\documentclass{article}
\usepackage{bookstyle}
\usepackage{cite}
\usepackage{graphicx}
\usepackage{amsfonts,amssymb,amsmath,amsthm}
\input{epsf}

\numberwithin{equation}{section}
\begin{document}
\title{Conformal Random Geometry}
\author{Bertrand Duplantier}
\address{Service de Physique Th\'{e}orique, \\ Orme des Merisiers, CEA/Saclay,\\ 91191 Gif-sur-Yvette Cedex, FRANCE}
\frontmatter
\maketitle
\mainmatter%
\section{\sc Preamble}
In these Lecture Notes\footnote{These Notes are based on my previous research survey article published in Ref. \cite{BDMan}, 
augmented by introductory sections, explanatory figures and some new material. Supplementary technical Appendices can be found
in Ref. \cite{BDMan}, or in
the forecoming extended version of the present Lectures on the 
Cornell University Library web site, arXiv.org.}, a comprehensive
description of the universal fractal geometry of conformally-invariant (CI) scaling 
curves or interfaces,
in the plane or half-plane, is given. They can be considered as complementary to those by Wendelin
 Werner.\footnote{W. Werner, {\it Some Recent Aspects of Random Conformally Invariant Systems} \cite{WW4}; see also
 \cite{stflour}.}

The present approach focuses on deriving critical exponents
associated with interacting random paths,
 by exploiting an underlying quantum gravity (QG) structure. The latter  
  relates exponents in the plane to those on a random lattice, i.e., in a 
fluctuating metric, using the so-called {\it 
Knizhnik, Polyakov and Zamolodchikov} (KPZ) map. This is accomplished within
the framework of random matrix theory and conformal field theory (CFT),
with applications to well-recognized geometrical critical models, like Brownian paths, self-avoiding walks, percolation, and
 more generally, the $O(N)$ or $Q$-state Potts 
models, and Schramm's Stochastic L\"owner Evolution (${\rm SLE}_{\kappa}$).\footnote{For an introduction, see the 
recent book by  G. F. Lawler \cite{lawlerbook}.}


Two fundamental ingredients of the QG construction are: the relation of bulk to 
Dirichlet boundary exponents, and
 additivity rules for QG {\it boundary} conformal dimensions associated with 
{\it mutual-avoidance} between sets of random paths. These relation and rules are
established  from the general
structure of
 correlation functions of arbitrary interacting random sets on a random lattice, 
as derived from {\it random matrix theory}.

The additivity of boundary exponents in quantum gravity for  mutually-avoiding paths is in contradistinction 
to the usual additivity of exponents in the  standard complex plane ${\mathbb C}$ or half-plane ${\mathbb H}$, where the latter 
additivity corresponds to the {\it statistical independence} of random 
processes, hence to possibly overlapping random paths. Therefore, with both additivities at hand, either in QG or in  
${\mathbb C}$ (or ${\mathbb H}$), and the possibility of 
multiple, direct or inverse, KPZ-maps between 
the random and the complex planes, any entangled structure made of interacting paths can be resolved and 
its exponents calculated, as explained in these Notes.

 From
this,
 non-intersection exponents for random walks or Brownian paths, self-avoiding 
walks (SAW's), or arbitrary
mixtures thereof are derived in particular.

Next, the multifractal 
function $f(\alpha,\,c)$ of the harmonic measure (i.e.,
electrostatic potential, or diffusion field) near any conformally
invariant fractal boundary or interface, is obtained as a function
of the central charge $c$ of the associated CFT.
It  gives the Hausdorff
dimension of the set of frontier points $w_{\alpha}$, where the potential varies with
distance  $r$ to the said point as $r^{\alpha}$. From an electrostatic point of view, this is equivalent to saying that
the frontier locally looks 
like a wedge of opening angle $0\leq \theta\leq 2\pi$, with a potential scaling like $r^{\pi/\theta}$, whence  
$\alpha=\pi/\theta$. Equivalently, the electrostatic charge contained in a ball of radius $r$ centered at $w_{\alpha}$, 
and the {\it harmonic measure}, i.e., the probability that an   
auxiliary  Brownian motion started at infinity, first hits the frontier in the same ball, both scale like $r^{\alpha}$.
 
In particular,  we shall see that Brownian paths,
SAW's in the scaling limit, and critical percolation clusters all have identical spectra corresponding
to the same central charge $c=0$. 
This result therefore states that the frontiers of a Brownian path or of the scaling limit of a critical
percolation cluster are just 
 identical with the scaling limit of a self-avoiding walk (or loop). 

 Higher multifractal
functions, like the double spectrum $f_2(\alpha,\alpha';c)$ of the double-sided harmonic measure 
on both sides of an SLE, are similarly
 obtained.

As a corollary, the Hausdorff 
dimension $D_{\rm H}$ of a {\it non-simple} scaling curve or
cluster {\it hull}, and the dimension $D_{\rm EP}={\rm sup}_{\alpha}
f(\alpha, c)$ of its {\it simple frontier}
 or {\it external
perimeter}, are shown to
obey the (superuniversal) {\it duality} equation $(D_{\rm H}-1)(D_{\rm
EP}-1)=\frac{1}{4}$, valid for any value of the central charge
$c$. 

For the ${\rm SLE}_{\kappa}$ process, this predicts the existence of a  
 $\kappa \to \kappa'=16/\kappa$ duality which associates simple
($\kappa' \leq 4)$ SLE paths as external frontiers of non-simple paths ($\kappa > 4)$ paths. 
  This duality is established via an algebraic
  symmetry  of the KPZ quantum gravity map. An extended {\it dual} KPZ
  relation is thus introduced for the SLE, 
  which commutes with the  $\kappa \to \kappa'=16/\kappa$ duality.

Quantum gravity allows one   to ``transmute'' random paths one into another, in particular
 Brownian paths into equivalent SLE paths. Combined
  with duality, this allows one to calculate  SLE exponents  
from simple QG fusion rules.
 
Besides  
the set of local singularity exponents $\alpha$ introduced above, the statistical description of the random geometry 
of a conformally invariant scaling curve or interface requires the introduction of   
 {\it logarithmic spirals}. These 
provide geometrical configurations of a scaling curve about a generic point that are conformally invariant, and correspond 
to the asymptotic logarithmic winding of the polar angle $\varphi$ at distance $r$, $\varphi=\lambda \ln r, r\to 0$, of 
the wedge (of opening
 angle $\theta=\pi/\alpha$)  
seen above.  

In complex analysis and probability theory, this  is best described by a new multifractal spectrum,    
the {\it mixed rotation harmonic spectrum} $f(\alpha,\lambda;c)$, 
which gives the Hausdorff dimension of the set of points possessing both a local logarithmic winding rate $\lambda$ 
and a local singularity exponent 
$\alpha$ with respect to the harmonic measure. 
 
The spectrum $f(\alpha,\lambda;c)$ of any conformally invariant scaling curve 
or interface is thus obtained as a function of the central charge $c$ labelling the associated CFT, or, equivalently, 
of the parameter $\kappa$ for the ${\rm SLE}_{\kappa}$ process. 
 Recently,  these results have been derived rigorously, including 
their various probabilistic senses, from first principle calculations within the SLE framework, thus vindicating the QG approach.  

The Lecture Notes by Wendelin Werner in this volume \cite{WW4}
  are based on the
rigorous construction of conformal ensembles of random curves using the SLE. Bridging the gap between these
physics and mathematics based approaches should constitute an interesting project for future studies.

A first step is the reformulation of the probabilistic SLE formalism in terms of standard conformal field
theory.\footnote{For an introduction, see M. Bauer and D. Bernard \cite{BB0}, and J. Cardy, {\it SLE for Theoretical Physicists},
\cite{cardySLE}.} A second one would be a more direct
 relation to standard models and methods of statistical mechanics in two dimensions like the Coulomb gas and Bethe Ansatz 
 ones.\footnote{See, e.g., W. Kager and B. Nienhuis \cite{NK}.} The natural emergence of quantum gravity 
in the SLE framework should be the next issue.
 
Let us start with a brief history of conformal invariance in statistical physics and probability theory.





\section{\sc{Introduction}}
\label{sec.intro}

\subsection{A Brief Conformal History}

\subsubsection*{Brownian Paths, Critical Phenomena, and Quantum Field Theory}

Brownian motion is the archetype of a random process,
hence its great importance in physics and probability
theory \cite{BBrownian}. The Brownian path is also the arche\-type of a scale invariant set, and in two dimensions is
a conformally-invariant one, as shown by P. L\'evy \cite{PLevy}. It is therefore perhaps the most natural random fractal \cite{mandelbrot}.  On the other hand, Brownian paths are intimately linked with quantum field theory (QFT).
 Intersections of Brownian paths indeed provide the random geometrical mechanism underlying QFT
  \cite{symanzyk}. In a Feynman diagram, any propagator line can be represented by a Brownian path,
  and the vertices are
  intersection points of the Brownian paths.
  This equivalence is widely used in polymer theory
  \cite{PGG,desC} and in rigorous studies of second-order phase
transitions and field theories  \cite{aizenman1}. Families of
universal critical exponents are in particular associated with {\it non-intersection} probabilities of collections of random
walks or Brownian paths, and these play an important role both in probability theory and quantum field theory
\cite{lawler1,fisher,aizenman2,duplantier1}.

A perhaps less known fact is the existence of highly non-trivial geometrical, actually {\it fractal}
(or {\it multifractal}), properties of Brownian paths or their subsets \cite{mandelbrot}. These types of geometrical
fractal properties generalize to all universality classes of, e.g., random walks (RW's), loop-erased random walks
(LERW's), self-avoiding walks (SAW's) or
polymers, Ising, percolation
and Potts models, $O(N)$ models, which are related in an essential way to standard critical phenomena and
field theory. The random fractal geometry is particularly rich in two dimensions.


\subsubsection*{Conformal Invariance and Coulomb Gas}
 In {\it two dimensions} (2D), the notion of {\it conformal invariance} \cite{BPZ,friedan,cardylebowitz}, and the
 introduction of the so-called
 ``Coulomb gas techniques'' and ``Bethe Ansatz''
 have brought a wealth of exact results in the Statistical Mechanics of critical models (see,
e.g., Refs.~\cite{NBRS} to 
\cite{JK3}).
Conformal field theory (CFT) has
lent strong support to the conjecture that statistical systems at
their critical point, in their scaling (continuum) limit, produce
{\it conformally-invariant} (CI) fractal structures, examples of
which are the continuum scaling limits of RW's, LERW's, SAW's, critical
Ising or Potts clusters.  A prominent role was played by Cardy's equation for the crossing probabilities in 2D percolation  \cite{cardy3}.
 To understand conformal invariance in a rigorous way  presented a mathematical challenge
(see, e.g.,  Refs. \cite{langlands,ai1,ben}). 
\begin{figure}[htb]
\begin{center}
\includegraphics[angle=0,width=.35\linewidth]{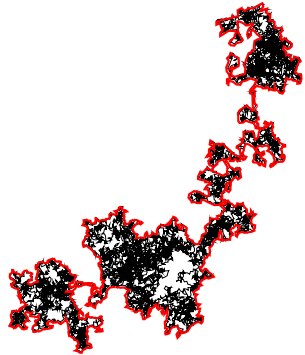}
\caption{{A planar Brownian path and its external frontier.}}
\end{center}
\label{fig.brownianfrontier}
\end{figure}
In the particular case of planar Brownian paths,  Beno\^{\i}t Mandelbrot  \cite{mandelbrot}
made the following famous conjecture in 1982: {\it in two dimensions, the external frontier of a planar
Brownian path has a Hausdorff dimension}
\begin{equation}
D_{\rm Brown.\, fr.}=\frac{4}{3}, \label{Mand}
\end{equation}
identical to that found by B. Nienhuis for
 a {\it planar self-avoiding walk} \cite{nien}. This identity has played an important role
in probability theory and theoretical physics in recent years, and will be a central theme in these Notes.
We shall understand this identity in the light of ``quantum gravity'', to which we turn now.
\begin{figure}[htb]
\begin{center}
\includegraphics[angle=-90,width=.7\linewidth]{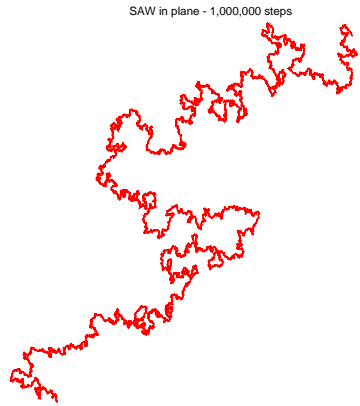}
\vskip.0cm
\caption{{A planar self-avoiding walk {\it (Courtesy of T. Kennedy)}.}}
\end{center}
\end{figure}
\subsubsection*{Quantum Gravity and the KPZ Relation}
Another breakthrough, not wide\-ly noticed at the time, was the introduction of ``2D quantum gravity'' (QG) in the
statistical mechanics of 2D critical systems.  V. A. Kazakov gave the solution of the Ising model on a random planar lattice \cite{kazakov}.
The  astounding discovery by Knizhnik, Polyakov, and Zamolodchikov of the ``KPZ'' map between critical
exponents in the standard plane and in a random 2D metric \cite{KPZ} led to the relation of the exponents found in Ref. \cite{kazakov} to those of Onsager
(see also \cite{david2}).
The first other explicit solutions and checks of KPZ were obtained for SAW's \cite{DK} and for the $O(N)$ model 
\cite{kostovgaudin,KS,KK}.

\subsubsection*{Multifractality}
The concepts of generalized dimensions and associated {\it multifractal} (MF)
 measures were developed in parallel two decades ago \cite
{mandelbrot2,hentschel,frisch,halsey}. It was later realized that
multifractals and field theory have deep connections, since
the algebras of their respective correlation functions reveal
striking similarities  \cite{cates}.

A particular example is given by classical potential theory, i.e., that of the electrostatic or
diffusion field near critical fractal boundaries,
or near diffusion
limited aggregates (DLA). The self-similarity of the fractal boundary is indeed reflected
in a multifractal behavior of the moments of the potential. In DLA,
the potential, also called harmonic measure, actually determines
the growth process \cite{halmeakproc,halseyerice,meak,BBE}.
For equilibrium statistical fractals, a first analytical example of multifractality was studied in ref.
  \cite{catesetwitten}, where the fractal boundary was chosen to be a simple RW, or a
SAW, both accessible to renormalization group
methods near four dimensions. In two dimensions, the existence of a multifractal spectrum for the Brownian path frontier
was established rigorously \cite{lawler97}.

In 2D, in analogy to the simplicity of the classical
method of conformal transforms to solve electrostatics of {
Euclidean} domains, a {\it universal}
solution could be expected  for the distribution of potential near any CI fractal in the plane.
It was clear that these multifractal spectra
should be linked with the conformal invariance classification, but outside the realm of the
usual {\it rational} exponents. That presented a second challenge to the theory.

\subsection{Conformal Geometrical Structures}
\subsubsection*{Brownian Intersection Exponents}
It was already envisioned in the
mid-eigh\-ties that the critical properties of planar Brownian paths, whose conformal invariance was
 well-established \cite{PLevy}, could be the opening gate to rigorous studies
of two-dimensional critical phenomena.\footnote{It is perhaps interesting to note that P.-G. de Gennes originally studied polymer
theory with the same hope of understanding from that perspective the broader class of critical phenomena. It turned out to be historically the converse:
the Wilson-Fisher renormalization group approach to spin models with $O(N)$ symmetry yielded in 1972 the polymer critical exponents
as the special case of the $N \to 0$ limit \cite{PGG}. Michael Aizenman, in a
seminar in the Probability Laboratory of the University of Paris VI
in 1984, insisted on the importance of the $\zeta_2$ exponent governing in two dimensions
the non-intersection
probability up to time $t$, $P_2(t) \approx t^{-\zeta_2}$, of two Brownian paths,
and promised a good bottle of Bordeaux wine for its resolution. 
A Ch\^ateau-Margaux 1982  was finally savored in company
of M. Aizenman, G. Lawler, O. Schramm, and W. Werner in 2001.} The precise values of the family $\zeta_L$
governing the similar non-intersection
properties of $L$ Brownian paths were later conjectured from conformal invariance and numerical studies in Ref. \cite{duplantier2}
(see also \cite{sokal,burdzy}). They correspond to a CFT with central charge $c=0$. Interestingly
enough, however, their analytic derivation  resisted  attempts by standard ``Coulomb-gas'' techniques.

\subsubsection*{Spanning Trees and LERW}
The related random process, the ``loop-erased random walk'', introduced in Ref. \cite{duke}, in which the loops of a simple RW are erased
sequentially, could also be expected to be accessible
to a rigorous approach. Indeed, it can be seen as the backbone of a spanning tree, and
the Coulomb gas predictions for the associated exponents \cite{D6,majumdar} were
obtained rigorously by  determinantal or Pfaffian techniques by R. Kenyon \cite{kenyon1}, in addition to the
conformal invariance of crossing probabilities \cite{kenyon2}. They correspond to a CFT with central charge $c=-2$.

\subsubsection*{Conformal Invariance and Brownian Cascade Relations}
The other route was followed by W. Werner \cite{werner}, joined later by G. F.  Lawler, who concentrated
on  Brownian path intersections, and on their general conformal invariance properties. They derived in particular important
``cascade relations'' between Brownian intersection exponents of packets of Brownian paths \cite{lawler2}, but still without a
derivation of the conjectured values of the latter.

\subsection{Quantum Gravity}
\subsubsection*{QG and Brownian Paths, SAW's and Percolation}
In the Brownian cascade structure of Lawler and Werner the author recognized the emergence of an underlying quantum gravity structure.
This led to an analytical derivation of
the (non-)intersection exponents for Brownian paths \cite{duplantier7}. The same QG structure, properly understood, also
gave access to exponents
of mixtures of RW's and SAW's, to the harmonic measure multifractal spectra of the latter two \cite{duplantier8},
 of a percolation cluster \cite{duplantier9}, and to the rederivation of
path-crossing exponents in percolation of Ref.~\cite{ADA}. Mandelbrot's conjecture (\ref{Mand})
also follows from
\begin{equation}
D_{\rm Brown.\, fr.}=2-2{\zeta}_{\frac{3}{2}}=\frac{4}{3}. \label{mandel}
\end{equation}
It was also observed there that the whole
class of Brownian paths, self-avoiding walks, and percolation clusters,
 possesses the same harmonic MF spectrum in two dimensions, corresponding to a unique central charge $c=0$.
 Higher MF spectra were also calculated \cite{duplantier10}.
 Related results were obtained in Refs. \cite{lawler3,cardy2}.

\subsubsection*{General CI Curves and Multifractality}
The general solution for the potential distribution near any conformal
fractal in 2D  was finally obtained from the same quantum gravity structure \cite{duplantier11}.
The exact multifractal spectra describing the singularities of the harmonic measure
 along the fractal boundary depend only on
the so-called
{\it central charge} $c$, the parameter which labels the universality class of the underlying CFT\footnote{Another 
intriguing quantum gravity structure was found in the classical combinatorial problem of {\it meanders} \cite{DIFEG}.}.

\subsubsection*{Duality}
A corollary  is the existence of a subtle geometrical {\it duality} structure
in boundaries of
random paths or clusters \cite{duplantier11}. For instance, in the Potts model, the {\it external perimeter}
(EP) of a Fortuin-Kasteleyn cluster, which bears the electrostatic charge and is a
{\it simple} (i.e., double point free) curve, differs from the full
cluster's  hull, which bounces onto itself in the scaling limit.  The EP's Hausdorff dimension $D_{\rm EP}$,
 and the hull's Hausdorff dimension $D_{\rm H}$
obey a duality relation:
\begin{equation}
(D_{\rm EP}-1)(D_{\rm H}-1)=\frac{1}{4}\, , \label{D-D}
\end{equation}
where $ D_{\rm EP} \leq D_{\rm H} $.
This generalizes the case of percolation hulls \cite{GA},
elucidated in Ref. \cite{ADA}, for which: $D_{\rm EP}=4/3, D_{\rm H}=7/4$. Notice
that the symmetric point of (\ref{D-D}), $D_{\rm EP}=D_{\rm H}=3/2$, gives the
maximum dimension of a  simple
conformally-invariant random curve in the plane.


\subsection{Stochastic L\"owner Evolution}
\subsubsection*{SLE and Brownian Paths}
In mathematics, O. Schramm, trying to reproduce by a continuum stochastic process both the conformal invariance and
Markov properties
of the scaling limit of loop-erased random walks, invented during the same period in 1999 the so-called ``Stochastic L\"owner Evolution'' (SLE)
\cite{schramm1},
a process parameterized by an auxiliary one-dimensional Brownian motion of variance or ``diffusion constant'' $\kappa$. It became quickly recognized as
a breakthrough since
it provided a powerful analytical tool to describe conformally-invariant scaling curves for various values of $\kappa$. The 
first identifications to standard critical models were proposed:
 LERW for $\kappa=2$, and
hulls of critical percolation
clusters for $\kappa=6$ \cite{schramm1}.

More generally, it was clear that the SLE described the continuum limit of
 hulls of critical cluster or loop models, and that the
 $\kappa$ parameter is actually in
one-to-one correspondence to the usual Coulomb gas
coupling constant $g$, $g=4/\kappa$. The easiest way \cite{histpoint} was to identify the Gaussian formula for
the windings about the tip of the SLE
given by Schramm in his original paper, with the similar one found earlier by H. Saleur and the author from Coulomb gas techniques
 for the windings in the $O(N)$ model \cite{BDHSwinding} (see, e.g., \cite{BDjsp} and section \ref{subsec.geoSLE} below).

Lawler, Schramm and Werner were then able to rigorously derive the Brownian intersection exponents \cite{lawler4}, as well
as Mandelbrot's conjecture \cite{lawler5} by relating them to the properties of ${\rm SLE}_{\kappa=6}$.\footnote{Wendelin Werner is being awarded the
Fields Medal on August 22nd, 2006,  at the International
 Congress of Mathematicians in Madrid, {\it ``for his contributions to the development of stochastic Loewner evolution,
 the geometry of two-dimensional Brownian motion, and conformal field theory.''}} S. Smirnov was able to
relate rigorously  the continuum limit of site percolation on the triangular lattice to
 the ${\rm SLE}_{\kappa=6}$ process \cite{smirnov1}, and derived Cardy's equation \cite{cardy3} from it. Other well-known percolation scaling behaviors
 follow from this \cite{lawler6,smirnov2}. The scaling limit of the LERW has also been rigorously shown to be the ${\rm SLE}_{\kappa=2}$
 \cite{LSWLERW}, as anticipated in Ref. \cite{schramm1}, while that of SAW's is expected to correspond to $\kappa=8/3$ \cite{BDjsp,SAWLSW,TGK}.

\subsubsection*{Duality for ${\rm SLE}_{\kappa}$}
The ${\rm SLE}_{\kappa}$ trace essentially describes boundaries of conformally-invariant random clusters.
For $\kappa \leq 4$, it is a simple path, while for $\kappa > 4$ it bounces onto itself. One can establish a dictionary between
the results obtained by quantum gravity and Coulomb gas techniques for Potts and $O(N)$ models \cite{duplantier11},
and those concerning the SLE \cite{BDjsp} (see below). The duality equation
(\ref{D-D}) then brings in a $\kappa\kappa'=16$ {\it duality} property \cite{duplantier11,BDjsp,duplantierdual} 
between Hausdorff dimensions:
\begin{equation}
\left[D_{\rm EP}(\kappa)-1\right] \left[ D_{\rm
H}(\kappa)-1\right]=\frac{1}{4},\ \kappa \geq 4\ ,
\label{dualione}
\end{equation}
where  $$D_{\rm EP}(\kappa)=D_{\rm H}(\kappa'=16/\kappa),\quad
\kappa'\leq 4 $$
gives the dimension of the (simple) frontier of
a non-simple ${\rm SLE}_{\kappa \geq 4}$ trace as the Hausdorff dimension of the simple
${\rm SLE}_{16/{\kappa}}$ trace. Actually, this extends to the whole multifractal spectrum of the harmonic measure near the
${\rm SLE}_{\kappa}$, which is identical to that of the ${\rm SLE}_{16/{\kappa}}$ \cite{duplantier11,BDjsp}. From that result was originally stated the duality prediction that
 the frontier of the non-simple ${\rm SLE}_{\kappa \geq 4}$ path
is locally a simple ${\rm SLE}_{16/{\kappa}}$ path \cite{duplantier11,BDjsp,duplantierdual}.

The SLE geometrical properties per se are an active subject of investigations \cite{RS}.  The value of the
 Hausdorff dimension of the SLE trace,
$D_{\rm H}(\kappa)=1+{\kappa}/{8}$, has been obtained rigorously by
V. Beffara, first in the case of  percolation ($\kappa=6$) \cite{beffara2}, and in general \cite{beffara},
 in agreement with the value predicted by the Coulomb gas approach
\cite{nien,SD,duplantier11, BDjsp}. The duality
(\ref{dualione}) predicts $D_{\rm EP}(\kappa)=1+({\kappa}/{8})\vartheta(4-\kappa)+({2}/{\kappa})\vartheta (\kappa -4)$ for the dimension of the SLE frontier \cite{duplantier11,BDjsp}.

The {\it mixed} multifractal spectrum
 describing the local rotations (windings) and singularities of the harmonic measure near a fractal boundary, 
introduced some time ago by Ilia Binder \cite{binder}, has been obtained for SLE, by a combination of Coulomb gas 
and quantum gravity methods \cite{DB}. 

\subsection{Recent Developments}
 At the same time, 
the relationship of $\rm{SLE}_{\kappa}$ to standard conformal field theory has been pointed out and developed, 
both in physics \cite{BB,BBH} and mathematics \cite{CRLSW,LF1,F}.  

A two-parameter family of Stochastic L\"owner Evolution processes, the so-called $\rm{SLE}(\kappa,\rho)$ processes,
introduced in Ref. \cite{CRLSW},  
  has been studied further \cite{WW}, in
particular in relation to the duality property mentioned above \cite{dubedat}.
 It  can be studied in the CFT framework \cite{cardyrho,kyto}, and we shall briefly describe it here from the QG point of view.
 Quite recently, $\rm{SLE}(\kappa,\rho)$ has also been described  in terms of randomly growing polygons \cite{BF}.
 
 A description of collections of SLE's in terms
of Dyson's circular ensembles has been proposed \cite{cardy2003}. Multiple SLE's are
 also studied in Refs. \cite{dubedat4,dubedat5,BBK}. 

Percolation remains a favorite model: Watts' crossing formula in percolation \cite{watts} has been derived 
 rigorously by J. Dub\'edat \cite{dubedat2,dubedat3}; the construction
 from ${\rm SLE}_6$ of the {\it full} 
 scaling limit of cluster loops in percolation has been recently achieved by F. Camia and C. Newman \cite{CN,CN1,CFN},
  V. Beffara has recently discovered a simplification of
 parts of Smirnov's original proof for the triangular lattice \cite{beffaraperc}, while
  trying to generalize it to other lattices \cite{beffaraotherlat}. It is also possible that the lines of zero
  vorticity in 2D turbulence are intimately related to percolation cluster boundaries
 \cite{2Dvort}.

Another proof has been given of the convergence of the scaling limit of loop-erased random walks to  $\rm{SLE}(\kappa=2)$ 
\cite{kozma}. The model of the ``harmonic explorer''
 has been shown to converge to $\rm{SLE}(\kappa=4)$ \cite{SS}. S. Smirnov seems to have been able very recently to prove that the critical Ising model corresponds to ${\rm SLE}_3$, as
  expected\footnote{Ilia Binder, private communication.}\cite{smirnovising}.

 Conformal loop ensembles have recently gained popularity. 
The  ``Brownian loop soup'' has been introduced \cite{WW1,WW2}, such that SLE curves 
are recovered as boundaries of clusters of such loops \cite{WW3,SW}.

Defining SLE or conformally invariant scaling curves on multiply-connected planar domains is an active subject of research 
\cite{konsevich,FKa,zhan,FB}. Correlation functions of the stress-energy tensor, a main object in CFT, has been described in terms of some probabilities 
 for the SLE process \cite{cardydoyonriva}. 
 
The Airy distribution for the area of self-avoiding loops has been found in theoretical physics by
J. Cardy \cite{cardyarea}, (see also \cite{cardyziff,richguttman,cardyloop}), while the 
 expected area of the regions of a given winding number inside the Brownian loop has been obtained recently by
 C. Garban and J. Trujillo Ferreras \cite{GTF} (see also \cite{richard}).
 
  The conformally invariant measure on self-avoiding loops has been constructed
recently \cite{WW5}, and is described in Werrner's lectures.

Gaussian free fields and their level sets, which play a fundamental role in the 
 Solid-On-Solid representation of 2D statistical models, are currently investigated in mathematics 
 \cite{sheffield}. The interface of the discrete Gaussian free field has been shown to converge to 
 ${\rm SLE}_4$ \cite{SS1}. When a relation between winding and height is imposed, reminiscent of a similar one
 in Ref. \cite{BDHSwinding}, other values of $\kappa$ are
 reached \cite{S}.

 The multifractal harmonic spectrum, originally derived in Ref. \cite{duplantier11} by QG, has been recovered 
by more standard CFT \cite{wiegmann}. The rigorous mathematical solution to the mixed multifractal spectrum of SLE has 
been obtained very recently in 
collaboration with Ilia Binder\cite{IABD} (see also \cite{Stockholm}).

On the side of quantum gravity and statistical mechanics, boundary correlators in 2D
QG, which were originally calculated via the Liouville
field theory \cite{FZZ,PT}, and are related to our quantum gravity approach, have been recovered from discrete models
on a random lattice \cite{KKK,KPS}. 
In mathematics, progress has been made towards a continuum theory of random planar graphs \cite{OAOS}, also in presence of 
percolation \cite{OA,OA2}. Recently, powerful combinatorial methods have entered the same 
field \cite{gschaeffer,BMS,DiFG,bouttier}. However, the KPZ relation has as yet eluded a rigorous approach.
 It would be worth studying further the relationship between SLE and Liouville theories.

 The Coulomb gas approach is still invaluable for discovering and analyzing the
 proper statistical models relevant to a given critical phenomenon. An example is that of the
 tricritical point of the $O(N)$ model, recently elucidated by Guo, Nienhuis and Bl\"ote \cite{GNB}. (See also \cite{NWB,JaS}.)
 
Readers interested in  general surveys of the SLE in relation
 to statistical mechanics are referred to Refs. \cite{NK,BDMan,cardySLE,BB0}.

\subsection{Synopsis}
The aim of the present Notes is to give a comprehensive description of conformal\-ly-invariant fractal geometry,
and of its underlying quantum gravity structure. In particular, we show how the repeated use of KPZ maps between
the critical exponents in the complex plane ${\mathbb C}$ and those in quantum gravity
allows the determination of a very large class of critical exponents arising in planar critical statistical systems, including the
multifractal ones, and their reduction to simple irreducible elements. Within this 
unifying perspective, we cover many well-recognized geometrical models, like RW's or SAW's and their intersection properties,
Potts and $O(N)$ models, and the multifractal properties thereof. 

We also adapt the quantum gravity
formalism to the ${\rm SLE}_{\kappa}$ process, revealing there a hidden algebraic duality in the KPZ map itself, which in turn
translates into the geometrical $\kappa \to \kappa'=16/\kappa$ duality between simple and non-simple SLE traces.
This KPZ algebraic duality also explains the duality which exists within the class of Potts and $O(N)$ models
between hulls and external frontiers.

In section \ref{sec.inter} we  first establish the values of the intersection exponents of random walks
or Brownian paths from quantum gravity. 
In section \ref{sec.mixing} we then move to the critical properties of arbitrary sets mixing simple random walks
or Brownian paths and self-avoiding walks, with arbitrary interactions thereof. 

Section \ref{sec.perco} deals with percolation. The QG method is illustrated in the case of path crossing exponents
and multifractal dimensions for percolation clusters. This completes the description of the universality
class of central charge $c=0$.

 We address in section \ref{sec.conform} the general
 solution for the multifractal potential distribution near any conformal
 fractal in 2D, which allows one to determine the Hausdorff dimension of the frontier.
 The multifractal spectra depend only on the central charge $c$, which labels the universality class of the underlying CFT.

 Another feature is the consideration in section \ref{sec.higher} of  higher multifractality, which occurs in a
 natural way in the joint distribution
 of potential on both sides of a random CI scaling path (or more generally, in the distribution of
 potential between the
 branches of a {\it star} made of an arbitrary number of CI paths). The associated universal multifractal spectrum
  then depends on several variables.

 Section \ref{sec.winding} describes the more subtle mixed multifractal spectrum associated with the local rotations
 and singularities along a conformally-invariant curve, as seen by the harmonic measure \cite{binder,DB}.
 Here quantum gravity and Coulomb gas techniques must be fused.

Section \ref{sec.geodual} focuses on the $O(N)$ and Potts models, on the ${\rm SLE}_{\kappa}$, and on the correspondence between
them. This is exemplified for the geometric duality existing between their cluster frontiers and hulls. The various Hausdorff
dimensions of $O(N)$ lines, Potts cluster boundaries, and SLE's traces are given. 

Conformally invariant paths have quite different critical properties and obey different quantum gravity rules, depending on
whether they are
{\it simple paths or not}. The next sections are devoted to the elucidation of this difference, and its treatment
 within a unified framework.

A fundamental algebraic duality which exists in the KPZ map is studied in section \ref{sec.duality}, and applied to the
construction rules for critical exponents associated with non-simple paths versus simple ones. These duality rules
are obtained from considerations of quantum gravity.

We then construct an extended KPZ formalism for the ${\rm SLE}_{\kappa}$ process,
which is valid for all values of the parameter $\kappa$.  It corresponds to the usual KPZ formalism for $\kappa \leq 4$ (simple paths), 
and to the algebraic dual one for $\kappa > 4$ (non-simple paths). The composition rules for calculating critical exponents 
involving multiple random paths in the SLE process are given, as well as short-distance expansion results where quantum gravity 
emerges in the complex plane.  The description of ${\rm SLE}(\kappa,\rho)$ in terms of quantum gravity is also given.  The  
exponents for multiple SLE's, and the equivalent ones for $O(N)$ and Potts models are listed.

Finally, the extended SLE quantum gravity formalism is applied to the calculation of all harmonic measure
exponents near multiple SLE traces, near a boundary or in open space.

Supplementary material can be found in a companion article \cite{BDMan}, or in the extended version of these Notes.
An Appendix there details the calculation, in quantum gravity, of non-intersection exponents for Brownian paths
 or self-avoiding walks.
Another Appendix establishes the
general relation between boundary and bulk exponents in quantum gravity, as well as the boundary additivity rules. They follow from a fairly
universal structure of correlation functions in quantum gravity. These QG relations are actually
sufficient to determine all exponents without further calculations. The example of the
 $O(N)$ model exponents is described in detail in Ref. \cite{BDMan} (Appendix B).

\medskip

The  quantum gravity techniques used here are perhaps not widely
known in the statistical mechanics community at-large, since they originally belonged to
string or random matrix theory. These techniques, moreover, are
not yet within the realm of rigorous mathematics. However, the
correspondence extensively used here, which exists between scaling
laws in the plane and on a random Riemann surface, appears to be 
fundamental and, in my opinion, illuminates many of the geometrical properties of
conformally-invariant random curves in the plane.
\section{\sc{Intersections of Random Walks}}
\label{sec.inter}

\subsection{Non-Intersection Probabilities}
\subsubsection*{Planar Case}
\begin{figure}[tb]
\begin{center}
\includegraphics[angle=0,width=.35\linewidth]{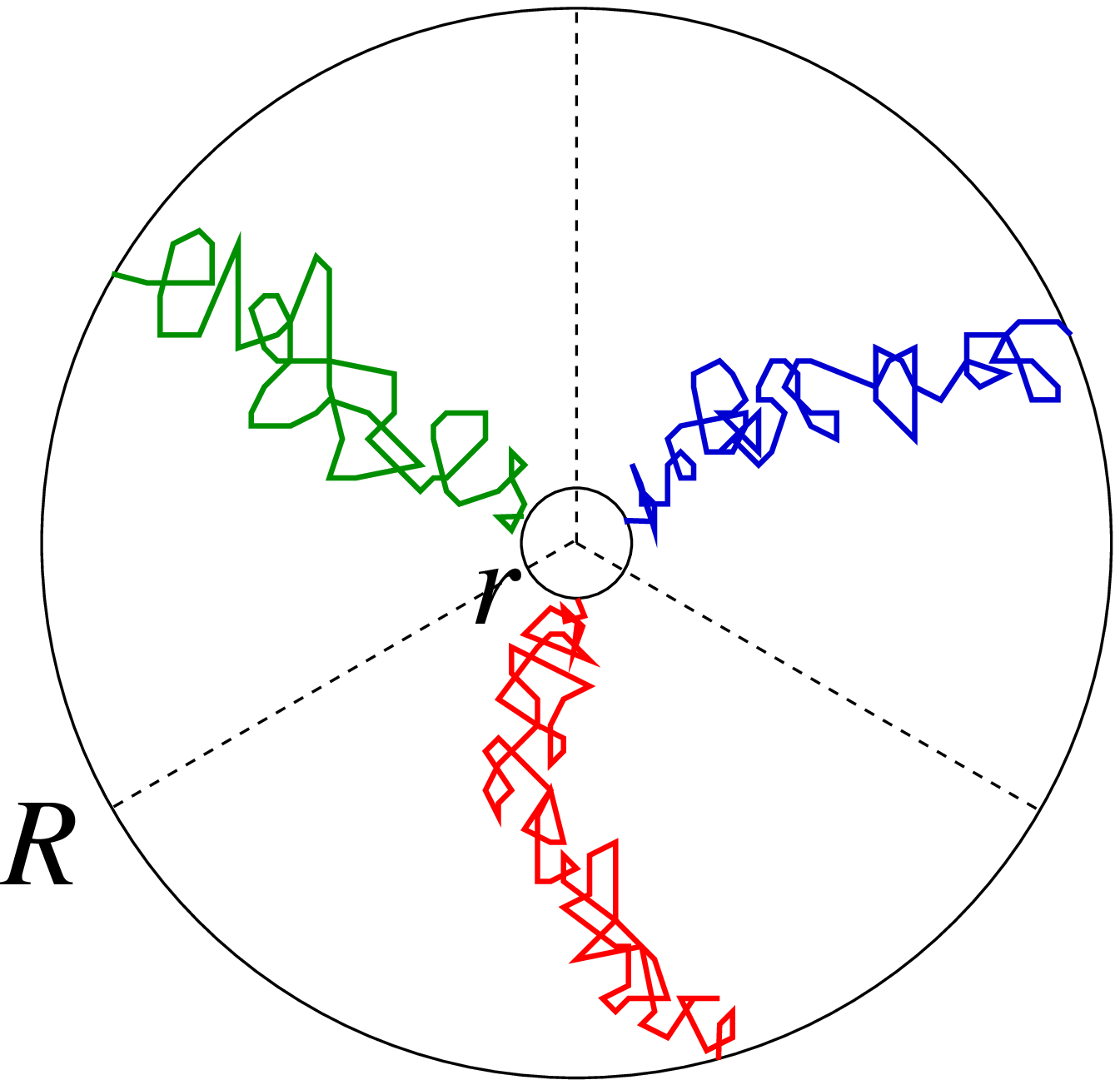}
\hskip.5cm
\includegraphics[angle=0,width=.40\linewidth]{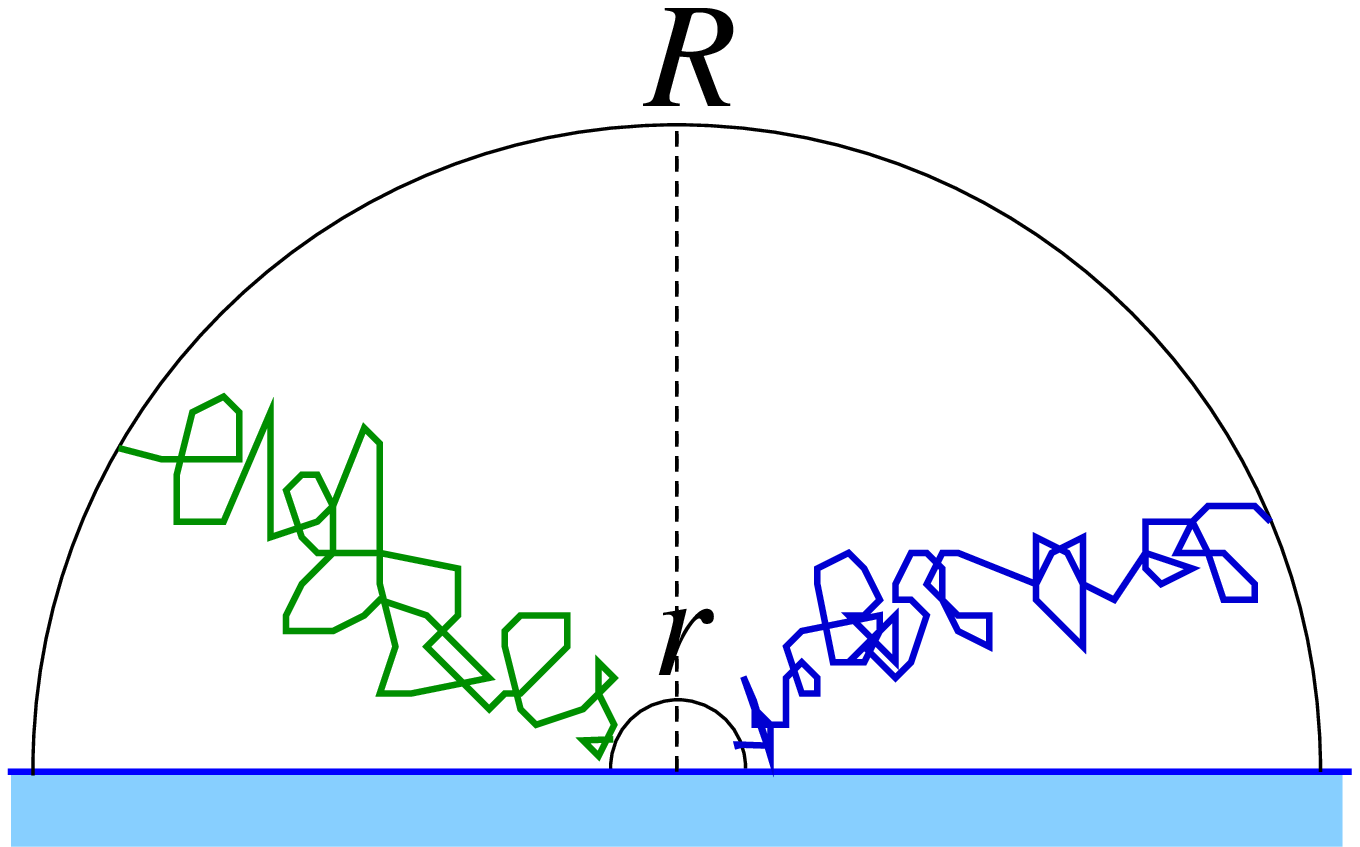}
\end{center}
\caption{{Non-intersecting planar random walks crossing an annulus from $r$ to $R$, or a half-annulus
in the half-plane ${\mathbb H}$.}}
\label{fig.rw111}
\end{figure}
Let us first define the so-called (non-){\it intersection exponents} for random walks or
Brownian motions. While simpler than the multifractal
exponents considered above, in fact they generate the latter. Consider
a number $L$ of independent random walks
$B^{(l)},l=1,\cdots,L$ in ${\mathbb Z} ^{2}$ (or Brownian paths in ${\mathbb R}^{2}={\mathbb C}),$
starting at fixed neighboring points, and the probability
\begin{equation}
{ P}_{L}\left( t\right) =P\left\{ \cup^{L}_{l,\, l'=1,\, l\neq l'} (B^{(l)}\lbrack
0,t\rbrack \cap B^{(l')}\lbrack 0,t\rbrack) =\emptyset \right\},
\label{pl}
\end{equation}
that the intersection of their paths up to time $t$ is
empty \cite{lawler1,duplantier1}. At large times one
expects this probability to decay as
\begin{equation}
{P}_{L}\left( t\right) \approx t^{-\zeta_{L}}, \label{zeta}
\end{equation}
where $\zeta _{L} $ is a {\it universal} exponent
depending only on $L$.
Similarly, the probability that the Brownian paths altogether traverse the
annulus ${\mathcal %
D}\left( r, R\right)$ in ${\mathbb C}$ from the inner boundary circle of radius $r$ to the outer
one at distance $R$ (Fig.~\ref{fig.rw111}) scales as
\begin{equation}
{P}_{L}\left( R\right) \approx \left({r}/{R}\right)^{2\zeta_{L}},
\label{zetaR}
\end{equation}
These exponents can be generalized to $d$ dimensions. Above the upper critical  dimension
$d=4$, RW's  almost surely do not intersect and $\zeta _{L}\left( d \geq 4\right)=0 $. The existence of
exponents $\zeta _{L}$ in $d=2, 3$ and their universality
have been proven \cite{burdzy}, and they can be calculated near $
d=4$ by renormalization theory  \cite{duplantier1}.

\subsubsection*{Boundary Case}

A generalization was introduced  for $L$ walks
constrained to stay in the half-plane ${\mathbb H}$ with Dirichlet boundary conditions on $\partial {\mathbb H}$
, and starting at neighboring
points near the boundary \cite{duplantier2}. The non-intersection probability
$\tilde{{P}_{L}}\left( t\right) $ of their paths is governed by a
boundary critical exponent $\tilde{\zeta}_{L}$ such that
\begin{equation}
\tilde{P} _{L}\left( t\right) \approx t^{-\tilde{\zeta}_{L}}.
\label{zetat}
\end{equation}
One can also consider the probability that the Brownian paths altogether traverse the
half-annulus ${\mathcal D}\left( r, R\right)$ in ${\mathbb H}$, centered on the boundary line $\partial {\mathbb H}$,
 from the inner boundary circle of radius $r$ to the outer
one at distance $R$ (Fig.~\ref{fig.rw111}). It scales as
\begin{equation}
{\tilde {P}}_{L}\left( R\right) \approx \left({r}/{R}\right)^{2\tilde \zeta_{L}}. \label{zetaRt}
\end{equation}
\subsubsection*{``Watermelon'' Correlations}
Another way to access these exponents consists in defining an {\it infinite measure} on mutually-avoiding Brownian paths. For definiteness,
let us first consider random walks on a lattice, and
{\it ``watermelon''} configurations in which $L$ walks ${B}^{({l})}_{{ij}}, l=1,...,L$, all started at point $i$,
are rejoined at the end at a point $j$, while staying mutually-avoiding in between. Their correlation
function is then defined as \cite{duplantier2}
\begin{equation}
\mathcal Z_{{L}}= \sum_{\scriptstyle
{\mathcal B}^{({\ell})}_{{ij}}\atop\scriptstyle {l=1,...,L}}
{\mu_{\scriptstyle{\rm RW}}}^{{-\left| {\mathcal B}\right|}}\propto \left|{i}-{j}\right|^{-4{\zeta}_{{L}}},
\label{watermelonB}
\end{equation}
where a critical fugacity $\mu_{\scriptstyle{\rm RW}}^{-1}$ is associated with the total number $\left|
{{\mathcal B}}\right| =\left|{\cup}^{{L}}_{{l=1}} {B}^{({l})}\right| $
of steps of the walks. When
$\mu_{\rm RW}$ is equal to the lattice connectivity constant (e.g., 4 for the square lattice ${\mathbb Z}^2$),
the corresponding term exactly counterbalances the exponential growth of the number of configurations.
The correlator then decays with distance
as a power law governed by the intersection exponent $\zeta_L$.

In the continuum limit
one has to let the paths start and end at distinct but neighboring points (otherwise they would immediately re-intersect),
 and  this correlation function then defines an
infinite measure on Brownian paths. (See the Lecture Notes by W. Werner.)

 An entirely  similar boundary correlator
$\tilde{\mathcal Z}_{{L}}$ can be defined, where
the $L$ paths are constrained to start and end near the Dirichlet boundary. It then decays as a power law:
$\tilde{\mathcal Z}_{{L}} \propto \left|{i}-{j}\right|^{-2{\tilde{\zeta}}_{{L}}},$ where now the boundary exponent
$\tilde{\zeta}_L$ appears.

\subsubsection*{Conformal Invariance and Weights}
It was first conjectured from conformal invariance arguments and
numerical simulations that in two dimensions  \cite{duplantier2}
\begin{equation}
\zeta _{L}=h_{0, L}^{\left( c=0\right) }=\frac{1}{24}\left(
4L^{2}-1\right), \label{Zeta}
\end{equation}
and for the half-plane
\begin{equation}
2\tilde{\zeta}_{L}=h_{1, 2L+2}^{\left( c=0\right)
}=\frac{1}{3}L\left( 1+2L\right), \label{zC2}
\end{equation}
where $h_{p,q}^{(c)}$ denotes the Ka\v {c} conformal weight
\begin{equation}
h_{p,q}^{(c)}=\frac{\left[ (m+1)p-mq\right] ^{2}-1}{4m\left(
m+1\right) }, \label{K}
\end{equation}
of a minimal conformal field theory of central charge
$c=1-6/[m\left(
m+1\right)] ,$ $m\in {\mathbb N} ^{*}$  \cite{friedan}. For Brownian motions
$c=0,$ and $m=2.$

\subsubsection*{Disconnection Exponent}
A discussion of the intersection exponents of random walks a priori requires a number $L \geq 2$ of them. Nonetheless, for $L=1$,
the exponent has a meaning: the non-trivial value $\zeta _{1}=1/8$ actually gives
the {\it disconnection exponent} governing the probability that an arbitrary point near
the origin of a single Brownian path remains accessible from infinity
without the path being crossed, hence stays connected to infinity. On a Dirichlet boundary, $\tilde \zeta_1$ retains its
standard value $\tilde \zeta_1=1$,
which can be derived directly, e.g., from the Green function formalism. It corresponds to
a path extremity located on the  boundary, which always stays accessible
due to Dirichlet boundary conditions.

\subsection{Quantum Gravity}
\subsubsection*{Preamble}
To derive the intersection exponents above, the idea
 \cite{duplantier7} is to map the original random walk problem in
the plane onto a random lattice with planar geometry, or, in other
words, in presence of two-dimensional {\it quantum gravity}
 \cite{KPZ}. The key point is that the random walk
intersection exponents on the random lattice are related to those
in the plane. Furthermore, the RW intersection problem can be
solved in quantum gravity. Thus, the exponents $\zeta_{L}$ (Eq.~(\ref{Zeta})) and $\tilde{\zeta}_{L}$ (Eq.~(\ref{zC2})) in the
standard complex plane or half-plane are derived from this mapping to a random
lattice or Riemann surface with fluctuating metric.

\subsubsection*{Introduction}
\begin{figure}[t]
\begin{center}
\includegraphics[angle=0,width=.45\linewidth]{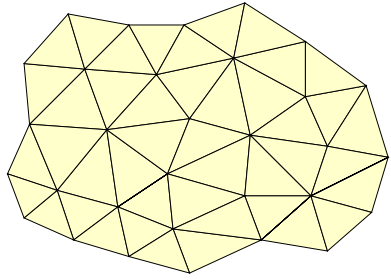}
\end{center}
\vskip.2cm
\caption{A random planar triangulated lattice. (Courtesy of Ivan Kostov.)}
\label{fig.largelattice}
\end{figure}

Random surfaces, in relation to string theory  \cite{see2}, have
been the subject and source of important developments in
statistical mechanics in two dimensions. In particular, the
discretization of string models led to the consideration of
abstract random lattices $G$, the connectivity fluctuations of
which represent those of the metric, i.e., pure 2D quantum gravity
 \cite{boulatov}. An example is given in figure \ref{fig.largelattice}.

  As is nowadays well-known, random (planar) graphs are in close relation to
   random (large) matrix models. Statistical ensembles of random matrices of large sizes have been introduced in
  1951 by E. Wigner in order to analyze the statistics of energy levels of heavy nuclei \cite{wigner}, leading to
  deep mathematical developments \cite{dyson,mehtagaudin,dysonmehta,mehtabook}.

  In 1974, G. 't Hooft discovered the so-called $1/N$ expansion in QCD \cite{thooft} and
  its representation in terms of planar diagrams. This opened the way to solving various
  combinatorial problems by using random matrix theory, the simplest of which is the enumeration of planar graphs
  \cite{bipz}, although this had been done earlier by W. T. Tutte by purely combinatorial methods \cite{tutte}.  Planarity then corresponds to the large-$N$ limit of a
 $N\times N$ Hermitian matrix theory.

\begin{figure}[t]
\begin{center}
\includegraphics[angle=0,width=.45\linewidth]{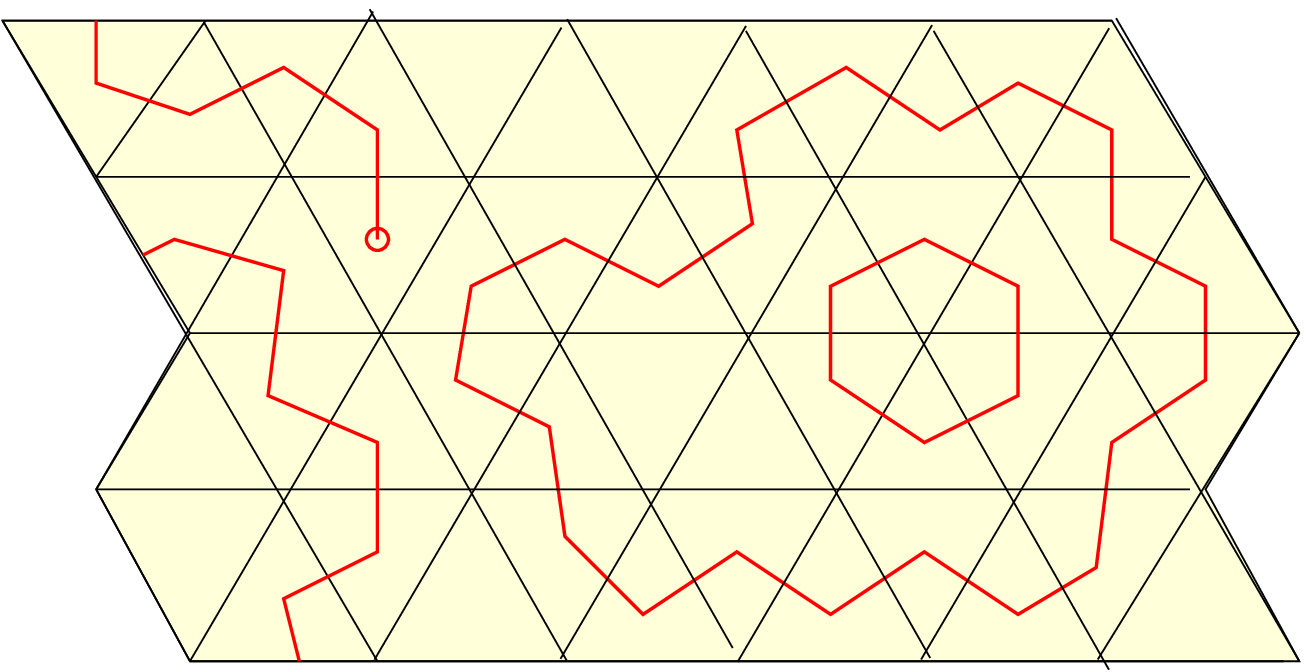}
\hskip.5cm
\includegraphics[angle=0,width=.45\linewidth]{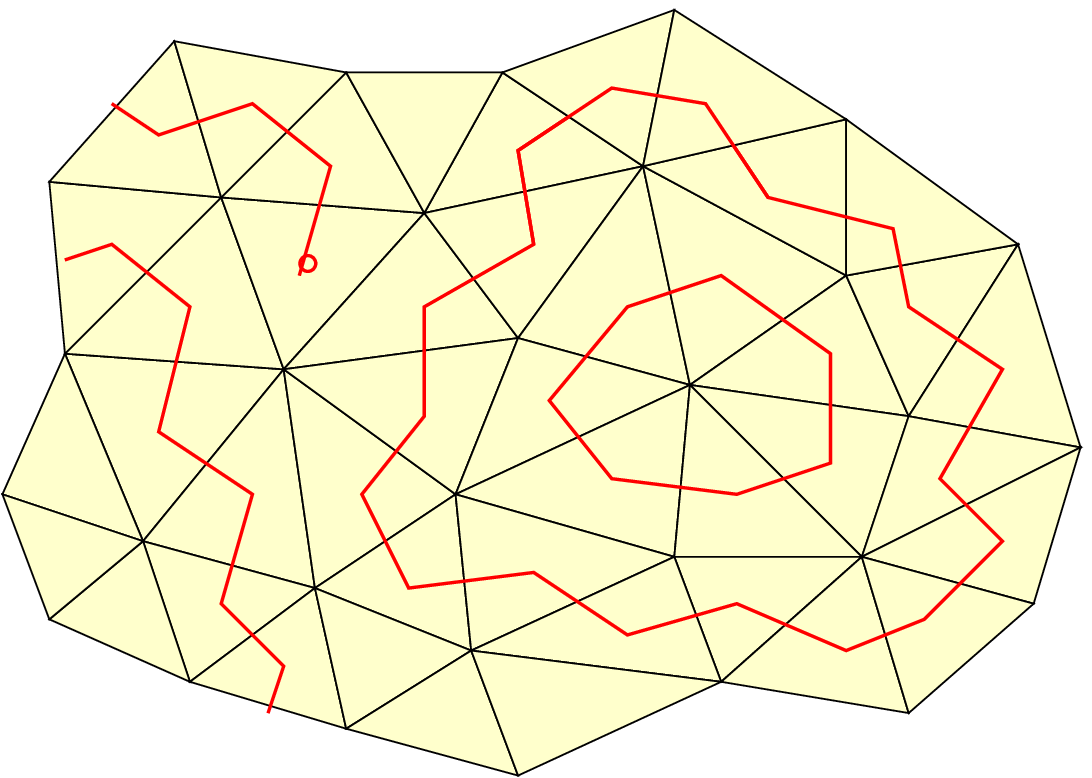}
\end{center}
\caption{A set of random lines on the regular triangular lattice and its counterpart on the
random triangular lattice. (Courtesy of I. K.)}
\label{fig.lgas}
\end{figure}

An further outstanding  idea was to redefine statistical mechanics {\it on} random planar lattices, instead
of doing statistical mechanics on  regular lattices \cite{kazakov}. One can indeed put any 2D statistical model (e.g.,
Ising model  \cite{kazakov}, self-avoiding walks  \cite{DK}, or $O(N)$ loop model \cite{kostovgaudin,KS,KK}) on a
random planar graph $G$ (figure \ref{fig.lgas}).  A new critical
behavior will emerge, corresponding to the confluence of the criticality of
the infinite random surface $G$ with the critical point of the original
model.

\begin{figure}[t]
\begin{center}
\includegraphics[angle=0,width=.5\linewidth]{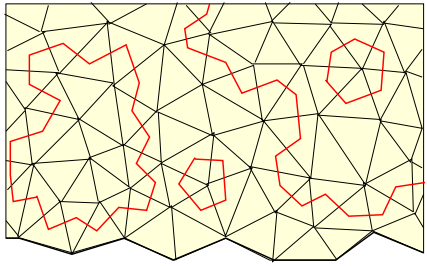}
\end{center}
\caption{A set of random loops near the bounday of a
randomly triangulated disk. (Courtesy of I. K.)}
\label{fig.dboundary}
\end{figure}

It is also natural to consider {\it boundary effects} by introducing random graphs with the {\it disk topology},
which may bear a statistical model
(e.g.,  a set of random loops as despicted in Figure \ref{fig.dboundary}). An interesting boundary (doubly) critical behavior of the statistical model
in presence of critical fluctuations of the metric can then be expected.

Another outstanding route was also to use 't Hooft's $1/N$ expansion of random matrices to generate the topological expansion
over random Riemann surfaces in terms of their genus \cite{99}.

All these developments led to a vast scientific literature, which of course can not be quoted here in its entirety!
 For a detailed introduction, the reader is referred to the 1993
 Les Houches or Altenberg lectures by F. David \cite{davidleshouches,davidaltenberg},
  to the 2001 Saclay lectures by B. Eynard \cite{eynard}, and to the monograph by J. Ambjorn {\it et al.} \cite{ambjornbook}.
 Among more specialized reviews, one can cite those by G. 't Hooft \cite{thooftplanar},
 by Di Francesco  {\it et al.} \cite{DiFGZJ} and by I. Kostov \cite{kostovsum}.

 The subject of random matrices is also widely
 studied in mathematics. In relation to the particular statistical mechanics purpose
 of describing (infinite) critical random planar surfaces, let us simply mention here
 the rigorous existence of a measure on  random planar graphs in the thermodynamical limit
  \cite{OAOS}.

 Let us finally mention that powerful combinatorial methods have
 been developped recently, where planar graph ensembles have been shown to be in bijection with random trees with various
adornments \cite{gschaeffer}, leading to an approach alternative to that by random matrices \cite{BMS,DiFG,bouttier}.

A brief tutorial on the statistical mechanics of random planar lattices and
their relation to
 random matrix theory, which contains the essentials required for understanding the statistical mechanics arguments
 presented here, can be found  in Refs. \cite{davidleshouches,davidaltenberg}.

\subsubsection*{KPZ Relation}
The critical system ``dressed by gravity'' has a larger
 symmetry under diffeomorphisms. This allowed Knizhnik, Polyakov, and
Zamolodchikov (KPZ)  \cite{KPZ} (see also  \cite{david2}) to establish the
existence of a fundamental relation between the conformal dimensions
$\Delta ^{\left( 0\right) }$ of scaling operators in the
plane and those in presence of gravity, $\Delta$:
\begin{equation}
\label{KPZg}
\Delta ^{\left( 0\right)}=U_{\gamma}(\Delta)=\Delta \frac{
\Delta -\gamma  }{1-\gamma},
\end{equation}
where $\gamma$, the {\it string susceptibility exponent}, is related to the central charge of the
statistical model in the plane:
\begin{equation}
\label{c(g)}
c=1-6\gamma^{2}/\left(1-\gamma\right),\;\; \gamma \leq 0.
\end{equation}
The same relation applies between conformal weights $\tilde \Delta ^{\left( 0\right)}$ in the half-plane ${\mathbb H}$ and
$\tilde \Delta $ near the boundary of a disk with fluctuating metric:
\begin{equation}
\label{KPZgb}
\tilde \Delta ^{\left( 0\right) }=U_{\gamma}\left(\tilde \Delta\right )=\tilde \Delta \frac{
\tilde \Delta -\gamma  }{1-\gamma}.
\end{equation}

For a minimal model of the series~(\ref{K}), $\gamma=-1/m$,
and the conformal weights in the plane ${\mathbb C}$ or half-plane ${\mathbb H}$
are $\Delta _{p,q}^{\left( 0\right) }:=
h_{p,q}^{\left( c\right) }.$

\subsubsection*{Random Walks in Quantum Gravity}
Let us now consider as a statistical model {\it random walks} on a
{\it random graph}. We know  \cite{duplantier2} that the central
charge $c=0$, whence $m=2$, $\gamma=-1/2.$ Thus the KPZ relation
becomes
\begin{equation}
\Delta ^{\left( 0\right) }={U}_{\gamma=-1/2}\left( \Delta
\right)=\frac{1}{3}\Delta \left( 1+2\Delta
\right):= U(\Delta), \label{KPZ}
\end{equation}
which has exactly the same analytical form as equation
(\ref{zC2})! Thus, from this KPZ equation one infers that the conjectured planar
Brownian intersection exponents  in the complex plane ${\mathbb C}$ (\ref{Zeta}) and  in ${\mathbb H}$ (\ref{zC2}) must
  be equivalent to the following Brownian intersection exponents in quantum gravity:
\begin{eqnarray}
 \label{delta}
{\Delta }_{L}&=&\frac{1}{2}\left(L-\frac{1}{2}\right),\\
\label{deltatilde}
\tilde{\Delta }_{L}&=&L.
\end{eqnarray}
Let us now sketch the derivation of these quantum gravity
exponents \cite{duplantier7}. A more detailed argument can be found in Ref. \cite{BDMan}.

\subsection{Random Walks on a Random Lattice}
\begin{figure}
\begin{center}
\includegraphics[angle=0,width=1.\linewidth]{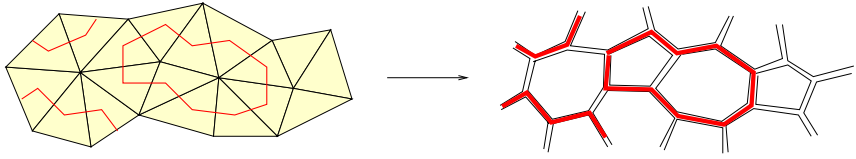}
\vskip.5cm
\end{center}
\caption{A randomly triangulated surface and its natural dual graph made of of {``$\varphi^3$''} trivalent vertices.
The set of random loops illustrates the possiblity to define an arbitrary statistical model on the trivalent graph.
(Courtesy of I. K.)}
\label{fig.loops}
\end{figure}
\subsubsection*{Random Graph Partition Function}
For definiteness, consider the set of planar random graphs $G$, built up with,
e.g., ``$\varphi^3$''-like  trivalent vertices tied together in a {\it random way} (Fig.~\ref{fig.loops}).
By duality, they form the set of dual graphs of the randomly triangulated planar lattices considered before.

The topology is fixed here to be that of a sphere $
\left( {\mathcal S}\right) $ or a disk $\left( \mathcal {D}\right)
$. The partition function of planar graphs is defined as
\begin{equation}
Z_{}(\beta,\chi )=\sum _{G(\chi)}{\frac{1}{S(G)}}e^{-\beta \left| G\right| },
\label{Zchi1}
\end{equation}
where $\chi $ denotes the fixed Euler characteristic of graph $G$; $\chi=2\;\left(
{\mathcal S}\right) ,1\;\left( {\mathcal D}\right)$; $\left| G\right|$
is the number of vertices of $G$, $S\left( G\right) $ its symmetry
factor (as an unlabelled graph).

The partition function of trivalent random planar graphs is generated in a Hermitian $M$-matrix theory
with a cubic interaction term  $e^{-\beta}{\rm Tr} M^3$. In
particular, the
combinatorial weights and symmetry factors involved in the definition of partition function (\ref{Zchi1})
can be properly understood from that matrix representation (see, e.g., \cite{davidleshouches,davidaltenberg}).

The partition sum converges for all values of the
parameter $\beta $ larger than some critical $\beta_c$. At $\beta
\rightarrow \beta_c^{+},$ a singularity appears due to the
presence of infinite graphs in (\ref{Zchi1})
\begin{equation}
Z_{}\left( \beta , \chi \right) \simeq \hbox{reg. part}+\left( \beta -\beta_c\right)
^{2-\gamma _{\rm str}(\chi)}, \label{Zchi2}
\end{equation}
where $\gamma _{\rm str}(\chi)$ is the string susceptibility
exponent, which depends on the topology of $G$ through the Euler characteristic.
 For pure gravity as described in (\ref{Zchi1}), the
embedding dimension $d=0$ coincides with the central charge $c=0,$
and  \cite{kostov}
\begin{equation}\gamma _{\rm str}(\chi)=2-\frac{5}{4} \chi , (c=0).
\label{gamchi}
\end{equation}
 In particular $\gamma_{\rm
str}(2)=-\frac{1}{2}$ for the spherical topology,  and $\gamma_{\rm
str}(1)=\frac{3}{4}$. The string susceptibility exponent appearing in KPZ formula (\ref{KPZg}) is the planar one
$$\gamma=\gamma _{\rm str}(\chi=2).$$

A particular partition function will play an important role later, that of the doubly punctured sphere. It is defined as
\begin{equation}
 Z\lbrack \hbox to 8.5mm{\hskip 0.5mm 
$\vcenter{\epsfysize=.45truecm\epsfbox{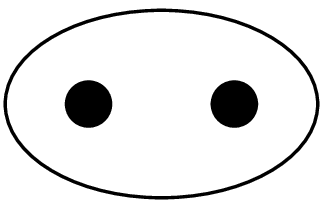}}$
              \hskip 10mm}
 \rbrack \
:= \frac{\partial^{2}}{\partial \beta ^{2}}
Z_{}(\beta,\chi =2)=\sum _{G(\chi=2)}{\frac{1}{S(G)}}\left| G\right|^2 e^{-\beta \left| G\right| }.
\label{Zdotdot}
\end{equation}
Owing to (\ref{Zchi2}) it scales as
\begin{equation}
 Z\lbrack \hbox to 8.5mm{\hskip 0.5mm
$\vcenter{\epsfysize=.45truecm\epsfbox{fig2.eps}}$
              \hskip 10mm}
 \rbrack \
\sim \left( \beta -\beta_c\right)
^{-\gamma _{\rm str}(\chi=2)}.
\label{Zdotdots}
\end{equation}
\begin{figure}[tb]
\begin{center}
\includegraphics[angle=0,width=.25\linewidth]{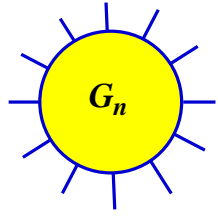}
\end{center}
\caption{A planar random disk with $n$ external legs.}
\label{Figuredisc1}
\end{figure}
The restricted partition function of a planar random
graph with the topology of a disk and a fixed number $n$ of external
vertices (Fig.~\ref{Figuredisc1}),
\begin{equation}
G_n(\beta ) = \sum _{n-{\rm leg}\ {\rm planar}\ G} e^{-\beta \left| G\right| },
\label{AGn}
\end{equation}
 can be calculated through the large$-{ N}$ limit in the random ${ N} \times { N}$
matrix theory \cite{bipz}. It has the integral representation
\begin{equation}
G_{n}\left( \beta \right) =\int^{b}_{a}d\lambda \, \rho\left( \lambda,\beta \right)
{\lambda }^n,
\label{AGnint}
\end{equation}
where $\rho \left( \lambda,\beta \right) $ is the spectral eigenvalue density of
the random matrix, for which the explicit expression is known as a function of
$\lambda ,\beta $
 \cite{bipz}. The {\it support} $[a, b]$ of the spectral density
depends on $\beta $. For the cubic potential  $e^{-\beta}{\rm Tr} M^3$ the explicit solution is
of the form (see, e.g., \cite{davidaltenberg})
\begin{equation}
\rho(\lambda,\beta) = \sqrt{[\lambda-a(\beta)][b(\beta)-\lambda]}\, (c(\beta)-\lambda)\quad
:\quad a(\beta)<b(\beta)\leq c(\beta),
\end{equation}
and is analytic in $\beta$ as long as ${\beta}$ is larger than the critical value ${\beta_c}$.
At this critical point $b(\beta_c)=c(\beta_c)$.
As long as $\beta>\beta_c$,
the density  vanishes like a square root at endpoint $b$:
 $\rho(\lambda,\beta)\propto[b(\beta)-\lambda]^{1/2}$. At $\beta_c$, the density has the {\it universal} critical behavior:
\begin{equation}
\label{rhocritique}
\rho(\lambda,\beta_c)\propto[b(\beta_c)-\lambda]^{3/2}.
\end{equation}
\subsubsection*{Random Walk Partition Functions}
\begin{figure}[tb]
\begin{center}
\includegraphics[angle=0,width=.35\linewidth]{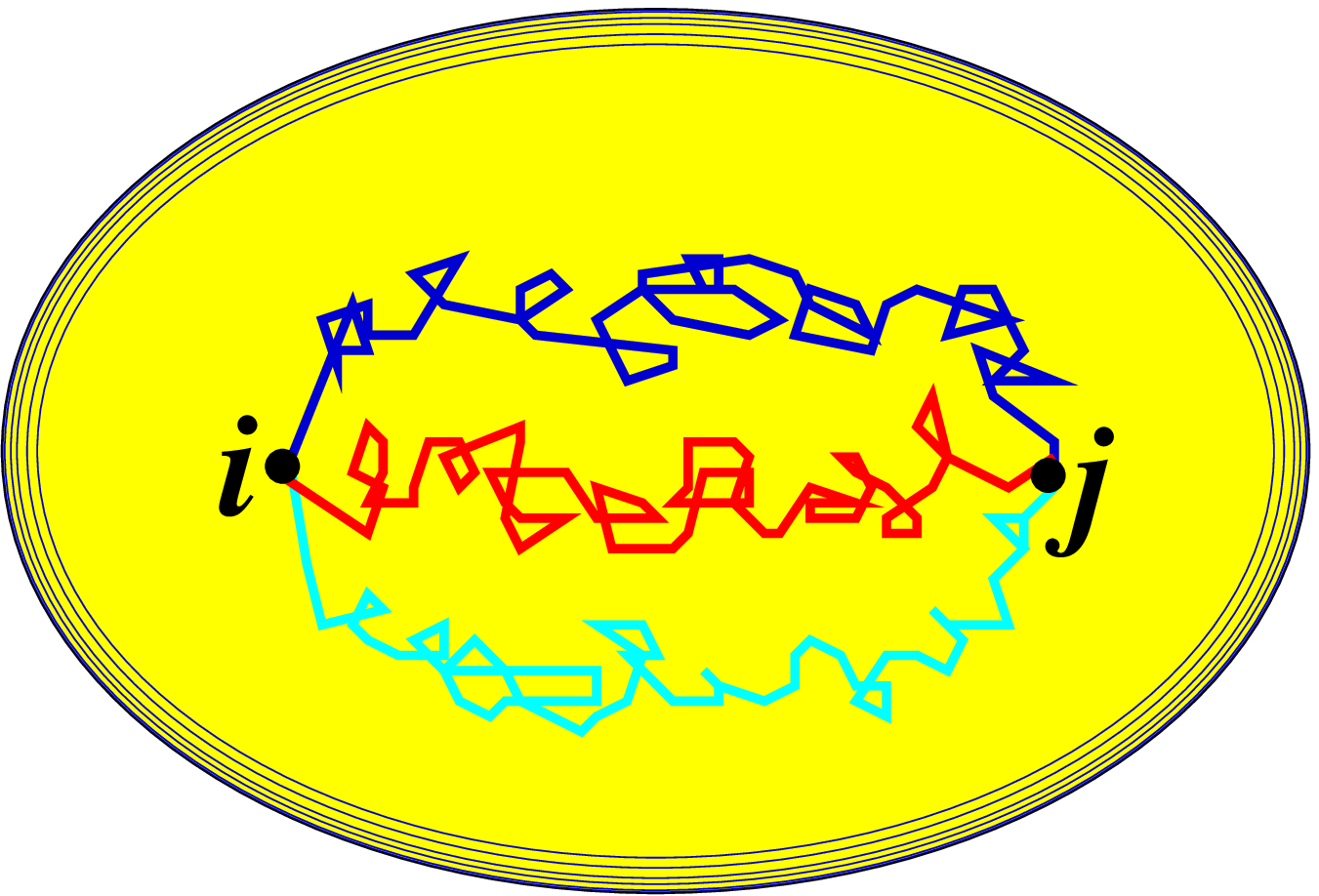}
\hskip.5cm
\includegraphics[angle=0,width=.35\linewidth]{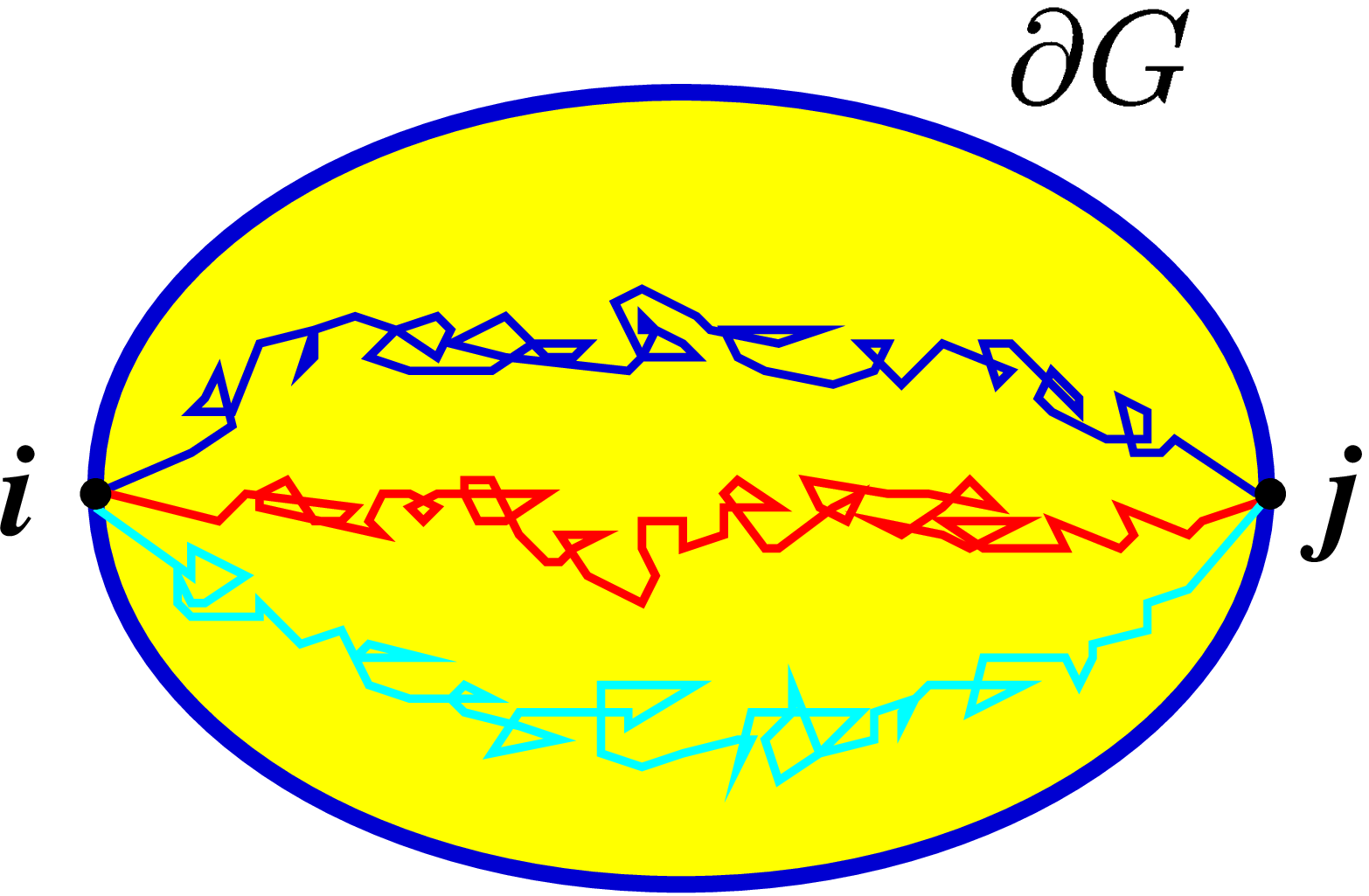}
\end{center}
\caption{$L=3$ mutually-avoiding random walks on a random sphere or traversing a random disk.}
\label{Fig.wmrw12}
\end{figure}

 Let us now consider a set of $L$ random walks ${\mathcal B}=\{{B}_{ij}^{\left( l\right)
},l=1,...,L\}$ on the {\it random graph} $G$ with the special
constraint that they start at the same vertex $i\in G,$ end at the
same vertex $j \in G$, and have no intersections in between. We
introduce the $L$-walk partition function on the random lattice
 \cite{duplantier7}:
\begin{equation}
Z_L(\beta ,z)=\sum _{{\rm planar}\ G}\frac{1}{ S(G)} e^{-\beta
\left| G\right| }\sum _{i,j\in G} \sum_{\scriptstyle
B^{(l)}_{ij}\atop\scriptstyle l=1,...,L} z^{\left| {\mathcal
B}\right| }, \label{Zl}
\end{equation}
where a fugacity $z$ is associated with the total number $\left|
{\mathcal B}\right| =\left| \cup^{L}_{l=1} B^{(l)}\right| $ of
vertices visited by the walks (Fig.~\ref{Fig.wmrw12}). This partition function is the quantum gravity analogue of the
correlator, or infinite measure (\ref{watermelonB}), defined in the standard plane.\\
\subsubsection*{RW Boundary Partition Functions}
We generalize this to the {\it boundary} case where $G$ now has
the topology of a disk and where the random walks connect two
sites $i$ and $j$ on the
boundary $\partial G:$%
\begin{equation}
\tilde Z_{L}(\beta , {\tilde \beta}, z)=\sum _{ {\rm disk}\ G}
e^{-\beta \left| G\right| }e^{-{\tilde \beta}{\left| \partial
G\right|}} \sum _{{i,j} \in \partial G}\sum_{{\scriptstyle
B^{(l)}_{ij}}\atop{\scriptstyle l=1,...,L}} z^{\left|{\mathcal
B}\right| },
\label{Ztilde}
\end{equation}
where $e^{-{\tilde \beta}}$ is the fugacity associated with the
boundary's length (Fig.~\ref{Fig.wmrw12}).

The double grand canonical partition functions (\ref{Zl}) and (\ref{Ztilde})
associated with non-intersecting RW's on a random lattice can be
calculated exactly  \cite{duplantier7}.
One in particular uses an equivalent representation of the random walks by their
{\it forward (or backward) trees}, which are trees uniformly spanning the sets of visited sites. This turns the
RW's problem into the solvable one of random trees on random graphs (see, e.g.,  \cite{DK}).
\subsubsection*{Random Walks and Representation by Trees}
Consider the set $%
B ^{\left( l\right) }\left[ i,j\right] $ of the points visited on the
random graph by a given walk $B^{\left( l\right) }$ between $i$ and $j$%
, and for each site $k \in B^{\left( l\right) }\left[ i,j\right] $ the
first entry, i.e., the edge of $G$ along which the walk $\left( l\right) $
reached $k$ for the first time. The union of these edges form a tree $%
T^{\left( l\right)}_{i,j}$ spanning all the sites of $B^{\left(
l\right) }\left[ i,j\right]$, called the forward tree. An important
property is that the measure on all the trees spanning a given set of points
visited by a RW is {\it uniform}  \cite{aldous}. This means that we can also
represent the path of a RW by its spanning tree taken with uniform
probability. Furthermore, the non-intersection property of the walks is by
definition equivalent to that of their spanning trees.

\subsubsection*{Bulk Tree Partition Function}
\begin{figure}[tb]
\begin{center}
\includegraphics[angle=0,width=.35\linewidth]{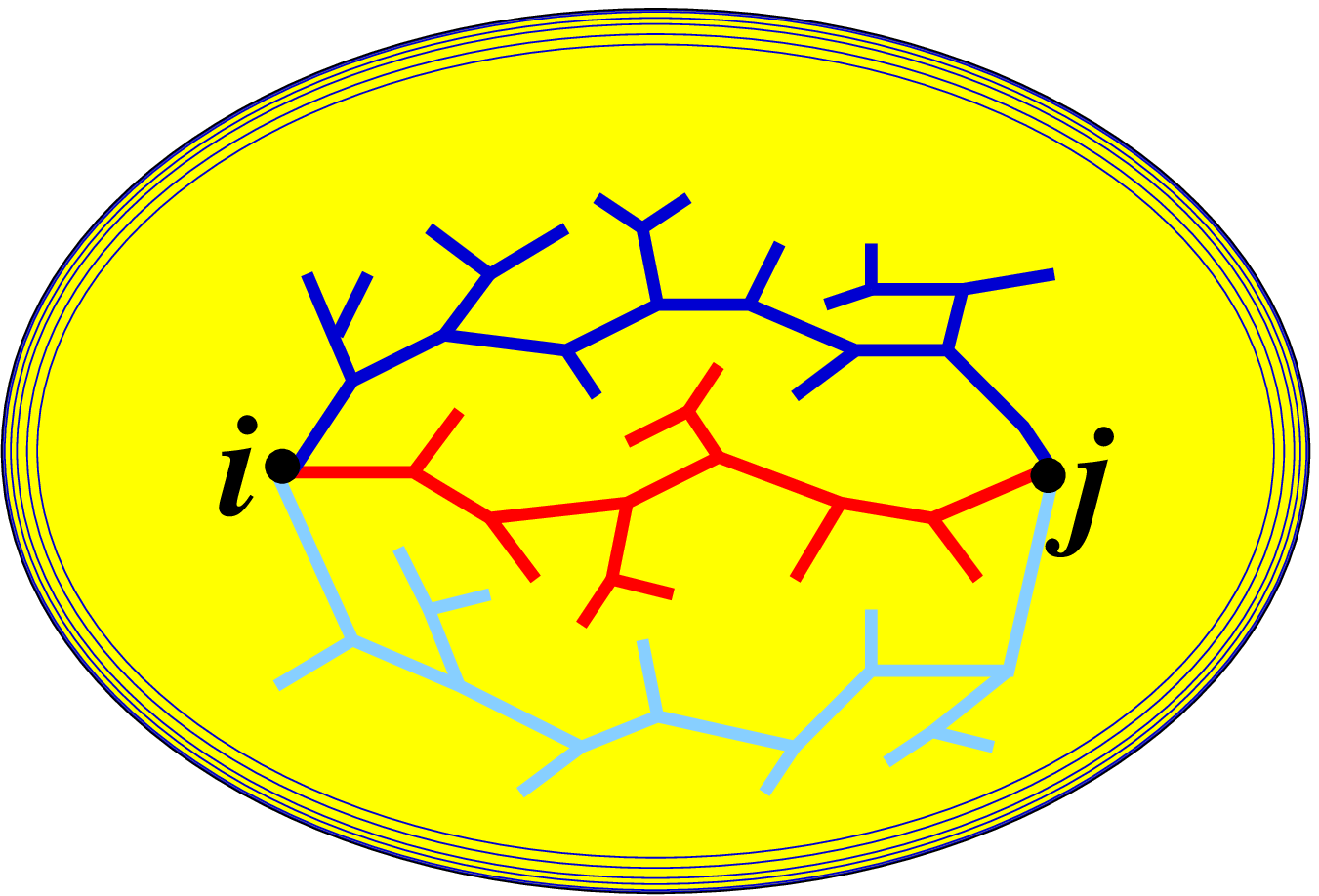}
\hskip.7cm
\includegraphics[angle=0,width=.35\linewidth]{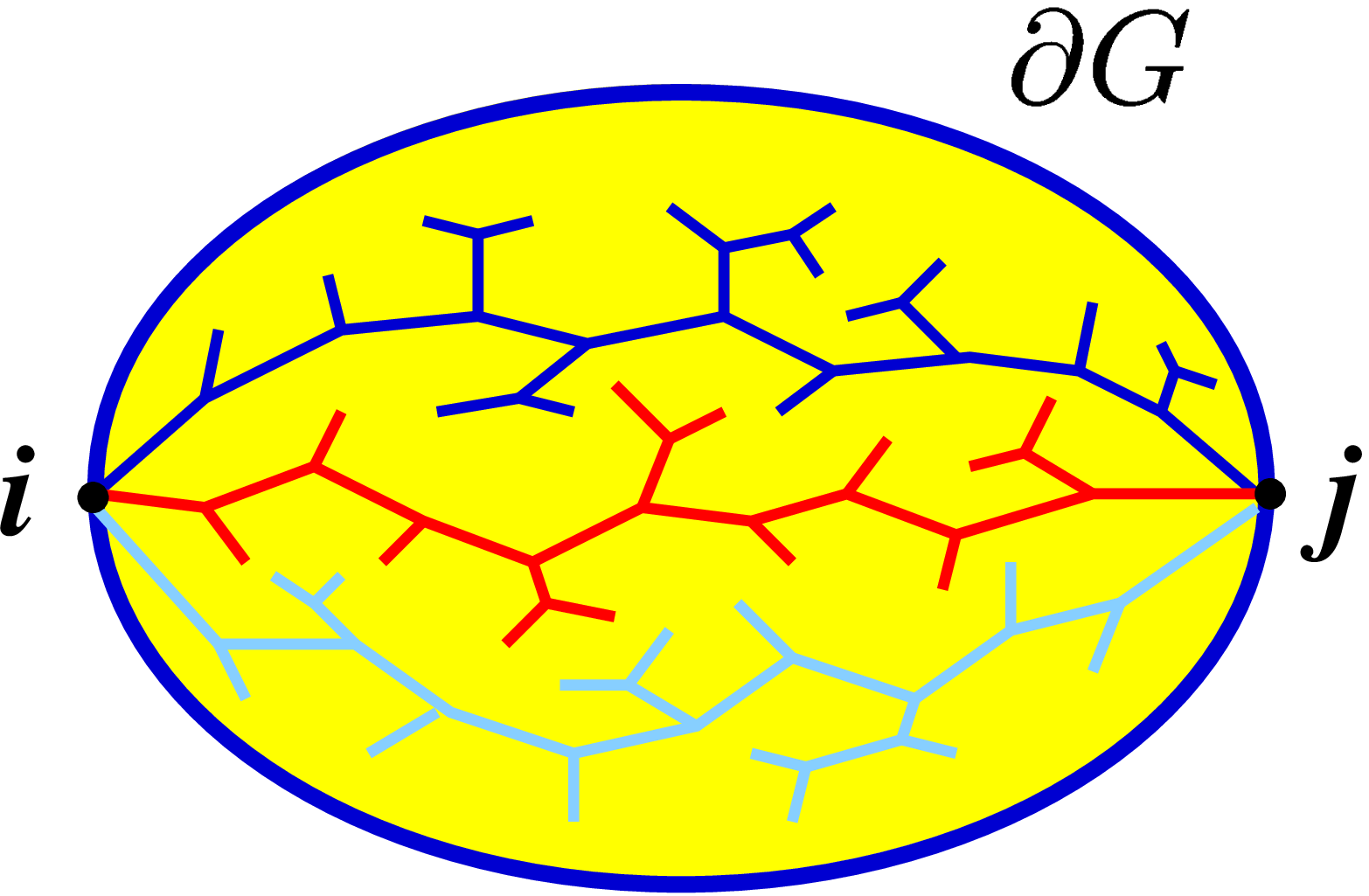}
\end{center}
\caption{$L=3$ mutually-avoiding random trees on a random sphere or traversing a random disk.}
\label{Fig.wmtree1}
\end{figure}

One
introduces the $L$-tree partition function on the random lattice (Fig.~\ref{Fig.wmtree1})
\begin{equation}
Z_L(\beta ,z)=\sum_{{\rm planar}\ G}{\frac{1}{ S(G)}}
e^{-\beta \left| G\right| }\sum _{i,j\in G}
\sum_{\scriptstyle T^{(l)}_{ij}\atop\scriptstyle l=1,...,L}
z^{\left| T\right| },
\label{AZl}
\end{equation}
where $\left\{ T_{ij}^{\left( l\right) }, l=1,\cdots, L \right\}$ is a set of $L$
trees,
all constrained to have sites $i$ and $j$ as end-points, and without mutual
intersections; a fugacity $z$ is in addition associated with the total
number $\left| T\right| =\left| \cup^{L}_{l=1} T^{(l)}\right| $
of vertices of the trees. In principle, the trees spanning the RW
paths can have divalent or trivalent vertices on $G$, but this is immaterial
to the critical behavior, as is the choice of purely trivalent graphs $G$,
so we restrict ourselves here to trivalent trees.
\subsubsection*{Boundary Partition Functions}
We generalize this to the {\it boundary} case where $G$ now has the topology of
a disk and where the trees connect two sites $i$ and $j$ on the
boundary $\partial G$ (Fig.~\ref{Fig.wmtree1})%
\begin{equation}
\tilde Z_{L}(\beta , z, \tilde z)=\sum _{ {\rm disk}\ G}
e^{-\beta \left| G\right| }\tilde z^{\left| \partial G\right|}
\sum _{{i,j} \in \partial G}\sum_{\scriptstyle
T^{(l)}_{ij}\atop\scriptstyle l=1,...,L}
z^{\left| T\right| },
\label{AZtilde}
\end{equation}
where $\tilde {z}\equiv e^{-\tilde\beta}$ is the fugacity associated with the boundary's length.

The partition function of the disk with two boundary punctures will play an important role. It is defined as
\begin{eqnarray}
\label{AZtilde0}
Z( \hbox to 9.5mm{\hskip 0.5mm
$\vcenter{\epsfysize=.45truecm\epsfbox{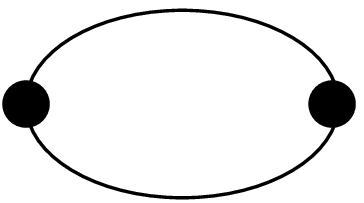}}$
              \hskip -80mm}
)&=&\sum _{ {\rm disk}\ G}
e^{-\beta \left| G\right| }\tilde z^{\left| \partial G\right|}\left| \partial G\right|^2\\
\nonumber
&=&\tilde Z_{L=0}(\beta , \tilde z),
\end{eqnarray}
and formally corresponds to the $L=0$ case of the $L$-tree boundary partition functions (\ref{AZtilde}).

\subsubsection*{Integral Representation}
\begin{figure}[tb]
\begin{center}
\includegraphics[angle=0,width=.4\linewidth]{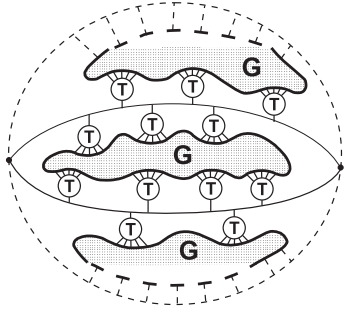}
\end{center}
\caption{Random trees on a random surface. The shaded areas represent portions
of random lattices $G$ with a disk topology (generating function
(\ref{AGn},\ref{AGnint}));
$L=2$ trees connect the end-points, each branch giving a generating
function $T$ (\ref{AT}). Two possible topologies are represented: for the disk,
the dashed lines represent the boundary, whereas for
the sphere the top and bottom dashed lines  should be identified with one another, as should
the upper and lower grey patches.}
\label{FigureA1}
\end{figure}

The partition function (\ref{AZl})
 has been calculated exactly   \cite{DK}, while (\ref{AZtilde}) was
 first considered in Ref. \cite{duplantier7}. The twofold
grand canonical partition function is calculated first by summing over the
abstract tree
configurations, and then gluing patches of random lattices in between these
trees. The rooted-tree generating function is defined as $T(x)
=\sum_{n\geq 1}x^{n}T_{n},$ where $T_{1}\equiv 1
$ and $T_{n}$ is the number of {\it rooted} planar trees with $n$ external
vertices (excluding the root). It reads
 \cite{DK}
\begin{equation}
T\left( x\right) =\frac{1}{2}(1-\sqrt{1-4x}).
\label{AT}
\end{equation}
The result for (\ref{AZl}) is then given by a multiple integral:
\begin{equation}
Z_{L}(\beta ,z)=\int_{a}^{b}\prod ^{L}_{l=1}
d\lambda_{l}\, \rho(\lambda_l,\beta)
\prod ^{L}_{l=1}{\mathcal T}(z\lambda _l,z\lambda _{l+1}),
\label{AZlint}
\end{equation}
with the cyclic condition  $\lambda _{L+1} \equiv \lambda _{1}$. The geometrical
interpretation is quite clear (Fig.~\ref{FigureA1}).
Each patch $l=1,\cdots,L$ of
random surface between trees $T^{\left( l-1\right) }$, $T^{\left( l\right) }$
contributes as a factor a spectral density $\rho \left( \lambda _{l}\right) $
as in Eq.~(\ref{AGnint}), while the backbone of
each tree $T^{\left( l\right) }$ contributes an inverse ``propagator'' $%
{\mathcal T}\left( z\lambda _{l},z\lambda _{l+1}\right) ,$ which couples the
 eigenvalues $\lambda _{l},\lambda _{l+1}$ associated with the two  patches adjacent to $T^{\left( l\right) }$:
\begin{equation}
{\mathcal T}(x,y):= [1-T(x)-T(y)]^{-1}.
\label{APropagator}
\end{equation}

The integral
representation of the boundary partition function (\ref{AZtilde}) is
\begin{eqnarray}
\tilde Z_{L}(\beta ,z , \tilde z)&=&\int_{a}^{b}%
\prod ^{L+1}_{l=1}d\lambda_{l}\, \rho(\lambda _{l},\beta)
\prod ^{L}_{l=1}{\mathcal T}(z\lambda _l,z\lambda
_{l+1}) \nonumber \\
& &\qquad\times (1-\tilde z\lambda_1)^{-1}
(1-\tilde z\lambda _{L+1})^{-1},
\label{AZtildeint}
\end{eqnarray}
with two extra propagators $\mathcal L$ describing the two boundary segments:
\begin{equation}
\label{AL}
\mathcal L(\tilde z\lambda):=(1-\tilde z\lambda)^{-1}.
\end{equation}
 This gives for the two-puncture disk partition function
(\ref{AZtilde0})
\begin{equation}
\label{AZtilde0int}
Z( \hbox to 9.5mm{\hskip 0.5mm
$\vcenter{\epsfysize=.45truecm\epsfbox{fig3.eps}}$
              \hskip -80mm}
)
=\int_{a}^{b} d\lambda \, \rho(\lambda,\beta)\,
(1-\tilde z\lambda )^{-2}.
\end{equation}
\subsubsection*{Symbolic Representation}
The structure of  $Z_{L}$ (\ref{AZlint}) and $\tilde Z_{L}$ (\ref{AZtildeint}) can be represented by using the suggestive
symbolic notation
\begin{eqnarray}
\label{AtZstar}
Z_{L} \sim \left(\int  d\lambda\, \rho \right)^{L} \star {\mathcal T}^{L},\,\,\,\,\,\,\,
\tilde Z_{L} \sim \left(\int  d\lambda\, \rho \right)^{L+1} \star {\mathcal T}^{L} \star \mathcal L^2,
\end{eqnarray}
where the $\star$ symbol represents both the factorized structure of the integrands and the convolution structure
of the integrals.
The formal powers also represent repeated $\star$ operations. This symbolic notation is
useful for the scaling analysis of the partition functions. Indeed the structure of the integrals reveals that
   each factorized component brings in its own contribution to the global scaling behavior \cite{BDMan}.
Hence the symbolic star notation directly translates into power counting,
in a way which is reminiscent of standard power counting for usual Feynman diagrams.

 One can thus write the formal power behavior
\begin{equation}
\label{AtZs}
Z_{L} \sim  \left(\int d\lambda \, \rho \star {\mathcal T}\right)^{L} ,\,\,\,\,\,\,\,
\tilde Z_{L}\sim \left(\int d\lambda \, \rho \star {\mathcal T}\right)^{L} \star \int  d\lambda \, \rho \star \mathcal L^2.
\end{equation}
This can be simply recast as
\begin{equation}
\label{AtZs'}
\tilde Z_{L}\sim Z_L \star \int d\lambda \, \rho\star \mathcal L^2.
\end{equation}
Notice that the last two factors precisely correspond to the scaling of the two-puncture boundary partition function
(\ref{AZtilde0int})
\begin{equation}
\label{AtZ0s}
Z( \hbox to 9.5mm{\hskip 0.5mm
$\vcenter{\epsfysize=.45truecm\epsfbox{fig3.eps}}$
              \hskip -80mm}
)=\tilde Z_{0}\sim  \int d\lambda \, \rho\star \mathcal L^2.
\end{equation}

\subsubsection*{Scaling Laws for Partition Functions}
The critical behavior of partition functions $Z_{L}$ and ${\tilde Z}_{L}$
is characterized by the existence of critical values of the parameters,
$\beta_c$ where the random lattice size diverges, $z_c$ where the number of
sites visited by the random walks also diverges, and ${\tilde z}_c$ where the boundary length diverges.

The analysis of singularities in the explicit expressions (\ref{AZlint}) and (\ref{AZtildeint}) can be performed by using
 the explicit propagators $\mathcal T$ (\ref{APropagator}) \& (\ref{AT}), $\mathcal L$ (\ref{AL}), and the critical behavior (\ref{rhocritique}) of the
 eigenvalue density $\rho(\lambda,\beta_c)$ of the random matrix theory representing the random lattice. One sees in
 particular that $z_c=1/2b(\beta_c)$ and ${\tilde z}_c=1/b(\beta_c)$.

 The critical behavior
of the bulk partition function $Z_{L}\left( \beta ,z\right)$ is then obtained by taking the
double
scaling limit $\beta \rightarrow \beta_c^{+}$ (infinite random surface) and $
z\rightarrow z_c^{-}$ (infinite RW's or trees), such that the average lattice and RW's sizes respectively scale as
\footnote{Hereafter
averages or expectation values like $\langle |G|\rangle$ are simply denoted by $|G|$.}
\begin{equation}
|G| \sim (\beta-\beta_c)^{-1},  \left| {\mathcal B}\right| \sim \left| { T}\right|\sim (z_c-z)^{-1}.
\label{sizes}
\end{equation}
We refer the reader to Appendix A in Ref. \cite{BDMan} for a detailed analysis of the singularities of multiple
integrals (\ref{AZlint}) and (\ref{AZtildeint}).  One observes in particular that the factorized structure
(\ref{AtZs}) corresponds precisely to the factorization of the various scaling components.

The analysis of the singular
behavior is  performed by using
{\it finite-size scaling} (FSS)  \cite{DK}, where one must have
$$\left| {\mathcal B}\right| \sim \left| {T}\right|\sim \left| G\right| ^\frac{1}{2} \iff
 z_c-z \sim (\beta-\beta_c)^{1/2}.$$
One obtains in this regime the global scaling of the full partition function
 \cite{BDMan,duplantier7}:
\begin{equation}
Z_{L}\left( \beta ,z\right) \sim \left( \beta -\beta_c\right) ^{L}
\sim {\left| G\right|} ^{-L}.
\label{Zll}
\end{equation}
Notice that the presence of a global power $L$ was expected from the factorized  structure (\ref{AtZs}).

The interpretation of partition function $Z_{L}$ in terms of conformal weights is the following:
It represents a random surface with two {\it
punctures} where two conformal operators, of conformal weights
$\Delta _{L}$, are located (here two vertices of $L$
non-intersecting RW's or trees). Using a graphical notation, it scales as
\begin{equation}
Z_{L}\sim Z\lbrack \hbox to 8.5mm{\hskip 0.5mm
$\vcenter{\epsfysize=.45truecm\epsfbox{fig2.eps}}$
              \hskip 10mm}
 \rbrack \ \times \left| G\right|
^{-2\Delta _{L}},\
\label{Zls}
\end{equation}
where the partition function of the doubly punctured surface is the
second derivative of $Z_{ }(\beta,\chi=2)$ (\ref{Zdotdot}):
\begin{equation}
 Z\lbrack \hbox to 8.5mm{\hskip 0.5mm
$\vcenter{\epsfysize=.45truecm\epsfbox{fig2.eps}}$
              \hskip 10mm}
 \rbrack \
= \frac{\partial^{2}}{\partial \beta ^{2}}
Z_{}(\beta,\chi =2).
\label{Zdots}
\end{equation}
From (\ref{Zdotdots}) we find
\begin{equation}
Z_{L}\sim  \left| G\right|
^{\gamma _{\rm str}(\chi =2)-2\Delta _{L}}.\
\label{Zlds}
\end{equation}
Comparing the latter to (\ref{Zll}) yields
\begin{equation}
2\Delta _{L}-\gamma _{\rm str}(\chi =2)=L, \label{deltal}
\end{equation}
where we recall that $\gamma _{\rm str}(\chi =2)=-1/2$. We thus get the first announced
result
\begin{equation}
\Delta _{L}=\frac{1}{2}\left(L-\frac{1}{2}\right).
\label{deltaL}
\end{equation}

\subsubsection*{Boundary Scaling \& Boundary Conformal Weights}
For the boundary partition function $\tilde Z_{L}$ (\ref{AZtildeint})
a similar analysis can be performed near the triple critical point $(\beta_c,{\tilde z}_c=1/b(\beta_c),z_c)$,
where the boundary length also diverges. One finds that the average boundary length $|\partial G|$
must scale with the area $|G|$ in a natural way (see Appendix A in Ref. \cite{BDMan})
\begin{equation}
|\partial G|\sim |G|^{1/2}.
\label{FSSdGG}
\end{equation}
The boundary partition
function $ \tilde Z_{L}$ corresponds to two boundary operators of
conformal weights ${\tilde \Delta}_{L},$ integrated over the boundary $\partial G,$
on a random surface with the topology of a disk. In terms of scaling behavior we write:
\begin{equation}
\tilde Z_{L}\sim Z( \hbox to 9.5mm{\hskip 0.5mm
$\vcenter{\epsfysize=.45truecm\epsfbox{fig3.eps}}$
              \hskip -80mm}
) \times \left|\partial G\right| ^{-2\tilde{\Delta }_{L}},
\label{Zlts}
\end{equation}
using the graphical representation of the two-puncture partition function (\ref{AZtilde0}).

\subsubsection*{Bulk-Boundary Relation}
The star representation in Eqs. (\ref{AtZs'}) and (\ref{AtZ0s}) is strongly suggestive of a
scaling relation between bulk and
boundary partition functions. From the exact expressions (\ref{AZlint}), (\ref{AZtildeint})  and (\ref{AZtilde0int})
of the various partition
functions, and the precise analysis of their singularities (see Appendix A in Ref. \cite{BDMan}), 
one indeed gets the further scaling equivalence:
\begin{equation}
Z_{L} \sim \frac{\tilde Z_{L}}{Z( \hbox to 9.5mm{\hskip 0.5mm
$\vcenter{\epsfysize=.45truecm\epsfbox{fig3.eps}}$
              \hskip -80mm}
)},
\label{ratio}
\end{equation}
where the equivalence holds true in terms of scaling behavior. It intuitively means that carving away from
the $L$-walk boundary partition function the contribution of one connected domain with two
boundary punctures brings one back to the $L$-walk bulk partition function.

Comparing Eqs.~(\ref{Zlts}), (\ref{ratio}), and~(\ref{Zlds}), and
using the FSS (\ref{FSSdGG}) gives
\begin{equation} {\tilde \Delta}_{L}=2\Delta_L-\gamma_{\rm str} (\chi=2).
\label{deltat}
\end{equation}
This relation between bulk and Dirichlet boundary behaviors in quantum gravity is quite general \cite{BDMan} and will also
play a
fundamental role
in the study of other critical systems in two dimensions. A general derivation can be found in Appendix C of 
Ref.~\cite{BDMan}.

From (\ref{deltaL}) we finally find the second announced result:
\begin{equation} {\tilde \Delta}_{L}=L.
\label{tdeltal}
\end{equation}

Applying the quadratic KPZ relation (\ref{KPZ}) to $\Delta
_{L}$ (\ref{deltaL})
and ${\tilde \Delta}_{L}$ (\ref{tdeltal}) above finally yields the values in
the plane ${{\mathbb C}}$ or half-plane ${\mathbb H}$
\begin{eqnarray}
\nonumber
\zeta _{L}&=&{U}_{\gamma=-1/2}\left( \Delta_L
\right)=\frac{1}{24} \left( 4L^2-1\right)\\
\nonumber
2\tilde \zeta _{L}&=&{U}_{\gamma=-1/2}\left( \tilde \Delta_L
\right)=\frac{1}{3}L \left( 1+2L \right),
\end{eqnarray}
as announced.

\subsection{Non-Intersections of Packets of Walks}
\subsubsection*{Definition}
\begin{figure}[htb]
\begin{center}
\centerline{\epsfxsize 5cm
\epsfbox{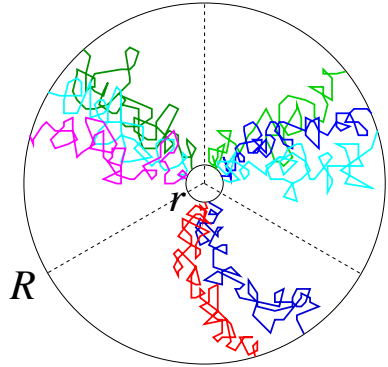}}
\caption{{Packets of $n_1=3, n_2=3$, and $n_3=2$ independent planar random walks, in a mutually-avoiding
star configuration, and crossing an annulus from $r$ to $R$.}}
\label{fig.rw332}
\end{center}
\end{figure}
\begin{figure}[tb]
\begin{center}
\includegraphics[angle=0,width=.5\linewidth]{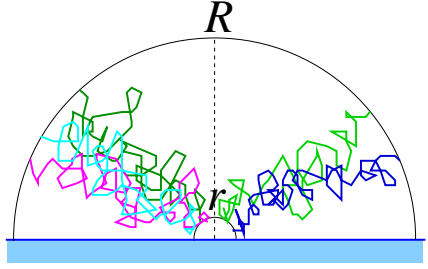}
\end{center}
\caption{Packets of $n_1=3$, and $n_2=2$ independent random walks, in a mutually-avoiding 
star configuration, and crossing the half-annulus from $r$ to $R$ in the half-plane ${\mathbb H}$.}
\label{fig.rwhalf332}
\end{figure}

Consider configurations made of $L$ mutually-avoiding bunches $l=1,\cdots,L$, each of them
made of $n_{l}$ walks {\it transparent} to each
other, i.e., $n_l$ independent RW's  \cite{werner}. All of them start at neighboring points (Fig.~\ref{fig.rw332}).
The probability of non-intersection of the $L$ packets up to time $t$
scales as
\begin{equation}
P_{n_{1},\cdots,n_{L}}(t) \approx t^{-\zeta
(n_{1},\cdots,n_{L})},
\label{Pbunch}
\end{equation}
and near a Dirichlet boundary (Fig.~\ref{fig.rwhalf332})
\begin{equation}
\tilde P_{n_{1},\cdots,n_{L}}(t) \approx t^{-\tilde \zeta
(n_{1},\cdots,n_{L})}.
\label{Ptbunch}
\end{equation}
The original case of $L$ mutually-avoiding RW's corresponds to $n_{1}=\cdots =n_{L}=1$.
Accordingly, the probability for the same $L$ Brownian path packets to cross the annulus $\mathcal D(r,R)$ in ${\mathbb C}$
(Fig.~\ref{fig.rw332})
scales as
\begin{equation}
P_{n_{1},\cdots,n_{L}}(r) \approx \left(r/R\right)^{2\zeta
(n_{1},\cdots,n_{L})},
\label{Pbunchr}
\end{equation}
and, near a Dirichlet boundary in ${\mathbb H}$ (Fig.~\ref{fig.rwhalf332}), as
\begin{equation}
\tilde P_{n_{1},\cdots,n_{L}}(r) \approx \left(r/R\right)^{2\tilde \zeta
(n_{1},\cdots,n_{L})}.
\label{Ptbunchr}
\end{equation}

The generalizations of former exponents $\zeta_L$, as well as ${\tilde \zeta}_{L}$, describing these $L$ packets
can be written as
conformal weights
$$\zeta (n_{1},\cdots,n_{L})=\Delta
^{(0)}
\left\{ n_{l}\right\}$$ in the plane ${{\mathbb C}}$, and $$2{\tilde \zeta}(n_{1},\cdots,n_{L}) ={\tilde %
\Delta}^{(0)}\left\{ n_{l}\right\} $$ in the half-plane ${\mathbb H}$. They can be calculated from quantum gravity, via
their conterparts $\Delta\left\{ n_{l}\right\}$ and $\tilde \Delta \left\{ n_{l}\right\}$. The details are given in
 \cite{BDMan} (Appendix A). We  sketch here the main steps.

\subsubsection*{Boundary Case}
One introduces the analogue
${\tilde Z}\left\{ n_{1},\cdots, n_{L}\right\}$ of partition function (\ref{Ztilde}) for the $L$ packets of walks.
In presence of
gravity each bunch contributes its own {\it normalized
boundary partition function} as a factor, and this yields a natural
generalization of the scaling equation~(\ref{ratio}) (see Appendix A in Ref.~\cite{BDMan})
\begin{equation}
\frac{{\tilde Z}\left\{ n_{1},\cdots, n_{L}\right\}}{
Z( \hbox to 9.5mm{\hskip 0.5mm
$\vcenter{\epsfysize=.45truecm\epsfbox{fig3.eps}}$
              \hskip -80mm}
)} \sim \prod _{l=1}^{L} \star \left\{\frac{{\tilde Z}\left( n_{l}\right)}{Z(
\hbox to 9.5mm{\hskip 0.5mm
$\vcenter{\epsfysize=.45truecm\epsfbox{fig3.eps}}$
              \hskip -80mm}
)}\right\}, \label{Zn}
\end{equation}
where the star product is to be understood as a scaling equivalence. Given the definition of boundary conformal weights
(see (\ref{Zlts})), the normalized left-hand fraction is to be identified
with $\left|\partial G\right| ^{-2{\tilde \Delta} \left\{n_{1},\cdots,n_{L}\right\}}$, while each normalized factor
${\tilde Z}\left( n_{l}\right)/{Z(
\hbox to 9.5mm{\hskip 0.5mm
$\vcenter{\epsfysize=.45truecm\epsfbox{fig3.eps}}$
              \hskip -80mm}
)}$ is to be identified
with $\left|\partial G\right| ^{-2{\tilde \Delta} \left( n_{l}\right)}$.
 Here ${\tilde \Delta}(n)$ is the boundary dimension of a
{\it single} packet of $n$ mutually transparent walks on the random surface.
 The {\it factorization} property
(\ref{Zn}) therefore immediately implies the {\it additivity of boundary
conformal dimensions in presence of gravity}
\begin{equation}
{\tilde \Delta}\left\{n_{1},\cdots,n_{L}\right\} =\sum ^{L}_{l=1}
{\tilde \Delta}(n_{l}). \label{deltan}
\end{equation}
\begin{figure}[htb]
\begin{center}
\includegraphics[angle=0,width=.9\linewidth]{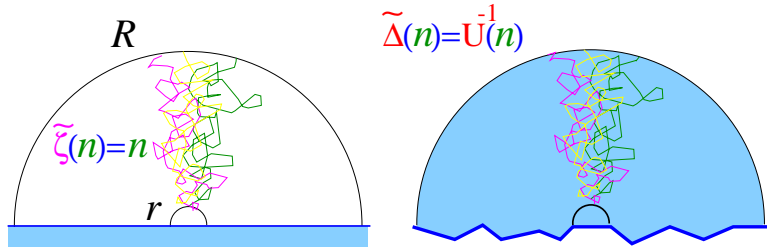}
\end{center}
\caption{{A packet of $n$ independent random walks and its boundary conformal dimensions in the
half-plane, ${\tilde \zeta}(n)\equiv {\tilde \Delta}^{(0)}(n)=n$, and in quantum gravity,
${\tilde \Delta}(n)={U}_{\gamma=-1/2}^{-1}(n).$}}
\label{fig.zetaQG}
\end{figure}
In the standard plane ${\mathbb C}$, a packet of $n$ independent random walks has a trivial boundary conformal dimension
${\tilde \Delta}^{(0)}(n)=n{\tilde \Delta}^{(0)}(1)=n,$ since for a single walk ${\tilde \Delta}^{(0)}(1)=1,$
as can be seen using the Green function formalism. We therefore know
${\tilde \Delta}(n)$ exactly, since it suffices to take the positive {\it inverse} of the KPZ map
(\ref{KPZ}) to get (figure \ref{fig.zetaQG})
\begin{equation}
{\tilde \Delta}(n)={U}_{\gamma=-1/2}^{-1}(n)=\frac{1}{4}(\sqrt{24n+1}-1).
\label{deltatn}
\end{equation}
One therefore finds:
\begin{equation}
{\tilde \Delta}\left\{n_{1},\cdots,n_{L}\right\} =
\sum ^{L}_{l=1}{U}_{\gamma=-1/2}^{-1}(n_l)=\sum ^{L}_{l=1}\frac{1}{4}(\sqrt{24n_l+1}-1).
\label{deltan'}
\end{equation}

\subsubsection*{Relation to the Bulk}
\begin{figure}[htb]
\begin{center}
\includegraphics[angle=0,width=.7\linewidth]{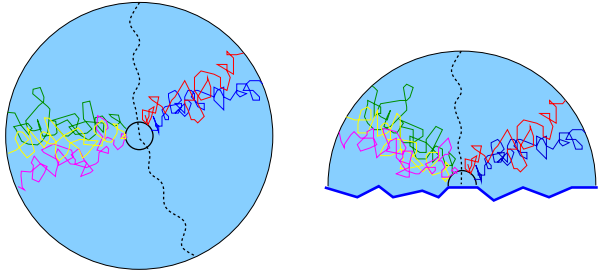}
\caption{In quantum gravity, conformal weights $\Delta \left\{  n_{1},\cdots, n_{L}\right\}$ for non-intersecting packets 
in the bulk (left) and ${\tilde \Delta}\left\{n_{1},\cdots,n_{L}\right\}$ near a boundary (right) are related by 
equation (\ref{deltanL}).}
\label{fig.rw32QG}
\end{center}
\end{figure}
One similarly defines for $L$ mutually-avoiding packets of $n_{1},\cdots,n_{L}$ independent walks
the generalization $Z\left\{  n_{1},\cdots, n_{L}\right\}$ of the bulk
partition function (\ref{Zl}) for $L$ walks on a random sphere. One then establishes on a random surface
the identification, similar to (\ref{ratio}), of
this bulk partition function with the normalized boundary one (see Ref. \cite{BDMan}, Appendix A):
\begin{equation}
Z\left\{  n_{1},\cdots, n_{L}\right\} \sim \frac{{\tilde Z}\left\{ n_{1},\cdots, n_{L}\right\}}{
Z( \hbox to 9.5mm{\hskip 0.5mm
$\vcenter{\epsfysize=.45truecm\epsfbox{fig3.eps}}$
              \hskip -80mm}
)} . \label{ZnZntilde}
\end{equation}
By definition of quantum conformal weights, the left-hand term of (\ref{ZnZntilde})
scales as $|G|^{-2\Delta \left\{n_{1},\cdots,n_{L}\right\}+\gamma _{\rm str}(\chi =2)}$, while the
right-hand term scales, as written above, as $\left|\partial G\right| ^{-2{\tilde \Delta}
\left\{n_{1},\cdots,n_{L}\right\}}$.
Using the area to perimeter scaling relation (\ref{FSSdGG}), we thus get the identity
existing in quantum gravity between bulk and boundary conformal weights, similar to (\ref{deltal}):
\begin{equation}
2\Delta \left\{  n_{1},\cdots, n_{L}\right\}-\gamma _{\rm str}(\chi =2)=
{\tilde \Delta}\left\{n_{1},\cdots,n_{L}\right\}, \label{deltanL}
\end{equation}
with   $\gamma _{\rm str}(\chi =2)=-\frac{1}{2}$ for pure gravity.

\subsubsection*{Back to the Complex Plane}
In the plane, using once again the KPZ relation (\ref{KPZ}) for
${\tilde \Delta}\left\{ n_{l}\right\}$  and ${\Delta}\left\{ n_{l}\right\}$, we obtain the general results
 \cite{duplantier7}
\begin{eqnarray}
\nonumber
2{\tilde
\zeta}(n_{1},\cdots,n_{L})&=&\tilde \Delta^{(0)}\{n_{1},\cdots,n_{L}\}=U\left(\tilde \Delta\left\{n_{1},\cdots,n_{L}
\right\}\right)\\
\nonumber
\zeta (n_{1},\cdots,n_{L})  &=&\Delta^{(0)}\{n_{1},\cdots,n_{L}\}=U\left(\Delta\left\{n_{1},\cdots,n_{L}\right\}\right),
\label{'ZetaL}
\end{eqnarray}
where we set $U:={U}_{\gamma=-1/2}$. One can finally write, using (\ref{deltatn}) and (\ref{deltan'})
\begin{eqnarray}
\label{Zetall}
2{\tilde \zeta}(n_{1},\cdots,n_{L})&=&U(x)=\frac{1}{3} x(1+2x)\\
\label{Zetal}
\zeta (n_{1},\cdots,n_{L})&=& V(x):=U\left[\frac{1}{2}\left(x-\frac{1}{2}\right)\right]=\frac{1}{24} (4x^{2}-1),\\
x&=&\sum_{l=1}^{L}{U}^{-1}(n_{l})=\sum_{l=1}^{L}\frac{1}{4}(\sqrt{24n_{l}+1}-1).
\label{ZetaL}
\end{eqnarray}
Lawler and Werner   first established
the existence of two functions $U$ and $V$ satisfying the ``cascade relations''
 (\ref{Zetall}-\ref{ZetaL}) by purely probabilistic means,
using the geometrical conformal invariance of Brownian motion \cite{lawler2}. The quantum gravity approach
 explained this structure by the linearity of boundary quantum gravity
(\ref{deltan}, \ref{deltan'}), and yielded the explicit functions $U$ and $V$
as KPZ maps (\ref{Zetall}--\ref{Zetal}) \cite{duplantier7}. The same expressions for these
functions have later been derived rigorously in probability
theory from the equivalence to ${\rm SLE}_6$  \cite{lawler4}.

\subsubsection*{Particular Values and Mandelbrot's Conjecture}
Let us introduce the notation $1^{(
L)}=\stackrel{L}{\overbrace{1,1,\cdots,1}}$ for $L$ mutually-avoiding walks in a star
configuration. Then the exponent $\zeta (2,1^{(
L)})$ describing a two-sided walk and $L$ one-sided walks, all
mutually-avoiding, has the value
\begin{eqnarray}
\nonumber
\zeta (2,1^{(L)})&=&V\left[{L {U}^{-1}(1)+{U}^{-1}(2)}\right]=V\left(L +\frac{3}{2}\right)\\
\nonumber
&=&\zeta_{L +\frac{3}{2}}=\frac{1}{6}(L+1)(L+2).
\end{eqnarray}
For $L=1$, $\zeta (2,1)=\zeta_{L=5/2}=1$ correctly gives
the exponent governing the escape probability of a RW from a given origin near another RW  \cite{lawlerisrael}.
(By construction the second one indeed appears as made of two independent RW's diffusing away from the origin.)

For $L=0$ one finds the non-trivial result
$$\zeta
(2,1^{(0)})=\zeta_{L=3/2}=1/3,$$
which describes the accessible points along a RW. It is formally related to the Hausdorff
dimension of the Brownian frontier by $D=2-2\zeta$ \cite{lawler}. Thus we
obtain for the dimension of the Brownian frontier  \cite{duplantier7}
\begin{equation}
D_{\rm Brown.\, fr.}=2-2{\zeta}_{\frac{3}{2}}=\frac{4}{3}, \label{mand}
\end{equation}
i.e., the famous {\it Mandelbrot conjecture}. Notice that the accessibility of a point on a
Brownian path is a statistical constraint equivalent to the non-intersection of ``$L=3/2$''
paths.\footnote{The understanding of the role played by
 exponent $\zeta_{3/2}=1/3$ emerged from a discussion in December 1997 at the
IAS at Princeton with M. Aizenman and R. Langlands about the meaning of half-integer
indices in critical percolation exponents.} The Mandelbrot conjecture was later
established rigorously in probability theory  by Lawler, Schramm and Werner \cite{lawler5}, using the
analytic properties of the non-intersection exponents derived from the
stochastic L\"owner evolution ${\rm SLE}_6$  \cite{schramm1}.


\section{\sc{Mixing Random \& Self-Avoiding Walks}}
\label{sec.mixing}
We now generalize the scaling structure obtained in the preceding
section to arbitrary sets of random or self-avoiding walks interacting together
 \cite{duplantier8} (see also  \cite{lawler2,lawler3}).

\subsection{General Star Configurations}

\subsubsection*{Star Algebra}
Consider a
general copolymer ${\mathcal S}$ in the plane ${{\mathbb C}}$ (or
in ${{\mathbb Z} }^{2}$), made of an arbitrary mixture of RW's or Brownian paths $%
\left( \makebox{set}{\rm \;}{\mathcal B}\right) ,$ and
SAW's or polymers $\left( \makebox{set}{\rm \;}{\mathcal P}\right)$, all
starting at neighboring points, and diffusing away, i.e., in a {\it star} configuration.
In the plane, any successive pair $\left( A,B\right) $ of
such paths, $A,B\in {\mathcal B}$ or ${\mathcal P},$ can be
constrained in a specific way: either they avoid each other
$\left( A\cap B=\emptyset ,\makebox{ denoted }A\wedge B\right) ,$ or
they are independent, i.e., ``transparent'' and can cross each
other (denoted $A\vee B)$ \cite{duplantier8,ferber}. This notation
allows any {\it nested} interaction structure  \cite{duplantier8}; 
for instance that the
branches $\left\{ {\mathcal P}_{\ell }\in {\mathcal P}\right\} _{\ell =1,...,L}$ of an
$L$-star
polymer, all mutually-avoiding, further avoid a collection of Brownian paths $%
\left\{ \mathcal B_{k}\in {\mathcal B}\right\} _{k=1,...,n},$ all
transparent to each other, which structure is represented by:
\begin{equation}
{\mathcal S}=\left( \bigwedge\nolimits_{\ell =1}^{L}{\mathcal P}_{\ell
}\right) \wedge \left( \bigvee\nolimits_{k=1}^{n}\mathcal B_{k}\right) .
\label{vw}
\end{equation}
A priori in 2D the order of the branches of the star polymer
may matter and is intrinsic to the $\left( \wedge ,\vee \right)$ notation.

 \subsubsection*{Conformal Operators and Scaling Dimensions}
To each {\it specific} star copo\-lymer center ${\mathcal S}$ is
attached a local conformal scaling operator $\Phi_{\mathcal S}$, which represents the presence of the
star vertex, with a scaling dimension
$x\left( {\mathcal S}\right)$  \cite{S1,duplantier4,DS2,duplantier8}. When the star is constrained to
stay in a {\it half-plane} ${\mathbb H}$, with Dirichlet boundary conditions, and
 its core placed near the {\it
boundary} $\partial {\mathbb H}$, a new boundary scaling operator ${\tilde \Phi}_{\mathcal S}$ appears, with a boundary
scaling dimension $\tilde{x}\left( {\mathcal S}\right)$  \cite{DS2}. To obtain proper scaling, one has to construct
 the partition functions of Brownian paths and polymers
having the same mean size $R$  \cite{duplantier4}. These partition functions then scale as powers of $R$,
with an exponent which mixes the
scaling dimension of the star core ($x\left( {\mathcal S}\right)$ or $\tilde{x}\left( {\mathcal S}\right)$),
with those of star dangling ends.

\subsubsection*{Partition Functions}
It is convenient to define for each star $%
{\mathcal  S}$ a grand canonical partition function  \cite
{duplantier4,DS2,ferber}, with fugacities $z$ and $z^{\prime }$ for
the total lengths $\left| {\mathcal  B}\right| $ and $\left| {\mathcal  P}\right| $ of
RW or SAW paths:
\begin{equation}
{\mathcal  Z}_{R}\left( {\mathcal  S}\right) =\sum_{{\mathcal  B},{\mathcal  P}\subset {\mathcal  S}%
}z^{\left| {\mathcal  B}\right| }z^{\prime \left| {\mathcal  P}\right| }\;{\bf 1}%
_{R}\left( {\mathcal  S}\right) ,  \label{zr}
\end{equation}
where one sums over all RW and SAW configurations respecting the mutual-avoi\-dance constraints built in star ${\mathcal  S}$
(as in (\ref{vw})),
further
constrained by the indicatrix $
\;{\bf 1}_{R}\left( {\mathcal  S}\right) $ to stay within a disk of radius $R$
centered on the star. At the critical values $z_{c}=\mu
_{\scriptstyle{\rm RW}}^{-1},z_{c}^{\prime }=\mu _{\scriptstyle{\rm SAW}}^{-1},$ where  $\mu _{\scriptstyle{\rm RW}}$ is
the coordination number of the underlying lattice for the RW's, and $\mu _{\scriptstyle{\rm SAW}}$ the
effective one for the SAW's, ${\mathcal  Z}_{R}$ obeys a power law decay  \cite
{duplantier4}
\begin{equation}
{\mathcal  Z}_{R}\left( {\mathcal  S}\right) \sim R^{-x\left( {\mathcal  S}\right) -x^{\bullet
}}.  \label{zrs}
\end{equation}
Here $x\left( {\mathcal  S}\right) $ is the scaling dimension of the operator $\Phi_{\mathcal S}$,
 associated only with the singularity
occurring at the center of the star where all critical paths meet, while $x^{\bullet}$
is the contribution of the independent dangling ends.
It reads $x^{\bullet   }=\left\| {\mathcal  B}\right\| x_{B,1}+\left\| {\mathcal  P}%
\right\| x_{P,1}-2{\mathcal  V},$ where $\left\| {\mathcal  B}\right\| $ and $\left\|
{\mathcal  P}\right\| $ are respectively the total numbers of Brownian or polymer
paths of the star; $x_{B,1}$ or $x_{P,1}$ are the scaling dimensions of the
extremities of a {\it single} RW ($x_{B,1}=0$) or SAW ($x_{P,1}=\frac{5}{48}$)%
 \cite{duplantier4,nien}. The last term in (\ref{zrs}), in which ${\mathcal  V}=\left\| {\mathcal  B}
\right\| +\left\| {\mathcal  P}\right\| $ is the number of dangling vertices, corresponds  to the
integration over the positions of the latter in the disk of radius $R$.

When the star is constrained to stay in a {\it half-plane} with its core
placed near the {\it boundary}, its partition function scales as  \cite
{duplantier4,duplantier2}
\begin{equation}
\tilde{{\mathcal  Z}}_{R}\left( {\mathcal  S}\right) \sim R^{-\tilde{x}\left( {\mathcal  S}%
\right) -x^{\bullet   }},  \label{zz}
\end{equation}
where $\tilde{x}\left( {\mathcal  S}\right) $ is the boundary scaling dimension, $%
x^{\bullet  }$ staying the same for star extremities in the bulk.

\subsubsection*{SAW Watermelon Exponents}
\begin{figure}[t]
\vskip-2cm
\epsfxsize=5.5truecm{\centerline{\epsfbox{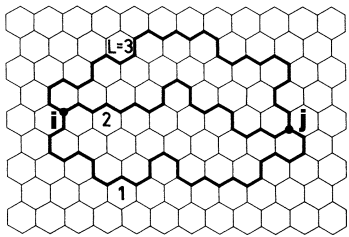}}}
\vskip-2cm
\caption{$L=3$ mutually- and self-avoiding walks on a regular (hexagonal) lattice.}
\label{fig.watermelonP}
\end{figure}

To illustrate the preceding section, let us consider the ``watermelon'' configurations of a set of
 $L$ mutually-avoiding SAW's ${\mathcal P}^{({\ell})}_{{ij}}, \ell=1,\cdots L$, all starting at the same point $i$,
and ending at the same point $j$ (Fig.~\ref{fig.watermelonP}) \cite{S1,duplantier4}. In a way similar to
(\ref{watermelonB}) for RW's, their
 correlator is defined as:
\begin{equation}
\mathcal Z_{{L}}= \sum_{\scriptstyle
{\mathcal P}^{({\ell})}_{{ij}}\atop\scriptstyle {\ell=1,...,L}}
{\mu_{\scriptstyle{\rm SAW}}}^{{-\left| {\mathcal P}\right|}}\propto \left|{i}-{j}\right|^{-2{x}_{{L}}},
\label{watermelonP}
\end{equation}
where the sum extends on all mutually- and self-avoiding configurations,
and where $\mu_{\rm SAW}$ is the effective growth constant of the SAW's on the lattice, associated with the total
polymer length  $\left| {\mathcal P}\right|$. Because of this choice, the correlator decays algebraically with a star exponent
$x_L\equiv x({\cal S}_L)$
corresponding, in the above notations, to the star
\begin{equation}
{\mathcal S}_L=\left( \bigwedge\nolimits_{\ell =1}^{L}{\mathcal P}_{\ell
}\right)
\label{vP}
\end{equation}
 made of $L$ mutually-avoiding polymers.

 A similar boundary watermelon correlator can be defined when points $i$
 and $j$ are both on the Dirichlet boundary \cite{DS2}, which decays with a boundary exponent
 ${\tilde x}_L\equiv \tilde{x}({\cal S}_L)$.
 The values of exponents ${x}_{{L}}$ and ${\tilde x}_L$ have been known since long ago in physics from the Coulomb gas or
 CFT approach \cite{S1,duplantier4,DS2}
 \begin{equation}
{x}_{{L}}=\frac{1}{12}\left(\frac{9}{4}L^2-1\right),\,\,\,\,\,\,\,{\tilde x}_L=\frac{1}{4}L\left(1+\frac{3}{2}L\right).
\label{expoSAW}
 \end{equation}
As we shall see, they provide a direct check of the KPZ relation in the quantum gravity approach \cite{DK}.
\subsection{Quantum Gravity for SAW's \& RW's}
As in section \ref{sec.inter}, the idea is to use the representation where the RW's or
SAW's are on a 2D random lattice, or a random Riemann surface,
i.e., in the presence of 2D quantum gravity  \cite{DK,KPZ}.
\subsubsection*{Example}
\begin{figure}[t]
\vskip-2cm
\epsfxsize=5.5truecm{\centerline{\epsfbox{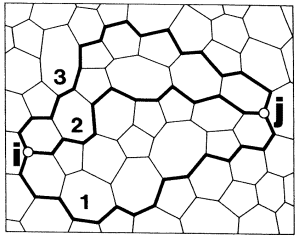}}}
\vskip-2cm
\caption{$L=3$ mutually- and self-avoiding walks on a trivalent random lattice.}
\label{fig.watermelonPQG}
\end{figure}
An example is given by the case of
 $L$ mutually- and self-avoiding walks, in the by now familiar ``watermelon'' configuration (Fig.~\ref{fig.watermelonPQG}).
In complete analogy to the random walk cases (\ref{Zl}) or (\ref{Ztilde}) seen in section \ref{sec.inter}, the quantum gravity partition
function is
 defined as
\begin{equation}
Z_{\scriptstyle{\rm SAW},L}(\beta ,z)=\sum _{{\rm planar}\ G}\frac{1}{ S(G)} e^{-\beta
\left| G\right| }\sum _{i,j\in G} \sum_{\scriptstyle
\Gamma^{(\ell)}_{ij}\atop\scriptstyle \ell=1,...,L} z^{\left| {\Gamma}\right| },
\label{ZlP}
\end{equation}
where the sum extends over all configurations of a set $\Gamma^{(\ell)}_{ij}, \ell=1,\cdots,L$ of $L$
mutually-avoiding SAW's with fugacity $z$ on a random planar lattice $G$ (f
ig. \ref{fig.watermelonPQG}).
A similar boundary partition function
is defined for multiple SAW's traversing a random disk $G$ with boundary $\partial G$
\begin{equation}
\tilde Z_{\scriptstyle{\rm SAW},L}(\beta , {\tilde \beta}, z)=\sum _{ {\rm disk}\ G}
e^{-\beta \left| G\right| }e^{-{\tilde \beta}{\left| \partial
G\right|}} \sum _{{i,j} \in \partial G}\sum_{{\scriptstyle
\Gamma^{(\ell)}_{ij}}\atop{\scriptstyle \ell=1,...,L}} z^{\left|{\Gamma}\right| }.
\label{ZtildeP}
\end{equation}
These partition functions, albeit non-trivial,  can be calculated exactly \cite{DK}.

Each path $\ell \in\{1,\cdots,L\}$ among the multiple SAW's can be represented topologically
by a line, which seperates on $G$ two successive planar domains with the disk topology,
labelled $\ell-1$ and $\ell$ (with the cyclic convention $0\equiv L$ on the sphere). For each polymer line $\ell$,
let us then call
 $m_{\ell}$ the number of edges coming from domain $\ell-1$ and
 $n_{\ell}$ the number of those coming from domain $\ell$, that are incident to line $\ell$.
Each disk-like planar domain $\ell$ has therefore a total number $n_{\ell}+m_{\ell+1}$ of outer edges, with an associated  generating function
(\ref{AGn}), (\ref{AGnint})
\begin{equation}
G_{n_{\ell}+m_{\ell+1}}\left( \beta \right) =\int^{b}_{a}d\lambda_{\ell} \, \rho\left(\lambda_{\ell},\beta \right)
{\lambda_{\ell}}^{n_{\ell}+m_{\ell+1}}.
\label{AGnintl}
\end{equation}
The combinatorial analysis of partition function (\ref{ZlP}) is easily seen to give, up to a coefficient \cite{DK}
\begin{eqnarray}
\nonumber
Z_{\scriptstyle{\rm SAW},L}(\beta ,z)=\sum_{m_{\ell},n_{\ell}=0}^{\infty}z^{\sum_{\ell=1}^{\infty}m_{\ell}+n_{\ell}}
  \prod_{\ell=1}^{L}\left(m_{\ell}\atop {m_{\ell}+n_{\ell}} \right) G_{n_{\ell}+m_{\ell+1}}(\beta),
\end{eqnarray}
where the combination numbers $\left(m_{\ell}\atop {m_{\ell}+n_{\ell}} \right)$ count the number of ways
to place along polymer line $\ell$ the sets of
$m_{\ell}$ and $n_{\ell}$ edges that are incident to that line. Inserting then for each planar domain $\ell$
the integral representation
(\ref{AGnintl}), and using Newton binomial formula for each line $\ell$ we arrive at ($L+1\equiv 1$)
\begin{equation}
Z_{\scriptstyle{\rm SAW},L}(\beta ,z)=\int^{b}_{a}\prod_{\ell=1}^{L}d\lambda_{\ell}\, \rho\left(\lambda_{\ell},\beta \right)
\prod_{\ell=1}^{L} \frac{1}{1-z\left(\lambda_{\ell}+\lambda_{\ell+1}\right)}.
\label{AZPfin}
\end{equation}
The combinatorial analysis of the boundary partition function (\ref{ZtildeP}) gives in a similar way
\begin{eqnarray}
\nonumber
\tilde Z_{\scriptstyle{\rm SAW},L}(\beta ,z)&=&\int^{b}_{a}\prod_{\ell=1}^{L+1}d\lambda_{\ell}\, \rho\left(\lambda_{\ell},\beta \right)
\prod_{\ell=1}^{L} \frac{1}{1-z\left(\lambda_{\ell}+\lambda_{\ell+1}\right)}\\
&\times&\frac{1}{1-{\tilde z}\left(\lambda_{1}\right)}\frac{1}{1-\tilde{z}\left(\lambda_{L+1}\right)},
\label{AZtildePfin}
\end{eqnarray}
where $\tilde z\equiv e^{-\tilde \beta}$, and  where the two last ``propagators'' account
for the presence of the two extra boundary lines.

\subsubsection*{From Trees to SAW's}
\label{remark}
At this stage, it is worth noting that the partition functions (\ref{AZPfin}) and (\ref{AZtildePfin})
for {\it self-avoiding walks} on a random lattice can be recovered in a simple way from the {\it tree} partition functions (\ref{AZlint})
and (\ref{AZtildeint}).

One observes indeed that
it is sufficient to replace in all integral expressions there each tree backbone propagator ${\mathcal T}(x, y)$ (\ref{APropagator}) by
 a SAW propagator
\begin{equation}
\label{calL}
{\cal L}(x,y):=(1-x-y)^{-1}.
\end{equation}
This corresponds to replace each {\it rooted tree} generating function $T(x)$ (\ref{AT}) building up
the propagator ${\mathcal T}(x, y)$,  by its small $x$
expansion, $T(x)=x+\cdots$. The reason is that the latter is the trivial generating function of a {\it rooted edge}. So each tree branching out of
each tree backbone line in Fig.~\ref{FigureA1} is replaced by a simple edge incident to the  backbone, which thus
becomes a simple SAW line.

\subsubsection*{SAW Quantum Gravity Exponents}
 The singular behavior of (\ref{AZPfin}) and (\ref{AZtildePfin})
arises when  the lattice fugacity $e^{-\beta}$, boundary's fugacity ${\tilde z}=e^{-\tilde \beta}$ and
polymer fugacity $z$ reach their
respective critical
values. The singularity analysis can be performed in a way entirely similar to
the analysis of the RWs' quantum gravity partition functions in
section \ref{sec.inter}.
 One uses the
remark
made above that each tree propagator $\cal T$ (\ref{APropagator}), with a square root singularity, is now
replaced by a SAW propagator $\cal L$ (\ref{calL}) with a simple singularity.

The result (\ref{Zll})
for $Z_L$ for trees is then simply replaced by \cite{DK}
\begin{equation}
Z_{{\scriptstyle SAW}, L} \sim \left( \beta -\beta_c\right) ^{3L/4},
\label{AZllP}
\end{equation}
which amounts to the {\it simple formal substitution} $L \to 3/4 \times L$ {\it for passing from RW's to SAW's.}
The rest of the analysis is exactly the same, and the fundamental result (\ref{deltal}) simply becomes
\begin{equation}
2\Delta_{{\scriptstyle SAW}, L}-\gamma _{\rm str}(\chi =2)=\frac{3}{4}L,
\label{AdeltalP}
\end{equation}
whith $\gamma _{\rm str}(\chi =2)=\gamma=-1/2,$ whence
\begin{equation}
\Delta_{{\scriptstyle SAW}, L}=\frac{1}{2}\left(\frac{3}{4}L-\frac{1}{2}\right).
\label{AdeltallP}
\end{equation}
The boundary-bulk relation (\ref{deltat}) remains valid:
\begin{equation}
{\tilde \Delta}_{{\scriptstyle SAW}, L}=2\Delta _{{\scriptstyle SAW}, L}-\gamma _{{\rm str}}(\chi =2),
\label{AdeltatP}
\end{equation}
so that one finds from the bulk conformal weight (\ref{AdeltallP})
\begin{equation}
{\tilde \Delta}_{{\scriptstyle SAW}, L}=\frac{3}{4}L.
\label{Adeltat'P}
\end{equation}
These are the quantum gravity conformal weights of a SAW $L$-star ${\cal S}_L$ \cite{DK}:
\begin{equation}
{\Delta}_{{\scriptstyle SAW},{L}}\equiv{\Delta}({{\cal S}_{L}})=\frac{1}{2}\left(\frac{3}{4}L-\frac{1}{2}\right),\,\,\,\,\,\,\,
{\tilde \Delta}_{{\scriptstyle SAW},L}\equiv {\tilde \Delta}({{\cal S}_{L}})=\frac{3}{4}L.
\label{expoSAWQG}
 \end{equation}
We now give the general formalism which allows the prediction of the complete family of conformal dimensions,
such as (\ref{expoSAWQG}) or (\ref{expoSAW}).
\subsubsection*{Scaling Dimensions, Conformal Weights, and KPZ Map}
Let us first recall that by definition any scaling dimension $x$ in the plane is twice the {\it conformal
weight} $\Delta ^{(0)}$ of the corresponding operator,
while near a boundary they are identical  \cite{BPZ,cardylebowitz}
\begin{equation}
x=2\Delta ^{\left( 0\right) },\quad
\tilde{x}=\tilde{\Delta}^{\left( 0\right) }.  \label{xdelta}
\end{equation}
The general relation (\ref{KPZ}) for Brownian paths depends only on the
central charge $c=0$, which also applies to self-avoiding walks or polymers.  For a critical
system with central charge $c=0$, the two universal functions:
\begin{eqnarray}
U\left( x\right) =U_{\gamma=-\frac{1}{2}}\left( x\right)=\frac{1}{3}x\left( 1+2x\right) , \hskip2mm
V\left( x\right) =\frac{1}{24}\left( 4x^{2}-1\right) ,  \label{U}
\end{eqnarray}
with $V\left( x\right) := U\left(\frac{1}{2}\left(x-\frac{1}{2}\right) \right)$,
generate all the scaling exponents. They transform the conformal weights in bulk quantum gravity,
${\Delta}$, or in boundary QG, $\tilde{\Delta}$, into the plane and half-plane ones (\ref{xdelta}):
\begin{eqnarray}
{\Delta}^{\left( 0\right)}=U({\Delta})=V(\tilde{\Delta}),\;
\tilde{\Delta}^{\left( 0\right)}=U(\tilde{\Delta}).
\label{KPZSAW}
\end{eqnarray}
These relations are for example satisfied by the dimensions (\ref{expoSAW}) and (\ref{expoSAWQG}).
\subsubsection*{Composition Rules}
Consider two  stars $A,B$ joined at their centers, and in a random {\it mutually-avoiding} star-configuration
$ A\wedge B$. Each star is made
of an arbitrary collection of Brownian paths and self-avoiding paths with arbitrary interactions of type (\ref{vw}).
Their respective bulk partition functions (\ref{zr}), (\ref{zrs}), or boundary partition functions
(\ref{zz}) have associated
planar scaling exponents $x\left( A\right)$, $x\left( B\right)$, or boundary exponents $\tilde{x}\left( A\right)$, 
$\tilde{x}\left( B\right)$. The corresponding scaling dimensions in {\it quantum gravity} are then,
for instance for $A$:
\begin{eqnarray}
\label{deltaA}
\tilde {\Delta}\left( A\right)=U^{-1}\left( \tilde{x}\left( A\right)\right),\;\;\;\;
{\Delta}\left( A\right)=U^{-1}\left[\frac{1}{2}{x}\left( A\right)\right],
\end{eqnarray}
where $U^{-1}\left( x\right) $ is the positive inverse of the KPZ map $U$
\begin{equation}
U^{-1}\left( x\right) =\frac{1}{4}\left( \sqrt{24x+1}-1\right) .
\label{u1}
\end{equation}
The key properties are given by the following propositions: \\
$\bullet$ {\it In $c=0$ quantum gravity the  boundary and bulk scaling
dimensions of a given random path set are related by:}
\begin{eqnarray}
\tilde {\Delta}( A)=2{\Delta}\left( A\right) -\gamma_{\rm str}(c=0)=2{\Delta}\left( A\right) +\frac{1}{2}.
\label{tdeltaA=deltaA}
\end{eqnarray}
This generalizes the relation (\ref{deltat}) for non-intersecting Brownian paths.\\
$\bullet$ {\it In quantum gravity the  boundary scaling
dimensions of two mutually-avoiding sets is the sum of their respective boundary scaling
dimensions:}
\begin{eqnarray}
\tilde {\Delta}\left( A\wedge B\right)=\tilde {\Delta}\left( A\right)
+\tilde {\Delta}\left(  B\right).
\label{deltaA+deltaB}
\end{eqnarray}
It generalizes identity (\ref{deltan}) for mutually-avoiding packets of Brownian paths.
The  boundary-bulk relation (\ref{tdeltaA=deltaA}) and the fusion rule (\ref{deltaA+deltaB})
 come from simple convolution  properties of partition functions on a random lattice
 \cite{duplantier7,duplantier8}. They are studied in detail in Ref. \cite{BDMan} (Appendices A \& C).

The planar scaling exponents $x\left( A\wedge B\right) $ in ${\mathbb C}$, and $\tilde{x}\left(
A\wedge B\right)$ in ${\mathbb H}$ of the two mutually-avoiding stars $A\wedge B$
are then given by the KPZ map (\ref{KPZSAW}) applied to Eq.~(\ref{deltaA+deltaB})
\begin{eqnarray}
x\left( A\wedge B\right) &=&2V\left[\tilde {\Delta}\left( A\wedge B\right) \right]
= 2V\left[\tilde {\Delta}\left( A\right)
+\tilde {\Delta}\left(  B\right)\right] \\
\tilde{x}\left( A\wedge B\right) &=&U\left[ \tilde {\Delta}\left( A\wedge B\right) \right]
=U\left[\tilde {\Delta}\left( A\right)
+\tilde {\Delta}\left(  B\right) \right]. \label{xprep}
\end{eqnarray}
Owing to (\ref{deltaA}), these scaling
exponents thus obey the {\it star algebra}
 \cite{duplantier7,duplantier8}
\begin{eqnarray}
x\left( A\wedge B\right) &=&2V\left[ U^{-1}\left( \tilde{x}\left(
A\right)
\right) +U^{-1}\left( \tilde{x}\left( B\right) \right) \right]  \label{x} \\
\tilde{x}\left( A\wedge B\right) &=&U\left[ U^{-1}\left(
\tilde{x}\left( A\right) \right) +U^{-1}\left( \tilde{x}\left(
B\right) \right) \right] . \label{xx}
\end{eqnarray}

These fusion rules (\ref{deltaA+deltaB}), (\ref{x}) and (\ref{xx}), which mix bulk and boundary exponents, are already
apparent in the derivation of non-intersection exponents for Brownian paths given in section \ref{sec.inter}. They also
apply to the $O(N)$ model, as shown in Ref. \cite{BDMan}, 
and are established in all generality in Appendix C there. They can also be seen as recurrence ``cascade'' relations
in ${{\mathbb C}}$ between successive conformal
Riemann maps of the frontiers of mutually-avoiding paths onto the half-plane boundary
 ${\partial {\mathbb H}}$, as in the original work  \cite{lawler2} on Brownian paths.

When the random sets $A$ and $B$ are {\it independent} and can
overlap, their scaling dimensions in the standard plane or half-plane are additive by trivial factorization of
partition functions or probabilities \cite{duplantier8}
\begin{eqnarray}
x\left( A\vee B\right)
=x\left( A\right) +x\left( B\right) ,\;\;\;\;
\tilde{x}\left( A\vee B\right) =
\tilde{x}\left( A\right) +\tilde{x}\left( B\right). \label{add}
\end{eqnarray}
This additivity no longer applies in quantum gravity, since overlapping 
paths get coupled by the fluctuations of the metric, and are no longer independent. In contrast, it is replaced by
 the additivity rule (\ref{deltaA+deltaB}) for mutually-avoiding paths (see Appendix C in Ref. \cite{BDMan} for a thorough discussion
 of this additivity property).

It is clear at this stage that the set of equations above is {\it
complete.} It allows for the calculation of any conformal
dimensions associated with a star
structure ${\mathcal S}$ of the most general type, as in (\ref{vw}),
involving $\left( \wedge ,\vee \right) $ operations
separated by nested{\it \ }parentheses  \cite{duplantier8}. Here follow some examples.

\subsection{RW-SAW Exponents }
The single extremity scaling
dimensions are for a RW or a SAW near a Dirichlet boundary 
$\partial {{\mathbb H}}$  \cite {cardy}\footnote{Hereafter we use a slightly different notation:
$\tilde{x}_{P}\left( 1\right)\equiv \tilde{x}_1$ in (\ref{expoSAW}), and $\tilde{\Delta}_{P}\left( 1\right)\equiv
{\tilde \Delta}_{{\scriptstyle SAW},1}$ in (\ref{expoSAWQG}).}
\begin{equation}
\tilde{x}_{B}\left( 1\right)=\tilde{\Delta}_{B}^{\left( 0\right)
}\left( 1\right)
=1,\;\tilde{x}_{P}\left( 1\right)=\tilde{\Delta}%
_{P}^{\left( 0\right) }\left( 1\right) =%
{{\frac{5}{8}}}%
,  \label{num}
\end{equation}
or in quantum gravity
\begin{equation}
\label{numbis}
\tilde{\Delta}_{B}\left( 1\right) =U^{-1}\left(
1\right) =1,\;\tilde{\Delta}_{P}\left( 1\right) =U^{-1}\left(\frac{5}{8}\right)=\frac{3}{4}.
\end{equation} 
Because of the star algebra described above these are the only
numerical seeds, i.e., generators, we need.

Consider  packets of $n$ copies of transparent RW's or $m$
transparent SAW's. Their boundary conformal dimensions in ${{\mathbb H}}$ are
respectively, by using (\ref{add}) and (\ref{num}),
$\tilde{\Delta}_{B}^{\left( 0\right) }\left( n\right) =n$ and
$\tilde{\Delta}_{P}^{\left( 0\right) }\left( m\right)
=\frac{5}{8}m$.  The inverse mapping to the random
surface yields the quantum gravity conformal weights $\tilde{\Delta}_{B}\left( n\right) =U^{-1}\left(
n\right) $ and $\tilde{\Delta}_{P}\left( m\right) =U^{-1}\left(\frac{5}{8}m\right).$ 
The star made of $L$ packets $\ell \in \left\{1,...,L\right\} $, each of them made of $n_{\ell }$ transparent
RW's and of $m_{\ell }$ transparent SAW's, with the $L$ packets
 mutually-avoiding, has planar scaling dimensions
\begin{eqnarray}
\tilde{\Delta}^{\left( 0\right) }\left\{ n_{\ell
},m_{\ell}\right\} &=&U\left( \tilde{\Delta}\left\{ n_{\ell },m_{\ell}\right\}\right) \\
\Delta
^{\left( 0\right)
}\left\{ n_{\ell },m_{\ell }\right\} &=&V\left( \tilde{\Delta}\left\{ n_{\ell },m_{\ell}\right\}%
\right) , \\
\label{gtDelta}
\tilde{\Delta}\left\{ n_{\ell },m_{\ell}\right\}
&=&\sum\nolimits_{\ell =1}^{L}U^{-1}\left( n_{\ell }+
{ {\frac{5}{8}}}%
m_{\ell }\right)\\
\nonumber
&=&\sum\nolimits_{\ell =1}^{L}\frac{1}{4}\left( \sqrt{24\left( n_{\ell}+ {\frac{5}{8}}m_{\ell}\right)+1}-1\right).
\end{eqnarray}
Take a copolymer star ${\mathcal S}_{L,L^{\prime }}$ made of $L$ RW's and $
L^{\prime }$ SAW's, all mutually-avoiding
$\left( \forall \ell=1,\cdots,L ,n_{\ell }=1, m_{\ell}=0;\;\;\forall \ell
^{\prime }=1,\cdots,L',n_{\ell^{\prime } }=0, m_{\ell ^{\prime }}=1\right)$. In QG the
linear boundary conformal weight (\ref{gtDelta}) is 
$\tilde{\Delta}\left({\mathcal S}_{L,L^{\prime }}\right) =L+\frac{3}{4}L^{\prime }$. By the $U$ and $V$ maps, it gives the 
scaling dimensions in
  ${{\mathbb H}}$ and ${{\mathbb C}}$
\begin{eqnarray*}
\tilde{\Delta}^{\left( 0\right) }\left( {\mathcal S}_{L,L^{\prime }}\right)&=&{\frac {1}{3}}
\left( L+{\frac {3 }{ 4}}L^{\prime }\right) \left( 1+2L+{\frac {3 }{ 2}} L^{\prime }\right)  \\
\Delta ^{\left( 0\right)}\left( {\mathcal S}_{L,L^{\prime }}\right)
&=&{\frac {1 }{ 24}}\left[ 4\left( L+{\frac {3 }{ 4}} L^{\prime }\right) ^{2}-1\right],
\end{eqnarray*}
recovering for $L=0$ the SAW star-exponents (\ref{expoSAW}) given above,  and for $L^{\prime}=0$ the RW non-intersection exponents
in ${{\mathbb H}}$ and ${{\mathbb C}}$ obtained in section \ref{sec.inter}
 \begin{eqnarray*}
2\tilde \zeta_L&=&\tilde{\Delta}^{\left( 0\right) }\left( {\mathcal S}_{L,L^{\prime }=0}\right)={\frac {1}{3}}
L \left( 1+2L\right)  \\
\zeta_L&=&\Delta ^{\left( 0\right)}\left( {\mathcal S}_{L,L^{\prime }=0}\right)
={\frac {1 }{ 24}}\left( 4 L ^{2}-1\right).
\end{eqnarray*}

Formula  (\ref{gtDelta}) encompasses all exponents previously known separately for RW's and SAW's
 \cite{duplantier4,DS2,duplantier2}. We arrive from it at a
striking {\it scaling equivalence: When overlapping with other paths in the standard plane, a self-avoiding walk is exactly
equivalent to $5/8$ of a random walk} \cite{duplantier8}.
Similar results were
later obtained in probability theory, based on the general
structure of ``completely conformally-invariant processes'', which correspond
  to  $c=0$ central charge conformal field theories  \cite {lawler3,lawler4}. Note that the construction of the scaling limit of SAW's still eludes a rigorous approach, although it is
 predicted to correspond to ``stochastic L\"owner evolution'' ${\rm SLE}_{\kappa}$ with
 $\kappa=8/3$, equivalent to a Coulomb gas with $g=4/\kappa=3/2$ (see  section \ref{sec.geodual} below).\\

From the point of view of {\it mutual-avoidance}, a ``transmutation'' formula  between SAW's and RW's
is obtained directly from
 the quantum gravity boundary additivity rule (\ref{deltaA+deltaB}) and the values (\ref{numbis}):
 {\it For mutual-avoidance, in quantum gravity, a self-avoiding walk is equivalent to $3/4$ of a random walk}. We shall now apply these rules to the
 determination of ``shadow'' or ``hiding'' exponents \cite{WW}.
\subsection{Brownian Hiding Exponents}
\begin{figure}[htb]
\begin{center}
\includegraphics[angle=0,width=.7\linewidth]{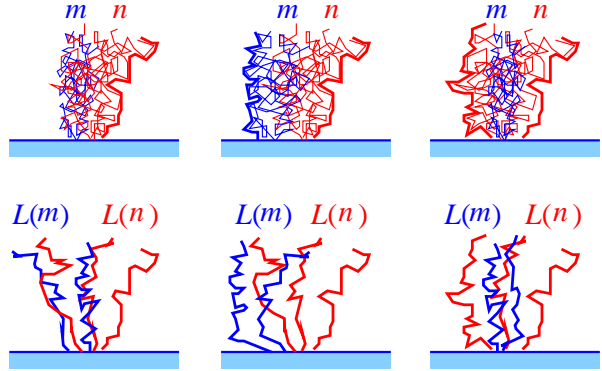}
\caption{Top: Two packets made of $m$ and $n$ independent Brownian paths, with three possible constraints for their
Brownian frontiers:
The right frontier is made only of paths of the $n$-packet; the left and right frontiers are
made exclusively of paths from the
$m$-packet and  $n$-packet, respectively; both Brownian frontiers are made exclusively from paths of the $n$-packet, i.e.,
the second $m$-packet is hidden by the former. Bottom: The conversion into equivalent problems for
two sets of multiple SAW's, made separately of $L(m)$ and $L(n)$ mutually-avoiding SAW's.}
\label{fig.sle4}
\end{center}
\end{figure}
Consider two packets made of $m$ and $n$ independent Brownian paths (or random walks)
diffusing in the half-plane away from the Dirichlet boundary $ \partial {\mathbb H}$, as
represented in figure \ref{fig.sle4}. Their left or right Brownian frontiers are selectively made of certain paths.

For instance,
 one can ask the following question, corresponding to the top left case in Fig.~\ref{fig.sle4}: What is the
  probability that the paths altogether diffuse up to a distance $R$ without the paths of the $m$ packet contributing to
  the Brownian frontier to the right? In other words, the $m$-packet stays in the left shadow of the other packet, i.e.,
  it is {\it hidden} from the outside to the right by the presence of this other packet.

  This probability decays with distance $R$ as a power law
  $$ P \approx R^{{-\tilde x}_{{m},{n}}},$$
  where ${\tilde x}_{{m},{n}}$ can be called a {\it shadow} or {\it hiding} exponent \cite{WW}.

By using quantum gravity, the exponent can be calculated immediately as the nested formula
$$
{\tilde x}_{{m},{n}}
={{U}}\,\left[{\frac{3}{4}}+{U^{-1}}\left[{m}+
{{U}}\left({U^{-1}}({n})-\frac{3}{4}\right)\right]\right].
$$
Let us explain briefly how this formula originates from the transmutation of Brownian
  paths into self-avoiding walks.

  First we transform separately each Brownian $m$ or $n$-packet
  into a packet of $L(m)$ or $L(n)$
  {\it mutually-avoiding}  SAW's
  (see figure \ref{fig.sle4}, bottom left). According to the quantum gravity theory established in the preceding section,
  one must have the exact equivalence of their quantum gravity boundary dimensions, namely:
  $$\frac{3}{4}L(n)=U^{-1}(n).$$
  Then one discards from the $L(n)$ SAW set its rightmost SAW,
  which will represent the right {\it frontier} of the original Brownian $n$-packet, since
   a Brownian frontier is a self-avoiding walk
(in the scaling limit). The resulting new set of $L(n)-1$ SAW's
  is now free to {\it overlap} with the other Brownian $m$-packet,
  so their boundary dimensions in the {\it standard} half-plane, $m$ and $U[\frac{3}{4}(L(n)-1)]$, do {\it add}. To finish,
   the rightmost SAW left aside should not intersect any other
   path. This corresponds {\it in QG} to an {\it additive} boundary dimension, equal to
   $\frac{3}{4}+U^{-1}[m+ U[\frac{3}{4}(L(n)-1)]]$.
   The latter is in turn transformed
  into a standard boundary exponent by a last application of KPZ map $U$, hence the formula above, {\bf QED}.

An explicit calculation then gives
$$
{\tilde x}_{{m},{n}}={m}+{n}+\frac{1}{4}\sqrt{24{m}+\left(\sqrt{1+24{n}}
-3\right)^2}
-\frac{1}{4}
\left(\sqrt{1+24{n}}-3\right),
$$
where the first term ${m}+{n}$ of course corresponds  to the simple boundary exponent of independent
Brownian paths,
while the two extra terms reflect the hidding constraint and cancel for $m=0$, as it must.

The other cases in Fig.~\ref{fig.sle4} can be treated in the same way and are left as exercizes.

\section{\sc{Percolation Clusters}}
\label{sec.perco}

\subsection{Cluster Hull and External Perimeter}
\begin{figure}[htbp]
\begin{center}
\includegraphics[angle=0,width=.6\linewidth]{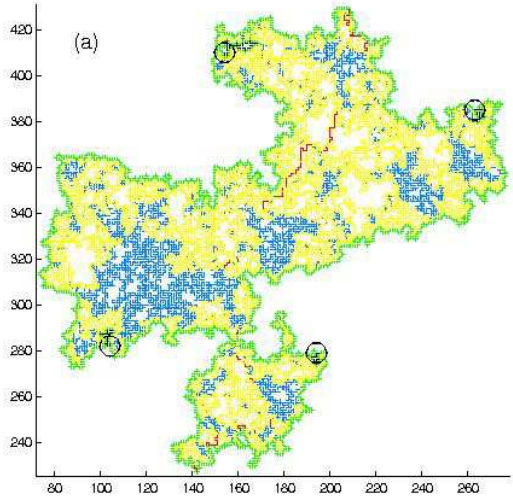}
\end{center}
\caption{A large percolation cluster, and its various scaling domains (Courtesy of J. Asikainen {\it et al.} \cite{aharony2}).}
\label{fig.ais1a}
\end{figure}

Let us consider, for definiteness, site percolation on the
2D triangular lattice. By universality, the results are expected to apply
to other 2D (e.g., bond) percolation models in the scaling limit. Consider then
a very large two-dimensional incipient cluster
${\mathcal C}$, at the percolation threshold $p_{c}=1/2$.
Figure \ref{fig.ais1a} depicts such a connected cluster.
\begin{figure}[htb]
\begin{center}
\includegraphics[angle=0,width=.6\linewidth]{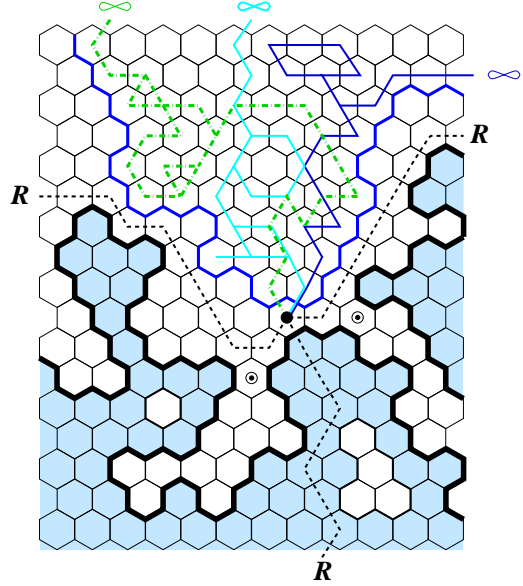}
\end{center}
\caption{An accessible site $(\bullet )$ on the external
perimeter for site percolation on the triangular lattice. It is
defined by the existence, in the {\it scaling limit}, of three non-intersecting,
and connected paths ${\mathcal S}_{3}$ (dotted
lines), one on the incipient cluster, the other two on
the dual empty sites. The entrances of fjords $\odot$ close in the scaling limit.
Point $(\bullet )$ is first reached by three
independent RW's, contributing to ${H}^3 (\bullet
)$. The hull of the incipient cluster (thick continuous line along hexagons) avoids the outer
frontier of the RW's (continuous line along hexagons). A Riemann map of the
latter onto the real line ${\partial {\mathbb H}}$ reveals the presence of an underlying
$\ell=3$ path-crossing {\it boundary}
operator, i.e,
a two-cluster boundary operator,
with dimension in the half-plane
$\tilde{x}_{\ell=3 }
=\tilde
{x}^{{\mathcal C}}_{k=2}=2.$ Both accessible hull and Brownian
paths have a frontier dimension $\frac{4}{3}$.}
\label{Figure4}
\end{figure}

\subsubsection*{Hull}
The boundary lines of a site
percolation cluster, i.e., of connected sets of occupied hexagons, form
random lines on the
dual hexagonal lattice (Fig.~\ref{Figure4}). (They are actually known to obey the statistics of random loops in the
${ O}\left( N=1\right)$ model,
where $N$ is the loop fugacity, in the so-called ``low-temperature phase'', or
of boundaries of Fortuin-Kasteleyn clusters in the $Q=1$ Potts model  \cite{SD}.)
Each critical connected cluster
thus possesses an external closed boundary, its {\it hull}, the fractal dimension of which is known to be
$D_{\rm H}=7/4$  \cite{SD}. (See also \cite{SRG}.)

In the scaling limit, however, the hull, which possesses many pairs of points at relative distances given by a
finite number of lattice meshes $a$, coils onto
itself to become a non-simple curve  \cite{GA}; it thus
develops a smoother outer (accessible) frontier ${\mathcal F}({\mathcal C})$ or {\it external perimeter} (EP).

\subsubsection*{External Perimeter and Crossing Paths}
The geometrical nature of this
external perimeter has recently been elucidated and its Hausdorff
dimension found to equal $D_{\rm EP}=4/3$  \cite{ADA}. For a site $w=\left( \bullet \right)$ to belong to the {\it
accessible} part of the hull, it must remain, in the {\it continuous
scaling limit},  the source of at least {\it three
non-intersecting crossing paths} , noted ${\mathcal S}_{3}=
{\mathcal P} \wedge {\bar {\mathcal P_1}} \wedge {\bar {\mathcal P_2}}$,
reaching to a (large) distance $R$ (Fig.~\ref{Figure4}). (Recall the notation
$A\wedge B$ for two sets, $A$, $B$, of random
paths, required to
be {\it mutually non-intersecting, }and{\it \ }$A\vee B$ for two {\it independent}, thus
possibly
intersecting, sets.) Each of these  paths is {``\it
monochromatic''}: one path  $\mathcal P$ runs only through occupied sites, which simply means that
$w$ belongs to a particular connected cluster; the other two
{\it dual} lines $ {\bar {\mathcal P}_{i=1,2}}$
run through  empty sites, and doubly
connect the external perimeter site $w$ to
``infinity'' in open space  \cite{ADA}.
The definition of the standard hull requires only the
origination, in the scaling limit, of a {\it ``bichromatic''} pair of lines
${\mathcal S}_2={\mathcal P} \wedge {\bar  {\mathcal P}}$, with one path running on occupied sites, and the dual one on empty ones.
Such hull points lacking a second dual line will not necessarily remain accessible from the outside
after the scaling limit is taken, because
their single exit path becomes a strait pinched by parts of the occupied cluster.
In the scaling limit,
the hull is thus a self-coiling and conformally-invariant (CI) scaling curve
which is not simple, while the external perimeter is a simple CI scaling curve.

The (bichromatic) set ${\mathcal S}_{3 }$ of three
non-intersecting connected paths in the percolation system is governed by a new critical exponent
$x\left( {\mathcal
S}_{3 } \right)(=2/3)$ such that $D_{\rm EP}=2-x\left( {\mathcal
S}_{3 }\right) $, while a  bichromatic pair of non-intersecting paths ${\mathcal
S}_2$ has an exponent $x\left( {\mathcal
S}_{2 }\right)(=1/4) $ such that $D_{\rm H}=2-x\left( {\mathcal
S}_{2}\right) $ (see below).

\subsection{Harmonic Measure of Percolation Frontiers}
Define $H\left( w,a\right):={H}\left({\mathcal F} \cap {B}(w,
a)\right) $
as the probability that a random walker,
launched from infinity, {\it first} hits the outer (accessible)
percolation hull's frontier or external perimeter ${\mathcal F}({\mathcal C})$ in the ball $B(w,a)$ centered at point
$w \in {\mathcal F}({\mathcal C})$. The moments $H^n$ of $H$ are
averaged over all realizations of RW's and ${\mathcal C}$
\begin{equation}
{\mathcal Z}_{n}=\left\langle \sum\limits_{w\in {\mathcal
F}/a}{H}^{n}\left({\mathcal F} \cap {B}(w,
a)\right) \right\rangle . \label{Zpe}
\end{equation}
For very large clusters ${\mathcal C}$ and
frontiers ${\mathcal F}\left( {\mathcal C}\right) $ of average
size $R,$ one expects  these moments to scale as: ${\mathcal Z}_{n}\approx \left( a/R\right) ^{\tau
\left( n\right) }$.

By the very definition of the $H$-measure, $n$ independent RW's diffusing away or towards
a neighborhood of a EP point $w$, give a geometric representation of the $n^{th}$
moment $H^{n}(w),$ for $n$ {\it integer}. The values so derived for $n\in
{\mathbb N}$ will be enough, by convexity arguments, to obtain the analytic
continuation for arbitrary $n$'s. Figure \ref{Figure4} depicts such $n$ independent random walks,
in a bunch, {\it
first} hitting the external frontier of a percolation cluster at a
site $w=\left( \bullet \right).$ The packet of independent RW's avoids
the occupied cluster, and defines its own envelope as a set of two
boundary lines separating it from the occupied part of the
lattice. The $n$ independent RW's, or Brownian
paths ${\mathcal B}$ in the scaling limit, in a bunch denoted $\left( \vee
{\mathcal B}\right) ^{n},$ thus  avoid the set ${\mathcal S}_{3 }$ of three {\it
non-intersecting} connected paths in the percolation system, and
this system is governed by a new family of critical exponents
 $x\left( {\mathcal
S}_{3 }\wedge n\right) $ depending on $n.$
The main lines of the derivation of the latter exponents by generalized conformal invariance are as follows.

\subsection{Harmonic and Path Crossing Exponents}

\subsubsection*{Generalized Harmonic Crossing Exponents}
The $n$ independent Brownian paths ${\mathcal B}$, in a bunch
 $\left(
\vee {\mathcal B}\right) ^{n},$ avoid a set ${\mathcal S}_{\ell }:= \left(
\wedge
{\mathcal P}\right) ^{\ell }$ of $\ell $  non-intersecting
crossing paths in
the
percolation system. The latter originate from the same hull site, and each
passes only through occupied sites, or only through empty ({\it dual}) ones
 \cite{ADA}.
The probability that the Brownian and percolation paths altogether traverse the
annulus ${\mathcal %
D}\left( a, R\right) $ from the inner boundary circle of radius $a$ to the outer
one at distance $R$, i.e., are in a ``star'' configuration ${\mathcal S}_{\ell
}\wedge
\left( \vee {\mathcal B}\right) ^{n}$, is expected to scale for $%
a/R\rightarrow 0 $ as
\begin{equation}
{\mathcal P}_{R}\left( {\mathcal S}_{\ell }\wedge n\right) \approx
\left( a/R\right) ^{x\left( {\mathcal S}_{\ell }\wedge n\right) },  \label{xppe}
\end{equation}
where we used ${\mathcal S}_{\ell }\wedge n = {\mathcal S}_{\ell }\wedge \left(
\vee
{\mathcal B}\right) ^{n}$ as a short hand notation, and where $x\left( {\mathcal
S}_{\ell
}\wedge n\right) $ is a new critical exponent
depending on $\ell $ and $n$.  It is convenient to introduce similar boundary
probabilities $\tilde{{\mathcal P}}_{R}\left( {\mathcal S}_{\ell }\wedge
n\right) \approx \left( a/R\right) ^{\tilde{x}\left( {\mathcal S}_{\ell }\wedge
n\right) }$ for the same star configuration of paths, now crossing through the
half-annulus $\tilde{{\mathcal D}}\left( a, R\right) $ in the half-plane ${\mathbb H}$.

\subsubsection*{Bichromatic Path Crossing Exponents}
For $n=0$, the probability ${\mathcal P}_{R}\left( {\mathcal S}_{\ell }\right)={\mathcal P}_{R}\left(
{\mathcal S}_{\ell}\wedge 0\right)\approx \left( a/R\right) ^{ x_{\ell}}$
[resp. $\tilde{{\mathcal P}}_{R}\left( {\mathcal S}_{\ell }\right)=\tilde{{\mathcal P}}_{R}\left(
{\mathcal S}_{\ell}\wedge 0\right)\approx \left( a/R\right) ^{\tilde x_{\ell}}$] is the probability
of having $\ell $ simultaneous non-intersec\-ting
path-crossings of the
annulus ${\mathcal %
D}\left( a, R\right) $ in the plane ${\mathbb C}$ [resp. half-plane ${\mathbb H}$], with associated exponents $x_{\ell
}:= x\left( {\mathcal S}_{\ell } \wedge
0\right) $ [resp.  $\tilde{x}_{\ell }:= \tilde{x}\left( {\mathcal S}_{\ell } \wedge
0\right)$]. Since these exponents are obtained from the limit $n\to 0$ of the harmonic measure
exponents, at least two paths run on occupied sites or empty sites, and these are
the {\it bichromatic} path crossing exponents \cite{ADA}. The {\it monochromatic} ones are different
in the bulk  \cite{ADA,JK}.

\subsection{Quantum Gravity for Percolation}
\subsubsection*{$c=0$ KPZ mapping}
Critical percolation is described
by a conformal field theory with the same vanishing central charge $c=0$ as RW's or SAW's
(see, e.g.,  \cite{cardylebowitz,cardyjapon}). Using again the fundamental mapping of this
conformal field theory (CFT) in the {\it
plane} $%
{{\mathbb C}}$, to the
CFT on a fluctuating random Riemann surface, i.e., in presence of {\it
quantum gravity}  \cite{KPZ},  the two
universal functions $U$ and $V$ only depend on the central charge $c$ of
the CFT, and  are the same as for RW's, and SAW's:
\begin{eqnarray}
U\left( x\right)=\frac{1}{3}x\left( 1+2x\right) , \hskip2mm V\left( x\right)
=U\left[\frac{1}{2}\left(x-\frac{1}{2}\right)\right]=\frac{1}{24}\left( 4x^{2}-1\right),  \label{Upe}
\end{eqnarray}
They suffice to generate all geometrical exponents involving
{\it mutual-avoidance} of random {\it star-shaped} sets of paths of the critical percolation
system. Consider  two arbitrary\ random sets $A,B,$ involving each
a collection of paths in a star configuration, with proper scaling crossing
exponents $x\left( A\right) ,x\left( B\right) ,$ or, in the half-plane, crossing
exponents $\tilde{x}\left( A\right) ,\tilde{x}\left(
B\right) .$ If one fuses the star centers and requires $A$ and $B$ to stay
mutually-avoiding, then the new crossing exponents, $x\left( A\wedge
B\right) $ and $\tilde{x}\left( A\wedge B\right) ,$ obey the same {\it star
fusion algebra} as in (\ref{x}) \cite{duplantier7,duplantier8}
\begin{eqnarray}
x\left( A\wedge B\right) &=&2V\left[ U^{-1}\left( \tilde{x}\left( A\right)
\right) +U^{-1}\left( \tilde{x}\left( B\right) \right) \right]  \nonumber \\
\tilde{x}\left( A\wedge B\right) &=&U\left[ U^{-1}\left( \tilde{x}\left(
A\right) \right) +U^{-1}\left( \tilde{x}\left( B\right) \right) \right] ,
\label{xpe}
\end{eqnarray}
where
$U^{-1}\left( x\right) $ is the inverse function
\begin{equation}
U^{-1}\left( x\right) =\frac{1}{4}\left( \sqrt{24x+1}-1\right) .
\label{u1pe}
\end{equation}

This structure immediately gives both the percolation crossing exponents
$x_{\ell}, \tilde {x}_{\ell}$  \cite{ADA}, and the harmonic crossing
exponents
$x\left( {\mathcal S}_{\ell }\wedge n\right) $ (\ref{xppe}).
\subsubsection*{Path Crossing Exponents}
First, for a
set ${\mathcal S}_{\ell }=\left( \wedge {\mathcal P}\right) ^{\ell }$ of $\ell
$ crossing paths, we have from the recurrent use of (\ref{xpe})
\begin{equation}
x_{\ell }=2V\left[ \ell\, U^{-1}\left( \tilde{x}_{1}\right) \right] ,\quad
\tilde{x}_{\ell }=U\left[ \ell\, U^{-1}\left( \tilde{x}_{1}\right) \right] .
\label{xl}
\end{equation}
For percolation, two values of half-plane crossing exponents $\tilde{x}_{\ell }$
are known by
{\it elementary} means: $\tilde{x}_{2}=1,\tilde{x}_{3}=2$  \cite{ai1,ADA}.
From (\ref{xl}) we thus find $U^{-1}\left( \tilde{x}_{1}\right)
=\frac{1}{2}U^{-1}\left(
\tilde{x}_{2}\right) =\frac{1}{3}U^{-1}\left( \tilde{x}_{3}\right) =\frac{1}{
2},$ (thus $ \tilde{x}_{1}=\frac{1}{3}$  \cite{cardy}), which in turn gives
\[
x_{\ell }=2V\left({\frac{1}{ 2}}\ell\right) =\frac{1}{12}\left(
{\ell}^{2}-1\right),
\tilde{x}_{\ell }=U\left({\frac{1}{ 2}} \ell\right)=\frac{\ell }{6}\left( \ell +1\right).
\]
We thus recover the identity  \cite{ADA} $x_{\ell }=x_{L=\ell }^{{O}\left(
N=1\right) }, \tilde{x}_{\ell }=\tilde{x}_{L=\ell +1}^{{O}\left( N=1\right)
}$ with the $L$-line exponents of the associated ${O}\left( N=1\right)$ loop
model, in the ``low-temperature phase''.
For $L$ {\it even}, these exponents also govern the existence of
$k=\frac{1}{2}L$
{\it spanning} clusters, with the identity $x_{k}^{\mathcal C}=x_{\ell =2k}=%
\frac{1}{12}\left( 4k^{2}-1\right) $ in the plane, and $\tilde{x}_{k}^{\mathcal C}=%
\tilde{x}_{\ell =2k-1}=\frac{1}{3}k\left( 2k-1\right) $ in the half-plane
 \cite{SD,D6,D7}.

\subsubsection*{Brownian Non-Intersection Exponents}
The non-intersection exponents (\ref{Zeta}) and (\ref{zC2})
of $L$ Brownian paths seen in section \ref{sec.inter} are identical to the percolation
path crossing exponents for
\begin{eqnarray}
2\zeta_L=x_{\ell},\;\; 2\tilde \zeta_L=\tilde{x}_{\ell},\;\;\;{\ell}=2L,
\end{eqnarray}
so we obtain a {\it complete scaling equivalence
between
a Brownian path
and {\it two} percolating crossing paths, in both the plane and half-plane}  \cite{duplantier9}.

\subsubsection*{Harmonic Crossing Exponents}
Finally, for the harmonic crossing exponents in (\ref{xppe}), we fuse the two objects
${\mathcal
S}_{\ell}$ and $\left( \vee {\mathcal B}\right) ^{n}$ into a new star ${\mathcal
S}_{\ell}\wedge n $, and use (\ref{xpe}). We just have seen that the
boundary $\ell$-crossing exponent of ${\mathcal S}_{\ell}$, $\tilde{x}_{\ell}$, obeys
$U^{-1}\left( \tilde{x}_{\ell}\right) =\frac{1}{2}\ell.$ The bunch of $n$
independent Brownian paths have their own  half-plane crossing exponent
$\tilde{x}\left( \left( \vee {\mathcal B}\right) ^{n}\right) =n\tilde{x}%
\left( {\mathcal B}\right) =n$ as above. Thus we obtain
\begin{equation}
x\left( {\mathcal S}_{\ell}\wedge n\right) =2V\left(\frac{1}{2}\ell
+U^{-1}\left( n\right) \right).  \label{fina}
\end{equation}
Specializing to the case $\ell=3$ finally gives from (\ref{Upe}-\ref{u1pe})
\[
x\left( {\mathcal S}_{3}\wedge n\right) =2+\frac{1}{2}\left( n-1\right) +\frac{5%
}{24}\left( \sqrt{24n+1}-5\right).
\]

\subsection{Multifractality of Percolation Clusters}
\subsubsection*{Multifractal Dimensions and Spectrum}
In terms of probability (\ref{xppe}), the harmonic measure moments (\ref{Zpe})
scale simply as ${\mathcal Z}_{n}\approx R^2{\mathcal P}_{R}\left( {\mathcal S}_{\ell =3}\wedge n\right)$
 \cite{cates},
which leads to
\begin{equation}
\tau \left( n\right) =x\left( {\mathcal S}_{3}\wedge n\right) -2.  \label{tt}
\end{equation}
Thus
\begin{equation}
\tau \left( n\right) =\frac{1}{2}\left( n-1\right) +\frac{5%
}{24}\left( \sqrt{24n+1}-5\right)
\label{taunperc}
\end{equation}
and the generalized dimensions $D\left( n\right)$
are:
\begin{equation}
D\left( n\right) =\frac{1}{n-1}\tau \left( n\right)=\frac{1}{2}+\frac{5}{\sqrt{24n+1}+5},\quad n\in \left[ -%
{\textstyle{\frac{1}{  24}}}%
,+\infty \right) ,  \label{dn}
\end{equation}
valid for all values of moment order $n,n\geq -\frac{1}{24}.$
We shall see in  section \ref{sec.conform} that these exponents $\tau(n)$  \cite{duplantier9} are {\it identical} to those
 obtained  for Brownian paths and self-avoiding walks.

\subsubsection*{Comparison to Numerical Results}
\begin{figure}[tb]
\begin{center}
\includegraphics[angle=0,width=.7\linewidth]{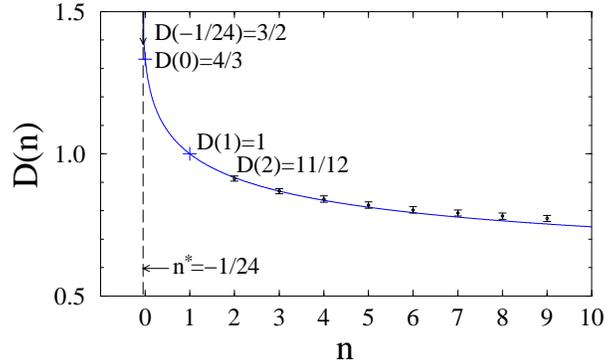}
\end{center}
\caption{Universal generalized dimensions $D(n)$ as a function of $n$, corresponding
to the harmonic measure near a percolation cluster, or to self-avoiding or
random walks, and comparison with 
the numerical data obtained by Meakin et al. (1988) for percolation.}
\label{Figure5}
\end{figure}

Only in the case of percolation has the harmonic measure been
systematically studied numerically, by Meakin et al.
 \cite{meakin}. We show in Figure \ref{Figure5} the exact curve $D\left(
n\right) $ (\ref{dn})
  \cite{duplantier9}, together with the
numerical results for $n\in \{2,...,9\} $  \cite{meakin}, showing
fairly good agreement.

Define now ${\mathcal N}\left(
H\right)$ as the number of boundary sites having a given
probability $H$ to be hit by a RW starting at infinity; the multifractal
formalism  yields, for $H\rightarrow 0,$ a power law behavior
\begin{equation}
{\mathcal N}\left( H\right)|_{H\rightarrow 0}\approx
H^{-(1+{n}^{\ast})}, \label{nh}
\end{equation}
with an exponent given by the lowest possible value of $n$, ${n^{\ast}=-1/24}$, where $D(n)$ reaches its maximal
value: $D(n^{\ast})=\frac{3}{2}$ (see section \ref{sec.conform}).

The average number ${\mathcal N}(H)$ (\ref{nh}) has been also
determined numerically for percolation clusters in Ref. \cite{MS}, and our prediction
$1+{n}^{\ast}=\frac{23}{24}=0.95833...$ compares very well with the numerical result
$0.951\pm 0.030$, obtained for $10^{-5}\leq H\leq 10^{-4}$.

The dimension of the measure's support is $D\left(
0\right)=\frac{4}{3} \neq D_{{\rm H}},$ where $D_{{\rm %
H}}=\frac{7}{4}$ is the Hausdorff dimension of the standard hull,
i.e., the complete outer boundary of critical percolating clusters
 \cite{SD}. The value $D_{{\rm EP}}=D(0)=\frac{4}{3}$ gives the dimension
of the {\it accessible external perimeter}. A direct derivation of
its exact value has ben first given by Aizenman {\it et al.} \cite{ADA}. The complement of the
accessible perimeter in the hull is made of deep fjords, which do
close in the scaling limit and are not probed by the harmonic
measure. This is in agreement with the instability phenomenon
observed on a lattice by Grossman-Aharony for the hull dimension  \cite{GA}.

A striking
fact is the complete identity of the multifractal dimensions  for
percolation with those for random walks and
self-avoiding walks, as we shall see in the next section.
 Seen from outside, these three scaling
curves are not distinguished by the harmonic measure. In fact they
are the same, and one of the main conclusions is
that {\it the external frontiers of a planar Brownian motion, or
of a critical percolation cluster are, in the scaling limit, identical to a critical
self-avoiding walk, with Hausdorff dimension $D=\frac{4}{3}$} \cite{duplantier8,duplantier9}.
In the same way,
the connected domain enclosed by a Brownian loop or by the frontier of a percolation cluster are the same
as the domain inside a closed SAW. (See also \cite{WW4}).

As we have seen, this fact is linked to the presence of a single
universal conformal field theory (with a vanishing central charge
$c=0$), and to the underlying presence of quantum gravity, which
organizes the associated conformal dimensions. S. Smirnov  \cite{smirnov1} proved that critical site
percolation on the triangular lattice has a conformally-invariant
scaling limit, and that the discrete cluster interfaces (hulls)
converge to the same stochastic L\"owner evolution process (${\rm SLE}_6$) as the
one involved for Brownian paths. This opened the way to a rigorous
derivation of percolation exponents  \cite{lawler6,smirnov2},
previously derived in the physics literature \cite{dennijs,nien,cardy}. V. Beffara  has thus been able to
derive
rigorously the values of percolation Hausdorff dimensions $D_{{\rm H}}$ \cite{beffara2}
and $D_{{\rm EP}}$ \cite{lawler5, beffara2}, already exactly known in physics \cite{SD,ADA}.
\subsubsection*{Double Layer Impedance}
Let us finally consider the different, but related, problem of the {\it
double layer impedance}
of a {\it rough} electrode. In some range of frequencies $\omega $, the
impedance contains an anomalous ``constant
phase angle'' (CPA) term $\left( i\omega
\right) ^{-\beta }$, where $\beta <1$.
From a
natural
RW representation of the impedance, a scaling law was 
proposed by Halsey and Leibig:
$\beta =\frac{D\left( 2\right) }{D\left( 0\right) }$
\label{beta}
(here in 2D), where $D\left( 2\right) $ and $D\left( 0\right)$ are the multifractal dimensions of the
$H$-measure on the rough electrode \cite{halsey7}. In the case of a
2D porous percolative electrode, our
results (\ref{dn}) give $D\left( 2\right)
= \frac{11}{12},$ $D\left( 0\right)=\frac{4%
}{3}$, whence $\beta =\frac{11}{16}=0.6875.$ This
compares very well with a numerical RW algorithm result \cite{MS}, which yields
an
effective CPA exponent $\beta \simeq 0.69,$
nicely vindicating the multifractal description \cite{halsey7}.\footnote{For a recent elaboration on the theory of 
Ref.~\cite{halsey7}, see also \cite{GFS}.}

\bigskip
\noindent
----------------------------------------------------

In the next sections, we consider arbitrary conformally-invariant curves and present a universal description of
multifractal functions for them. They are derived
from conformal field theory and
quantum gravity. The geometrical findings are described in
detail, including the cases of Brownian paths, self-avoiding walks, Ising clusters, and $Q=4$ Potts
Fortuin-Kasteleyn clusters, which are of particular interest.  We also make explicit the relation between
 a conformally-invariant scaling curve with CFT central charge $c$  \cite{duplantier11}, and the
 stochastic L\"owner process ${\rm SLE}_{\kappa}$  \cite{schramm1}. A fundamental geometric duality property
 for the external boundaries in $O(N)$ and Potts models, and SLE is obtained.
 
\section{\sc{Conformally Invariant Frontiers and Quantum Gravity}}
\label{sec.conform}

\subsection{Harmonic Measure and Potential near a Fractal Frontier}
\subsubsection*{Introduction}
The {\it harmonic measure}, i.e., the diffusion or electrostatic potential
field near an equipotential fractal boundary \cite{BBE}, or,
equivalently, the electric charge appearing on the frontier of a
perfectly conducting fractal, possesses a self-similarity
property, which is reflected in a {\it multifractal}
behavior. Cates and Witten  \cite{catesetwitten} considered the
case of the Laplacian diffusion field near a simple random walk,
or near a self-avoiding walk, using renomalization group arguments near $d=4$ dimensions. The associated exponents can be
recast as those of star copolymers made of a bunch of independent
RW's diffusing away from a generic point of the absorber, similar to those introduced in section \ref{sec.mixing}.

For a
Brownian path, the very existence of a harmonic multifractal spectrum
has been first rigorously established in Ref.~\cite{lawler97}. The
exact solution to this problem in two dimensions was given in Ref. \cite{duplantier8}. From a mathematical point of view, it could in principle
 be derived from the results of refs.
 \cite{lawler2,lawler3,lawler4,lawler5} taken altogether. Here we consider the general case of a conformally invariant
 scaling curve, using QG \cite{duplantier11}, while a rigorous approach is also possible \cite{IABD, Stockholm}.

\subsubsection*{Harmonic Measure}
Consider a two-dimensional very large ``absorber'',
a conformally-invariant critical random cluster,
hereafter generically called ${\mathcal C}$. It can be for instance a percolation cluster, a random walk, a SAW, a Fortuin-Kasteleyn
cluster in the Potts model, etc.
(The figures illustrate the case of a random walk or Brownian path.)

One  defines the harmonic measure $H\left( w\right) $ as the
probability that a random walker launched from infinity, {\it
first} hits the outer ``hull's frontier'' or accessible frontier $\mathcal F:= \partial
\mathcal C$ of $\mathcal C$ at point $w \in \partial \mathcal C$.
For a given point $w \in \partial
\mathcal C$, let $B(w,r)$ be the ball (i.e., disk)
of radius $r$ centered at $w$. Then $H({\partial\mathcal C} \cap B(w,r))$ is  the total
harmonic measure of the points of the frontier inside the ball
$B(w,r)$.

\subsubsection*{Potential Theory}
\begin{figure}[tb]
\begin{center}
\includegraphics[angle=0,width=.45\linewidth]{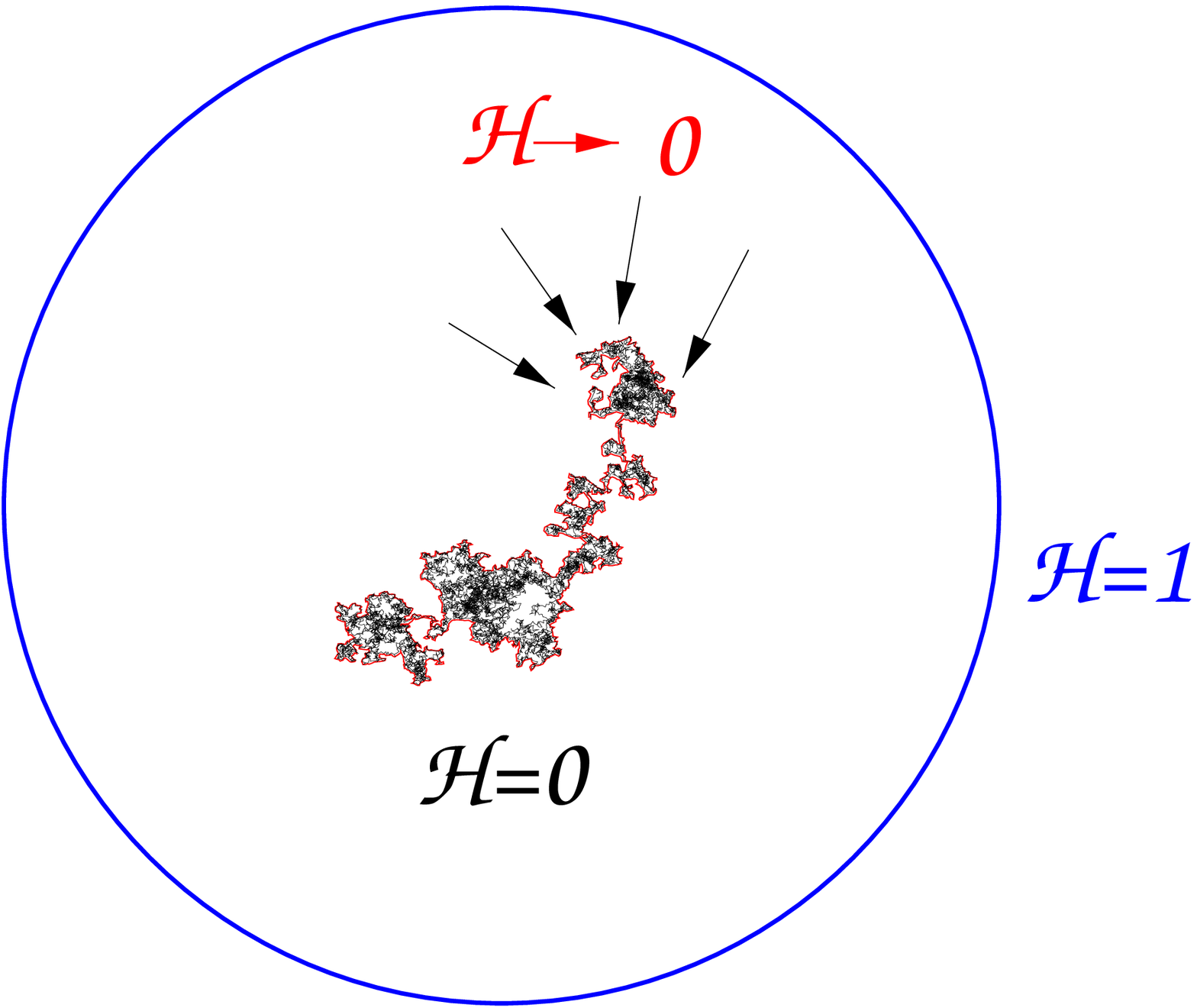}
\hskip.7cm
\includegraphics[angle=0,width=.45\linewidth]{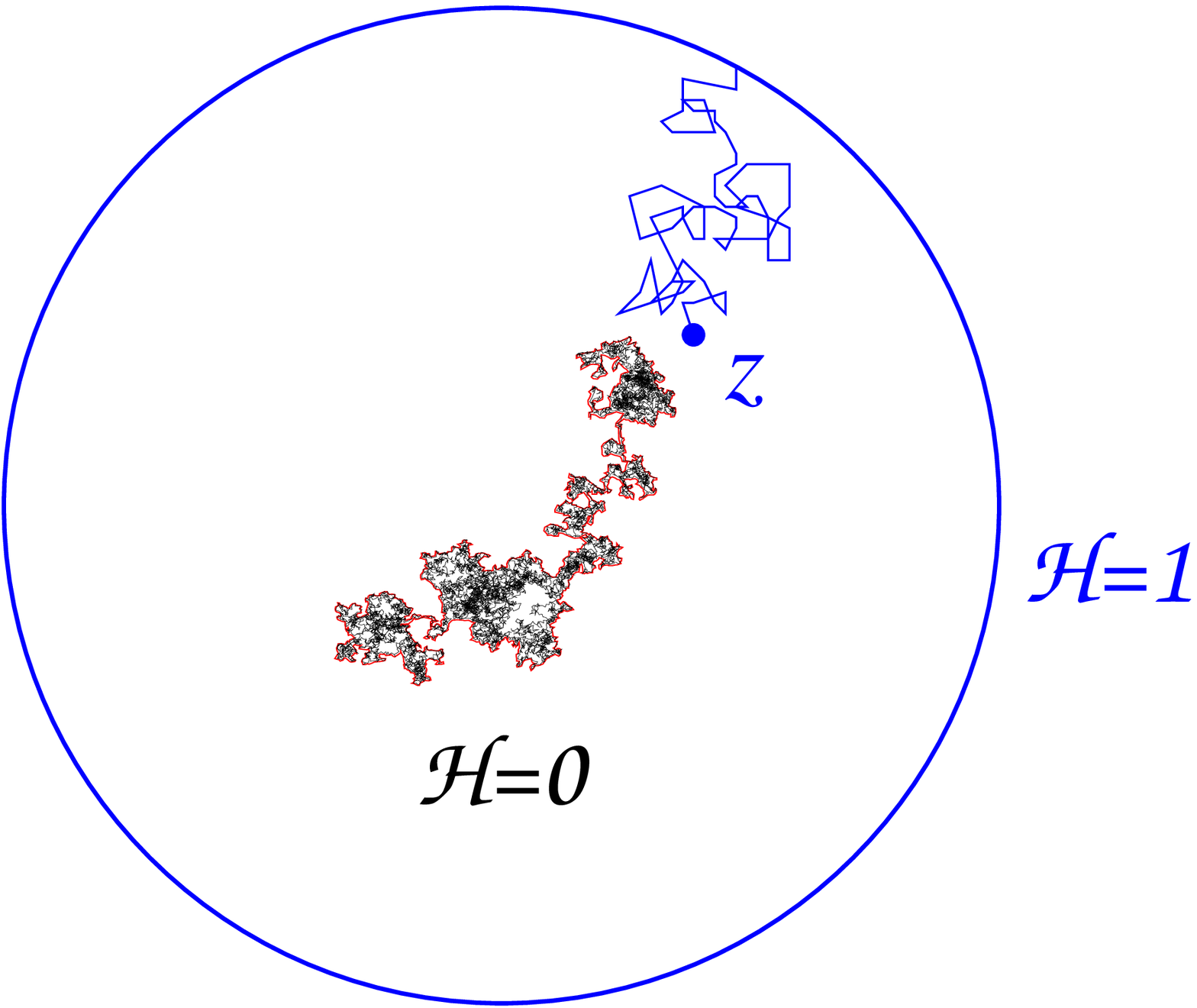}
\end{center}
\caption{Potential near a charged Brownian set and the equivalent Kakutani's diffusion process.}
\label{fig.potentialH}
\end{figure}
One can also consider potential theory near the same
fractal boundary, now charged. One assumes the absorber to be
perfectly conducting, and introduces the harmonic potential
 ${\mathcal H}\left( z\right) $ at an
exterior point $z \in {{\mathbb C}}$, with Dirichlet boundary
conditions ${\mathcal H}\left({w \in {\partial \mathcal C}}\right)=0$ on the outer
(simply connected)  frontier $ \partial \mathcal C$, and
${\mathcal H}(w)=1$ on a circle ``at $\infty$'', i.e., of a large radius
scaling like the average size $R$ of $ \partial \mathcal C$ (Fig.~\ref{fig.potentialH}). As is well-known from a theorem due to Kakutani \cite{kakutani},
${\mathcal H}\left( z\right)$ is identical to the probability that a  random walker (more
precisely, a Brownian motion)
started at $z$ escapes to ``$\infty$'' without having hit
${\partial \mathcal C}$ (Fig.~\ref{fig.potentialH}).

The  harmonic measure $H\left({\partial \mathcal C} \cap B(w,r)\right)$ defined above
 then also appears as the integral of the Laplacian of
${\mathcal H}$ in the disk $B(w,r)$, i.e., the {\it boundary
charge} contained in that disk.

\subsubsection*{Multifractal Local Behavior}
The multifractal formalism
  \cite{mandelbrot2,hentschel,frisch,halsey} further involves
characterizing subsets ${\partial\mathcal C}_{\alpha }$ of sites of the
 frontier ${\partial \mathcal C}$ by a H\"{o}lder exponent $\alpha ,$
such that the $H$-measure of the frontier points in the ball
$B(w,r)$ of radius $r$ centered at $w_{\alpha}\in
{\partial\mathcal C}_{\alpha }$ scales as
\begin{equation}
{H}\left({\partial \mathcal C} \cap { B}(w_{\alpha}, r) \right)
\approx \left( r/R\right) ^{\alpha }. \label{ha'}
\end{equation}
The Hausdorff or ``fractal dimension'' $f\left( \alpha \right) $
of the set ${\partial\mathcal C}_{\alpha }$ is such that
\begin{equation}
{\rm Card}\, {\partial\mathcal C}_{\alpha } \approx R^{f(\alpha)},
\label{ca'}
\end{equation}
and defines the {\it multifractal spectrum} of the harmonic measure.
\subsubsection*{Local Behavior of the Potential}
Similarly, one can consider the local behavior of the
potential near point $w_{\alpha} \in {\partial\mathcal C}_{\alpha }$,
\begin{equation}
{\mathcal H}\left( z \to w\in {\partial\mathcal C}_{\alpha }\right) \approx
\left( |z-w|/R\right) ^{\alpha }, \label{ha''}
\end{equation}
in the scaling limit $a \ll r=|z-w| \ll R$  (with $a$ the
underlying lattice constant if one starts from a lattice description before
taking the scaling limit $a \to 0$).

\begin{figure}[tb]
\begin{center}
\includegraphics[angle=0,width=.450\linewidth]{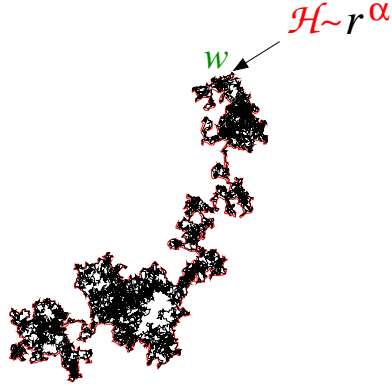}
\end{center}
\caption{Multifractal scaling of the potential (or of the harmonic measure) near a charged Brownian set.}
\label{fig.brownianpotentialH}
\end{figure}

Thus the potential scales with the same $\alpha$-exponent as the harmonic measure (\ref{ha'})
around point $w_{\alpha},$ and
$f(\alpha)={\rm dim}\, {\partial\mathcal C}_{\alpha }$ thus appears as the Hausdorff
dimension of boundary points inducing the local behavior
(\ref{ha''}) (Fig.~\ref{fig.brownianpotentialH}).\footnote{The local definitions of the exponent $\alpha$ and
of $f(\alpha)$ as given in (\ref{ha'}) and (\ref{ca'}), or (\ref{ha''}), are only
heuristic, since the way of taking limits was not explained.
For any given point $w$ on the boundary of a random fractal object, in
general no stable local exponents $\alpha$ exist, such that they are
obtained by a ``simple limit'' to the point. One then proceeds in
another way (see, e.g., \cite{binder}). Define the set ${\partial\mathcal C}_{\alpha,
\eta}$ of points on the boundary $\partial\mathcal C$, $w=w_{\alpha,
\eta}$, for which there exists a decreasing series of radii $r_j,
j\in \mathbb N$ tending towards $0$, such that $r_j^{\alpha+\eta}\leq
\omega(w, r_j)\leq r_j^{\alpha-\eta}$.  The multifractal spectrum
$f(\alpha)$ is then globally defined as the limit  $\eta \to 0$
of the  Hausdorff dimension of the set ${\partial\mathcal
C}_{\alpha, \eta}$, i.e.,
$$f(\alpha)=\lim_{\eta \to 0}{\rm dim}
\left\{w\, : \exists\
\{r_{j}\to 0,\, j\in \mathbb N\} : r_j^{\alpha+\eta}\leq \omega(w,
r_j)\leq r_j^{\alpha-\eta}\right\}.$$}


\subsubsection*{Equivalent Wedge Angle}
\begin{figure}[htb]
\begin{center}
\includegraphics[angle=0,width=.4\linewidth]{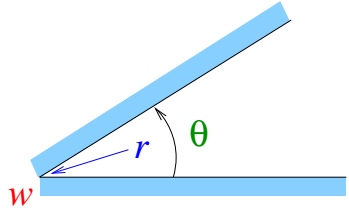}
\caption{Wedge of angle $\theta$.}
\label{fig.angle}
\end{center}
\end{figure}
In  2D the {\it complex} potential
$\varphi(z)$ (such that the electrostatic potential ${\mathcal H}(z)=\Im
\varphi(z)$ and the field's modulus $|{\bf E}(z)|=|\varphi'(z)|$) for a
{\it wedge} of angle $\theta$, centered at $w$ (Fig.~\ref{fig.angle}), is
\begin{equation}
\varphi(z) = (z-w)^{{\pi}/{\theta}}.
\end{equation}
 By Eq.~(\ref{ha''}) a
H\"older exponent $\alpha$ thus defines a local equivalent
``electrostatic'' angle $\theta={\pi}/{\alpha},$ and the MF
dimension $\hat f(\theta)$ of the boundary subset with such
$\theta$ is
\begin{equation}
\hat f(\theta) = f(\alpha={\pi}/{\theta}). \label{fchapeau}
\end{equation}
\subsubsection*{Harmonic Moments}
One then considers a covering of the frontier $\partial \mathcal C$
by balls $B(w,r)$ of radius $r$, and centered at
points $w$ forming a discrete subset $  {\partial \mathcal C}/r$ of $\partial \mathcal C$.
We are interested in the moments of the harmonic measure content $H(w,r):=H(\partial \mathcal C \cap B(w,r))$ of those balls,
averaged over all realizations of ${\mathcal C}$
\begin{equation}
{\mathcal Z}_{n}=\left\langle \sum\limits_{z\in {\partial
{\mathcal C}/r}}{H}^{n}\left(w,r \right) \right\rangle , \label{Za}
\end{equation}
 where $n$ is, { a priori},
a real number. For very large absorbers ${\mathcal C}$ and
frontiers $\partial \mathcal C$ of average
size $R,$ one expects these moments to scale as
\begin{equation}
{\mathcal Z}_{n}\approx \left( r/R\right) ^{\tau \left( n\right)
}, \label{Z2a'}
\end{equation}
 where the multifractal scaling
exponents $\tau \left( n\right) $ encode {\it generalized
dimensions}
\begin{equation}
D\left( n\right) =\frac{\tau\left( n\right)}{n-1} , \label{dn'}
\end{equation}
which vary in a non-linear way with
$n$  \cite{mandelbrot2,hentschel,frisch,halsey}. Several {\it a
priori} results are known. $D(0)$ is the Hausdorff dimension of
the accessible frontier of the fractal. By construction, $H$ is a
normalized probability measure, so that $\tau (1)=0.$ Makarov's
theorem  \cite{makarov}, here applied to the H\"{o}lder regular
curve describing the frontier  \cite{ai2}, gives the so-called
information dimension $\tau ^{\prime }\left( 1\right)
=D\left( 1\right) =1$.

The multifractal spectrum $f\left( \alpha \right)$
appearing in (\ref{ca'})
is given by the symmetric Legendre transform of $\tau \left( n\right)$:
\begin{equation}
\alpha =\frac{d\tau }{dn}\left( n\right) ,\quad \tau \left(
n\right) +f\left( \alpha \right) =\alpha n,\quad
n=\frac{df}{d\alpha }\left( \alpha \right) .  \label{alpha}
\end{equation}
Because of the statistical ensemble average (\ref{Za}), values of $%
f\left( \alpha \right) $ can become negative for some domains of
$\alpha $  \cite{catesetwitten}.


\subsection{Calculation of Multifractal Exponents from Quantum Gravity}
Let us now give the main lines of the derivation of exponents $\tau\left(
n\right) $, hence $f(\alpha)$,  via {\it conformal invariance} and {\it quantum gravity} \cite{duplantier11}. The
recent joint work with I. A. Binder \cite{IABD} on  harmonic (mixed) spectra for SLE establishes rigorously
these multifractal results. (See also \cite{Stockholm}.)
\begin{figure}[tb]
\begin{center}
\includegraphics[angle=0,width=.6\linewidth]{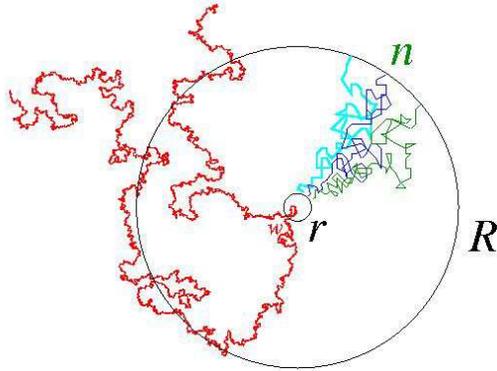}
\end{center}
\caption{Representation of moments (\ref{Za}) by a packet of $n$
independent Brownian paths diffusing away from a SAW, or equivalently from a Brownian frontier, from a short distance $r$  to
a large distance $R$.}
\label{diffuse}
\end{figure}
\subsubsection*{Representation of Moments by Random Walks}
By the very definition of the $H$-measure, $n$ independent RW's or Brownian motions 
diffusing away from the absorber, i.e., from the cluster's
hull's frontier $\partial \mathcal C$, and diffusing without
hitting $\partial \mathcal C$, give a geometric representation
of the $n^{th}$ moment ${H}^{n},$ in Eq.~(\ref{Za}) for $n$ {\it integer} (Fig.~\ref{diffuse}).
Convexity arguments yield the analytic continuation to arbitrary
$n$'s.

Recall the notation $A\wedge B$ for two random
sets required to traverse, {\it without mutual intersection},
the
annulus ${\mathcal%
D}\left( r, R\right) $ from the inner boundary circle of radius
$r$ to the outer one at distance $R$, and{\it \ }$A\vee B$ for two
{\it independent}, thus possibly intersecting, sets. With this notation, one can define, as in Eq.~(\ref{zrs}), a grand canonical partition function
 which describes the star configuration
of the Brownian paths ${\mathcal B}$ and cluster $\mathcal C$:
${\partial\mathcal C}\wedge {n}:={\partial\mathcal C}\wedge\left( \vee {\mathcal B}\right) ^{n} $.
At the critical point, it
is expected to scale for $%
r/R\rightarrow 0$ as
\begin{equation}
{\mathcal Z}_{R}\left( {\partial\mathcal C}\wedge n\right) \approx
\left( r/R\right) ^{x\left( n\right)+\cdots }, \label{xp}
\end{equation}
where the scaling exponent
\begin{equation}
\label{xnwedge}
x\left(n\right):=x\left({\partial\mathcal C}\wedge n\right)
\end{equation}
depends on $n$ and is associated
with the conformal operator creating the star vertex ${\partial\mathcal C}\wedge {n}$.
The dots after exponent $x(n)$ express the fact that there may be an
additional contribution to the exponent,
independent of $n$, corresponding to the entropy associated with the extremities of the random frontier
(see, e.g., Eq.~(\ref{zrs})).

By normalization, this contribution actually does not appear in the multifractal moments.
Since $H$ is a probability measure, the sum (\ref{Za}) is indeed normalized as
\begin{equation}
{\mathcal Z}_{n=1}=1, \label{Za1}
\end{equation}
or in terms of star partition functions:
\begin{equation}
{\mathcal Z}_{n}={\mathcal Z}_{R}\left( {\partial\mathcal C}\wedge n\right) /{\mathcal Z}_{R}\left(  {\partial\mathcal C}\wedge 1\right).
\end{equation}
The scaling behavior (\ref{xp}) thus gives
\begin{equation}
\label{ZnZ1a}
 {\mathcal Z}_{n} \approx
(r/R)^{x\left( n\right)-x\left(1\right)}.
\end{equation}
The last exponent actually obeys the identity $x(1)=x\left(  {\partial\mathcal C}\wedge 1\right)=2$, which
will be obtained directly, and can also be seen as a consequence of Gauss's theorem in two
dimensions \cite{catesetwitten}. Thus we can also write as in (\ref{xppe})
\begin{equation}
{\mathcal Z}_{n} = (R/r)^2\ {\mathcal P}_{R}\left(  {\partial\mathcal C}\wedge n\right),
\end{equation}
where ${\mathcal P}_{R}\left( {\partial\mathcal C}\wedge n\right)$ is a (grand-canonical) excursion measure
from $r$ to $R$ for the random set $ {\partial\mathcal C}\wedge n$,
with proper scaling ${\mathcal P}_{R}\approx
\left( r/R\right) ^{x\left( n\right)}$. The factor $(R/r)^2$
is the area scaling factor of the annulus $\mathcal D (r,R)$.

Owing to Eqs.~(\ref{Z2a'}) (\ref{ZnZ1a}) we get
\begin{equation}
\tau \left( n\right) =x(n)
-x\left(1\right)=x\left(n\right) -2.
\label{tt'}
\end{equation}

\subsubsection*{Proper Scaling Dimensions}
\begin{figure}[htb]
\begin{center}
\includegraphics[angle=0,width=.7\linewidth]{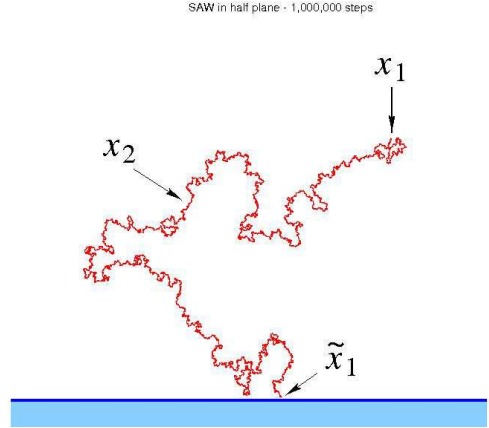}
\caption{{Scaling dimensions along a single conformally invariant curve.}}
\label{fig.halfa}
\end{center}
\end{figure}
 In the absence of diffusing Brownian paths, conformally invariant scaling curves possess their own scaling dimensions. Typically, for a 
single scaling curve, like, e.g., a self-avoiding path, there are three possible environments, 
corresponding to the neighborhoods of the {\it tip}, with scaling dimension $x_1$, of a point {\it inside} the curve ($x_2$),  
or of a {\it boundary} point (${\tilde x}_1$). In these notations, the subscript obviously corresponds to the number of path components attached 
to the considered point (Fig.~\ref{fig.halfa}). Generalizations are given by star exponents $x_L$ and ${\tilde x}_L$, associated with multiple paths, 
as in sections \ref{sec.mixing} or 
\ref{subsec.multiline} below.

The Hausdorff  dimension of the curve is related to the scaling dimension $x_2$ in a well-known way:
$D_{\rm H}=2-x_2.$

\subsubsection*{Quantum Gravity}
To calculate exponents, we again use the fundamental mapping between
the conformal field theory, describing a critical statistical system in the
plane ${\mathbb C}$ or half-plane ${\mathbb H}$, and
the same CFT  in
presence of quantum gravity  \cite{KPZ,DK,david2}. Two
universal functions $U$ and $V$, which now depend on the central
charge $c$ of the CFT, describe the KPZ map between conformal dimensions in bulk or boundary QG and those in
the standard plane or half-plane:
\begin{eqnarray}
U\left( x\right)=U_{\gamma}\left( x\right):=x\frac{x-\gamma}{1-\gamma} , \hskip2mm V\left(
x\right)= V_{\gamma}\left( x\right)=\frac{1}{4}\frac{x^{2}-\gamma^2}{1-\gamma},  \label{Ua}
\end{eqnarray}
with
\begin{equation}
\label{UVa}
V_{\gamma}\left( x\right):= U_{\gamma}\left(\frac{1}{2}\left(x+\gamma
\right) \right).
\end{equation}
The parameter $\gamma$ is the {\it string susceptibility exponent}
of the random 2D surface (of genus zero), bearing the CFT of
central charge $c$ \cite{KPZ}; $\gamma$ is the solution of
\begin{equation}
c=1-6{\gamma}^2(1-\gamma)^{-1}, \gamma \leq 0. \label{cgamma}
\end{equation}
In order to simplify the notation, we shall hereafter in this section
drop the subscript $\gamma$ from functions $U$ and $V$.

The function $U$ maps quantum gravity conformal weights, whether in the bulk or on a boundary,
into their counterparts in ${\mathbb C}$ or ${\mathbb H}$, as in (\ref{KPZg}) (\ref{KPZgb}). The function $V$ has been
tailored to map quantum gravity
{\it boundary} dimensions to the corresponding conformal dimensions in the full plane ${\mathbb C}$,
as in (\ref{Zetal}) (\ref{ZetaL}).
The {\it positive} inverse
function of $U$, $U^{-1}$, is
\begin{equation}
U^{-1}\left( x\right)
=\frac{1}{2}\left(\sqrt{4(1-\gamma)x+\gamma^2}+\gamma\right),
\label{U1a}
\end{equation}
and transforms the conformal weights of a conformal operator in ${\mathbb C}$ or ${\mathbb H}$ into
the conformal weights of the same operator in quantum gravity, in the bulk or on the boundary.

\subsubsection*{Boundary Additivity Rule}
Consider two arbitrary  random sets $A,B,$ with boundary scaling
exponents $\tilde{x}\left( A\right)$,

$\tilde{x}\left( B\right)$ in the {\it half-plane} ${\mathbb H}$ with Dirichlet boundary conditions.
When these two sets are mutually-avoiding, the scaling exponent $x\left( A\wedge
B\right)$ in ${\mathbb C}$, as in (\ref{xnwedge}), or $\tilde x\left( A\wedge
B\right)$ in ${\mathbb H}$ have the universal structure
 \cite{duplantier8,duplantier9,duplantier11}
\begin{eqnarray}\label{xa}
x\left( A\wedge B\right) &=&2V\left[ U^{-1}\left( \tilde{x}\left(
A\right)
\right) +U^{-1}\left( \tilde{x}\left( B\right) \right) \right],  
\\
\tilde x\left( A\wedge B\right) &=&U\left[ U^{-1}\left( \tilde{x}\left(
A\right)
\right) +U^{-1}\left( \tilde{x}\left( B\right) \right) \right].  
\label{xb}
\end{eqnarray}
We have seen these fundamental relations in the $c=0$ case above; they are established
for the general case in Ref. \cite{BDMan}. $U^{-1}\left( \tilde{x} \right)$ is, on the random disk
with Dirichlet boundary conditions, the boundary scaling dimension corresponding to
$\tilde{x}$ in the half-plane  ${\mathbb H}$, and  in Eqs.~(\ref{xa}) (\ref{xb})
\begin{eqnarray}
U^{-1}\left(\tilde x\left( A\wedge B\right)\right) &=&U^{-1}\left( \tilde{x}\left(
A\right)
\right) +U^{-1}\left( \tilde{x}\left( B\right) \right)  
\label{xc}
\end{eqnarray}
is a
{\it linear} boundary exponent corresponding to the fusion of two ``boundary
operators'' on the random disk, under the Dirichlet mutual avoidance condition $A \wedge B$.
This quantum boundary conformal dimension is mapped back by $V$ to the scaling
dimension in ${\mathbb C}$, or by $U$ to the boundary scaling dimension in ${\mathbb H}$ \cite{duplantier11}.  

\subsubsection*{Exponent Construction}
\begin{figure}[tb]
\begin{center}
\includegraphics[angle=0,width=.7\linewidth]{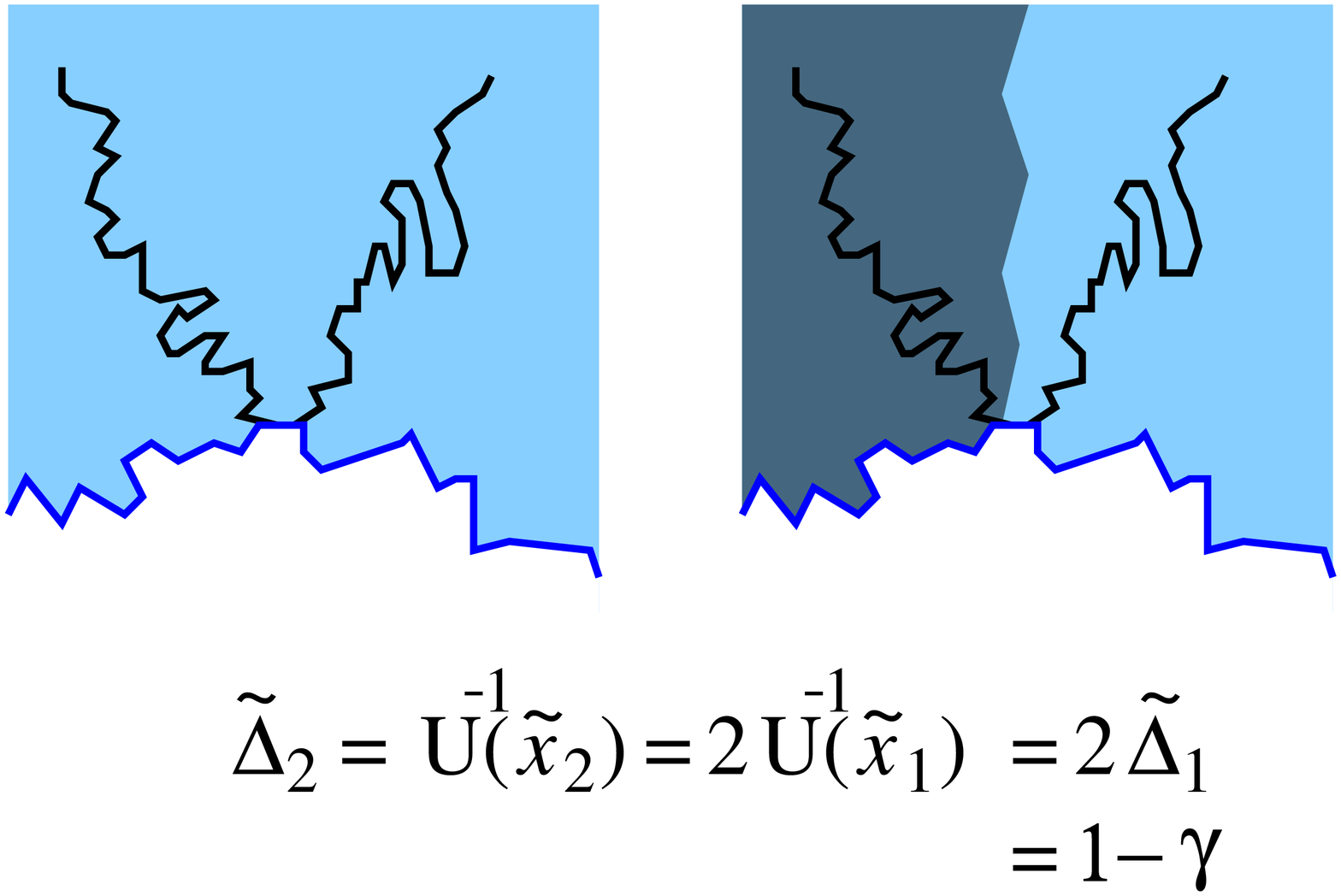}
\end{center}
\caption{Illustration of the additivity rule (\ref{xtilde}): each of the two non-intersecting strands of
a simple random path
defines its own boundary sector of the random disk near the Dirichlet boundary.}
\label{split}
\end{figure}
\begin{figure}[tb]
\begin{center}
\includegraphics[angle=0,width=.48\linewidth]{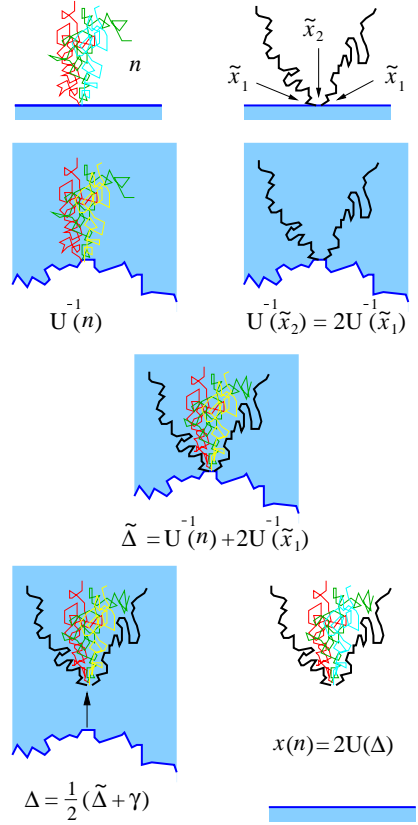}
\end{center}
\caption{The quantum gravity construction (\ref{xtilde}) (\ref{deltatildea}) of exponents (\ref{finab}).}
\label{sumsmall}
\end{figure}

For determining the harmonic exponents $x(n)$ (\ref{xnwedge}),
we use (\ref{xa}) for $A={\partial \mathcal C}$ and $B=\left( \vee {\mathcal B}\right) ^{n}$.

\noindent $\bullet$ We first need the {\it boundary} (conformal) scaling dimension (b.s.d.)
$\tilde x_2:=\tilde{x}\left({\partial \mathcal C}\right)$ associated with the presence of the random frontier
near the Dirichlet boundary ${\mathbb H}$. Since
this frontier is simple, it can be seen as made of two non-intersecting semi-infinite strands (Fig.~\ref{split}).
Its b.s.d. in quantum gravity thus obeys
(\ref{xc})
\begin{equation}
U^{-1}\left( \tilde{x_2}\right) =2U^{-1}\left( \tilde{x_1}\right),
\label{xtilde}
\end{equation}
where $\tilde x_1$ is the boundary scaling dimension of a semi-infinite frontier path
 originating at the boundary of ${\mathbb H}$.

\noindent $\bullet$ As before, the packet of $n$ independent Brownian paths has a boundary scaling dimension
$\tilde{x}\left( \left( \vee {\mathcal B}\right) ^{n}\right)=n.$

\noindent $\bullet$ From (\ref{xc}) the QG boundary dimension of the whole set is (see Fig.~\ref{sumsmall}):
\begin{equation}
\tilde \Delta:=U^{-1}\left[\tilde x\left({\partial \mathcal C} \wedge n\right)\right]=2U^{-1}\left( \tilde{x_1}\right)
 +U^{-1}\left( n\right).
\label{deltatildea}
\end{equation}
Its associated QG bulk conformal dimension is therefore
$\Delta=\frac{1}{2}(\tilde \Delta+\gamma)$. From Eqs.~(\ref{UVa}) or (\ref{xa})
we finally find
\begin{eqnarray}
\nonumber
 x\left( n\right) &=&2U(\Delta)=2V(\tilde\Delta)\\
\label{finab}
 &=&2V\left[2U^{-1}\left( \tilde{x_1}\right)+U^{-1}\left( n\right) \right].
\end{eqnarray}
The whole construction is illustrated in Fig.~\ref{sumsmall}.

\noindent $\bullet$ The value of the QG b.s.d. of a {\it simple} semi-infinite random path is
\begin{equation}
\label{tildex1}
U^{-1}\left( \tilde{x_1}\right) =\frac{1}{2}(1-\gamma).
\end{equation}
It is derived in section \ref{sec.SLEKPZ} below 
from the exponents of the $O(N)$ model, or of the SLE. It can be directly derived from Makarov's theorem:
\begin{equation}
\alpha(n=1)=\tau'(n=1)=\frac{ dx}{ d n}(n=1)=1,
\end{equation}
which, applied to (\ref{finab}), leads to the same result. We thus finally get
\begin{equation}
x\left( n\right) =2V\left(1-\gamma
 +U^{-1}\left( n\right) \right)=2U\left( \frac{1}{2}+\frac{1}{2}
 U^{-1}\left( n\right) \right).  \label{finaa}
\end{equation}
This result satisfies the identity: $x(1)=2U(1)=2$, which is related to Gauss's theorem, as mentioned above.

\subsubsection*{Multifractal Exponents}

\noindent $\bullet$ The multifractal exponents $\tau(n)$ (\ref{tt'}) are obtained from
(\ref{Ua}-\ref{U1a}) as \cite{duplantier11}
\begin{eqnarray}
\nonumber
\tau\left( n\right)&=&x(n)-2\\
&=&\frac{1}{2}(n-1)+\frac{1}{4}\frac{2-\gamma}{1-\gamma}
[\sqrt{4(1-\gamma)n+{\gamma}^2}-(2-\gamma)]\ .
\label{tauoriginal}
\end{eqnarray}
Similar exponents, but
associated with moments taken at the tip,
later appeared in the context of the ${\rm SLE}$ process
(see II in Ref. \cite{lawler4}, and  \cite{lawleresi}; see also  \cite{hastings} for Laplacian random walks.)
The  whole family will be given in section \ref{sec.multifSLE}.

\noindent $\bullet$ The Legendre transform is easily performed to yield:
\begin{eqnarray}
\alpha &=&\frac{d{\tau} }{dn}\left( n\right)=\frac{1}{2}
+\frac{1}{2} \frac{2-\gamma}{\sqrt{4(1-\gamma)n+{\gamma}^2}};
\label{a'}
\\
\nonumber
\\
f\left( \alpha \right)&=&
\frac{1}{8}\frac{(2-\gamma)^2}{1-\gamma}\left(3- \frac{1}{2\alpha
-1}\right) -\frac{1}{4}\frac{\gamma^2}{1-\gamma}\alpha,
\label{foriginal}
\\
\quad \alpha &\in& \left(
{\textstyle{ \frac{1}{ 2}}}%
,+\infty \right) . \nonumber
\end{eqnarray}

\noindent It is convenient to express the results in terms of the central charge $c$ with the help of:
\begin{eqnarray}
\label{c,gamma}
\frac{1}{4}\frac{(2-\gamma)^2}{1-\gamma}=\frac{25-c}{24};\;\;\;\;
\frac{1}{4}\frac{\gamma^2}{1-\gamma}=\frac{1-c}{24}.
\end{eqnarray}
We finally find the\\
\noindent $\bullet$ {\it Multifractal Exponents}
\begin{eqnarray}
\label{taunc}
\tau\left( n\right) &=&\frac{1}{2}(n-1)+\frac{25-c}{24}
\left(\sqrt{\frac{24n+1-c}{25-c}}-1\right),
\\
\nonumber
\\
D\left( n\right) &=&\frac{\tau\left( n\right)}{n-1}=\frac{1}{2}+
{\left(\sqrt{\frac{24n+1-c}{25-c}}+1\right)}^{-1},
\label{D''}\\
\quad n&\in& \left[ n^{\ast}= -\frac{1-c}{24} ,+\infty \right)\ ;
\nonumber
\end{eqnarray}
$\bullet$ {\it Multifractal Spectrum}
\begin{eqnarray}
\alpha &=&\frac{d{\tau} }{dn}\left( n\right)=\frac{1}{2}
+\frac{1}{2} \sqrt{\frac{25-c}{24n+1-c}}; \label{a'a}
\\
\nonumber
\\
\label{foriginalbis} f\left( \alpha \right)&=& \frac{25-c}{48}\left(3-
\frac{1}{2\alpha -1}\right) -\frac{1-c}{24}\alpha,
\\
\quad \alpha &\in& \left(
\frac{1}{2}
,+\infty \right) . \nonumber
\end{eqnarray}

\subsubsection*{Other Multifractal Exponents}
This formalism immediately allows generalizations. For instance,
in place of a packet of $n$ independent random walks, one can consider a packet of $n$ {\it
independent self-avoiding walks} $\mathcal P$, which avoid the fractal boundary.
The associated multifractal exponents $ x\left( {\partial\mathcal
C}\wedge \left( \vee {\mathcal P}\right)^{n} \right)$ are given by
(\ref{finaa}), with the argument $n$ in $U^{-1}(n)$ simply
replaced by  ${\tilde x}\left( \left( \vee {\mathcal P}\right)
^{n}\right) =n{\tilde x} \left( {\mathcal P}\right) =\frac{5}{8}n
$  \cite{duplantier8}. These exponents govern the universal
multifractal behavior of the moments of the probability that a SAW
escapes from $\mathcal C$. One then gets a spectrum $\bar f\left(\bar\alpha\right)$ such
that $${\bar f}\left(\bar\alpha=\tilde{x} \left( {\mathcal
P}\right)\alpha \right) =
f\left(\alpha=\pi/\theta\right)={\hat f}(\theta),$$ thus unveiling
the  same invariant
underlying wedge distribution as the harmonic measure (see also \cite{cardy2}).\\

\subsection{{Geometrical Analysis of Multifractal Spectra}}
\label{subsec.geometry}

\subsubsection*{Makarov's Theorem}
The generalized dimensions $D(n)$ satisfy, for any
$c$, $\tau'(n=1)=D(n=1)=1$, or equivalently $f(\alpha=1)=1$, i.e.,
{\it Makarov's theorem}  \cite{makarov}, valid for any simply
connected boundary curve. From (\ref{D''}), (\ref{a'a}) we also
note a fundamental relation, independent of $c$:
\begin{equation}
3-2D(n)=1/\alpha=\theta/\pi. \label{Dtheta}
\end{equation}
We also have the {\it superuniversal} bounds: $\forall c, \forall
n,\frac{1}{2}=D(\infty) \leq D(n) \leq D(n^{\ast})=\frac{3}{2}$,
corresponding to $0 \leq \theta\leq 2\pi$.
\subsubsection*{An Invariance Property of $f(\alpha)$}
\label{subsec.inv}
It is interesting to note that the general multifractal function
(\ref{foriginalbis}) can also be written as
\begin{eqnarray}
f\left( \alpha \right)-\alpha&=& \frac{25-c}{24}
\left[1-\frac{1}{2}\left(2\alpha -1 + \frac{1}{2\alpha
-1}\right)\right]. \label{f-aa}
\end{eqnarray}
Thus the multifractal function possesses the invariance symmetry
\begin{eqnarray}
f\left( \alpha \right)-\alpha=f\left( {\alpha}^\prime
\right)-{\alpha}^{\prime}, \label{inv}
\end{eqnarray}
for $\alpha$ and ${\alpha}^{\prime}$ satisfying the duality
relation:
\begin{eqnarray}
(2\alpha-1)(2{\alpha}^{\prime}-1)=1,
\end{eqnarray}
or, equivalently
${\alpha}^{-1}+{{\alpha}^{\prime}}^{-1}=2.$ 
When associating an equivalent electrostatic wedge angle $\theta=\pi / \alpha$ to each local
singularity exponent $\alpha$, one gets the complementary rule
for angles in the plane
\begin{eqnarray}
\theta+{\theta}^{\prime}=\frac{\pi}{\alpha}+\frac{\pi}{{\alpha}^{\prime}}=2\pi.
\label{tetateta'}
\end{eqnarray}
Notice that by definition of the multifractal dimension $f\left( \alpha \right)$,
$R^{f\left( \alpha \right)-\alpha}$ is the total harmonic measure content of points of type $\alpha$
or equivalent angle $\theta=\pi/\alpha$
along the multifractal frontier. The symmetry (\ref{inv}) thus means that this harmonic content is invariant when
taken at the complementary angle in the plane $2\pi-\theta$. The basic symmetry (\ref{inv}) thus  reflects 
 that of the external frontier itself under the {\it exchange of
interior and exterior domains}.

It is also interesting to note that, owing to the explicit forms (\ref{D''}) of $D(n)$
and (\ref{a'a}) of $\alpha$, the
condition (\ref{tetateta'}) becomes, after a little algebra,
\begin{equation}
D(n)+D(n')=2. \label{DD'}
\end{equation}
This basic interior-exterior
symmetry, first observed  \cite{BDH} for the $c=0$
result of  \cite{duplantier8}, is valid for {\it any} conformally
invariant boundary.


\subsubsection*{Equivalent Wedge Distribution}
The geometrical
multifractal distribution of wedges $\theta$ along the boundary takes the form:
\begin{eqnarray}
\hat
f(\theta)=f\left(\frac{\pi}{\theta}\right)=\frac{\pi}{\theta}-\frac{25-c}{12}
 \frac{(\pi-\theta)^2}{\theta (2\pi -\theta)}\ .
\label{fchap}
\end{eqnarray}
Remarkably enough, the second term also describes the contribution
by a wedge to the density of electromagnetic modes in a cavity
 \cite{BD}. The simple shift in (\ref{fchap}), $25 \to 25 -c$, from
the $c=0$ case to general values of $c$, can then be related to
results of conformal invariance in a wedge  \cite{DuCa}. The
partition function for the two sides of a wedge of angle $\theta$
and size $R$, in a CFT of central charge $c$, indeed scales as
 \cite{Ca}
\begin{equation}
 \hat {\mathcal Z} (\theta,c) \approx R^{-
 {c(\pi-\theta)^2}/12\,{\theta (2\pi -\theta)}}\ .
\label{hatZ}
\end{equation}
Thus, one can view the $c$ dependance of result (\ref{fchap}) as follows: the
number of sites, $R^{\hat f(\theta,c)}$,
 with local wedge angle $\theta$ along a random path with central charge $c$, is the same
 as the number of sites, $R^{\hat f(\theta,c=0)}$, with wedge angle $\theta$
along a {\it self-avoiding walk} ($c=0$), renormalized  by the partition function $\hat
{\mathcal Z}(\theta,c)$ representing the presence of a
$c$-CFT along such wedges:
$$R^{\hat f(\theta,c)} \propto R^{\hat f(\theta,c=0)}/\hat {\mathcal Z} (\theta,c).$$

\subsubsection*{Hausdorff Dimension of the External Perimeter}
The maximum of $f(\alpha)$ corresponds
to $n=0$, and gives the Hausdorff dimension $D_{\rm EP}$ of
 the support of the measure, i.e.,
the {\it accessible} or {\it external perimeter} as:
\begin{eqnarray}
D_{\rm
EP}&=&{\sup}_{\alpha}f(\alpha)=f(\alpha(n=0))\\&=&D(0)=\frac{3-2\gamma}{2(1-\gamma)}
=\frac{3}{2}-\frac{1}{24}\sqrt{1-c}\left(\sqrt{25-c}-\sqrt{1-c}\right).
\label{D(c)}
\end{eqnarray}
This corresponds to a {\it typical} sigularity exponent
\begin{equation}
\hat\alpha={\alpha(0)}=1-\frac{1}{\gamma}=\left(\frac{1}{12}\sqrt{1-c}\left(\sqrt{25-
c}-\sqrt{1-c}\right)\right)^{-1}=(3-2D_{\rm EP})^{-1}\ ,
\label{halpha}
\end{equation}
and to a  typical wedge angle
\begin{equation}
\hat\theta={\pi}/{\hat \alpha}=\pi(3-2D_{\rm EP})\ .
\label{htheta}
\end{equation}

\subsubsection*{Probability Densities}
The
probability $P(\alpha)$ to find a singularity exponent $\alpha$
or, equivalently, $\hat P (\theta)$ to find an equivalent opening
angle $\theta$ along the frontier is
\begin{equation}
P(\alpha)=\hat P(\theta)\propto R^{f(\alpha)- f(\hat\alpha)} \ .
\end{equation}
Using the values found above, one can recast this probability as
(see also  \cite{cardy2})
\begin{equation}
P(\alpha)=\hat P(\theta)\propto \exp\left[-\frac{1}{24}\ln
R\left(\sqrt{1-c}\sqrt{\omega}-\frac{\sqrt{25- c}}{2 \sqrt \omega}
\right)^2\right]\ , \label{prob}
\end{equation}
where  $$\omega:=\alpha-\frac{1}{2}=\frac{\pi}{\theta}-\frac{1}{2}\
.$$

\subsubsection*{Universal Multifractal Data}
\begin{figure}[tb]
\begin{center}
\includegraphics[angle=0,width=.7\linewidth]{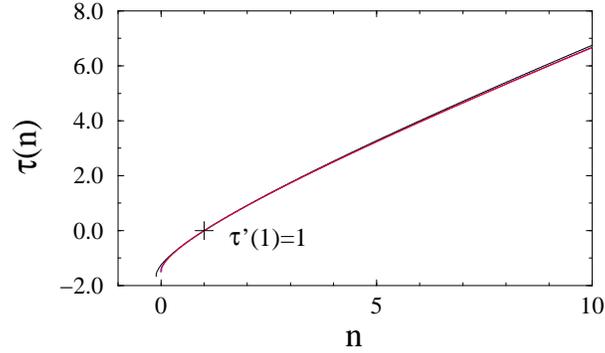}
\end{center}
\caption{Universal multifractal exponents
$\tau(n)$ (\ref{taunc}). The curves are indexed by the central charge $c$: 2D spanning trees ($c=-2$); 
 self-avoiding or random walks, and
percolation ($c=0$); 
Ising clusters or $Q=2$ Potts clusters ($c=\frac{1}{2}$); $N=2$ loops, or $Q=4$
Potts clusters
($c=1$). The curves are almost
indistinguishable at the scale shown.}
\label{Figure6}
\end{figure}

\begin{figure}[tb]
\begin{center}
\includegraphics[angle=0,width=.7\linewidth]{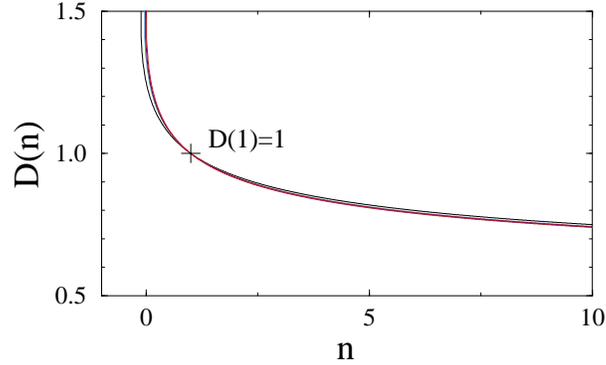}
\end{center}
\caption{Universal generalized dimensions
$D(n)$ (\ref{D''}). The curves are indexed as in Fig.~\ref{Figure6} and are almost
indistinguishable at the scale shown.}
\label{Figure7}
\end{figure}

\begin{figure}[tb]
\begin{center}
\vskip.3cm
\includegraphics[angle=0,width=.75\linewidth]{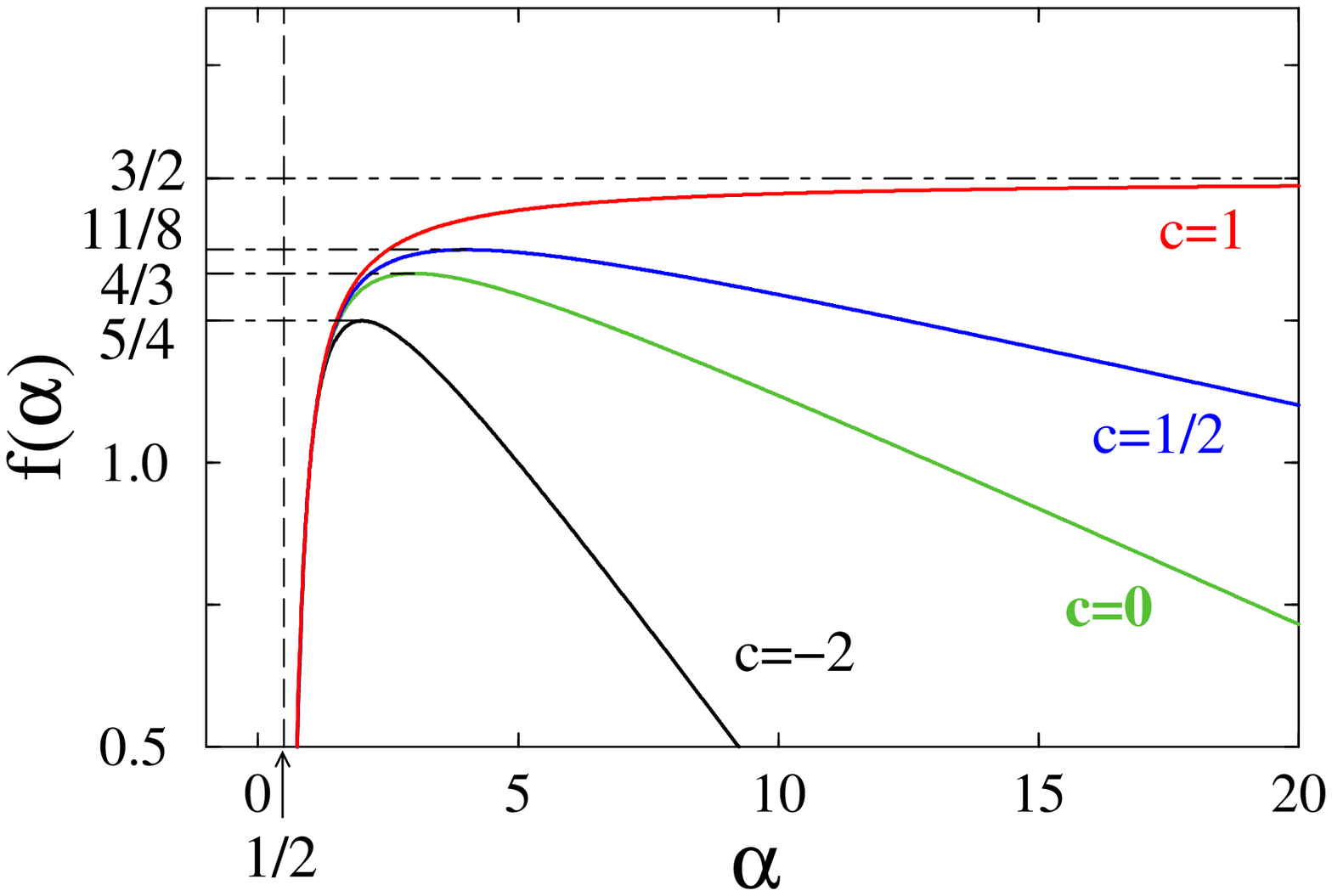}
\end{center}
\caption{Universal harmonic multifractal spectra
$f(\alpha)$ (\ref{foriginalbis}). The curves are indexed by the central charge $c$, and correspond
 to: 2D spanning trees ($c=-2$); self-avoiding or random walks, and
percolation ($c=0$);
Ising clusters or $Q=2$ Potts clusters ($c=\frac{1}{2}$); $N=2$ loops, or $Q=4$
Potts clusters
($c=1$). The maximal dimensions are those of the {\it
accessible} frontiers. The left branches of the various $f(\alpha)$ curves are largely
indistinguishable, while their right branches split for large $\alpha$,
corresponding
 to negative values
of $n$.}
\label{Figure8}
\end{figure}

The multifractal exponents $\tau(n)$ (Fig.~\ref{Figure6}) or the generalized
dimensions $D(n)$ (Fig.~\ref{Figure7}) appear quite similar for various
values of $c$, and a numerical simulation would hardly distinguish
the different universality classes, while the $f(\alpha)$
functions, as we see in Fig.~\ref{Figure8}, do distinguish these classes,
especially for negative $n$, i.e. large $\alpha$. In Figure \ref{Figure8}
 we display the multifractal functions $f$, Eq.~(\ref{foriginalbis}),
corresponding to various values of $-2 \leq c \leq 1$, or,
equivalently, to a number of components $N \in [0, 2]$, and $Q \in
[0,4]$ in the $O(N)$ or Potts models (see below).

\subsubsection*{Needles}
The singularity at $\alpha=\frac{1}{2}$, or $\theta=2\pi$, in the
multifractal functions $f$, or $\hat f$, corresponds to boundary
points with a needle local geometry, and Beurling's theorem
 \cite{Beur} indeed insures that the H{\"o}lder exponents $\alpha$ are
 bounded below by $\frac{1}{2}$. This corresponds to large
values of $n$, where, asymptotically, for {\it any}
 universality class,
\begin{equation}
\forall c, \lim_{n \to \infty} D(n)=\frac{1}{2}. \label{1/2}
\end{equation}

\subsubsection*{Fjords}
The right branch of $f\left( \alpha \right) $ has a linear
asymptote
\begin{equation}
\lim_{\alpha \rightarrow \infty} f\left(\alpha \right)/{\alpha} =
n^{\ast}=-(1-c)/24.
\end{equation}
The $\alpha \to \infty$ behavior corresponds to moments of lowest
order $n\rightarrow {n^{\ast}}$, where $D(n)$ reaches its maximal
value: $\forall c, D(n^{\ast})=\frac{3}{2}$, common to {\it all}
simply connected, conformally-invariant, boundaries.
Its linear shape is quite reminiscent of that of the multifractal
function of the growth probability as in the case of a 2D DLA
cluster  \cite {ball}. 
This describes almost inaccessible sites: Define ${\mathcal N}\left(
H\right)$ as the number of boundary sites having a given
probability $H$ to be hit by a RW starting at infinity; the MF
formalism yields, for $H\rightarrow 0,$ a power law behavior
\begin{equation}
{\mathcal N}\left( H\right)|_{H\rightarrow 0}\approx
H^{-(1+{n}^{\ast})} \label{nha}
\end{equation}
with an exponent
\begin{equation}
1+n^{\ast}=\frac{23+c}{24}<1.
\end{equation}

\subsubsection*{RW's, SAW's and Percolation}
Brownian paths, scaling self-avoiding walks and critical percolation clusters all correspond to CFT's with $c=0$,
for which we find
\begin{eqnarray}
\tau \left( n\right)&=&\frac{1}{2}\left( n-1\right)
+\frac{5}{24}\left(
\sqrt{24n+1}-5\right) ,\\ \label{tauf}
D\left( n\right)&=&\frac{1}{2}+\frac{5}{\sqrt{24n+1}+5},\quad n\in \left[ -%
{{\frac{1}{24}}}%
,+\infty \right),  \label{dna}\\
f\left( \alpha \right)&=&\frac{25}{48}\left( 3-\frac{1}{2\alpha -1}\right) -%
\frac{\alpha }{24},\quad \alpha \in \left(
{{\frac{1}{ 2}}}%
,+\infty \right) , \label{mf}
\end{eqnarray}
where we recognize in particular the percolation exponents (\ref{taunperc}, \ref{dn}). We thus have the general result:

\noindent {\it In two dimensions,   the harmonic multifractal exponents $\tau(n)$ and spectra} $f\left(
\alpha \right) $ {\it of a random walk, a critical percolation cluster, and a self-avoiding walk are
identical  in the scaling limit.}\\
{\it The  external frontier of a Brownian path and the accessible perimeter of a percolation cluster are
identical to a self-avoiding walk in the scaling limit, with Hausdorff dimension
$
D_{\rm EP}={\rm sup}_{\alpha}f(\alpha,
c=0)={4}/{3},
$
i.e., the Mandelbrot conjecture.}
\subsubsection*{Ising Clusters}
 A critical  Ising
cluster ($c=\frac{1}{2}$) possesses a multifractal spectrum with respect to the
harmonic measure:
\begin{eqnarray}
\tau \left( n\right)&=&\frac{1}{2}\left( n-1\right)
+\frac{7}{48}\left(
\sqrt{48n+1}-7\right) , \label{taufis}\\
f\left( \alpha \right) &=&\frac{49}{96}\left( 3-\frac{1}{2\alpha -1}\right) -%
\frac{\alpha }{48},\quad \alpha \in \left(
\frac{1}{2}
,+\infty \right), \label{fis}
\end{eqnarray}
with the dimension of the accessible perimeter
\begin{equation}
D_{\rm EP}={\rm sup}_{\alpha}f(\alpha,
c={1}/{2})=\frac{11}{8}.
\end{equation}

\subsubsection*{$Q=4$ Potts Clusters, and ``Ultimate Norway''}
The {\it limit} multifractal spectrum is obtained for $c=1$,  which is an upper
 or lower bound for all $c$'s, depending on the position of $\alpha$ with respect to $1$:
\begin{eqnarray}
\nonumber
f(\alpha,c<1) &<& f(\alpha,c=1),\;  1 < \alpha,\\
\nonumber
f(\alpha=1,c)&=&1,\; \forall c,\\
\nonumber
f(\alpha,c<1) &>& f(\alpha,c=1),\;  \alpha <1.
\end{eqnarray}
This MF spectrum provides an exact example of a {\it left-sided} MF spectrum,
with an asymptote $f\left(\alpha \to \infty, c=1\right)\to
\frac{3}{2}$ (Fig.~\ref{Figure8}). It corresponds to singular boundaries
where ${\hat f}\left(\theta \to 0, c=1\right)=\frac{3}{2}=D_{\rm
EP}$, i.e., where the external perimeter is everywhere dominated
by ``{\it fjords}'', with typical angle $\hat \theta =0$. It is
tempting to call it the ``ultimate Norway''.

The frontier of a $Q=4$ Potts Fortuin-Kasteleyn cluster, or the ${\rm SLE}_{\kappa=4}$ provide such an example for  this {\it
left-handed} multifractal spectrum ($c=1$) (see section \ref{sec.geodual}). The MF data are:
\begin{eqnarray}
\tau \left( n\right)&=&\frac{1}{2}\left( n-1\right) +
\sqrt{n}-1, \label{taufc}\\
f\left( \alpha \right) &=&\frac{1}{2}\left( 3-\frac{1}{2\alpha
-1}\right),\quad \alpha \in \left(
\frac{1}{2}
,+\infty \right), \label{fc}
\end{eqnarray}
with accessible sites forming a set of Hausdorff dimension
\begin{equation}
D_{\rm EP}={\rm sup}_{\alpha}f(\alpha,c=1)=\frac{3}{2},
\end{equation}
which is also the maximal value common to all multifractal
generalized dimensions $D(n)=\frac{1}{n-1}\tau(n)$. The external perimeter
which bears the electrostatic charge is a non-intersecting
{\it simple} path. We therefore arrive at the
striking conclusion that in the plane, a conformally-invariant
scaling curve which is simple has a Hausdorff dimension at
most equal to $D_{\rm EP}=3/2$ \cite{duplantier11}. The
corresponding $Q=4$ Potts frontier, while still possessing a set
of double points of dimension $0$, actually develops
a logarithmically growing number of double points  \cite{aharony}. The values
of the various Hausdorff dimensions predicted for Potts clusters have been verified
in a nice numerical study \cite{aharony,aharony2}.
\section{\sc{Higher Multifractal Spectra}}
\label{sec.higher}
It is interesting to note that one can define {\it higher
multifractal} spectra as those depending on several $\alpha$
variables  \cite{duplantier10}. A first example is given by the
double moments of the harmonic measure on {\it both} sides of a
random path.
\subsection{Double-Sided Spectra}
\label{subsec.double}
\begin{figure}[tb]
\begin{center}
\includegraphics[angle=0,width=.9\linewidth]{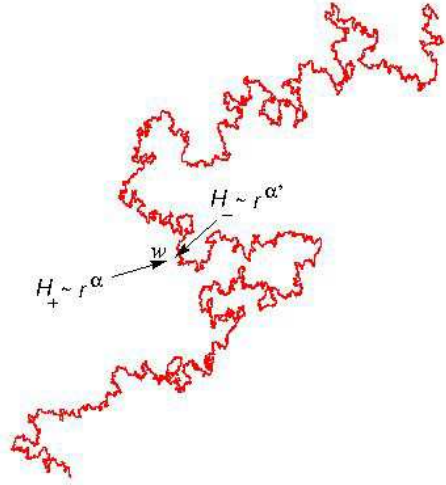}
\end{center}
\caption{Double distribution of potential ${\mathcal H}$ on both sides of a simple scaling curve
(here a SAW). The local exponents on both sides of point $w=w_{\alpha,\alpha'}$ are
$\alpha$ and $\alpha'$. The Hausdorff dimension of such points along the SAW is $f_2(\alpha,\alpha')$.}
\label{Figure1}
\end{figure}

\subsubsection*{Double-Sided Potential}
 When it is {\it simple}, i.e., double point free,
a conformally scaling curve $\mathcal C$ can be reached from
both sides.  Notice, however, that one
can also address the case of non-simple random paths, by concentrating
on the double-sided potential near {\it cut-points}. For a Brownian path for instance, one can
consider the subset of pinching or cut-points, of
Hausdorff dimension $D=2-2\zeta_{2}=3/4$, where the path splits
into two non-intersecting parts.  The path  is then locally
accessible from
both directions. 

Taking Dirichlet boundary conditions on the random curve, one can then
consider  the joint distribution of potential on both sides, such that the
potential scales as
\begin{equation}
{\mathcal H}_{+}\left( z \to w\in {\partial\mathcal C}_{\alpha,\alpha'
}\right) \approx |z-w|^{\alpha}, \label{ha+}
\end{equation}
when approaching $w$ on one side of the scaling curve, while scaling as
\begin{equation}
{\mathcal H}_{-}\left( z \to w\in {\partial\mathcal C}_{\alpha,\alpha'
}\right) \approx |z-w|^{\alpha'}, \label{ha-}
\end{equation}
on the other side (Fig.~\ref{Figure1}). The multifractal
formalism now characterizes subsets ${\mathcal C}_{\alpha,\alpha' }$
of boundary sites $w$ with two such H\"{o}lder exponents, $\alpha
,\alpha'$, by their Hausdorff dimension
$f_2\left( \alpha,\alpha' \right) :={\rm dim}\left({\mathcal
C}_{\alpha,\alpha' }\right)$. The standard one-sided multifractal spectrum
$f(\alpha)$ is then recovered as the supremum:
\begin{equation}
f(\alpha)={\rm sup_{\alpha'}}f_2\left( \alpha,\alpha' \right).
\label{sup}
\end{equation}

\subsubsection*{Equivalent Wedges}
As above, one can also define two equivalent ``electrostatic''
angles from singularity exponents $\alpha,\alpha'$, as
$\theta={\pi}/{\alpha},\theta'={\pi}/{\alpha'}$ and the 
dimension $\hat f_2(\theta,\theta')$ of the boundary subset with
such $\theta,\theta'$ is then
\begin{equation}
\hat f_2(\theta,\theta') :=
f_2(\alpha={\pi}/{\theta},\alpha'={\pi}/{\theta'}).
\label{fchapeau'}
\end{equation}

\subsubsection*{Double Harmonic Moments}
As before, instead of considering directly the potential $\mathcal H$, one can consider
equivalently the harmonic measure. Let  ${H}\left( w,r\right):={H}({\mathcal C} \cap { B}(w, r))$
be the harmonic measure (as seen from ``infinity'') of the intersection of $\mathcal C$ and the ball
${B}(w, r)$ centered at point $w \in {\mathcal C}.$
Let us consider a
covering of the path by such balls centered at points
forming a discrete subset ${\mathcal C}/r$ of $\mathcal C$.

Define the double moments of the harmonic measure:
\begin{equation}
{\mathcal Z}_{n,n'}=\left\langle \sum\limits_{w\in {\mathcal
C}/r}\left[{H}_{+}(w,r)\right]^{n} \left[{H}_{-}(w,r)\right]^{n'}\right\rangle, \label{ZZ''}
\end{equation}
where ${H}_{+}(w,r)$
and ${H}_{-}(w,r)$
are
respectively the harmonic measures on the ``left'' or ``right''
sides of the random path.
\begin{figure}[tb]
\begin{center}
\includegraphics[angle=0,width=.9\linewidth]{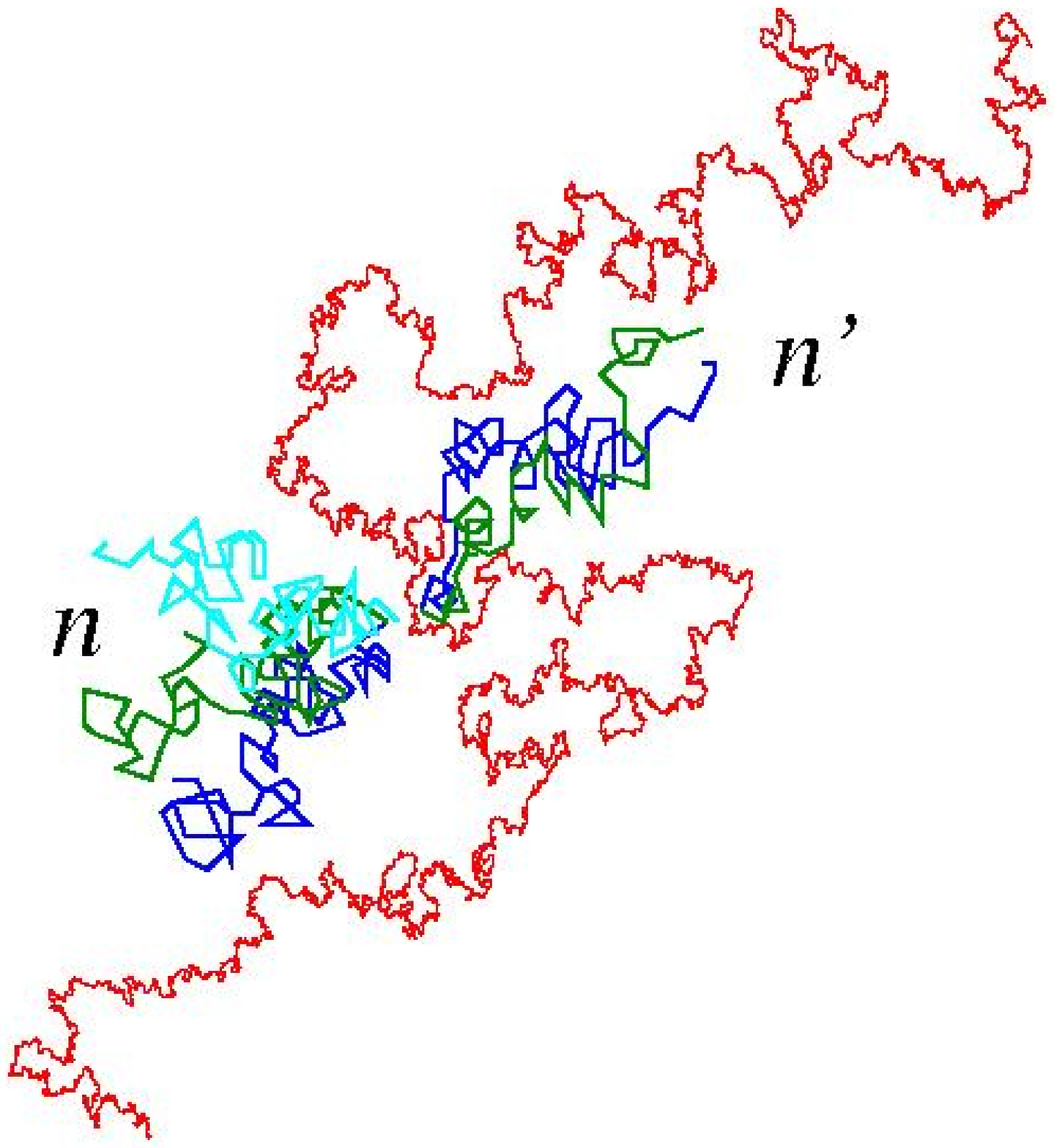}
\end{center}
\caption{Representation of the double moments (\ref{ZZ''}) by two packets of $n$ and $n'$
independent Brownian paths diffusing away from a SAW.}
\label{sawbrown}
\end{figure}
These moments are represented by two packets of $n$ and $n'$
independent Brownian paths
diffusing away from the fractal path (Fig.~\ref{sawbrown}).

They have a
multifractal scaling behavior
 \begin{equation}
{\mathcal Z}_{n,n'}\approx \left( r/R\right) ^{\tau_2 \left(
n,n'\right) },  \label{ZZ2''}
\end{equation}
where the exponent $\tau_2 \left(n,n'\right)$ now depends on two
moment orders $n,n'$.  A Hausdorff dimension
is given by the double Legendre transform:
\begin{eqnarray}
\nonumber
\alpha &=&\frac{\partial\tau_2 }{\partial n}\left( n,n'\right) ,
\quad \alpha' =\frac{\partial\tau_2 }{\partial n'}\left( n,n'\right), \\
f_2\left( \alpha, \alpha' \right) &=&\alpha n+\alpha' n'-\tau_2 \left( n,n'\right), \label{alpha'''}\\
\nonumber
n&=&\frac{\partial f_2}{\partial \alpha}\left( \alpha, \alpha'
\right) , \quad n'=\frac{\partial f_2}{\partial \alpha' }\left(
\alpha, \alpha' \right).
\end{eqnarray}
It yields the dimension  of the
subset $\mathcal C_{\alpha,\alpha'}$ of frontier points
$w_{\alpha,\alpha'}$, where the potential $\mathcal H$ scales as
in Eqs.~(\ref{ha+}-\ref{ha-}), or where the harmonic content of a ball $B(w_{\alpha,\alpha'},r)$ scales
as $(r/R)^{\alpha}$ on one side, and $(r/R)^{\alpha'}$ on the other.

From definition (\ref{ZZ''}) and Eq.~(\ref{ZZ2''}), we recover for $n'=0$ the
one-sided multifractal exponents
\begin{equation}
\tau \left( n\right)=\tau_2 \left( n,n'=0\right),
\end{equation}
and putting these values in the Legendre transform  (\ref{alpha'''}) yields identity (\ref{sup}),
as it must.
\subsubsection*{One and Two-Sided Cases}
In analogy to Eqs.~(\ref{tt'}), (\ref{finab}), the exponent
$\tau_2(n,n')$ is associated with a scaling dimension $x_2(n,n')$, calculated in the quantum gravity
formalism in a way similar to (\ref{finaa}) \cite{BDjsp,BDMan}:
\begin{eqnarray}
\nonumber
\tau_2(n,n')&=&x_2(n,n')-2 \\
x_2(n,n')&=&2V\left[
1-\gamma +U^{-1}\left( n\right)+U^{-1}\left( n'\right) \right].
\label{finaa'}
\end{eqnarray}
The two-sided multifractal spectrum
 is then obtained by a double Legendre transform as \cite{BDMan,BDjsp}
\begin{eqnarray}
f_2\left( \alpha, \alpha'
\right)&=&\frac{25-c}{12}-\frac{1}{2(1-\gamma)}
{\left[1-\frac{1}{2}\left(\frac{1}{\alpha}+\frac{1}{\alpha'}\right)\right]}^{-1} \nonumber \\
& &-\frac{1-c}{24}\left(\alpha+\alpha'\right), \label{f_2c}
\end{eqnarray}
\begin{equation}
{\alpha}=\frac{1}{\sqrt{4(1-\gamma)n+\gamma^2}}\left[1
+\frac{1}{2}\left(\sqrt{4(1-\gamma)n+\gamma^2}+\sqrt{4(1-\gamma)n'+\gamma^2}\right)\right],
\label{alphann'c}
\end{equation}
with a similar symmetric equation for $\alpha'$. This doubly multifractal spectrum
possesses the desired property $${\rm sup}_{\alpha'}
f_2(\alpha, \alpha')=f(\alpha),$$ where $f(\alpha)$ is
(\ref{foriginalbis}) above.
The domain of definition of the doubly multifractal function $f_2$ is
independent of $c$ and given by
\begin{equation}
1-\frac{1}{2}\left({\alpha}_{}^{-1}+{\alpha'}_{}^{-1}\right) \geq 0,
\label{domain}
\end{equation}
in accordance to Eq.~(\ref{f_2c}).
The domain of definition of distribution $\hat f_2$ is the image
of domain (\ref{domain}) in $\theta$-variables:
 \begin{equation}
\theta +\theta' \leq 2\pi. \label{domaint}
\end{equation}
The total {\it electrostatic} angle is thus less than $2\pi$,
which simply accounts for the electrostatic screening of local
wedges by fractal randomness, as expected.\\

Notice finally that there also exists a single-sided distribution \cite{BDMan}
\begin{eqnarray}
f_{1}(\alpha)=\frac{25-c}{12}-
\frac{1}{8(1-\gamma)}\left(1-\frac{1}{2 \alpha}\right)
-\frac{1-c}{24}{\alpha}, \label{faic1}
\end{eqnarray}
which corresponds to the potential distribution in the
{\it vicinity of the tip} of a conformally-invariant scaling path, and naturally
differs from the usual $f(\alpha)={\rm sup}_{\alpha'}
f_2(\alpha,\alpha'))$ spectrum, which describes the potential on
one side of the scaling path.

\subsubsection*{Brownian and SAW's Double Spectra}

In the case of a Brownian path or a self-avoiding walk, one obtains
 \cite{BDMan,BDjsp}
\begin{eqnarray}
\nonumber
f_2\left( \alpha, \alpha'
\right)&=&\frac{25}{12}-\frac{1}{3}{a_{B,P}^2}
{\left[1-\frac{1}{2}\left(\frac{1}{\alpha}+\frac{1}{\alpha'}\right)\right]}^{-1}
-\frac{1}{24}\left(\alpha+\alpha'\right), \label{faa'}\\
\nonumber
a_B&=&\frac{3}{2}\ ({\rm RW}),\
\ a_P=1\ ({\rm SAW}). \label{a''}
\end{eqnarray}
These doubly multifractal spectra thus are different for RW's and SAW's. The SAW spectrum corresponds to
(\ref{f_2c}) for $c=0, \gamma=-1/2$, and  possesses the required
property $$f_P(\alpha):={\rm sup}_{\alpha'} f_{2,P}(\alpha,
\alpha')=f(\alpha),$$ where $f(\alpha)$ is (\ref{mf}) above. For a Brownian path,
the one-sided spectrum $$ f_{B}(\alpha):={\rm sup}_{\alpha'} f_{2,B}(\alpha,
\alpha')=\frac{51}{48}-\frac{49}{48}\frac{1}{2\alpha-1}-\frac{\alpha}{24},$$ such that
$f_B(\alpha) < f(\alpha)$,
gives the spectrum of
cut-points along the Brownian frontier. This set of Hausdorff dimension $\frac{3}{4} < 1$ is disconnected, and
$f_B(\alpha=1)= 0$, in contrast to Makarov's theorem, $f(\alpha=1)= 1 $, for any
connected set in the plane.

\subsection{Higher Multifractality of Multiple Path Vertices}
One can consider a star configuration ${\mathcal S}_L$ of a number $L$
 of {\it similar simple scaling paths},
all originating at the same vertex $w$. Higher moments can then be defined by looking at the joint
distribution of the harmonic measure contents in each sector between the arms. We shall not describe
this general case here, which can be found in full detail in Ref. \cite{BDMan}.

\section{\sc{Winding of Conformally Invariant Curves}}
\label{sec.winding}

\begin{figure}[htb]
\epsfxsize=5.5truecm{\centerline{\epsfbox{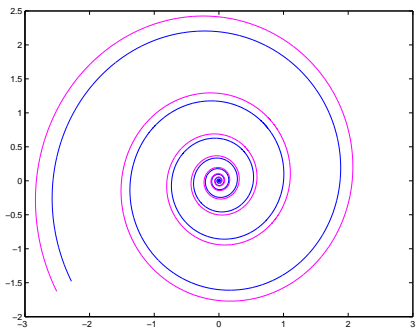}}}
\caption{A double logarithmic spiral mimicking the local geometry of the two strands of the conformally-invariant
frontier path.}
\label{spiral2}
\end{figure}

Another important question arises concerning the {\it   geometry
 of the  equipotential lines} near a random (CI)
fractal curve. These lines are expected to  rotate wildly, or wind, in a spiralling motion that closely follows
the boundary itself. The key geometrical object is here the {\it  logarithmic spiral},
which is conformally invariant (Fig.~\ref{spiral2}).
The MF description should generalize to a
{\it   mixed} multifractal spectrum, accounting for {\it   both scaling
and winding} of the equipotentials  \cite{binder}.

In this
section, we describe the exact solution to this mixed MF spectrum
for any random CI curve \cite{DB}. In particular, it is shown
to be related by a scaling law to the usual 
harmonic MF spectrum. We use the same conformal tools as before, fusing quantum
gravity and Coulomb gas methods, which allow the
description of Brownian paths interacting and winding with CI curves,
thereby providing a probabilistic description of the potential map near any CI curve.
With I. A. Binder, we have also obtained recently a rigorous derivation of this spectrum
for the SLE \cite{IABD}.
\\

\subsection{Harmonic Measure and Rotations}

Consider again a
(CI) critical random cluster, or scaling curve, generically
called ${\mathcal C}$. Let ${\mathcal H}\left( z\right) $ be the potential at an 
exterior point $z \in {  {\mathbb C}}$, with Dirichlet boundary
conditions ${\mathcal H}\left({w \in \partial \mathcal C}\right)=0$ on the outer
(simply connected) boundary $\partial \mathcal C$ of $\mathcal C$, and
${\mathcal H}(w)=1$ on a circle ``at $\infty$'', i.e., of a large radius
scaling like the average size $R$ of $ \mathcal C$. As we have seen, the potential
${\mathcal H}\left( z\right)$ is identical to the probability that a Brownian path
started at $z$ escapes to ``$\infty$'' without having hit
${\mathcal C}$. 

Let us now consider the {\it  degree with which the
curves wind in the complex plane about point} $w$ and call
$\varphi(z)={\rm  arg}\,(z-w)$. In the scaling limit, the multifractal formalism, here generalized to take into
account rotations  \cite{binder}, characterizes subsets
${\partial\mathcal C}_{\alpha,\lambda}$ of boundary sites by a
H\"{o}lder exponent $\alpha$, and a rotation rate $\lambda$,
such that their potential lines respectively scale and {\it logarithmically spiral} as
\begin{eqnarray}
\nonumber
{\mathcal H}\left( z \to w\in {\partial\mathcal C}_{\alpha,\lambda }\right) &\approx& r ^{\alpha },\\
\varphi\left( z \to w\in {\partial\mathcal C}_{\alpha,\lambda
}\right) &\approx& \lambda \ln\, r\ , 
\label{ha}
\end{eqnarray}
in the limit  $r=|z-w| \to 0$. The Hausdorff dimension
${\rm  dim}\left({\partial\mathcal C}_{\alpha,\lambda }\right)=f\left(
\alpha, \lambda\right)$ defines
 the mixed MF spectrum, which is CI since {\it under a conformal map
 both $\alpha$ and $\lambda$ are locally invariant}.

\begin{figure}[htb]
\begin{center}
\includegraphics[angle=0,width=.4\linewidth]{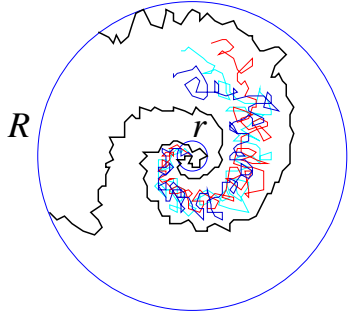}
\end{center}
\caption{Two-sided boundary curve $\partial \mathcal C$ and Brownian $n$-packet winding together from the disk of
radius $r$ up to distances of order $R$, as measured by the winding angle 
$\varphi (w,r)={\rm  arg}({\partial\mathcal C}\wedge n)$ as in (\ref{ZDB}) and in (\ref{Phinp}).}
\label{Fig.escape}
\end{figure}

As above, we 
consider the  harmonic measure $H\left(w,r\right)$, which is the integral of the Laplacian of 
${\mathcal H}$ in a disk $B(w,r)$ of radius $r$ centered at $w \in \partial \mathcal C$, i.e., the boundary 
charge in that disk. It scales as $r^{\alpha}$ with the same exponent as in (\ref{ha}), and 
is also the probability that a Brownian path started at large distance $R$ first hits the boundary at a point inside 
 $B(w,r)$. Let $\varphi (w,r)$ be the associated winding angle of the path down
to distance $r$ from $w$. The {\it   mixed} moments of $H$ and
$e^{\varphi}$, averaged over all realizations of ${\mathcal C}$, are defined as
\begin{equation}
{\mathcal Z}_{n,p}=\left\langle \sum\limits_{w\in {\partial {\mathcal C}}/r} 
H^{n}\left(w,r\right) \exp\, (p\,\varphi (w,r)) \right\rangle
\approx \left( r/R\right) ^{\tau \left(
n,p\right) }, \label{ZDB}
\end{equation}
where the sum runs over the centers of a covering of the boundary by disks of radius $r$, and where $n$ and $p$ are 
real numbers. As before, the $n^{\rm th}$ moment of $H\left(w,r\right)$ is the probability that $n$ independent Brownian 
paths diffuse 
along the boundary and all first hit it at points inside the disk $B(w,r)$. The angle $\varphi (w,r)$ is then their common winding angle down to 
distance $r$ (Fig.~\ref{Fig.escape}). 
 
The scaling limit in (\ref{ZDB}) involves 
multifractal scaling exponents $\tau \left( n,p\right)$
which vary in a non-linear way with $n$ and $p$.
 They give the multifractal spectrum $f\left(
\alpha, \lambda\right)$ via a symmetric double Legendre transform:
\begin{eqnarray}
\nonumber
\alpha &=&\frac{\partial\tau }{\partial n}\left( n,p\right) ,
\quad \lambda =\frac{\partial\tau }{\partial p}\left( n,p\right), \\ \nonumber
f\left( \alpha, \lambda \right)&=&\alpha n+\lambda p-\tau \left( n,p\right) ,\\
n&=&\frac{\partial f}{\partial \alpha}\left( \alpha, \lambda
\right) , \quad p=\frac{\partial f}{\partial \lambda }\left(
\alpha, \lambda \right).
\label{legendre}
\end{eqnarray}
Because of the ensemble average (\ref{ZDB}), $f\left(
\alpha,\lambda \right)$ can become negative for some
$\alpha,\lambda$.

\subsection{Exact Mixed Multifractal Spectra}

 The 2D conformally
invariant random statistical system
 is labelled by its {\it   central charge} $c$, $c\leq 1$  \cite{BPZ}. 
 The main result is the following exact scaling law  \cite{DB}:
\begin{eqnarray}
\label{scalinglaw}
 f(\alpha,\lambda)&=&(1+\lambda^2) f\left(\frac{\alpha}{1+\lambda^2}\right)-b \lambda^2\ ,\\
\nonumber
 b&:=&\frac{25-c}{12}\geq 2\ ,
\end{eqnarray}
where $f\left({\alpha}\right)=
f\left({\alpha},\lambda=0\right)$ is the usual harmonic
MF spectrum in the absence of prescribed winding, first
obtained in Ref. \cite{duplantier11}, and described in section \ref{sec.conform},  Eq.~(\ref{foriginalbis}). It can be recast as:
\begin{eqnarray}
 f(\alpha)&=&\alpha+b-\frac{b\alpha^2}{2\alpha-1},\\ \nonumber b&=&\frac{25-c}{12}.
\label{falpha}
\end{eqnarray}
We thus arrive at the very simple formula for the mixed spectrum:
\begin{eqnarray}
 f(\alpha,\lambda)=\alpha+b-\frac{b\alpha^2}{2\alpha-1-\lambda^2}\ .
\label{falphalambda}
\end{eqnarray}
Notice that by conformal symmetry $${\sup}_{\lambda}f(\alpha,\lambda)=f(\alpha,\lambda=0),$$
 i.e.,
the most likely situation in the absence of prescribed rotation
is the same as $\lambda=0$, i.e. {\it  winding-free}.
The domain of definition of the usual $f(\alpha)$ (\ref{falpha}) is $1/2 \leq \alpha
$  \cite{Beur}, thus for $\lambda$-spiralling points Eq.~(\ref{scalinglaw}) gives
\begin{eqnarray}
  \frac{1}{2}({1+\lambda^2}) \leq {\alpha}\ ,
\label{alpha'}
\end{eqnarray}
in agreement with a theorem by Beurling  \cite{Beur,binder}.

We have seen in section  \ref{subsec.geometry} the geometrical meaning to the exponent $\alpha$:
For an angle with opening $\theta$, $\alpha={\pi}/{\theta}$, and the quantity ${\pi}/{\alpha}$ can be regarded as
 a local generalized angle with respect to the harmonic measure. The geometrical MF spectrum
of the boundary subset with such opening angle $\theta$ and spiralling rate
$\lambda$ reads from (\ref{falphalambda})
\begin{eqnarray}
\hat f(\theta,\lambda)\equiv f\left(\alpha=\frac{\pi}{\theta},\lambda\right)=\frac{\pi}{\theta}+b-b\frac{\pi}{2}
\left(\frac{1}{\theta}+
 \frac{1}{\frac{2\pi}{1+\lambda^2} -\theta}\right).
 \nonumber
\label{fthetalambda}
\end{eqnarray}
As in (\ref{alpha'}), the domain of definition in the
$\theta$ variable is
\begin{eqnarray}
\nonumber
0 \le \theta \le \theta(\lambda),\;\;\;
\theta(\lambda)={2\pi}/({1+\lambda^2}).
\end{eqnarray}
The maximum is reached when the two frontier strands about point $w$ locally collapse into a single
$\lambda$-spiral, whose inner opening angle is $\theta(\lambda)$  \cite{Beur}.

In the absence of prescribed winding ($\lambda=0$), the maximum
$D_{\rm  EP}:= D_{\rm  EP}(0)={\sup}_{\alpha}f(\alpha,\lambda=0)$
gives the dimension of the {\it   external perimeter} of the fractal
cluster, which is a {\it simple} curve without double points, and may differ from the full hull  \cite{duplantier11,ADA}.
Its dimension (\ref{D(c)}) reads in this notation
$$D_{\rm  EP}=\frac{1}{2}(1+b)-\frac{1}{2}\sqrt{b(b-2)},\;\;\;b=\frac{25-c}{12}.$$
It corresponds to typical values $\hat
\alpha=\alpha(n=0,p=0)$ and $\hat\theta={\pi}/{\hat \alpha}=\pi(3-2D_{\rm  EP}).$

For spirals, the maximum value
$D_{\rm  EP}(\lambda)={\sup}_{\alpha}f(\alpha,\lambda)$ still
corresponds  in the Legendre transform (\ref{legendre}) to $n=0$,
and gives the dimension of the {\it   subset of the  external
perimeter made of logarithmic spirals of type $\lambda$}. Owing to (\ref{scalinglaw})
we immediately get
\begin{eqnarray}
D_{\rm  EP}(\lambda)=(1+\lambda^2)D_{\rm  EP} -b \lambda^2 \ .
\label{supf}
\end{eqnarray}
This corresponds to typical scaled values $$\hat
\alpha(\lambda)=(1+\lambda^2)\hat \alpha, \;\; \hat
\theta(\lambda)=\hat \theta/(1+\lambda^2).$$
Since $b \geq 2$ and
$D_{\rm  EP} \leq 3/2$, the EP dimension decreases
with spiralling rate, in a simple parabolic way.
\begin{figure}[t]
\begin{center}
\includegraphics[angle=0,width=0.81\linewidth]{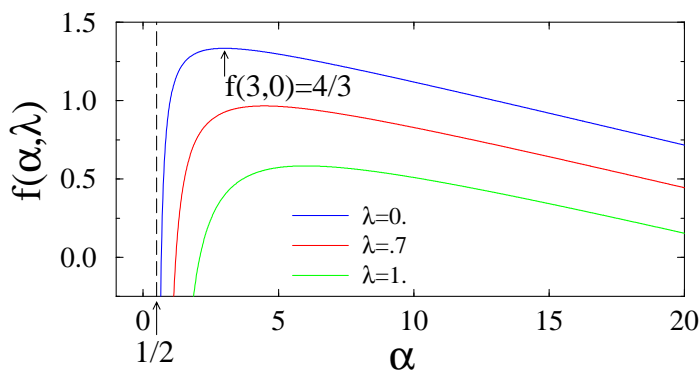}
\end{center}
\caption{Universal multifractal spectrum $f(\alpha,\lambda)$ for
$c=0$ (Brownian frontier, percolation EP and
SAW), and for three different values of the
spiralling rate $\lambda$. The maximum $f(3,0)=4/3$ is the Hausdorff dimension of the frontier.}
\label{Figure1DB}
\end{figure}
\begin{figure}[t]
\begin{center}
\includegraphics[angle=0,width=0.81\linewidth]{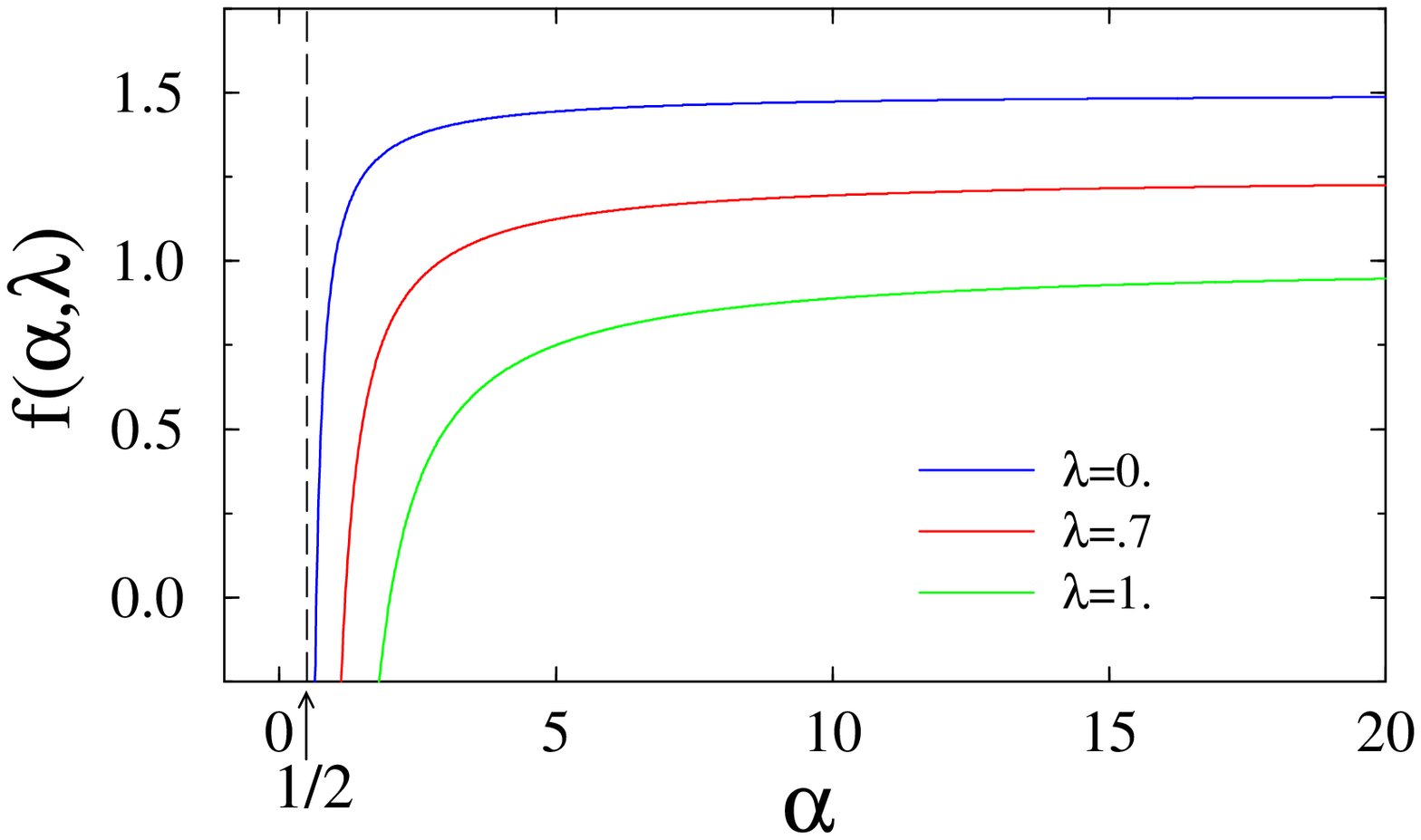}
\end{center}
\caption{Left-sided multifractal spectra $f(\alpha,\lambda)$ for
the limit case $c=1$, the ``ultimate Norway'' (frontier of a $Q=4$ Potts cluster or ${\rm  {\rm SLE}}_{\kappa=4}$).}
\label{Figure2DB}
\end{figure}

Fig.~\ref{Figure1DB}  displays typical multifractal functions
$f(\alpha,\lambda;c)$. The example choosen, $c=0$,
corresponds to the cases of a SAW,
or of a percolation EP, the scaling limits of which both coincide
with the Brownian frontier  \cite{duplantier8,duplantier9,lawler3}. The original
singularity at $\alpha=\frac{1}{2}$ in the
rotation free MF functions $f(\alpha,0)$, which describes boundary points with a needle local
geometry, is shifted for $\lambda \ne 0$ towards the minimal value (\ref{alpha'}). The right branch of
$f\left( \alpha,\lambda\right) $ has a linear asymptote
$\lim_{\alpha \rightarrow +\infty} f\left(\alpha,\lambda
\right)/{\alpha} =-(1-c)/24.$ Thus the $\lambda$-curves all become parallel for $\alpha \to
+\infty$, i.e., $\theta \to 0^{+}$, corresponding to deep fjords
where winding is easiest.

Limit multifractal spectra are
obtained for $c=1$, which exhibit examples of {\it
left-sided} spectra, with a horizontal asymptote
$f\left(\alpha\to +\infty,\lambda; c=1\right)=
\frac{3}{2}-\frac{1}{2}\lambda^2$ (Fig.~\ref{Figure2DB}). This corresponds to
the frontier of a $Q=4$ Potts cluster (i.e., the
${\rm  {\rm SLE}}_{\kappa=4}$), a universal random scaling curve, with the
maximum value $D_{\rm  EP}=3/2$, and a vanishing typical opening
angle $\hat \theta=0$, i.e., the ``ultimate Norway'' where the EP
is dominated by ``{\it   fjords}'' everywhere
 \cite{duplantier11,BDjsp}.
\begin{figure}[htbp]
\begin{center}
\includegraphics[angle=0,width=0.81\linewidth]{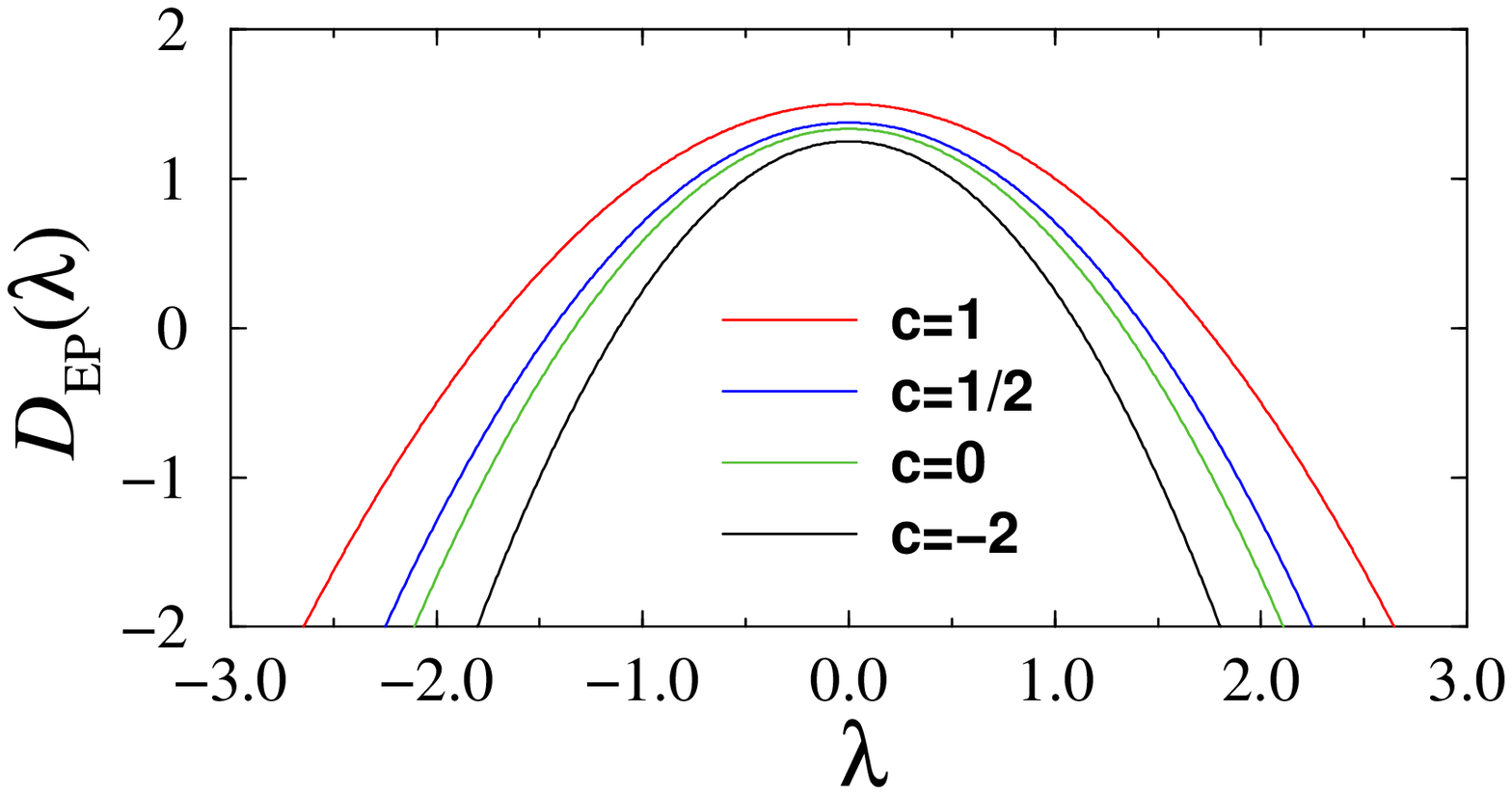}
\end{center}
\caption{Dimensions $D_{\rm  EP}(\lambda)$ of the external frontiers as a function of rotation rate.  The curves
are indexed by increasing central charge $c$, and correspond
respectively to: loop-erased RW ($c=-2; {\rm {\rm SLE}}_{2}$); Brownian or
percolation external frontiers, and self-avoiding walk ($c=0; {\rm SLE}_{8/3}$);
Ising clusters ($c=\frac{1}{2}; {\rm SLE}_{3}$); $Q=4$ Potts clusters ($c=1; {\rm SLE}_{4}$).}
\label{Figure3DB}
\end{figure}
Fig. \ref{Figure3DB} displays the dimension $D_{\rm
EP}(\lambda)$ as a function of the rotation
rate $\lambda$, for various values of $ c \leq 1$,
corresponding to different statistical systems. Again, the $c=1$
case shows the least decay with $\lambda$, as expected from the
predominence of fjords there.


\subsection{Conformal Invariance and Quantum Gravity}

We now give the main lines of the derivation of exponents $\tau\left(
n,p\right)$, hence $f(\alpha,\lambda)$  \cite{DB}. As usual, $n$ {\it
independent} Brownian paths ${\mathcal B}$, starting a small distance $r$ away from
a point $w$ on the frontier $\partial \mathcal C$, and
diffusing without hitting $\partial \mathcal C$, give a geometric
representation of the $n^{th}$ moment, $ H^{n}$, of the harmonic measure in Eq.~(\ref{ZDB})
for {\it integer} $n$ (Fig.~\ref{Fig.escape}), extended by convexity  to
arbitrary $n$'s. Let us introduce an abstract (conformal) field
operator $\Phi_{{\partial\mathcal C}\wedge {n}}$ characterizing the
presence of a vertex where $n$ such Brownian paths and the cluster's
frontier diffuse away from each other in the {\it   mutually-avoiding} configuration
${\partial\mathcal
C}\wedge {n}$  \cite{duplantier8,duplantier9}; to this operator is associated a scaling
dimension $x(n)$. To measure rotations using the moments (\ref{ZDB}) we have to consider
expectation values with insertion of the mixed operator
\begin{equation}
\Phi_{{\partial\mathcal C}\wedge n} e^{p\,{\rm  arg}({\partial\mathcal
C}\wedge n)} \longrightarrow x\left( n,p\right) , \label{Phinp}
\end{equation}
where ${\rm  arg}({\partial\mathcal C}\wedge n)$ is the winding angle
common to the frontier and to the Brownian paths (see Fig.~(\ref{Fig.escape})), and where  $x(n,p)$ is the
{\it scaling dimension} of the operator $\Phi_{{\partial\mathcal C}\wedge n} e^{p\,{\rm  arg}({\partial\mathcal
C}\wedge n)}$. It is directly related to $\tau(n,p)$ by  
\begin{equation}
x\left( n,p\right)=\tau \left( n,p\right)+2. \label{xnp}
\end{equation}
For $n=0$, one recovers the previous scaling dimension
\begin{eqnarray}
\nonumber
x(n,p=0)=x(n),\,\,
\tau(n,p=0) =\tau \left( n\right)=x\left(n\right) -2.
\end{eqnarray}
As in section \ref{sec.conform}, we use the fundamental KPZ mapping of the
CFT in the {\it   plane} ${{\mathbb C}}$  to the CFT on a
 random Riemann surface, i.e., in presence of 2D quantum gravity  \cite{KPZ}, and the
 universal functions $U$
and $V$, acting on conformal weights, which describe the map:
\begin{eqnarray}
U\left( x\right) &=&x\frac{x-\gamma}{1-\gamma} , \hskip2mm
V\left( x\right) =\frac{1}{4}\frac{x^{2}-\gamma^2}{1-\gamma}.
\label{UV}
\end{eqnarray}
with $V\left( x\right) = U\left(
\frac{1}{2}
\left( x+\gamma \right) \right)$. As before, the parameter $\gamma$
 is the solution of
$c=1-6{\gamma}^2(1-\gamma)^{-1}, \gamma \leq 0.$

For the purely harmonic exponents $x(n)$,
describing the mutually-avoiding set ${\partial\mathcal C}\wedge n$, we have seen in
Eqs.~(\ref{finaa}) and (\ref{xtilde}) that
\begin{eqnarray}
x(n)&=&2V\left[2 U^{-1}\left(
\tilde{x_1}\right)
 +U^{-1}\left( n \right) \right],  
\label{xcn}
\end{eqnarray}
where $U^{-1}\left( x\right)$ is the positive inverse of $U$,
\begin{eqnarray}
\nonumber
2 U^{-1}\left( x\right)
=\sqrt{4(1-\gamma)x+\gamma^2}+\gamma\, .
\label{u1DB}
\end{eqnarray}
In (\ref{xcn}), we recall that the arguments $\tilde{x_1}$ and $n$ are respectively the {\it   boundary} scaling dimensions
(b.s.d.) (\ref{xtilde}) of the simple path ${\mathcal S}_1$ representing a
semi-infinite random frontier (such that ${\partial\mathcal C}= {\mathcal
S}_1\wedge{\mathcal S}_1$),
 and of the packet of $n$ Brownian paths, both diffusing into the upper {\it   half-plane} ${\mathbb H}$.
The function $U^{-1}$ transforms these half-plane b.s.d's into the corresponding b.s.d.'s in quantum
gravity, the {\it linear combination} of which gives, still in QG, the
b.s.d. of the mutually-avoiding set ${\partial\mathcal
C}\wedge n=(\wedge{\mathcal S}_1)^2\wedge n$. The function $V$ finally
maps the latter b.s.d. into the scaling dimension in ${\mathbb C}$.
The path b.s.d. $\tilde{x_1}$ (\ref{xtilde}) obeys $U^{-1}\left(
\tilde{x_1}\right) =(1-\gamma)/2$.

\begin{figure}[tb]
\begin{center}
\vskip.3cm
\includegraphics[angle=0,width=.9\linewidth]{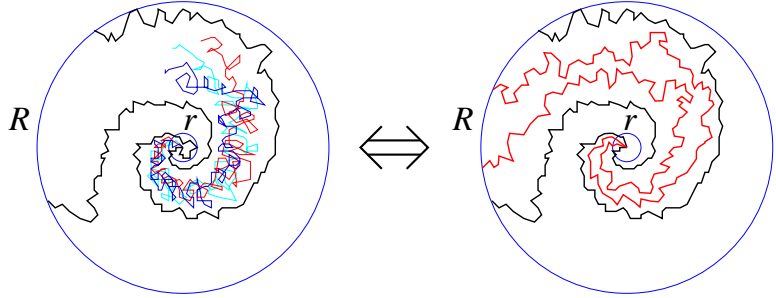}
\end{center}
\caption{Equivalence (\ref{kn}) between two simple paths in a mutually-avoiding configuration
${\mathcal S}_2={\mathcal S}_1 \wedge {\mathcal S}_1$, further avoided by a packet of $n$ 
independent Brownian motions, and $k(n)$ simple paths in a mutually-avoiding star configuration  ${\mathcal S}_{k(n)}$.}
\label{Fig.equiv}
\end{figure}

It is now useful to consider $k$ semi-infinite
random paths ${\mathcal S}_1$, joined at a single vertex in a
{\it   mutually-avoiding star} configuration ${\mathcal
S}_k=\stackrel{k}{\overbrace{{\mathcal S}_1\wedge{\mathcal
S}_1\wedge\cdots{\mathcal S}_1}}=(\wedge{\mathcal S}_1)^k$.
(In this notation the frontier near any of its points is a two-star
${\partial\mathcal C}={\mathcal S}_2$.)
The scaling dimension of ${\mathcal
S}_k$ can be obtained from the same b.s.d. additivity rule in quantum gravity, as in
(\ref{xa}) or (\ref{xcn})  \cite{duplantier11}
\begin{eqnarray}
x({\mathcal S}_k)&=&2V\left[ k\, U^{-1}\left( \tilde{x_1}\right) \right]\ .  
\label{xk}
\end{eqnarray}
The scaling dimensions (\ref{xcn}) and (\ref{xk}) coincide when
\begin{eqnarray}
\label{equal}
x(n)&=&x({\mathcal S}_{k(n)})\\
\label{kn}
k(n)&=& 2+\frac{U^{-1}\left( n \right)}{U^{-1}\left( \tilde{x_1}\right)}.
\end{eqnarray}
Thus we state the {\it   scaling star-equivalence}
\begin{eqnarray}
{\partial\mathcal C}\wedge n \Longleftrightarrow {\mathcal S}_{k(n)},
\label{equiv}
\end{eqnarray}
{\it   of two mutually-avoiding simple paths ${\partial\mathcal C}={\mathcal S}_2={\mathcal S}_1 \wedge {\mathcal S}_1$, 
further avoiding $n$ Brownian motions, to $k(n)$
simple paths in a mutually-avoiding star configuration ${\mathcal S}_{k(n)}$} (Fig.~\ref{Fig.equiv}). This 
equivalence plays an essential role in the computation of the complete rotation
spectrum (\ref{Phinp}).

\subsection{Rotation Scaling Exponents}
\label{rotationscaling}
 The Gaussian distribution of
the winding angle about the {\it   extremity} of a scaling path,
like ${\mathcal S}_1$, was derived in Ref. \cite{BDHSwinding}, using exact Coulomb
gas methods. The argument can be generalized to the winding angle
of a star ${\mathcal S}_k$ about its center  \cite{BDtopub}, where
one finds that the angular variance is reduced by a factor
$1/k^2$ (see also  \cite{wilson}). The
scaling dimension associated with the rotation scaling operator
$\Phi_{{\mathcal S}_k}e^{p\, {\rm  arg }\left({\mathcal S}_k\right)}$ is found
by analytic continuation of the Fourier transforms evaluated there  \cite{DB}:
\begin{eqnarray}
x({\mathcal S}_k;p)=x({\mathcal
S}_k)-\frac{2}{1-\gamma}\frac{p^2}{k^2}  \ , \nonumber 
\end{eqnarray}
i.e., is given by a quadratic shift in the star scaling exponent. To calculate
the scaling dimension (\ref{xnp}), it is 
sufficient to use the star-equivalence (\ref{kn}) above to
conclude that
\begin{eqnarray}
x(n,p)=x({\mathcal
S}_{k(n)};p)=x(n)-\frac{2}{1-\gamma}\frac{p^2}{k^2(n)}  \ ,
\nonumber
\end{eqnarray}
which is the key to our problem. Using Eqs.~(\ref{kn}),
(\ref{xcn}), and (\ref{UV}) gives the
useful identity:
\begin{eqnarray}
\frac{1}{8}({1-\gamma})k^2(n)=x(n)-2+b \ ,
\nonumber
\end{eqnarray}
with $b=\frac{1}{2}\frac{(2-\gamma)^2}{1-\gamma}=\frac{25-c}{12}$.
Recalling (\ref{xnp}), we arrive at the multifractal result:
\begin{eqnarray}
\tau(n,p)=\tau(n)-\frac{1}{4}\frac{p^2}{\tau(n)+b}  \ ,
\label{taunp}
\end{eqnarray}
where $\tau(n)=x(n)-2$ corresponds to the purely harmonic
spectrum with no prescribed rotation.

\subsection{Legendre Transform} 

The structure of the full
$\tau$-function (\ref{taunp}) leads by a formal Legendre transform
(\ref{legendre}) directly to the identity
\begin{eqnarray}
 f(\alpha,\lambda)&=&(1+\lambda^2)f(\bar{\alpha})-b\lambda^2 \ ,
 \nonumber
\end{eqnarray}
where $f(\bar{\alpha})\equiv\bar{\alpha}n-\tau(n)$, with $\bar \alpha={d{\tau}(n)
}/{dn}$, is the purely
harmonic MF function. It depends on the natural reduced variable $\bar{\alpha}$
{\it   \`a la} Beurling ($\bar{\alpha} \in \left[\frac{1}{2},+\infty \right)$)
\begin{eqnarray}
\nonumber
\bar{\alpha}&:=&\frac{\alpha}{1+\lambda^2}=\frac{d{x}
}{dn}\left( n\right)=\frac{1}{2} +\frac{1}{2}
\sqrt{\frac{b}{2n+b-2}}\, ,
\end{eqnarray}
whose expression emerges explicitly from (\ref{xcn}). Whence Eq.(\ref{scalinglaw}), {\bf QED}.

It is interesting to consider also higher multifractal spectra
 \cite{BDjsp}. For a confor\-mally-invariant scaling curve which is simple,
i.e., without double points, like the external frontier $\partial
\mathcal C$, here taken alone, define the universal function
$f_2(\alpha,\alpha',\lambda)$ which gives the Hausdorff dimension of the
points where the potential varies jointly with distance $r$ as
$r^{\alpha}$ on one side of the curve, and as $r^{\alpha'}$ on
the other, given a winding at rate $\lambda$. This function is
\begin{eqnarray}
f_2\left( \alpha, \alpha'; \lambda
\right)&=&b-\frac{1}{2(1-\gamma)}
{\left(\frac{1}{1+\lambda^2}-\frac{1}{2\alpha}-\frac{1}{2\alpha'}\right)}^{-1} \nonumber \\
& &-\frac{b-2}{2}\left(\alpha+\alpha'\right), \label{f_2twistexp}
\end{eqnarray}
and satisfies the generalization of scaling relation
(\ref{scalinglaw})
\begin{eqnarray}
 {f}_2(\alpha,\alpha';\lambda)&=&(1+\lambda^2){f}_2(\bar{\alpha},{\bar\alpha'};0)-b\lambda^2 \ .
\label{f_2twist}
\end{eqnarray}

\section{\sc{$O(N)$ \& Potts Models and the Stochastic L\"owner Evolution}}
\label{sec.geodual}
\subsection{{Geometric Duality in $O(N)$ and Potts Cluster Frontiers}}

\subsubsection*{$O(N)$ Model}
The $O(N)$ model partition function is that of a gas $\mathcal G$
of self- and mutually-avoiding {\it loops} on a given lattice,
e.g., the hexagonal lattice  \cite{nien}:
\begin{equation}
{Z}_{O(N)} = \sum_{\mathcal G }K^{{\mathcal N}_{B}}N^{{\mathcal
N}_{P}},
\label{ZON'}
\end{equation}
where $K$ and $N$ are two fugacities, associated respectively with the
total number of occupied bonds ${\mathcal N}_{B}$, and with the
total number of loops ${\mathcal N}_{P}$,
i.e., polygons drawn on the lattice. 
For $N \in [-2,2]$, this model possesses a critical point (CP),
$K_c$, while the whole {\it ``low-temperature''} (low-$T$) phase,
i.e., ${K}_c < K$, has critical universal properties, where the
loops are {\it denser} than those at the critical
point \cite{nien}.

\subsubsection*{Potts Model}
The partition function of the $Q$-state Potts model \cite{Potts} on, e.g., the
square lattice, with a second order critical point for $Q \in
[0,4]$, has a Fortuin-Kasteleyn (FK) representation  {\it at} the CP \cite{FK}: $
Z_{\rm Potts}=\sum_{\cup (\mathcal C)}Q^{\frac{1}{2}{\mathcal
N}_{P}},$ where the configurations $\cup (\mathcal C)$ are those
of unions of FK clusters on
the square lattice, with 
a total number ${\mathcal N}_{P}$ of polygons encircling all
clusters, and filling the medial square lattice of the original
lattice  \cite{dennijs,nien}. Thus the critical Potts model becomes
a {\it dense} loop model, with loop fugacity
$N=Q^{\frac{1}{2}}$, while one can show that its {\it tricritical}
point with site dilution \cite{NienPT} corresponds to the $O(N)$ CP \cite{D6,D7}.  
The {\it geometrical} clusters,  made of like spins in the 
(critical) Potts model,  have been argued to also correspond to the tricritical branch of the Potts model with dilution, 
\cite{JS}. Their frontiers are then the 
critical loops of the 
corresponding 
$O(N)$ model with the same central charge, in agreement with RefS. \cite{D6,D7}.

\subsubsection*{Coulomb Gas}
The $O(N)$ and Potts models thus possess the same ``Coulomb gas''
representations  \cite{dennijs,nien,D6,D7}:
\begin{equation}
N=\sqrt{Q}=-2 \cos \pi g, \nonumber
\end{equation}
with $g \in [1,\frac{3}{2}]$ for the $O(N)$ CP or tricritical Potts model, and $ g \in
[\frac{1}{2},1]$ for the low-$T$ $O(N)$ or critical Potts
models; the coupling constant $g$ of the Coulomb gas also yields
the central charge:
\begin{equation}
c=1-6{(1-g)^2}/{g}. \label{cg}
\end{equation}
Notice that from the expression (\ref{cgamma}) of $c$ in
terms of $\gamma \leq 0$ one arrives at the simple relation:
\begin{equation}
\gamma=1-g,\ g \geq 1;\ \gamma=1-1/g,\ g\leq 1. \label{ggamma}
\end{equation}
The above representation for $N=\sqrt Q \in [0,2]$ gives a range
of values $- 2 \leq c \leq 1$; our results also apply for $c
\in(-\infty, -2]$, corresponding, e.g., to the $O\left(N\in
[-2,0]\right)$ branch, with a low-$T$ phase for $g \in
[0,\frac{1}{2}]$, and CP for $g \in [\frac{3}{2},2].$

\subsubsection*{Hausdorff Dimensions of  Hull Subsets}
The fractal dimension $D_{\rm EP}$ of the accessible perimeter,
Eq.~(\ref{D(c)}), is, like $c(g)=c(g^{-1})$, a symmetric function of $g$ and $g^{-1}$ once
 rewritten in terms of $g$:
\begin{equation}
D_{\rm EP}=1+ \frac{1}{2}g^{-1}\vartheta(1-g^{-1})+\frac{1}{2}g
\vartheta(1-g), \label{DEP}
\end{equation}
where $\vartheta$ is the Heaviside distribution. Thus $D_{\rm EP}$ is given by two
different analytic expressions on either side of the separatrix
$g=1$. The dimension of the full hull, i.e., the
complete set of outer boundary sites of a cluster, has been
determined for $O(N)$ and Potts clusters  \cite{SD}, and is
\begin{equation}
D_{\rm H}=1+\frac{1}{2}g^{-1}, \label{DH}
\end{equation}
for the {\it entire} range of the coupling constant $g \in
[\frac{1}{2},2]$. Comparing to Eq.~(\ref{DEP}), we therefore see
that the accessible perimeter and hull Hausdorff dimensions {\it coincide}
for $g\ge 1$, i.e., at the $O(N)$ CP (or for tricritical Potts
clusters), whereas they {\it differ}, namely $D_{\rm EP} < D_H$,
for $g < 1$, i.e., in the $O(N)$ low-$T$ phase, or for critical
Potts clusters. This is the generalization to any Potts model of
the effect originally found in percolation  \cite{GA}. This can be
directly understood in terms of the {\it singly connected} sites
(or bonds) where fjords close in the scaling limit. Their
dimension is given by \cite{SD}
\begin{equation}
D_{\rm SC}=1+\frac{1}{2}g^{-1}-\frac{3}{2}g. \label{DSC}
\end{equation}
For critical $O(N)$ loops, $g \in (1,2]$, so that $D_{\rm SC} <
0,$ hence there exist no closing fjords, thereby explaining the identity:
\begin{equation}
D_{\rm EP} = D_{\rm H}. \label{id}
\end{equation}
In contrast, one has  $g \in [\frac{1}{2},1)$ and $D_{\rm SC} > 0$ for critical Potts
clusters and for the $O(N)$ low-$T$ phase. In this case, pinching points of
positive dimension appear in the scaling limit,
so that $D_{\rm EP} < D_{\rm H}$ (Table 1).\\

\begin{table}[tb]
\begin{tabular}{| c | c | c | c | c | c |}
\hline
$Q$          &     0     &    1     &      2      &      3      &  4          \\
\hline
$c$          &     -2    &    0     &     1/2     &      4/5       &  1        \\
\hline
$D_{\rm EP}$ & ${5}/{4}$ &${4}/{3}$ & ${11}/{8}$  & ${17}/{12}$ & ${3}/{2}$ \\
\hline
$D_{\rm H}$  & $2$       &${7}/{4}$ & ${5}/{3}$   & ${8}/{5}$   & ${3}/{2}$ \\
\hline
$D_{\rm SC}$ & ${5}/{4}$ &${3}/{4}$ &$ {13}/{24}$ & ${7}/{20}$  &   $ 0$        \\
\hline
\end{tabular}
\vskip.5cm
\caption{{Dimensions for the critical $Q$-state Potts
model; $Q=0,1,2$ correspond  to spanning trees,
percolation and Ising clusters, respectively.}}
\end{table}

\subsubsection*{Duality}
We then find from Eq.~(\ref{DEP}), with $g\leq 1$:
\begin{equation}
\left(D_{\rm EP}-1\right) \left( D_{\rm H}-1\right)=\frac{1}{4}.
\label{duali}
\end{equation}
The symmetry point $D_{\rm EP} = D_{\rm H}=\frac{3}{2}$
corresponds to $g=1$, $N=2$, or $Q=4$, where, as expected, the
dimension $D_{\rm SC}=0$ of the pinching points vanishes.

For percolation, described either by $Q=1$, or by the low-$T$
$O(N=1)$ model with $g=\frac{2}{3}$, we recover the result
$D_{\rm EP}=\frac{4}{3}$, recently derived in Ref. \cite{ADA}. For the
Ising model, described either by $Q=2, g=\frac{3}{4}$, or by the
$O(N=1)$ CP with $g'=g^{-1}=\frac{4}{3}$, we observe that the EP
dimension $D_{\rm EP}=\frac{11}{8}$ coincides, as expected, with
that of critical $O(N=1)$ loops, which in fact appear as EP's.
This is a particular case of a further duality relation between
the critical Potts and CP $O(N)$ models:
\begin{equation}
D_{\rm EP}\left(Q(g)\right)= D_{\rm
H}\left[O\left(N(g')\right)\right],\makebox{\rm for}\; g'=g^{-1}, g
\le 1\ .
\end{equation}
In terms of this duality, the central charge takes the simple
expression:
\begin{equation}
c=(3-2g)(3-2g'). \label{cdual}
\end{equation}
Exactly the same duality exists between the frontiers of Potts FK and geometrical clusters, as studied in Ref. \cite{JS}.

\subsection{{Geometric Duality of  ${\rm SLE}_{\kappa}$}}
\label{subsec.geoSLE}

\subsubsection*{Stochastic L\"owner Evolution}

\begin{figure}[htb]
\begin{center}
\includegraphics[angle=0,width=.72\linewidth]{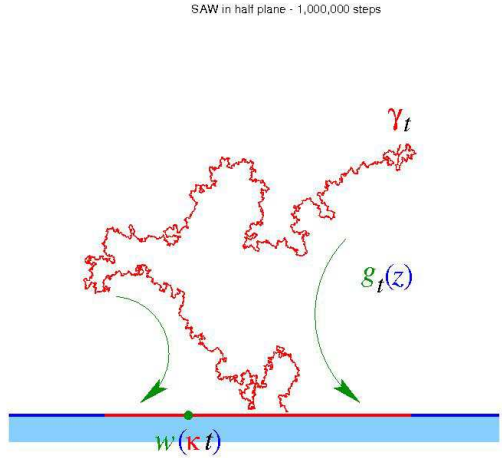}
\caption{The trace $\gamma{[0,t ]}$ of the chordal SLE process $\gamma_t$ up to time $t$,  and the Riemann map ${g}_{{t}}(z)$ 
which maps the slit half-plane ${\mathbb H}\backslash \gamma{[0,t ]}$ to ${\mathbb H}$. The image of 
$\gamma_t$ is the Brownian motion ${w}({\kappa}\, {t})$, ($w$ is standard one-dimensional Brownian motion).}
\end{center}
\label{fig.newhalfa2}
\end{figure}

  An introduction to the stochastic L\"owner evolution process (${\rm SLE}_{\kappa}$)
 can be found in  Refs. \cite{lawlerbook}, \cite{lawleresi}, \cite{stflour}. Here we consider the so-called {\it chordal} SLE in the complex 
 half-plane. A similar definition of radial SLE exists for the complex plane.
  
 The trace  $\gamma{[0,t ]}$ of this process  $\gamma_t $ is a conformally-invariant random path in the half-plane 
 ${\mathbb H}$. The Riemann conformal map ${g}_{{t}}(z): {\mathbb H}\backslash  \gamma{[0,t ]}\to {\mathbb H}$, from the slit 
 half-plane to  ${\mathbb H}$ itself, obeys the stochastic L\"owner equation \cite{schramm1}
 $$\partial_{{t}}
{g}_{{t}}(z)=\frac{2}{{g}_{{t}}(z)-{w}({\kappa}\, {t})},$$
where $w(\kappa t)$ is a one-dimensional Brownian motion on the real line ${\mathbb R}=\partial  {\mathbb H}$, with diffusion constant 
$\kappa \in [0,\infty ).$  The map is normalized (``hydrodynamic'' normalization) and the curve parameterized 
with respect to time $t$ (half-plane capacity para\-metrization),  
 so that ${g}_{{t}}(z)$ has 
the asymptotic behavior at infinity:
$${g}_{{t}}(z)=z+\frac{2t}{z} +{\mathcal O} (1/z^2), \,\,\, z\to \infty,$$
for all $t>0$.

The random path can be a simple or a non-simple path
 with self-contacts.
 The ${\rm SLE}_{\kappa}$ is parameterized by $\kappa$, which describes the rate of the auxiliary Brownian
 motion along the boundary, which is the source for the process. When $\kappa \in [0,4]$, the random curve is simple,
 while for $\kappa \in (4,8)$, the curve is a self-coiling path \cite{RS}.
 For $\kappa \geq 8$ the path is space filling.
 
The trace of this SLE process 
 essentially describes the boundaries of (Potts) clusters or {\it hulls}
 we have introduced above, or the random lines of the $O(N)$ model. 
 The correspondence to the previous parameters,
 the central charge $c$, the string susceptibility exponent $\gamma$, or
 the Coulomb gas constant $g$, is as follows.

 In the original work by Schramm  \cite{schramm1}, the variance
  of the Gaussian winding angle $\vartheta$ of the single extremity of a ${\rm SLE}_{\kappa}$ of size $R$ was calculated,
 and found to be $$\langle \vartheta^2\rangle ={\kappa} \,{\rm ln} R.$$ In  \cite{BDHSwinding} we found, for instance
 for the extremity of a random line in the $O(N)$ model,
 the corresponding angular variance $$\langle \vartheta^2\rangle=(4/g)\, {\rm ln} R,$$ from which we immediately infer the identity
 \begin{equation}
 \kappa=\frac{4}{g}\ .
 \label{k}
 \end{equation}

The low-temperature branch $g \in [\frac{1}{2},1)$ of the $O(N)$
model, for $N\in [0,2)$, indeed corresponds to $\kappa \in (4,8]$
and describes non-simple curves, while $N\in [-2,0], g\in
[0,\frac{1}{2}]$ corresponds to $\kappa \geq 8$. The critical
point branch $g \in [1,\frac{3}{2}], N\in [0,2]$
 gives $\kappa \in [\frac{8}{3},4]$, while $g \in [\frac{3}{2},2], N\in [-2,0]$
 gives $\kappa \in [2,\frac{8}{3}]$.
The range $\kappa \in [0,2)$ probably corresponds to higher
multicritical points with $g>2$. Owing to Eq.~(\ref{ggamma}) for
$\gamma$, we have
\begin{eqnarray}
\label{gk}
\gamma&=&1-\frac{4}{\kappa},\ \kappa \leq 4\ ;\\
\label{gkd}
\gamma&=&1-\frac{\kappa}{4},\ \kappa \geq 4\ .
\end{eqnarray}

\subsubsection*{Duality}
 The central charge (\ref{cgamma}) or (\ref{cg}) is accordingly:
\begin{equation}
c=1-24{\left(\frac{\kappa}{4}-1\right)^2}/{\kappa}\ , \label{ck}
\end{equation}
an expression which of course is symmetric under the {\it duality}
$\kappa/4 \to 4/\kappa=\kappa'$, or
\begin{equation}
\kappa \kappa'=16\ ,
\label{duality}
\end{equation}
 reflecting the symmetry under $gg'=1$  \cite{duplantier11}.
The self-dual form of the central charge is accordingly:
\begin{equation}
c=\frac{1}{4}(6-\kappa)(6-\kappa'). \label{cdualSLE}
\end{equation}
From Eqs.~(\ref{DH}) and (\ref{DEP}) we respectively find \cite{duplantier11}
\begin{equation}
D_{\rm H}=1+\frac{1}{8}\kappa\ , \label{DHs}
\end{equation}
\begin{equation}
D_{\rm EP}=1+ \frac{2}{\kappa}\vartheta(\kappa-4)+\frac{\kappa}{8}
\vartheta(4-\kappa)\ , \label{DEPs}
\end{equation}
the first result being later derived rigorously in probability theory
 \cite{RS,beffara}.

For $\kappa \leq 4$, we have $D_{\rm
EP}(\kappa)=D_{\rm H}(\kappa)$. For $\kappa \geq 4$, the
self-coiling scaling paths obey the duality equation
(\ref{duali}) derived above, recast here in the context of the
${\rm SLE}_{\kappa}$ process:
\begin{equation}
\left[D_{\rm EP}(\kappa)-1\right] \left[ D_{\rm
H}(\kappa)-1\right]=\frac{1}{4},\ \kappa \geq 4\ ,
\label{dualibis}
\end{equation}
where now $$D_{\rm EP}(\kappa)=D_{\rm H}(\kappa'=16/\kappa)\quad
\kappa'\leq 4\ .$$ Thus we predict that the external perimeter of
a self-coiling ${\rm SLE}_{\kappa \geq 4}$ process is, by {\it
duality}, the simple path of the ${\rm SLE}_{(16/{\kappa})=\kappa' \leq
4}$ process.

The symmetric point $\kappa=4$ corresponds to the $O(N=2)$ model,
or $Q=4$ Potts model, with $c=1$. The value $\kappa=8/3, c=0$
corresponds to a self-avoiding walk, which thus appears
 \cite{duplantier9,ADA} as the external frontier of a $\kappa=6$
process, namely that of a percolation hull
 \cite{schramm1,smirnov1}.

 Let us now study more of the ${\rm SLE}$'s random geometry
 using the quantum gravity method described here.

Up to now, we have described general conformally-invariant curves in the plane in terms of the
universal parameters $c$ (central charge) or $\gamma$ (string susceptibility). The multifractal
results described in the
sections above thus apply to the ${\rm SLE}$ after substituting $\kappa$ for $\gamma$ or $c$. Care should be taken,
however, in such a substitution since two dual values of $\kappa$ (\ref{duality}) correspond to a same value of $\gamma$.
The reason is that up to now we have considered boundary geometrical properties which actually were
{\it self-dual}. An exemple is the harmonic multifractal spectrum of the ${\rm SLE}_{\kappa \geq 4}$ frontier, which
is identical to that of the smoother (simple) ${\rm SLE}_{(16/{\kappa})=\kappa' \leq 4}$ path. So we actually saw only
the set of simple SLE traces with $\kappa \leq 4$. When dealing with higher multifractality, we assumed the random
curves to be simple. When dealing with non-simple random paths, boundary quantum gravity rules are to be
modified as explained now.

\section{\sc{Quantum Gravity Duality and SLE}}
\label{sec.duality}
\subsection{Dual Dimensions}
\label{sec.dualityQG}
It will be convenient to introduce the following notations. The standard KPZ map reads:
\begin{equation}
x=U_{\gamma}(\Delta)=\Delta \frac{\Delta-\gamma}{1-\gamma}\ ,
\label{KPZS}
\end{equation}
where $x$ is a planar conformal dimension and $\Delta$ its quantum gravity counterpart, and
where we recall that $\gamma$ is the negative root of
\begin{equation}
c=1-6{\gamma}^2(1-\gamma)^{-1}, \gamma \leq 0. \label{cgammaa}
\end{equation}
We introduce the {\it dual quantum dimension} of $\Delta$, $\Delta'$ such that:
\begin{equation}
\Delta' := \frac{\Delta-\gamma}{1-\gamma}\ ,
\label{dualdelta}
\end{equation}
and \begin{equation}
x=U_{\gamma}(\Delta)=\Delta \Delta'\ .
\label{dd'}
\end{equation}
Similarly, let us define the variable $\gamma'$, dual of susceptibility exponent $\gamma$, by:
\begin{equation}
(1-\gamma)(1-\gamma')=1\ ,
\label{dualgamma}
\end{equation}
which is simply the (``non-physical'') positive root of Eq.~(\ref{cgammaa}):
\begin{equation}
c=1-6{\gamma'}^2(1-\gamma')^{-1}, \gamma' \geq 0. \label{cgammaa'}
\end{equation}
The dual equation of (\ref{dualdelta}) is then:
\begin{equation}
\Delta = \frac{\Delta'-\gamma'}{1-\gamma'}\ ,
\label{dual'delta}
\end{equation}
By construction we have the simultaneous equations:
\begin{equation}
\Delta=U_{\gamma}^{-1}(x),\, \Delta'=\frac{U_{\gamma}^{-1}(x)-\gamma}{1-\gamma},
\label{dualU}
\end{equation}
with the positive solution
\begin{equation}
U_{\gamma}^{-1}\left( x\right)
=\frac{1}{2}\left(\sqrt{4(1-\gamma)x+\gamma^2}+\gamma\right) .
\label{U1aa}
\end{equation}

We define a dual KPZ map $U_{\gamma'}$ by the same equation as (\ref{KPZS}), with ${\gamma'}$ substituted
for ${\gamma}$.
It has the following properties\footnote{It generalizes to any operator the so-called ``wrong'' 
KPZ gravitational dressing  
of the boundary identity operator \cite{KS}.}:
\begin{eqnarray}
\label{KPZdual}
x&=&U_{\gamma}(\Delta)=U_{\gamma'}(\Delta')\ ,\\
\label{KPZdualinv}
\Delta'&=&U_{\gamma'}^{-1}(x)=\frac{U_{\gamma}^{-1}(x)-\gamma}{1-\gamma}\ ,\\
\label{KPZdualinv'}
\Delta&=&U_{\gamma}^{-1}(x)=\frac{U_{\gamma'}^{-1}(x)-\gamma'}{1-\gamma'}\ .
\end{eqnarray}

\subsubsection*{Boundary KPZ for Non Simple Paths}
The additivity rules in quantum gravity for the boundary scaling dimensions of mutually-avoiding random paths $A$ and $B$
 are:
\begin{eqnarray}
\label{sa}
\tilde\Delta\left( A\wedge B\right)&=&\tilde \Delta(A)+\tilde\Delta(B)\,\,\,\,\,\,\, {\rm{(simple\,\, paths),}}\\
\label{nsa}
\tilde\Delta'\left( A\wedge B\right)&=&\tilde\Delta'(A)+\tilde\Delta'(B)\,\,\,\, {\rm{(non-simple\,\, paths).}}
\end{eqnarray}
For simple paths, like random lines in the $O(N)$ model at its critical point, or the SLE trace for $\kappa \leq 4$ 
the boundary dimensions are additive 
in quantum gravity, a fundamental fact repeatedly used 
above. On the other hand, for non-simple paths, the {\it dual dimensions are additive} in boundary quantum gravity.
 This is the case of random lines in the dense phase of the $O(N)$ model, or, equivalently, of hulls of
 Fortuin-Kasteleyn clusters in the Potts model, or of the ${\rm SLE}_{\kappa \geq 4}$ trace.
 These additivity rules are derived
  from the consideration of partition functions on a random surface in the dilute or dense phases.
 (See \cite{BDMan}, Appendices B \& C.)

The composition rules for non-simple paths are different
from the ones for simple paths, when written in terms of the standard string
susceptibility exponent $\gamma$, but they are formally identical in terms of the dual exponent $\gamma'$.

\subsubsection*{Bulk KPZ for Non-Simple Paths}
For determining the complete set of scaling dimensions, it remains to relate bulk and boundary dimensions.
In the dilute phase, i.e., for simple paths, we have seen the simple relation in a random metric
(see Appendix C in Ref. \cite{BDMan}):
\begin{equation}
2\Delta -\gamma=\tilde \Delta \ .
\label{bulkboundary}
\end{equation}
 The KPZ map from boundary dimension in quantum gravity to
bulk dimension in the plane reads accordingly
\begin{eqnarray}
x=2 U_{\gamma}(\Delta)=2 U_{\gamma}\left(\frac{1}{2}{(\tilde\Delta+\gamma)}\right)
=2V_{\gamma}(\tilde\Delta),
\label{VD}
\end{eqnarray}
where
\begin{eqnarray}
\label{VV}
V_{\gamma}(x)&=&\frac{1}{4}\frac{x^2-\gamma^2}{1-\gamma}\ ,
\end{eqnarray}
an expression repeatedly used above. When dealing with non-simple paths,
these relations have to be changed to:
\begin{equation}
2\Delta =\tilde \Delta \ ,
\label{bulkboundarydense}
\end{equation}
as shown in detail in Ref. \cite{BDMan}. At this stage, the reader will
not be surprised that this relation is just identical to the dual of (\ref{bulkboundary})
\begin{equation}
2\Delta' -\gamma'=\tilde \Delta' \ ,
\label{bulkboundarydual}
\end{equation}
when now written in terms of both dual dimensions and susceptibility exponent.
As a consequence, the scaling dimension of a bulk operator in a dense system reads:
\begin{eqnarray}
x=2 U_{\gamma}(\Delta)=2 U_{\gamma}\left(\frac{1}{2}{\tilde\Delta}\right)
=\frac{1}{2}\tilde\Delta\frac{\tilde\Delta -2\gamma}{1-\gamma},
\label{UD}
\end{eqnarray}
which by duality can necessarily be written as:
\begin{eqnarray}
\label{xdensedual}
x&=&2 V_{\gamma'}(\tilde\Delta'),\\
\nonumber
V_{\gamma'}(x)&=&\frac{1}{4}\frac{x^2-{\gamma'}^2}{1-\gamma'}\ ,
\end{eqnarray}
as can be easily checked. This QG duality is analyzed in greater detail in Ref.~\cite{BDMan}.

In summary, the composition rules for scaling dimensions,
whether on a boundary or in the bulk, take a unique analytic form for both phases (simple or non-simple paths), 
provided one replaces the string susceptibility exponent $\gamma$ in the simple case 
by its dual variable $\gamma'$ in the non-simple case, and QG dimensions by their duals. This
 applies to the dense phase of the $O(N)$ model, or to
Potts cluster boundaries, and in particular to ${\rm SLE}_{\kappa \geq 4}$.


\subsection{\sc{KPZ for SLE}}
\label{sec.SLEKPZ}

\subsubsection*{QG Duality for SLE}
The QG duality is perfectly adapted to the parametrization of the ${\rm SLE}_{\kappa}$ process.
Indeed we have from (\ref{gk}) and (\ref{gkd})
\begin{eqnarray}
\label{gg'k}
\gamma&=&1-\frac{4}{\kappa},\ \gamma'=1-\frac{\kappa}{4},\, \kappa \leq 4;\\
\label{gg'kd}
\gamma&=&1-\frac{\kappa}{4},\ \gamma'=1-\frac{4}{\kappa} \ ,\kappa \geq 4,
\end{eqnarray}
so that the analytical forms of $\gamma$ and its dual $\gamma'$ are simply exchanged when passing from simple paths
($\kappa \leq 4$)
to non-simple ones ($\kappa > 4$). Because of the equivalent dual equations (\ref{KPZdual}), by choosing either the 
$\gamma$-solution or the $\gamma'$-solution, depending whether $\kappa \leq 4$ or $\kappa \geq 4$, we can write 
\begin{equation}
x=\left\{\begin{array}{ll}
U_{\gamma(\kappa \leq 4)}(\Delta)={\mathcal U}_{\kappa}(\Delta) & \mbox{$\kappa \leq 4$}\\
U_{\gamma'(\kappa \geq 4)}(\Delta')={\mathcal U}_{\kappa}(\Delta') & \mbox{$\kappa \geq 4$,}
\end{array}
\right.
\label{KPZSLE}
\end{equation}
 with now a single function, valid for all values of parameter $\kappa$
\begin{eqnarray}
{\mathcal U}_{\kappa}(\Delta)=\frac{1}{4}\Delta\left({\kappa}\Delta +4-{\kappa}\right).
\label{USLE}
\end{eqnarray}
Similarly, the inverse KPZ map (\ref{U1aa}) reads, according to (\ref{KPZdualinv}) or (\ref{KPZdualinv'}):
\begin{eqnarray}
\nonumber
\Delta&=&U_{\gamma(\kappa \leq 4)}^{-1}\left( x\right)={\mathcal U}_{\kappa}^{-1}\left( x\right),\,\,\,\kappa \leq 4,\\
\label{KPZSLEinv}
\Delta'&=&U_{\gamma'(\kappa \geq 4)}^{-1}\left( x\right)={\mathcal U}_{\kappa}^{-1}\left( x\right),\,\kappa \geq 4,
\end{eqnarray}
again with a single expression of the inverse function, valid for any $\kappa$
\begin{eqnarray}
{\mathcal U}_{\kappa}^{-1}\left( x\right)=
\frac{1}{2\kappa}\left(\sqrt{16\kappa x+(\kappa-4)^2}+\kappa-4\right).
\label{U-1SLE}
\end{eqnarray}
I emphasize that ${\mathcal U}_{\kappa}$ coincides with the KPZ map for $\kappa \leq 4$, while it
 represents the dual of the latter when $\kappa \geq 4$ and then acts on the dual dimension $\Delta'$. For instance,
  we have the important result at the origin
\begin{equation}
{\mathcal U}_{\kappa}^{-1}\left( 0\right)=\frac{1}{2\kappa}\left[\,|\kappa-4| +\kappa -4\right]
=\left(1-\frac{4}{\kappa}\right) \vartheta (\kappa  -4),
\label{U01}
\end{equation}
which vanishes for simple paths, and is non-trivial for non-simple ones.

 It remains to define
the analogue of the $V$ function (\ref{VV}) or its dual (\ref{xdensedual}):
\begin{eqnarray}
x=\left\{\begin{array}{ll}
2V_{\gamma(\kappa \leq 4)}(\tilde\Delta)=2{\mathcal V}_{\kappa}(\tilde\Delta) & \mbox{$\kappa \leq 4$}\\
2V_{\gamma'(\kappa \geq 4)}(\tilde\Delta')=2{\mathcal V}_{\kappa}(\tilde\Delta') & \mbox{$\kappa \geq 4$,}
\end{array}
\right.
\label{KPZSLEV}
\end{eqnarray}
with again a single function, valid for all values of parameter $\kappa$
\begin{eqnarray}
\nonumber
{\mathcal V}_{\kappa}(\Delta)&=&{\mathcal U}_{\kappa}\left[\frac{1}{2}\left(\Delta+1-\frac{4}{\kappa}\right)\right]\\
&=&\frac{1}{16\kappa}\left[\kappa^2\Delta^2-(\kappa-4)^2\right],
\label{VSLE}
\end{eqnarray}
but acting on the boundary dimension in quantum gravity or on its dual, depending on whether 
$\kappa \leq 4$ or $\kappa \geq 4$.

\subsubsection*{Composition Rules for SLE}
Finally we can conclude with general composition rules for the SLE process. Indeed, the boundary rule in ${\mathbb H}$
(\ref{sa}) or
its dual (\ref{nsa}), owing to Eqs.~(\ref{KPZSLE}) and (\ref{KPZSLEinv}), read in a unified way
in terms of parameter $\kappa$:
\begin{eqnarray}
\label{tildexABk}
\tilde x(A\wedge B)&=&{\mathcal U}_{\kappa}\left[ {\mathcal U}_{\kappa}
^{-1}\left( \tilde{x}\left(A\right)\right)
+{\mathcal U}_{\kappa}^{-1}\left( \tilde{x}\left( B\right)\right)  \right],
\end{eqnarray}
valid for the entire range of $\kappa$. Similarly, the composition rules for SLE's in the plane ${\mathbb C}$ are found
from Eqs.~(\ref{bulkboundary}) or (\ref{bulkboundarydual}), and recast according to (\ref{KPZSLEV}) and (\ref{KPZSLEinv})
into a unified formula, valid for any $\kappa$
\begin{eqnarray}
\label{xABk}
x(A\wedge B)=2{\mathcal V}_{\kappa}\left[ {\mathcal U}_{\kappa}^{-1}\left( \tilde{x}\left(A\right)\right)
+{\mathcal U}_{\kappa}^{-1}\left( \tilde{x}\left( B\right)\right) \right].
\end{eqnarray}
Thus we see that by introducing dual equations we have been able to unify the composition rules for the SLE in a unique way, 
which no longer depends explicitly on the range of $\kappa$.
\subsection{Short Distance Expansion}
\begin{figure}[htb]
\begin{center}
\includegraphics[angle=0,width=.8\linewidth]{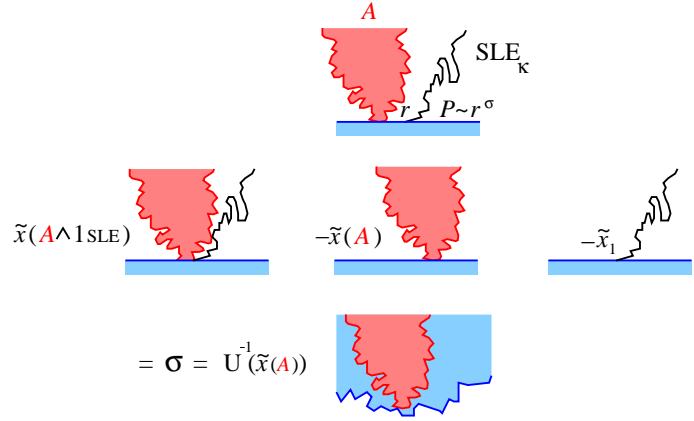}
\caption{Boundary contact exponent $\sigma$ between an arbitrary conformally invariant set $A$ and a chordal SLE. It
is given by fusion rules, and because
of the properties of the KPZ map it is identical with the QG boundary conformal dimension $U^{-1}({\tilde x}_A)$
of the set $A$ alone.}
\label{fig.sle2}
\end{center}
\end{figure}
\subsubsection*{Boundary SDE}
Consider the power law governing the behavior of two mutually-avoiding random paths $A$ and $B$ anchored at the Dirichlet boundary line, and 
approaching each other at short distance $r$ along the line. The probability of such an event scales like 
\begin{equation}
\tilde{\mathcal P}_{A, B}(r) \propto r^{\tilde x_{A,B}}, \,\, r \to 0,
\label{SDE}
\end{equation}
where the short-distance  exponent reads  \cite{BPZ, duplantier4}:
\begin{equation}
{\tilde x_{A,B}}= \tilde x(A\wedge B)-\tilde x(A)-\tilde x(B).
\label{SDEexp}
\end{equation}
We simply use the fusion rule (\ref{tildexABk}) and the quadratic map (\ref{USLE}) to immediately get
\begin{equation}
{\tilde x_{A,B}}= \frac{\kappa}{2}{\mathcal U}_{\kappa}^{-1}
\left( \tilde{x}_A\right)\,{\mathcal U}_{\kappa}^{-1}\left(\tilde{x}_B\right),
\label{SDEU}
\end{equation}
where we use $\tilde{x}_A=\tilde{x}\left(A\right)$ as a short-hand notation.
In terms of quantum gravity boundary dimensions, or their dual, this SDE exponent splits into
\begin{eqnarray}
{\tilde x_{A,B}}=\left\{\begin{array}{ll} \frac{\kappa}{2} \tilde\Delta_A\tilde\Delta_B  & \mbox{$\kappa \leq 4$} \\
\frac{\kappa}{2} \tilde\Delta'_A\tilde\Delta'_B & \mbox{$\kappa \geq 4$.}
\end{array}
\right.
\label{SDED}
\end{eqnarray}
So we see that the short-distance expansion (SDE) along the boundary of ${\mathbb H}$ is governed by the product of 
the quantum boundary dimensions, or of their duals, 
depending on the phase we are in. In particular, if one chooses the set $B$ to be the chordal SLE trace itself,
its boundary dimension $\tilde x_1=(6-\kappa)/2\kappa$ is such that
$\tilde \Delta_1=U_{\gamma}^{-1}(\tilde x_1)=\frac{1}{2}(1-\gamma)$ in the dilute phase, or $\tilde \Delta_1=U_{\gamma}^{-1}(\tilde x_1)=\frac{1}{2}+\gamma$ 
in the dense phase. That corresponds to the single expression ${\mathcal U}_{\kappa}^{-1}(\tilde x_1)={2}/{\kappa}$, 
 which is $\tilde \Delta_1$ for $\kappa \leq 4$ or  $\tilde \Delta_1^{'}$ for $\kappa \geq 4$. 
 In this case, the expressions (\ref{SDEU}) or (\ref{SDED}) simplify to
\begin{eqnarray}
\label{SDESLE}
{\tilde x_{A,1}}&=&{\mathcal U}_{\kappa}^{-1}
\left( \tilde{x}_A\right)=\frac{1}{2\kappa}\left(\sqrt{16\kappa \tilde{x}_A +(\kappa-4)^2}+\kappa-4\right)\\
\nonumber
&=&\left\{\begin{array}{ll}\tilde\Delta_A & \mbox{$\kappa \leq 4$}\\
\tilde\Delta'_A& \mbox{$\kappa \geq 4$.}
\end{array}
\right.
\end{eqnarray}
{\it The boundary contact exponent between an SLE and an arbitrary conformally invariant scaling set $A$ in the standard
(half-) plane
is therefore identical with the purely
gravitational boundary exponent of the set $A$!}

In a way, {in the standard plane, the local randomness of the SLE
acts exactly as quantum gravity for approaching scaling sets like $A$, when the latter have to avoid the SLE.}

 This explains the observation made in Ref. \cite{BB} that the boundary SDE of any operator with the SLE trace
might be seen as exhibiting (boundary) quantum gravity. However, we see that if for $\kappa \leq 4$
the SDE exponent (\ref{SDESLE}) is 
indeed the KPZ solution $\tilde \Delta$, for $\kappa \geq 4$ it necessarily transforms to the dual dimension 
$\tilde\Delta'$ introduced above 
in (\ref{dualdelta}) . The appearance of the simple quantum gravity dimension
results from the consideration of the SDE with a boundary SLE, since the 
general structure 
of SDE exponent (\ref{SDED}) is clearly still quadratic and given by the product of quantum gravity dimensions or their dual.

\subsubsection*{Bulk SDE}
One can also consider the SDE for random paths in the full plane, corresponding to the so-called radial SLE. 
Consider the power law governing the behavior of two mutually-avoiding random paths $A$ and $B$  
approaching each other at short distance $r$ in the plane, with probability  
\begin{equation}
{\mathcal P}_{A, B}(r) \propto r^{x_{A,B}}, \,\, r \to 0,
\label{SDEbulk}
\end{equation}
where the short-distance exponent now reads:
\begin{equation}
{x_{A,B}}= x(A\wedge B)-x(A)-x(B).
\label{SDEbexp}
\end{equation}
 In the case where the set $B$ is chosen to be the radial SLE trace itself, taken at a typical medial point,
 the expression simplify into
\begin{eqnarray}
{x_{A,2}}={\mathcal U}_{\kappa}^{-1}
\left( \tilde{x}_A\right) +\frac{(\kappa -4)^2}{8\kappa}
=\left\{\begin{array}{ll} \tilde\Delta_A +\frac{(\kappa -4)^2}{8\kappa} & \mbox{$\kappa \leq 4$}\\
 \tilde\Delta'_A +\frac{(\kappa -4)^2}{8\kappa} & \mbox{$\kappa \geq 4$.}
\end{array}
\right.
\label{SDESLEb}
\end{eqnarray}
So the SDE of the SLE trace with any operator $A$ in the plane again generates the boundary dimension of $A$
in quantum gravity or its dual, modulo a constant shift. Notice that this shift is self-dual 
with respect to $\kappa\kappa'=16$ and reads also
$\frac{(\kappa -4)^2}{8\kappa}=\frac{1-c}{12}$.

\subsection{Multiple Paths in $O(N)$, Potts Models and SLE}
\label{subsec.multiline}
\begin{figure}[htb]
\begin{center}
\includegraphics[angle=0,width=.35\linewidth]{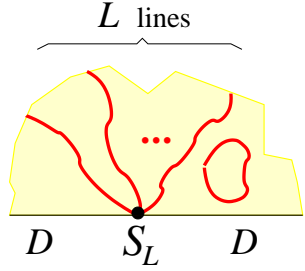}
\caption{{A boundary star ${\cal S}_L$ made of $L$ random lines in the $O(N)$ loop model with Dirichlet
boundary conditions. It can also be considered as an $L$-multiple SLE. (Courtesy of I. Kostov.)}}
\label{fig.SL}
\end{center}
\end{figure}
Let us consider the scaling dimensions associated with several (mutually-avoiding) random paths starting from a same small
neighborhood, also called star exponents in the above (Figure \ref{fig.SL}). It is simpler to first give them for the $O(N)$ model, before
transferring them to the SLE. These exponents can be derived explicitly from the quantum gravity
approach, in particular in presence of a boundary (see Appendix B in Ref. \cite{BDMan}). (See also refs.
\cite{DK,kostovgaudin,KK,cardy2003}.)

\subsubsection*{Multiple SLE's in QG}
Near the boundary of a random surface with Dirichlet conditions, the conformal dimensions read
\begin{eqnarray}
\label{tdeltaL}
\tilde\Delta_L=\frac{L}{2}(1-\gamma)\, ,\,\,\,\,\, \tilde\Delta_L^D&=&\frac{L}{2}+\gamma\, ,
\end{eqnarray}
 where the ``$D$'' superscript stands for the dense phase. The quantum bulk dimensions read similarly
\begin{eqnarray}
\label{deltaLON}
\Delta_L=\frac{L}{4}(1-\gamma)+\frac{\gamma}{2}\, ,\,\,\,\,\, \Delta_L^D&=&\frac{L}{4}+\frac{\gamma}{2}\, .
\end{eqnarray}
The dilute phase corresponds to (\ref{gk}) for $\kappa \leq 4$, while the dense one covers
(\ref{gkd}) with $\kappa \geq 4$:
\begin{eqnarray}
\label{tdeltaLk}
\tilde\Delta_L&=&\frac{2L}{\kappa},\;\;\;\;\;\;\;\;\;\;\;\;\;\; \Delta_L=\frac{1}{2\kappa}(2L+\kappa-4),\;\;  \kappa \leq 4 \\
\label{tdeltaLDk}
\tilde\Delta_L^D&=&\frac{L}{2}+1-\frac{\kappa}{4},\;\; \Delta_L^D=\frac{1}{8}\left(2L+4-\kappa\right),\;\; \kappa \geq 4.
\end{eqnarray}
By using dual dimensions (\ref{dualdelta}) for the dense phase, these results are unified into
\begin{eqnarray}
\label{tdeltaLkk}
\tilde\Delta_L&=&\frac{2L}{\kappa}\, ,\;\;\;\;\;\;\;\;\;\;\;\;\;\;\;\;\;\;\;\;\;\;\;\;\;\;\;\;{\kappa \leq 4} \\
\label{deltaLkk}
\Delta_L&=&\frac{1}{2\kappa}(2L+\kappa-4)\, ,\;\;\;\;\;\;\;\;\;\kappa \leq 4 \\
\label{tdeltaLDkd}
{{\tilde{\Delta}}_L}^D{}'&=&\frac{2L}{\kappa}\, , \;\;\;\;\;\;\;\;\;\;\;\;\;\;\;\;\;\;\;\;\;\;\;\;\;\;\;\;\kappa \geq 4\\
\label{deltaLDkd}
{\Delta}_L^D{}'&=&\frac{1}{2\kappa}(2L+\kappa-4)\, ,\;\;\;\;\;\;\;\;\; \kappa \geq 4.
\end{eqnarray}
Hence we observe that in the dense phase the dual dimensions play the role of the original ones in the dilute phase.

\subsubsection*{Multiple SLE's in ${\mathbb H}$ and ${\mathbb C}$}
The scaling dimensions $\tilde x_L$ in the standard complex half-plane ${\mathbb H}$, or $x_L$ in the complex
plane ${\mathbb C}$, can now be obtained
from the quantum gravity ones by the KPZ $U$-map (\ref{KPZS}), or, in the SLE formalism, from the $\mathcal U_\kappa$ (\ref{KPZSLE}) or
$\mathcal V_\kappa$ (\ref{KPZSLEV}) adapted KPZ maps. From the last equations (\ref{tdeltaLkk}) to (\ref{deltaLDkd}),
it is clear that by duality the analytic form of the dimensions stays the same in the two phases $\kappa \leq 4$, and
$\kappa \geq 4$. Indeed we get:
\begin{eqnarray}
\label{txLkk}
\tilde x_L&=&\mathcal U_\kappa(\tilde\Delta_L)=\frac{L}{2\kappa}(2L+4-\kappa)\, ,\;\;\;\;\;\;\;\;\;\;\;\;\;\;\;\;\;\;{\kappa \leq 4} \\
\label{xLkk}
 x_L&=&2\mathcal V_\kappa(\tilde\Delta_L)=
 \frac{1}{8\kappa}\left[4L^2-(4-\kappa)^2\right]\, ,\,\;\;\;\;\;\;\;\;\kappa \leq 4 \\
\label{txDkd}
\tilde x_L&=&\mathcal U_\kappa({{\tilde{\Delta}}_L}^D{}')=\frac{L}{2\kappa}(2L+4-\kappa)\, , \;\;\;\;\;\;\;\;\;\;\;\;\;\;\;\;\kappa \geq 4\\
\label{xLDkd}
x_L&=&2\mathcal V_\kappa({{\tilde{\Delta}}_L}^D{}')=\frac{1}{8\kappa}\left[4L^2-(4-\kappa)^2\right]
\, ,\;\;\;\;\;\;\; \kappa \geq 4.
\end{eqnarray}

\subsection{${\rm SLE}(\kappa,\rho)$ and Quantum Gravity}
\begin{figure}[htb]
\begin{center}
\includegraphics[angle=0,width=.95\linewidth]{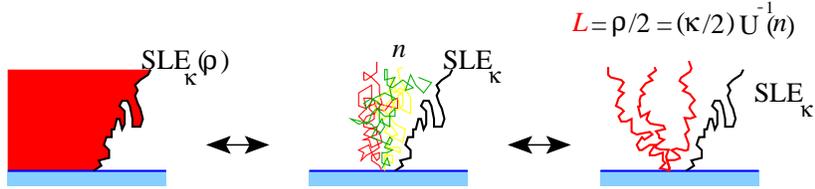}
\caption{{Left: The drift $\rho$ in ${\rm SLE}(\kappa,\rho)$ introduces a pressure that pushes the path further away from the left part
of the half-plane boundary. Middle: The $n$-Brownian packet, equivalent to the drift $\rho$, that is avoided by the
standard ${\rm SLE}_{\kappa}$. Right: The equivalence to the avoidance of a number $L=\rho/2$ of multiple
${\rm SLE}_{\kappa}$s.}}
\label{fig.sle1}
\end{center}
\end{figure}
An extension of the original SLE, the ${\rm SLE}(\kappa,\rho)$ stochastic process, has been introduced in Ref. \cite{CRLSW} 
(see also \cite{WW,dubedat,cardyrho,kyto}).
A drift term of strength $\rho$ is added to the boundary Brownian process that appears in
the L\"owner equation driving the uniformizing
Riemann map of the ${\rm SLE}_\kappa$ trace. For $\rho=0$, one recovers the
usual SLE process: ${\rm SLE}(\kappa,\rho=0)={\rm SLE}_{\kappa}.$
As a consequence, the chordal ${\rm SLE}(\kappa,\rho)$ feels an asymmetrical
``pressure'' that tends to push it away from one side of the Dirichlet boundary (Figure \ref{fig.sle1}).

As shown in Ref. \cite{WW}, the ${\rm SLE}(\kappa,\rho)$ is completely equivalent, in terms of conformal properties or critical exponents,
to a standard
 ${\rm SLE}_{\kappa}$ in the presence, on the same side of the boundary, of a packet of $n$ independent Brownian paths
 which are avoided by the
 ${\rm SLE}_{\kappa}$ trace, and exert a ``conformal pressure'' (Fig.~\ref{fig.sle1}). The value of $n$ is given by the formula
\begin{equation}
\rho=\kappa\, \mathcal U_\kappa^{-1}(n).
\label{pressure}
\end{equation}
We can then use the QG formalism to give yet another representation of the ${\rm SLE}(\kappa,\rho)$ process and
 give a simple meaning to parameter $\rho$ (\ref{pressure}).
 The equivalent Brownian packet associated with the ${\rm SLE}(\kappa,\rho)$ process can indeed be replaced by
 multiple SLE's. Multiple ${\rm SLE}_{\kappa}$s and a Brownian packet are conformally equivalent if and only if their
 boundary QG dimensions (for ${\kappa \leq 4}$), or their dual boundary QG dimensions (for ${\kappa \geq 4}$), coincide:
 \begin{eqnarray}
 \nonumber
\tilde\Delta_L&=&\frac{2L}{\kappa}=\mathcal U_\kappa^{-1}(n)\, ,\,\,\, {\kappa \leq 4} \\
\nonumber
{{\tilde{\Delta}}_L}^D{}'&=&\frac{2L}{\kappa}=\mathcal U_\kappa^{-1}(n)\, , \;\;\kappa \geq 4;
\end{eqnarray}
both cases yield naturally the same anaytical result. Therefore the parameter $\rho/2\equiv L$ {\it simply appears as  the number
$L$ of equivalent multiple ${\rm SLE}_{\kappa}$'s avoided by the original one} (Fig.~\ref{fig.sle1}) (See also \cite{kyto}.)

\subsubsection*{Contact Exponents for ${\rm SLE}(\kappa,\rho)$}
\begin{figure}[htb]
\begin{center}
\includegraphics[angle=0,width=.8\linewidth]{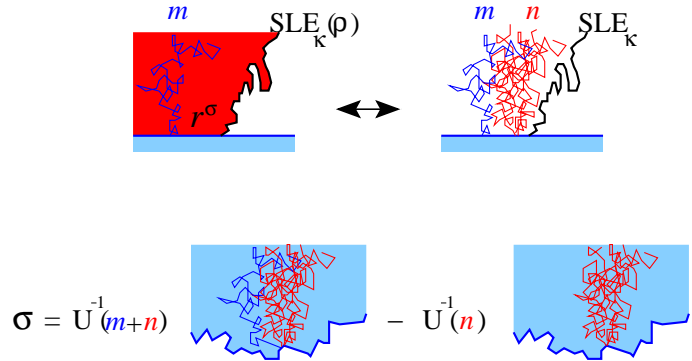}
\caption{{Top: Contact exponent of a packet of $m$ Brownian paths that avoids the trace of an
${\rm SLE}(\kappa,\rho)$. The $m$-packet overlaps with the equivalent $n$-packet associated with the drift parameter
$\rho$
and adds to the pressure exerted by the
latter onto the trace of ${\rm SLE}_\kappa.$
Bottom: The QG representation of exponent $\sigma$.}}
\label{fig.sle3}
\end{center}
\end{figure}
One can use this QG conformal equivalence to predict other properties of the composite ${\rm SLE}(\kappa,\rho)$ process.

A first question is that of the minimal value of the drift or pressure parameter $\rho$ such that the trace of
${\rm SLE}(\kappa,\rho)$ completely avoids the negative part $\partial {\mathbb H}^{-}$ of the half-plane boundary. For $\kappa \leq 4 $, the trace
of ${\rm SLE}_\kappa$  always avoids $\partial {\mathbb H}$, while for $\kappa > 4$ it always  bounces onto it.
The minimal value of $\rho$ simply corresponds to a minimal number $n=0$ of equivalent Brownian paths, whence:
\begin{equation}
\rho=\kappa\,{\mathcal U}_{\kappa}^{-1}\left( 0\right)
=\left({\kappa}-{4}\right) \vartheta (\kappa  -4),
\label{U0rho}
\end{equation}
where we used  (\ref{U01}). As expected, this minimal value for $\rho$ \cite{dubedat}
is non vanishing only for $\kappa > 4$.

Consider also the probability $P(r)$ that a packet of $m$ independent Brownian paths avoids a chordal
${\rm SLE}(\kappa,\rho)$, while starting at distance $r$ from it on the boundary (Fig.~\ref{fig.sle3}).
This probability scales as
$P(r) \approx r^{\sigma(m,\rho)}$, and the
contact exponent $\sigma(m,\rho)$ can be calculated with the help of the Brownian equivalence and of the
contact exponent (\ref{SDESLE}) for standard SLE (see figures \ref{fig.sle1} and \ref{fig.sle2}). One finds
$$\sigma(m,\rho)=\mathcal U_\kappa^{-1}(m+n)-\mathcal U_\kappa^{-1}(n),$$
where $n$ is given by $\rho=\kappa\,\mathcal U_\kappa^{-1}(n)$. Again a contact exponent, $\sigma$, acting in the
{\it standard} (half-) plane, actually {\it is} a quantum gravity exponent! (Fig.~\ref{fig.sle3}).

\subsection{{Multifractal Exponents for Multiple SLE's}}
\label{sec.multifSLE}

In section \ref{sec.conform} above we have
studied in detail the multifractal spectrum associated with the harmonic measure near a
conformally-invariant frontier, generalized to the mixed rotation spectrum in section \ref{sec.winding}.
We also looked at the double-sided distribution of potential near a simple fractal curve.
We have seen in previous
sections \ref{sec.dualityQG} and \ref{sec.SLEKPZ} how to
extend the formalism to non-simple curves, by using duality.
We now briefly apply it to some other spectra associated with the harmonic measure near multiple paths.
They include the so-called
SLE derivative exponents \cite{lawler4}.

\subsubsection*{Boundary Multifractal Exponents}
Let us start with geometrical properties of CI curves near the boundary
of ${\mathbb H}$ ({\it chordal} SLE).  We specifically look at
the scaling behavior of moments of the harmonic measure, or in SLE terms, of powers (of the modulus of) the derivative
of the Riemann map that maps the SLE trace back to the half-line ${\mathbb R}= \partial {\mathbb H}$
 \cite{lawler4,lawleresi,stflour}.

\begin{figure}[tb]
\begin{center}
\includegraphics[angle=0,width=0.3\linewidth]{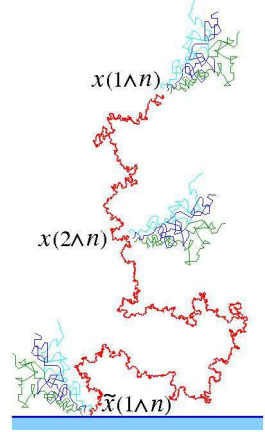}
\end{center}
\caption{Representation of harmonic moments by packets of
independent Brownian paths diffusing away from a single SLE trace, hence a $L=1$ star $\mathcal S_1$.
There are three locations to probe the harmonic measure:
at the SLE origin on the boundary, at the SLE tip in the plane, or along the fractal curve itself.
The corresponding scaling exponents are respectively $\tilde x(1\wedge n)$, $x(1\wedge n)$ and $x(2\wedge n)$.}
\label{halcbis}
\end{figure}

Consider the $L$-leg boundary operator
${\tilde\Phi}_{{\mathcal S_L}}$ creating a star made of $L$ semi-infinite
random paths $\tilde {\mathcal S}_1$, diffusing in the upper half-plane ${\mathbb H}$ and started at
a single vertex on the real line $\partial {\mathbb H}$ in a
{\it   mutually-avoiding star} configuration
${\mathcal S}_L=(\wedge{\mathcal S}_1)^L,$  as seen in section \ref{subsec.multiline} (Fig~\ref{fig.SL}). Its boundary
scaling dimension $\tilde x_L$ is given by Eqs.
(\ref{txLkk}) or (\ref{txDkd}):
\begin{eqnarray}
\label{txL}
\tilde x({\mathcal S}_L)=\tilde x_L=\frac{L}{2\kappa}(2L+4-\kappa)\, , \,\,{\forall \kappa}
\end{eqnarray}
with the inversion formula:
\begin{eqnarray}
\label{uk-1}
\mathcal U_\kappa^{-1}(\tilde x_L)=L\,\mathcal U_\kappa^{-1}(\tilde x_1)=\frac{2L}{\kappa}
\, , \,\, {\forall \kappa}.
\end{eqnarray}
We thus dress this $L$-star ${\mathcal S_L}$ by a packet of $n$ independent Brownian paths diffusing
away from the apex of the star,
located on the boundary, while avoiding the star's paths (Fig.~\ref{halcbis}).
In our standard notation, this reads:
$${\mathcal S}_L \wedge\{\stackrel{n}{\overbrace{{\mathcal B}\vee{\mathcal B}\vee\cdots{\mathcal B}}}
\} =(\wedge{\mathcal S}_1)^L\wedge (\vee{\mathcal B})^n\equiv L\wedge n.$$
The corresponding boundary scaling dimension $\tilde x(L\wedge n)$ in ${\mathbb H}$
is given by the KPZ structure relations (\ref{tildexABk}, \ref{uk-1}):
\begin{eqnarray}
\label{tildexLn}
\tilde x(L\wedge n)
={\mathcal U}_{\kappa}\left[ L\,\frac{2}{\kappa}
+{\mathcal U}_{\kappa}^{-1}\left( n\right) \right].
\end{eqnarray}
\subsubsection*{Boundary Derivative Exponents}
It is interesting to isolate in these exponents the contribution $\tilde x_L$ (\ref{txL})
coming from the $L$ SLE paths, and
 define a subtracted exponent, the (boundary) derivative exponent \cite{lawler4}, which is obtained after simplication as
\begin{eqnarray}
\label{tildelambda}
\nonumber
\tilde \lambda_{\kappa}(L\wedge n)&:=&\tilde x(L\wedge n)-\tilde x_L
=n+ L\, {\mathcal U}_{\kappa}^{-1}\left( n\right)\\ \nonumber
{\mathcal U}_{\kappa}^{-1}\left( n\right)&=&\frac{1}{2\kappa}\left[\sqrt{16\kappa n+(\kappa-4)^2} +\kappa -4\right].
\end{eqnarray}
For $L=1$, one recovers the result of Ref.~ \cite{lawler4}.
The linear structure so obtained is in agreement with the short-distance expansion results (\ref{SDEU}) and (\ref{SDED});
 mutual-avoidance between  SLE and  Brownian paths
enhances the independent path exponent $\tilde x_L+n$ by $L$ times a typical boundary QG term.

\subsubsection*{Boundary Disconnection Exponents}
Notice that for $n=0$ the exponent is not necessarily trivial:
\begin{eqnarray}
\label{tildelambda0}
\nonumber
\tilde \lambda_{\kappa}(L\wedge 0)
= L\, {\mathcal U}_{\kappa}^{-1}\left( 0\right)=L\, \left(1-\frac{4}{\kappa}\right) \vartheta (\kappa  -4).
\end{eqnarray}
So this exponent takes non-zero values for $\kappa > 4$, i.e. for
self-coiling CI curves. This is typical of a {\it disconnection exponent}: Consider a point
$z$ located along the boundary $\partial {\mathbb H}$ at short distance  $r=|z-w|$ from the origin
$w$ where all paths of the SLE star
$\mathcal S_L$ are started. The probability $\tilde{\mathcal P}_{L\wedge 0}$ that point $z$ stays connected to
infinity without being encircled by the collection of
SLE traces scales like
\begin{equation}
\nonumber
\tilde{\mathcal P}_{L\wedge 0}(z) \propto r^{\tilde \lambda_{\kappa}(L\wedge 0)}=r^{\,L(1-{4}/{\kappa})},\; r\to 0, \;\kappa \geq 4.
\label{tildePL0}
\end{equation}
If $\kappa \leq 4$, the probability that the SLE paths return to the boundary is zero, and any point
$w \ne 0$ stays connected to infinity, hence a vanishing disconnection exponent
$\tilde\lambda_{\kappa \leq 4}(L,0)=0$.

\subsubsection*{Planar Multifractal Exponents}
Let us consider now the scaling exponent $x(L\wedge n)$ associated with the $n^{\rm th}$ moment
of the harmonic measure near
the tip of a collection of $L$ radial SLE paths in the plane.
It suffices to use the bulk general composition formula (\ref{xABk}) in place of the boundary one (\ref{tildexABk}) 
   in (\ref{tildexLn}) above, to immediately get:
\begin{eqnarray}
\label{xLn}
\nonumber
x(L\wedge n)
=2{\mathcal V}_{\kappa}\left[ L\,\frac{2}{\kappa}
+{\mathcal U}_{\kappa}^{-1}\left( n\right) \right].
\end{eqnarray}
It is useful to
separate the contribution $x_L$ of the tip of star $\mathcal S_L$
\begin{eqnarray}
\nonumber
\label{xL}
x_L=\frac{1}{8\kappa}\left[4L^2-(4-\kappa)^2\right]\, ,
\end{eqnarray}
and define a {\it bulk derivative exponent}
\begin{eqnarray}
\label{lambda}
\nonumber
\lambda_{\kappa}(L\wedge n):= x(L\wedge n)-x_L=\frac{n}{2}
+\frac{1}{2}\left(L+ \frac{\kappa}{4}-1 \right){\mathcal U}_{\kappa}^{-1}\left( n\right),
\end{eqnarray}
which
 generalizes the $L=1$ case considered in Ref. \cite{lawler4}.

\subsubsection*{Planar Disconnection Exponents}
 For  $n=0$ this yields the set of disconnection exponents
\begin{eqnarray}
\nonumber
\lambda_{\kappa}(L\wedge 0)=\left\{\begin{array}{ll} 0 & \mbox{$\kappa \leq 4$}\\
\nonumber
 \frac{1}{2}\left(L+ \frac{\kappa}{4}-1 \right)\left(1-\frac{4}{\kappa}\right)& \mbox{$\kappa \geq 4$,}
\end{array}
\right.
\label{lambdaL0exp}
\end{eqnarray}
which governs the probability $\mathcal P_{L\wedge 0}(r)$ that a point
$z \in {\mathbb C}$, located  at distance  $r=|z-w|$ from the star's tip $w$,
 stays connected to
infinity without being encircled by the collection of
SLE traces:
\begin{equation}
\nonumber
\mathcal P_{L\wedge 0}(r) \propto r^{\lambda_{\kappa}(L\wedge 0)}, \; r\to 0.
\label{PL0}
\end{equation}
Here again, it
 strongly depends on whether
the random paths are simple or not, respectively for $\kappa \leq 4$ and $\kappa > 4$.
If $\kappa \leq 4$, the SLE paths are simple curves that cannot encircle any exterior point; the latter
 therefore stays connected to infinity, hence a vanishing disconnection exponent.

For $L=1$, we recover the disconnection exponent associated with the tip of a single radial SLE trace, or,
equivalently, with the end of an open line in the $O(N)$ model,
a result appearing in  \cite{lawler4,RS}.

\subsubsection*{Double-Sided Exponents}
Let us mention that boundary double-sided exponents can be defined, corresponding to
double moments of the harmonic measure on both sides of a multiple SLE trace, or, equivalently, to
double-sided derivative exponents  \cite{lawler4,BDMan}.  We have in mind  configurations where two packets
of $n_1$ and $n_2$ Brownian paths diffuse on both sides of a boundary star $\mathcal S_L$.
 They are easily computed at level $L$ from the QG method, and the interested reader is referred to  \cite{BDMan}.

\subsubsection*{Winding of Multiple SLE's}
Let us finally return to the winding angle variance at points where $k$ strands come together in a star configuration
$\mathcal S_k$. We have  seen in section \ref{sec.winding} that the variance of $k$ paths up to distance $R$ is
reduced by a factor $1/k^2$ with respect to the $k=1$ single path case, namely:
\begin{equation}
\nonumber
\langle \vartheta^2 \rangle_k=\frac{\kappa}{k^2}\, \ln R.
\label{variance-k}
\end{equation}
In the case of {\it non-simple} paths ($\kappa >4$), one can further consider the winding at points where $k$ strands
meet together, amongst which $j$
 adjacent pairs (with $2j\leq k$) are conditioned not to hit each other \cite{wilson}. In each pair the two strands,
 which otherwise would bounce on each other, are disconnected from each other, and that corresponds, in our notations, to
 a star configuration:
\begin{equation}
{\mathcal
S}_{k,j}=\stackrel{k-2j}{\overbrace{{\mathcal S}_1\wedge{\mathcal
S}_1\wedge\cdots{\mathcal S}_1}}\wedge\stackrel{j}{\overbrace{({\mathcal S}_1\wedge 0\wedge{\mathcal S}_1)
\wedge\cdots \wedge({\mathcal S}_1\wedge 0\wedge{\mathcal S}_1)}}.
\label{configuration}
\end{equation}
 Wieland and Wilson  made
 the interesting conjecture that
 in this case the winding angle variance grows like \cite{wilson}
\begin{equation}
\langle \vartheta^2 \rangle_{k,j}=\frac{\kappa}{(k+j\,{\rm max}(0,\kappa/2-2))^2}\, \ln R.
\label{variance-kj}
\end{equation}
This can be derived from the quantum gravity formalism as follows.
A generalization of Eq.~(\ref{kn}) gives the  number of paths, $k(j)$, which is equivalent to $k$ strands in a star configuration
${\mathcal
S}_{k,j}$ (\ref{configuration}), as
\begin{eqnarray}
\nonumber
\label{kj}
k(j)&=& k+j\frac{{\mathcal U}_{\kappa}^{-1}\left( 0\right)}{{\mathcal U}_{\kappa}^{-1}\left( \tilde{x_1}\right)}.
\end{eqnarray}
Indeed, one simply has to gauge the  extra (quantum gravity) conformal weight $j \times {\mathcal U}_{\kappa}^{-1}\left( 0\right)$,
associated with the $j$ disconnected pairs, by the (QG) boundary conformal weight
${\mathcal U}_{\kappa}^{-1}\left( \tilde{x_1}\right)$ of a single path extremity.
Because of the value (\ref{U01})
and the value (\ref{uk-1})
we find
\begin{eqnarray}
\nonumber
\label{kj1}
k(j)&=& k+j\left(\frac{\kappa}{2}-2\right) \vartheta (\kappa  -4),
\end{eqnarray}
which gives a variance
\begin{equation}
\nonumber
\langle \vartheta^2 \rangle_{k,j}=\langle \vartheta^2 \rangle_{k(j)}=\frac{\kappa}{k^2(j)}\, \ln R,
\label{variance-kjk}
\end{equation}
which is just the conjecture (\ref{variance-kj}), {\bf QED}.

\section*{{Acknowledgements}}

It is a pleasure to thank Michael Aizenman for his collaboration on path-crossing exponents in percolation
(section \ref{sec.perco}), after a seminal discussion together with Bob Langlands, and for
many enjoyable and fruitful discussions over time;
Amnon Aharony for the same shared collaboration;
Ilia A.  Binder for his collaboration on the mixed multifractal rotation spectrum in section \ref{sec.winding};
 Peter Jones and Beno\^{\i}t Mandelbrot for invitations to the Department of Mathematics at Yale University and to the Mittag-Leffler Institute,
and many stimulating discussions; Fran\c{c}ois David for precious help with his Lectures; Emmanuel Guitter for generously preparing the figures;
Jean-Marc Luck for
extensive help with \LaTeX; Bernard Nienhuis and
Vincent Pasquier for friendly and interesting discussions; David A. Kosower for a
careful reading of the manuscript; Ivan K. Kostov for our early collaboration and intensive discussions;
Ugo Moschella for friendly help while preparing the file; and Thomas C. Halsey for many ``multifractal''
discussions over time, and a
careful reading of the manuscript.

Last but not least, Anton Bovier, Fran\c{c}ois Dunlop, Frank den Hollander and Aernout van Enter are especially thanked for having
organized this successful Les Houches Session LXXXIII in such a
  friendly and efficient way, and for their care in  editing this volume.


\begin{thebibliography}{99}
\bibitem{BDMan} B. Duplantier, {\it Conformal Fractal Geometry \& Boundary Quantum Gravity}, in:
{\it Fractal Geometry and Applications: A Jubilee of Beno\^{i}t Mandelbrot} (M. L. Lapidus and M.
van Frankenhuysen, eds.), Proc.
 Symposia  Pure Math. vol. 72, Part 2, 365-482 (AMS, Providence, R.I., 2004), arXiv:math-ph/0303034.

\bibitem{WW4} W. Werner, {\it Some Recent Aspects of Conformally Invariant Systems}, in: Les
Houches Summer School LXIII {\it Mathematical Statistical Physics}, July 4 - 29, 2005, this volume, arXiv:math.PR/0511268.

\bibitem{stflour} W. Werner, Lecture Notes from the 2002 Saint-Flour Summer School, Springer, L. N. Math. {\bf 1840}, 107-195,
(2004), arXiv:math.PR/0303354.

\bibitem{lawlerbook} G. F. Lawler, {\it Conformally Invariant Processes in the Plane}, Mathematical Surveys and Monographs, AMS,
Vol. {\bf 114} (2005).

\bibitem{BB0} M. Bauer and D. Bernard, {\it 2D growth processes: SLE and Loewner chains}, Phys. Rep., to appear,
arXimath-ph/0602049.

\bibitem{cardySLE} J. Cardy, {\it {\rm SLE} for Theoretical Physicists}, Ann. Physics {\bf 318},  81-118 (2005),
cond-mat/0503313.


\bibitem{NK}  W. Kager and B. Nienhuis,  J. Stat. Phys. {\bf 115}, 1149-1229 (2004), arXiv:math-ph/0312056.

\bibitem{BBrownian} B. Duplantier, {\it Brownian Motion, ``Diverse and Undulating''}, in: {\it Einstein, 1905-2005},
Poincar\'e Seminar 2005, Eds. T. Damour, O. Darrigol, B. Duplantier and V. Rivasseau (Birkha\"user Verlag, Basel, 2006).

\bibitem{PLevy} P. L\'evy, {\it Processus stochastiques et mouvement brownien} (Gauthier-Villars, Paris, 1965).

\bibitem{mandelbrot}  B. B. Mandelbrot, {\it The Fractal Geometry of Nature}
(Freeman, New-York, 1982).


\bibitem{symanzyk}  K. Symanzyk, in{\it \ Local Quantum Theory}, edited by
R. Jost (Academic Press, London, New-York, 1969).

\bibitem{PGG} P.-G. de Gennes, Phys.
Lett. {\bf A38},  339-340 (1972).

\bibitem{desC} J. des Cloizeaux and G. Jannink,
{\it Polymers in Solution, their Modeling and Structure}
 (Clarendon, Oxford, 1989).

\bibitem{aizenman1}  M. Aizenman, Phys. Rev. Lett. {\bf 47},  1-4, 886 (1981);
Commun.  Math. Phys. {\bf 86},  1-48 (1982); D. C. Brydges, J.
Fr\"{o}hlich, and T. Spencer, Commun.  Math. Phys. {\bf 83},  123-150 (1982).


\bibitem{lawler1}  G. F.  Lawler, Commun.  Math. Phys. {\bf 86},  539-554 (1982); {\it
Intersection of Random Walks} (Birkh\"{a}user, Boston, 1991).

\bibitem{fisher} M. E. Fisher, J. Stat. Phys. {\bf 34},  667-729 (1984).

\bibitem{aizenman2}  M. Aizenman, Commun.  Math. Phys. {\bf 97}, 91-110 (1985);
G. Felder and J. Fr\"{o}hlich, {\it ibid.}, 111-124; G. F.  Lawler, {\it ibid.},
583-594.
\bibitem{duplantier1}  B. Duplantier, Commun.  Math. Phys. {\bf 117}, 279-329
(1987).

\bibitem{BPZ}  A. A.  Belavin, A. M.  Polyakov and A.~ B.  Zamolodchikov, Nucl.
Phys. {\bf B241}, 333-380 (1984).

\bibitem{friedan}  D. Friedan,
J. Qiu, and S. Shenker, Phys. Rev. Lett. {\bf
52}, 1575-1578 (1984).

\bibitem{cardylebowitz}
J. L. Cardy, in {\it   Phase Transitions and Critical Phenomena}, edited by C.
  Domb and J. L. Lebowitz, (Academic Press, London, 1987), Vol. 11.

\bibitem{NBRS} B. Nienhuis, A.N. Berker, E.K. Riedel and M. Schick, Phys. Rev. Lett. {\bf 43}, 737 (1979).


\bibitem{dennijs} M. den Nijs, J. Phys. A {\bf 12}, 1857-1868 (1979);
Phys. Rev. B {\bf 27}, 1674-1679 (1983).

\bibitem{nien} B. Nienhuis, Phys. Rev. Lett. {\bf 49}, 1062-1065 (1982); J. Stat. Phys. {\bf 34}, 731-761 (1984);
in {\it Phase Transitions and Critical Phenomena}, edited by C.
  Domb and J. L. Lebowitz, (Academic Press, London, 1987), Vol. 11.

\bibitem{NienPT} B. Nienhuis, J. Phys. A {\bf 15}, 199-213 (1982).

\bibitem{cardy}  J. L. Cardy, Nucl. Phys. {\bf B240} [FS12], 514-532 (1984).

\bibitem{S1} H. Saleur, J. Phys. A {\bf 19}, L807-L810 (1986); {\it ibid.} {\bf 20}, 455-470 (1987).

\bibitem{duplantier4}  B. Duplantier, Phys. Rev. Lett. {\bf 57}, 941-944, 2332 (1986);
J. Stat. Phys. {\bf 54}, 581-680 (1989); [see also
L. Sch\"afer, C. von Ferber, U. Lehr, and B. Duplantier, Nucl. Phys. {\bf B374}, 473-495 (1992)].

\bibitem{DS2}  B. Duplantier and H. Saleur, Phys. Rev. Lett. {\bf 57}, 3179-3182 (1986).

\bibitem{BDD} B. Duplantier, J. Phys. A {\bf 19}, L1009-L1014 (1986).

\bibitem{DSd}
B. Duplantier and H. Saleur, Nucl. Phys. {\bf B290} [FS20], 291-326 (1987).


\bibitem{SD}  H. Saleur and B. Duplantier, Phys. Rev. Lett. {\bf 58}, 2325-2328
(1987).

\bibitem{DStheta} B. Duplantier and H. Saleur, Phys. Rev. Lett. {\bf 59}, 539-542 (1987).

\bibitem{BDHSwinding} B. Duplantier and H. Saleur,
Phys. Rev. Lett. {\bf 60}, 2343-2346 (1988).

\bibitem{Ca1} J. L. Cardy,
J. Phys. A {\bf 21}, L797-L799 (1988); {\it ibid.} {\bf 31}, L105-L110 (1998) [see also
M. Aizenman, Nucl. Phys. {\bf B485} [FS], 551-582 (1997)];  {\it ibid.} {\bf 47}, L665-L672 (2001);
J. L. Cardy and H. Saleur,
J. Phys. A {\bf 22}, L601-L604 (1989); J. L. Cardy and R. M. Ziff, J. Stat. Phys. {\bf 110}, 1-33 (2003).



\bibitem{BBl}  M. T. Batchelor and H. W. J. Bl\"ote, Phys. Rev. Lett. {\bf 61}, 138-140 (1988); 1042 (1988).

\bibitem{BNW} M. T. Batchelor, B. Nienhuis, and S. O. Warnaar,
       Phys. Rev. Lett. {\bf 62}, 2425-2428 (1989).

\bibitem{nien1} B. Nienhuis,
Int. J. Mod. Phys. {\bf B4}, 929-942 (1990).

\bibitem{WBN}
S.~O.~Warnaar, M.~T.~Batchelor and B.~Nienhuis,
J. Phys. {\bf A 25}, 3077-3095 (1992).
\bibitem{NWB}
 B. Nienhuis, S.~O. Warnaar and H.~W. J. Bl\"ote,
J. Phys. A {\bf 26}, 477-493 (1993).
\bibitem{WNS} S.~O.~Warnaar, B.~Nienhuis and K.~A.~Seaton,
Phys. Rev. Lett. {\bf 69}, 710 (1992).

\bibitem{BNW1}
 V.~V. Bazhanov, B. Nienhuis and S.~O. Warnaar,
Phys. Lett. B. {\bf 322}, 198-206 (1994).

\bibitem{cardy3} J. L. Cardy, J. Phys. A {\bf 25}, L201-L206 (1992).

\bibitem{S3} H. Saleur, Nucl. Phys. {\bf B382}, 486-531 (1992), arXiv:hep-th/9111007.


\bibitem{BSY} M. T. Batchelor, J. Suzuki, and C. M. Yung,
       Phys. Rev. Lett. {\bf 73}, 2646-2649 (1994);
       M. T. Batchelor and C. M. Yung, Phys. Rev. Lett. {\bf 74}, 2026-2029 (1995);
       J. Phys. A {\bf 28}  L421-L426 (1995), arXiv:cond-mat/9507010;
       Nucl. Phys. {\bf B453}, 552-580 (1995), arXiv:hep-th/9506074.

 \bibitem{KGN}
J. Kondev, J.C. de Gier and B. Nienhuis, 
J. Phys. A {\bf 29}, 6489
(1996).

\bibitem{Nientardif} B. Nienhuis,
Physica A {\bf 251}, 104-114 (1998).


 \bibitem{JK1} J. L. Jacobsen and J. Kondev, Phys. Rev. Lett. {\bf 81}, 2922-2925 (1998), arXiv:cond-mat/9805178; Nucl. Phys.
{\bf B532}, 635-688 (1998), arXiv:cond-mat/9804048;
 J. Stat. Phys. {\bf 96}, 21-48 (1999), arXiv:cond-mat/9811085.

\bibitem{DJP} V. S. Dotsenko, J. L. Jacobsen, and M. Picco, Nucl. Phys. {\bf B618}, 523-550 (2001), arXiv:hep-th/0105287.

\bibitem{JRS} J. L. Jacobsen, N. Read, and H. Saleur,
Phys. Rev. Lett. {\bf 90}, 090601 (2003).

\bibitem{JK3} J. L. Jacobsen and J. Kondev, Phys. Rev. Lett. {\bf 92}, 210601 (2004), arXiv:cond-mat/0401504.

\bibitem{langlands}  R. Langlands, P. Pouliot, and Y. Saint-Aubin, Bull. Amer. Math. Soc.
{\bf 30}, 1-61 (1994).

\bibitem{ai1}  M. Aizenman, in {\it Mathematics of Multiscale Materials};
the IMA Volumes in Mathematics and its Applications, K. M. Golden
et al. eds, Springer-Verlag (1998).

\bibitem{ben}  I. Benjamini and O. Schramm, Commun.  Math. Phys. {\bf 197},
75-107 (1998).

\bibitem{kazakov}  V. A.  Kazakov, Phys. Lett. {\bf A119}, 140-144 (1986).

\bibitem{KPZ}  A. M.  Polyakov, Mod. Phys. Lett. {\bf A 2}, 893-898 (1987); V. G.
Knizhnik, A. M.  Polyakov and A.~ B.  Zamolodchikov, Mod. Phys. Lett.
{\bf A 3}, 819-826 (1988).

\bibitem{david2} F. David, Mod. Phys. Lett.  {\bf A 3}, 1651-1656 (1988); J.
Distler and H. Kawai, Nucl. Phys. {\bf B321}, 509-527 (1988).


\bibitem{DK}  B. Duplantier and I. K. Kostov, Phys. Rev. Lett. {\bf
61}, 1433-1436 (1988); Nucl. Phys. {\bf B340}, 491-541 (1990).

\bibitem{kostovgaudin} I. K. Kostov, Mod. Phys. Lett. {\bf A 4}, 217-226 (1989);
M. Gaudin and I. K. Kostov, Phys. Lett. {\bf B220}, 200-206 (1989); I. K. Kostov, Nucl. Phys. {\bf B376}, 539-598 (1992),
arXiv:hep-th/9112059.

\bibitem{KS} I. K. Kostov and M. Staudacher, Nucl. Phys. {\bf B384}, 459-483 (1992), arXiv:hep-th/9203030.

\bibitem{KK} V. Kazakov and I. K. Kostov, Nucl. Phys. {\bf B386}, 520-557 (1992), arXiv:hep-th/9205059.


\bibitem{mandelbrot2}  B. B. Mandelbrot, J. Fluid.\ Mech. {\bf 62}, 331-358 (1974).

\bibitem{hentschel}  H. G. E.  Hentschel and I. Procaccia, Physica D
{\bf 8}, 435-444 (1983).

\bibitem{frisch}  U. Frisch and G. Parisi, in Proceedings of the
International School of Physics ``Enrico Fermi'', course LXXXVIII,
edited by M. Ghil (North-Holland, New York, 1985) p. 84.

\bibitem{halsey}  T. C. Halsey, M. H. Jensen, L. P. Kadanoff, I. Procaccia, and
B. I. Shraiman, Phys. Rev. \ A {\bf 33}, 1141-1151 (1986); {\it ibid.} {\bf 34}, 1601 (1986).


\bibitem{cates}  M. Cates and J. M. Deutsch, Phys. Rev. A {\bf 35}, 4907-4910
(1987); B. Duplantier and A. Ludwig, Phys. Rev. Lett. {\bf 66},
247-251 (1991); C. von Ferber, Nucl. Phys. {\bf B490}, 511-542 (1997).

\bibitem{halmeakproc}
T. C. Halsey, P. Meakin, and I. Procaccia, Phys. Rev. Lett. {\bf 56}, 854-857 (1986).

\bibitem{halseyerice}
T. C. Halsey, {\it Multifractality, Scaling, and Diffusive Growth} in
Fractals: Physical Origin and Properties, L. Pietronero, ed. (Plenum
Publishing Co., New York, 1989).

\bibitem{meak}
P. Meakin, {\it Fractals, Scaling and Growth Far from
Equilibrium}, Cambridge Nonlinear Sc. Series {\bf 5} (1999).

\bibitem{BBE}  B. B.  Mandelbrot and C. J. G.. Evertsz, Nature {\bf 348}, 143-145 (1990);
C. J. G. Evertsz and B. B. Mandelbrot, Physica A {\bf 177}, 589-592 (1991).


\bibitem{catesetwitten}  M. E. Cates and T. A. Witten, Phys. Rev. Lett. {\bf %
56}, 2497-2500 (1986); Phys. Rev. A {\bf 35}, 1809-1824 (1987).

\bibitem{lawler97}
G. F.  Lawler, {\it The frontier of a Brownian path is
multifractal}, preprint (1998).

\bibitem{duplantier2}  B. Duplantier and K.-H. Kwon, Phys. Rev. Lett. {\bf 61}, 2514-2517 (1988).

\bibitem{sokal} B. Li and A. D. Sokal, J. Stat. Phys. {\bf 61}, 723-748 (1990); E.
E. Puckette and W. Werner, Elect. Commun.  in Probab. {\bf 1}, 49-64
(1996).

\bibitem{burdzy}  K. Burdzy and G. F.  Lawler, Probab. Th. Rel. Fields {\bf 84}%
, 393-410 (1990); Ann. Probab. {\bf 18}, 981-1009 (1990).


\bibitem{duke} G. F. Lawler, Duke Math. J. {\bf 47}, 655-694 (1980); {\bf 53}, 249-270 (1986).

\bibitem{D6} B. Duplantier, J. Stat. Phys. {\bf 49}, 411-431 (1987); B. Duplantier and F. David,
J. Stat. Phys. {\bf 51}, 327-434 (1988).

\bibitem{majumdar} S. Majumdar, Phys. Rev. Lett. {\bf 68}, 2329-2331 (1992);
 B. Duplantier, Physica A {\bf 191}, 516-522 (1992).

\bibitem{kenyon1} R. W. Kenyon, Acta Math. {\bf 185}, 239-286 (2000).

\bibitem{kenyon2} R. W. Kenyon, J. Math. Phys. {\bf 41}, 1338-1363 (2000).

\bibitem{werner}  W. Werner, Probab. Th. Rel. Fields {\bf 108}, 131-152 (1997).

\bibitem{lawler2}  G. F.  Lawler and W. Werner, Ann.  Probab. {\bf 27}, 1601-1642 (1999).


\bibitem{duplantier7}  B. Duplantier, Phys. Rev. Lett. {\bf 81}, 5489-5492 (1998).

\bibitem{duplantier8}  B. Duplantier, Phys. Rev. Lett. {\bf 82}, 880-883 (1999), arXiv:cond-mat/9812439.

\bibitem{duplantier9}  B. Duplantier, Phys. Rev. Lett. {\bf 82}, 3940-3943 (1999), arXiv:cond-mat/9901008.


\bibitem{ADA} M. Aizenman, B. Duplantier and A. Aharony, Phys. Rev. Lett. {\bf 83}, 1359-1362 (1999), arXiv:cond-mat/9901018 .

\bibitem{duplantier10}
B. Duplantier, in {\it Fractals: Theory and Applications in
Engineering}, M. Dekking {\it et al.} eds., pp. 185-206
(Springer-Verlag, 1999).



\bibitem{lawler3}  G. F.  Lawler and W. Werner, J. European Math. Soc. {\bf 2}, 291-328 (2000).


\bibitem{cardy2} J. L. Cardy, J. Phys. A {\bf 32}, L177-L182 (1999), arXiv:cond-mat/9812416.

\bibitem{duplantier11} B. Duplantier, Phys. Rev. Lett. {\bf 84}, 1363-1367 (2000), arXiv:cond-mat/9908314.

\bibitem{DIFEG} P. Di Francesco,  O. Golinelli, and E. Guitter,
Nucl. Phys. {\bf B570}, 699-712 (2000), arxiv:cond-mat/9910453; P. Di Francesco, E. Guitter, and  J. L. Jacobsen,
 Nucl. Phys. {\bf B580}, 757-795 (2000), arXiv:cond-mat/0003008.


\bibitem{GA} T. Grossman and A. Aharony, J. Phys. A {\bf 19}, L745-L751 (1986); {\it ibid.} {\bf 20}, L1193-L1201 (1987).


\bibitem{schramm1} O. Schramm, Israel Jour. Math. {\bf 118}, 221-288 (2000).

\bibitem{histpoint} For the identification of $\kappa=2$ to LERW, Schramm mentions that he used the Gaussian
winding result found by R. Kenyon for spanning trees and LERW \cite{kenyon1,kenyon2}. The earlier general formula found in Ref. \cite{BDHSwinding} was actually immediately giving
the SLE parameter $\kappa$ in terms of the Coulomb gas constant $g=4/\kappa$, hence for the $O(N)$ model, $N=-2\cos \frac{4\pi}{\kappa}$.

\bibitem{BDjsp} B. Duplantier, J. Stat. Phys. {\bf 110}, 691-738 (2003), arXiv:cond-mat/0207743.


\bibitem{lawler4}  G. F.  Lawler, O. Schramm, and W. Werner, Acta Math.  {\bf 187}, (I) 237-273, (II) 275-308 (2001),
arXiv:math.PR/9911084,  arXiv:math.PR/0003156; Ann. Inst. Henri Poincar\'e
PR {\bf 38}, 109-123 (2002), arXiv:math.PR/0005294.
\bibitem{lawler5} G. F.  Lawler, O. Schramm, and W. Werner, Acta Math.  {\bf 189}, 179-201 (2002), arXiv:math.PR/0005295; Math. Res. Lett. {\bf 8}, 401-411
(2001), math.PR/0010165.

\bibitem{smirnov1} S. Smirnov, C. R. Acad. Sci. Paris S\'er. I Math. {\bf 333}, 239-244 (2001).
\bibitem{lawler6} G. F.  Lawler, O. Schramm, and W. Werner,
Electronic J. of Probability {\bf 7}, 2002, paper no.2,
arXiv:math.PR/0108211.

\bibitem{smirnov2} S. Smirnov and W. Werner, Math. Res. Lett. {\bf 8}, 729-744 (2001), arXiv:math.PR/0109120.

\bibitem{LSWLERW} G. F.  Lawler, O. Schramm, and W. Werner, Ann. Probab. {\bf 32}, 939-995 (2004), arXiv:math.PR/0112234.


\bibitem{SAWLSW} G. F.  Lawler, O. Schramm, and W. Werner, {\it On the Scaling limit of Planar Self-Avoiding Walks}, in:
{\it Fractal Geometry and Applications: A Jubilee of Beno\^{i}t Mandelbrot} (M. L. Lapidus and M.
van Frankenhuysen, eds.), Proc.
 Symposia  Pure Math. vol. 72, Part 2, 339-364 (AMS, Providence, R.I., 2004), arXiv:math.PR/0204277.


\bibitem{TGK} T. G. Kennedy, Phys. Rev. Lett. {\bf 88}, 130601 (2002).

\bibitem{duplantierdual} The duality prediction follows immediately from the $g \to 1/g$ duality results of \cite{duplantier11},
and the $\kappa=4/g$ correspondence between the SLE parameter $\kappa$ and the Coulomb gas constant $g$.


\bibitem{RS} S. Rohde and O. Schramm, Ann. Math. {\bf 161}, 879-920 (2005), arXiv:math.PR/0106036.

\bibitem{beffara2} V. Beffara, Ann. Probab. {\bf 32}, 2606-2629 (2004), arXiv:math.PR/0204208; see also
 V. Beffara and V. Sidoravicius, {\it Percolation Theory}, in: {\it Encyclopedia of
Mathematical Physics} (Elsevier, Amsterdam, 2006), arXiv:math.PR/0507220.

\bibitem{beffara}  V. Beffara, {\it The Dimension of the SLE Curves}, arXiv:math.PR/\-0211322.

\bibitem{binder} I. A. Binder, {\it   Harmonic Measure and Rotation of Simply Connected Planar Domains},
PhD Thesis, Caltech (1997).


\bibitem{DB}
B. Duplantier and I. A. Binder, Phys. Rev. Lett. {\bf 89}, 264101 (2002); arXiv:cond-mat/0208045.





\bibitem{BB}

M. Bauer and D. Bernard, Phys. Lett. {\bf B543}, 135-138 (2002), arXiv:math-ph/0206028; Commun. Math. Phys. {\bf 239}, 493-521 (2003),
arXiv:hep-th/0210015; Phys. Lett. {\bf B557}, 309-316 (2003), arXiv:hep-th/0301064; Ann. Henri Poincar\'e {\bf 5}, 289-326 (2004), arXiv:math-ph/0305061;
Phys. Lett. {\bf B583}, 324-330 (2004), arXiv:math-ph/0310032;  Proceedings of the conference
{\it Conformal Invariance and Random Spatial Processes}, Edinburgh, July 2003, arXiv:math-ph/0401019.


\bibitem{BBH} M. Bauer, D. Bernard, J. Houdayer, J. Stat. Mech. P03001 (2004), arXiv:math-ph/0411038.

\bibitem{CRLSW} G. F.  Lawler, O. Schramm, and W. Werner, J. Amer. Math. Soc. {\bf 16}, 917-955 (2003), arXiv:math.PR/0209343.

\bibitem{LF1} R. Friedrich and W. Werner, C. R. Acad. Sci. Paris S\'er. I Math. {\bf 335}, 947-952 (2002), arXiv:math.PR/0209382;
  Commun.  Math. Phys.,  {\bf 243}, 105-122 (2003), arXiv:math-ph/0301018.

\bibitem{F} R. Friedrich, {\it On Connections of Conformal Field Theory and Stochastic L{\oe}wner Evolution}, arXiv:math-ph/0410029.


\bibitem{WW} W. Werner, Ann. Fac. Sci. Toulouse, {\bf 13}, 121-147  (2004), arXiv:math.PR/0302115.

\bibitem{dubedat} J. Dub\'edat,  Ann. Probab. {\bf 33}, 223-243 (2005), arXiv:math.PR/0303128.

\bibitem{cardyrho} J. Cardy,  {\it ${\rm SLE}(\kappa,\rho)$ and Conformal Field Theory}, arXiv:math-ph/0412033.
\bibitem{kyto} K. Kyt\"ol\"a, {\it On Conformal Field Theory of ${\rm SLE}(\kappa,\rho)$}, arXiv:math-ph/0504057.


\bibitem{BF} R. O. Bauer, R. M. Friedrich, {\it Diffusing polygons and SLE($\kappa,\rho$)}, arXiv:math.PR/0506062.






\bibitem{cardy2003} J. L. Cardy,  J. Phys. A {\bf 36}, L379-L386 (2003); erratum J. Phys. A {\bf 36}, 12343 (2003),
arXiv:math-ph/0301039; Phys. Lett. {\bf B582}, 121-126 (2004),
arXiv:hep-th/0310291.

\bibitem{dubedat4} J. Dub\'edat, {\it Commutation Relations for SLE}, arXiv:math.PR/0411299.
\bibitem{dubedat5} J. Dub\'edat, {\it  Euler Integrals for Commuting SLEs}, arXiv:math.PR/0507276.


\bibitem{BBK} M. Bauer, D. Bernard, K. Kyt\"ol\"a, J. Stat. Phys. {\bf 120}, 1125-1163 (2005), arXiv:math-ph/0503024.

\bibitem{watts} G. M. T. Watts, J. Phys. A {\bf 29}, L363-L368 (1996).


\bibitem{dubedat2} J. Dub\'edat, {\it Excursion Decompositions for SLE and Watts' Crossing Formula}, arXiv:math.PR/0405074.

\bibitem{dubedat3} J. Dub\'edat, Commun. Math. Phys. {\bf 245}, 627-637 (2004), arXiv:math.PR/0306056.


\bibitem{CN} F. Camia and C. M. Newman,  {\it Continuum Nonsimple Loops and 2D Critical Percolation}, arXiv:math.PR/0308122.

\bibitem{CN1} F. Camia and C. M. Newman,  {\it The Full Scaling Limit of Two-Dimensional Critical Percolation},
 arXiv:math.PR/0504036.

\bibitem{CFN}  F. Camia, L. R. G. Fontes and C. M. Newman, {\it The Scaling Limit Geometry of Near-Critical 2D Percolation},
arXiv:cond-mat/0510740.
\bibitem{beffaraperc} V. Beffara, {\it Cardy's Formula on the Triangular Lattice, the Easy Way}, Preprint,
http://www.umpa.ens-lyon.fr/~vbeffara/files/Proceedings-Toronto.pdf (2005).
\bibitem{beffaraotherlat} V. Beffara, {\it Critical Percolation on Other Lattices}, Talk at the Fields Institute,
http://www.fields.utoronto.ca/audio/05-06/ (2005).
\bibitem{2Dvort} D. Bernard, G. Boffetta, A. Celani and G. Falkovich, Nature Physics {\bf 2}, 124-128 (2006).

\bibitem{kozma} G. Kozma, {\it Scaling limit of loop-erased random walk - a naive approach}, arXiv:math.PR/0212338.

\bibitem{SS} O. Schramm and S. Sheffield, Ann.  Probab. {\bf 33}, 2127-2148 (2005), arXiv:math.PR/0310210.

\bibitem{smirnovising} S. Smirnov, in preparation (2006).

\bibitem{WW1} G. F. Lawler and W. Werner, Probab. Th. Rel. Fields {\bf 128}, 565-588 (2004), arXiv:math.PR/0304419.


\bibitem{WW2} W. Werner, {\it Conformal Restriction and Related Questions},  Proceedings of the conference
{\it Conformal Invariance and Random Spatial Processes}, Edinburgh, July 2003, Probability Surveys 2, 145-190
(2005), arXiv:math.PR/0307353.

\bibitem{WW3} W. Werner,  C. R. Acad. Sci. Paris S\'er. I Math. {\bf 337}, 481-486 (2003), arXiv:math.PR/0308164.
\bibitem{SW} S. Sheffield and W. Werner, in preparation (2006).

\bibitem{konsevich} M. Konsevich, {\it CFT, SLE and Phase Boundaries}, Preprint of the Max Planck Institute
(Arbeitstagung 2003), 2003-60a.

\bibitem{FKa} R. Friedrich and J. Kalkkinen, Nucl. Phys. {\bf B687},  279-302 (2004), arXiv:hep-th/0308020.

\bibitem{zhan} D. Zhan, Probab. Theory Related Fields {\bf 129}, 340-380 (2004), arXiv:math.PR/0310350.


\bibitem{FB} R. O. Bauer and R. M. Friedrich, C. R. Math. Acad. Sci. Paris {\bf 339}, 579-584 (2004), arXiv:math.PR/0408157;
{\it On radial stochastic Loewner evolution in multiply connected domains},
arXiv:math.PR/0412060;
  {\it On Chordal and Bilateral SLE in multiply connected domains}, arXiv:math.PR/0503178.

\bibitem{cardydoyonriva}  B. Doyon, V. Riva and J. Cardy, {\it Identification of the stress-energy
tensor through conformal restriction in SLE and related processes}, arXiv:math-ph/0511054.

\bibitem{cardyarea} J. Cardy, J. Phys. A {\bf 34}, L665-672 (2001), arXiv:cond-mat/0107223.


\bibitem{richguttman} C. Richard, A. J. Guttmann and I. Jensen, J. Phys. A {\bf 34}, L495-501 (2001), arXiv:cond-mat/0107329.
\bibitem{cardyloop} J. Cardy, Phys. Rev. Lett. {\bf 72}, 1580-1583 (1994).
\bibitem{cardyziff} J. Cardy and R. Ziff, J. Stat. Phys. {\bf 101}, 1-33 (2003), arXiv:cond-mat/0205404.



\bibitem{GTF} C. Garban and J. A. Trujillo Ferreras,
{\it The expected area of the filled planar Brownian loop is $\pi/5$}, arXiv:math.PR/0504496.

\bibitem{richard} C. Richard, J. Phys. A {\bf 37}, 4493-4500 (2004), arXiv:cond-mat/0311346.
\bibitem{WW5} W. Werner, {\it The Conformally Invariant Measure on Self-Avoiding Loops}, arXiv:math.PR/0511605.

\bibitem{sheffield} S. Sheffield, {\it Gaussian Free Fields for Mathematicians}, arXiv:math.PR/0312099.

\bibitem{SS1} O. Schramm and S. Sheffield, {\it Contour Lines of the 2D Gaussian Free Field}, in preparation (2006).

\bibitem{S} S. Sheffield, in preparation (2006).





\bibitem{wiegmann}
E. Bettelheim, I. Rushkin, I. A. Gruzberg and P. Wiegmann, Phys. Rev. Lett. {\bf 95},  170602 (2005), arXiv:hep-th/0507115.

\bibitem{IABD}
I. A. Binder and B. Duplantier, {\it Multifractal Properties of Harmonic Measure and Rotations
for Stochastic Loewner Evolution}, in preparation (2006).

\bibitem{Stockholm} D. Beliaev, {\it Harmonic Measure on Random
Fractals}, PhD thesis, KTH, Stockholm, Sweden (2005).






\bibitem{FZZ} V. Fateev, A. B. Zamolodchikov, and Al. Zamolodchikov, arXiv:hep-th/0001012.

\bibitem{PT} B. Ponsot and J. Techner, Nucl. Phys. {\bf B622}, 309-327 (2002), arXiv:hep-th/0110244.

\bibitem{KKK} I. K. Kostov,  Nucl.\ Phys.\ {\bf B658}, 397-416 (2003), arXiv:hep-th/0212194;
Nucl.\ Phys.\  {\bf B689}, 3-36 (2004), arXiv:hep-th/0312301.

\bibitem{KPS} I. K. Kostov, B. Ponsot, and D. Serban, Nucl. Phys. {\bf B683}, 309-362 (2004), arXiv:hep-th/0307189.


\bibitem{OAOS} O. Angel and O. Schramm,  Commun. Math. Phys. {\bf 241}, 191--213 (2003), arXiv:math.PR/0207153.

\bibitem{OA} O. Angel, Geom. Funct. Anal. {\bf 13}, 935-974 (2003),
           arXiv: math.PR/0208123.

\bibitem{OA2} O. Angel {\it Scaling of Percolation on Infinite Planar Maps, I}, arXiv:math.PR/0501006.

\bibitem{gschaeffer}  G. Schaeffer, {\it Bijective census and random generation of {E}ulerian planar
              maps with prescribed vertex degrees}, Electron. J. Combin. {\bf 4 } (1997) \# 1; G. Schaeffer,
     {\it Conjugaison d'arbres et cartes combinatoires al\'eatoires}, PhD Thesis (1998),
      Universit\'e Bordeaux 1.


\bibitem{BMS} M. Bousquet-M\'elou and G. Schaeffer, {\it The degree distribution in bipartite planar maps:
        applications to the Ising model}, arXiv:math.CO/0211070.

\bibitem{DiFG} Ph. Di Francesco and E. Guitter, Phys. Rep. {\bf 415}, 1-88 (2005).

\bibitem{bouttier} J. Bouttier, {\it Physique statistique des surfaces al\'eatoires et combinatoire bijective
des cartes planaires}, PhD Thesis (2005), Universit\'e Paris 6.

\bibitem{GNB}
W. N. Guo, B. Nienhuis, H. W. J. Bl\"ote,
Phys. Rev. Lett. {\bf 96}, 045704 (2006), arXiv:cond-mat/0511277.

\bibitem{JaS} W. Janke and A. M. J. Schakel, Phys. Rev. Lett. {\bf 95}, 135702 (2005).

\bibitem{see2}  See, e.g., A. M.  Polyakov, {\it Gauge Fields and Strings}{\bf
\ }(Harwood-Academic, Chur, 1987).

\bibitem{boulatov}  D. V. Boulatov, V. A. Kazakov, I. K. Kostov and A. A. Migdal, Nucl. Phys. {\bf B275}, 641-686 (1986);
F. David, Nucl. Phys. {\bf B257}, 45-58 (1985); {\it ibid.} 543-576; J. Ambjorn, B.
Durhuus, and J. Fr\"{o}hlich, {\it ibid.}, 433-449.


\bibitem{wigner} E. P. Wigner, Proc. \ Cambridge \ Philos.\ Soc.\ {\bf 47},  790 (1951).

\bibitem{dyson} F. Dyson, J. Math. Phys. {\bf 3},  140, 157, 166, 1199 (1962); Commun. Math. Phys. {\bf 19}, 235 (1970).

\bibitem{mehtagaudin} M. L. Mehta and M. Gaudin, Nucl. Phys. {\bf A 18},  395 (1960); {\bf A 25},  447 (1961).

\bibitem{dysonmehta} F. Dyson and M. L. Mehta, J. Math. Phys. {\bf 4},  701, 713 (1963).

\bibitem{mehtabook} M. L. Mehta, {\it Random Matrices}, 2nd edition (Academic Press, New York, 1991).

\bibitem{thooft} G. 't Hooft, Nucl. Phys. {\bf B72},  461 (1974).

\bibitem{bipz}  E. Br\'{e}zin, C. Itzykson, G. Parisi, J.-B. Zuber,
Commun. \ Math.\ Phys.\ {\bf 59}, 35-51 (1978).

\bibitem{tutte} W. T. Tutte, Can. J. Math. {\bf 14},  21 (1962).

\bibitem{99} \'E. Br\'ezin and V. A. Kazakov, Phys. Lett. {\bf B236} (1990) 144; M. Douglas and S. Shenker,
Nucl. Phys. {\bf B335} (1990) 635; D. Gross and A. A. Migdal, Phys. Rev. Lett. {\bf 64} (1990) 127;
Nucl. Phys. {\bf B340} (1990) 333.

\bibitem{davidleshouches} F. David, {\it Simplicial Quantum Gravity and Random Lattices}, in:
{\it Gravitation and Quantizations}, in: Les Houches Summer School, Session LVII, July 5 - August 1, 1992
(Elsevier, Amsterdam, 1993).

\bibitem{davidaltenberg} F. David, {\it Random Matrices and Two-Dimensional Gravity}, in:
{\sl Fundamental Problems in Statistical Mechanics VIII}, Altenberg (Germany), June 28
-- July 10, 1993 (Elsevier, Amsterdam, 1994).

\bibitem{eynard} B. Eynard, {\it Random Matrices}, Saclay Lectures in Theoretical Physics, SPhT-T01/014 (2001).

\bibitem{ambjornbook} J. Ambjorn, B. Durhuus and T. Jonsson, {\it Quantum Geometry, a Statistical Field Theory Approach}, Cambridge Monographs in
Mathematical Physics (Cambridge University Press, Cambridge, 1997).

\bibitem{thooftplanar} G. 't Hooft, {\it Counting Planar Diagrams with Various Restrictions}, arXiv:hep-th/9808113.

\bibitem{DiFGZJ} Ph. Di Francesco, P. Ginsparg, and J. Zinn-Justin, Phys. Rep. {\bf 254},  1-133 (1995).

\bibitem{kostovsum} I. K. Kostov, {\it Conformal Field Theory Techniques in Random Matrix Models}, arXiv:hep-th/9907060.



\bibitem{kostov}  I. K. Kostov and M. L. Mehta, Phys. Lett. {\bf B189}, 118-124
(1987).

\bibitem{aldous}  D. Aldous, SIAM J. Disc. Math. {\bf 3}, 450 (1990); A. Broder,
in {\it 30th Annual Symp. Foundations Computer Sci.} 442, (IEEE, New York,
1989).

\bibitem{lawlerisrael}  G. F.  Lawler, Israel Jour. Math. {\bf 65}, 113-132 (1989).

\bibitem{lawler}  G. F.  Lawler, Elect. Commun.  in Probab. {\bf 1}, 29-47 (1996).


\bibitem{ferber}  C. von Ferber and Y. Holovatch, Europhys. Lett. {\bf 39},
31-36 (1997); Phys. Rev. E {\bf 56}, 6370-6386 (1997); {\it ibid.} {\bf 59}, 6914-6923 (1999); {\it ibid.} {\bf 65},
092801 (2002); Physica A {\bf 249},  327-331 (1998).

\bibitem{aharony2} J. Asikainen, A. Aharony, B. B. Mandelbrot, E. M. Rauch, J.-P. Hovi,
 Eur. Phys. J. B {\bf 34}, 479-487 (2003), arXiv:cond-mat/0212216.


\bibitem{SRG} B. Sapoval, M. Rosso, and J.-F. Gouyet, J. Physique Lett. {\bf 46}, 149-156 (1985).

\bibitem{JK} J. L. Jacobsen and P. Zinn-Justin, Phys. Rev. E {\bf 66}, 055102 (2002),
arXiv:cond-mat/0207063.

\bibitem{cardyjapon} J. L. Cardy, {\it   Lectures on Conformal Invariance and Percolation},
Chuo University, Tokyo (2001), arXiv:math-ph/0103018.


\bibitem{D7}  B. Duplantier, Phys. Rep. {\bf 184}, 229-257 (1989).

\bibitem{meakin}  P. Meakin {\it et al.},
Phys. Rev. A {\bf 34}, 3325-3340 (1986); see also: P. Meakin, {\it
{\it ibid.}} {\bf 33}, 1365-1371 (1986); in {\it Phase Transitions and
Critical Phenomena}, vol. 12, edited by C. Domb and J.L. Lebowitz
(Academic Press, London, 1988).

\bibitem{MS}  P. Meakin and B. Sapoval, Phys. Rev. A {\bf 46}, 1022-1034 (1992).

\bibitem{halsey7}  T. C. Halsey and M. Leibig, Ann. Phys. (N.Y.) {\bf 219}, 109-147
(1992).

\bibitem{GFS}
D. S. Grebenkov, M. Filoche, and B. Sapoval, Eur. Phys. J. B {\bf 36}, 221-231 (2003).



\bibitem{kakutani} S. Kakutani, Proc. Acad. Japan. {\bf 20}, 706-714 (1944).

\bibitem{makarov}  N. G. Makarov, Proc.\ London Math. Soc. {\bf 51}, 369-384 (1985); see also
P. W. Jones and T. H. Wolff, Acta Math. {\bf 161}, 131-144 (1988);
C. J. Bishop and P. W. Jones,
Ann. Math. {\bf 132}, 511-547 (1990).

\bibitem{ai2}  M. Aizenman and A. Burchard, Duke Math. J. {\bf 99}, 419-453 (1999).

\bibitem{BDH} R. C. Ball, B. Duplantier, and T. C. Halsey, unpublished (1999).








\bibitem{lawleresi} G. F.  Lawler, {\it An Introduction to the Stochastic L\"owner Evolution}, in: {\it Random Walks and Geometry},
V. Kaimonovich ed., 263-293 (de Gruyter, Berlin, 2004).



\bibitem{hastings} M. B. Hastings, Phys. Rev. Lett. {\bf 88}, 055506 (2002).

\bibitem{BD} R. Balian and B. Duplantier, Ann. Physics {\bf 112}, 165-208 (1978), p.183.

\bibitem{DuCa} B. Duplantier and J. L. Cardy, private discussion (Aspen, 1999).

\bibitem{Ca} J. L. Cardy, Nucl. Phys. {\bf B300}, 377-392 (1988).

\bibitem{Beur}
A. Beurling, \emph{The Collected Works of {A}rne {B}eurling. {V}ol.
1},
  Contemporary Mathematicians, Birkh\"auser Boston Inc., Boston, MA,
1989,
  Complex analysis, edited by L. Carleson {\it et al}.
\bibitem{ball}  R. C. Ball and R. Blumenfeld, Phys. Rev. A {\bf 44}, R828-R831
(1991).


\bibitem{aharony} A. Aharony and J. Asikainen, Fractals {\bf 11} (Suppl.), 3-7 (2003), arXiv:cond-mat/0206367.















\bibitem{BDtopub}
B. Duplantier, unpublished.

\bibitem{wilson}
B. Wieland and D. B. Wilson, Phys. Rev. E {\bf 68}, 056101 (2003).


\bibitem{Potts} R. B. Potts, Proc. Camb. Phil. Soc. {\bf 48}, 106-109 (1952);
 F. Y. Wu, Rev. Mod. Phys. {\bf 54}, 235-268 (1982), erratum {\bf 55}, 315 (1983).

\bibitem{FK} C. M. Fortuin and P. W. Kasteleyn, Physica {\bf 57}, 536-564 (1972).



\bibitem{JS} W. Janke and A. M. J. Schakel, Nucl. Phys. {\bf B700}, 385 (2004), arXiv:cond-mat/0311624. [See also A. Coniglio and W. Klein,
J. Phys. A {\bf 13}, 2775-2780 (1980); A. L. Stella and C. Vanderzande,
Phys. Rev. Lett. {\bf 62}, 1067-1070 (1989); {\it ibid.} {\bf 63}, 2537;
B. Duplantier and H. Saleur, {\it ibid.} {\bf 63}, 2536.]









\end{thebibliography}
\end{document}